\documentclass[12pt,twoside]{article}
\usepackage[xetex,a4paper,hmargin={1.0in,1.0in},top=1.0in,bottom=1.0in]{geometry}
\usepackage{setspace} 
\usepackage[T1]{fontenc}
\usepackage[utf8]{inputenc}  
\usepackage{newtxtext} 
\usepackage[varg]{newtxmath} 
\def\cal{\scr}
\usepackage{graphicx} 
\usepackage{ntheorem}  
\usepackage{sansmath}
\usepackage{marvosym}
\usepackage{stmaryrd} 
\usepackage{mathtools,amsmath}
\usepackage{hypernat}
\usepackage[labelfont={bf,footnotesize},textfont={footnotesize,singlespacing}]{caption}
\usepackage{footmisc}

\usepackage{subcaption}
\usepackage[colorlinks,citecolor=red,linkcolor=red,anchorcolor=blue,urlcolor=blue]{hyperref}
\usepackage[toc]{appendix} 
\theorembodyfont{\upshape}
\theorempreskip{0.0pt plus 1.0pt}
\theorempostskip{0.0pt plus 1.0pt}
\newcounter{nitnum}[section] 
\def\nitskip{\vskip 1.75pt plus 0.5pt minus 0.5pt}
\renewcommand{\thenitnum}{\textbf{\arabic{section}.\arabic{nitnum}}}  
\newenvironment{nit}[1]{
\removelastskip\nitskip\par\noindent\refstepcounter{nitnum}%
\textbf{\thenitnum~{#1}:\enspace}\ignorespaces%
}%
{%
\ignorespacesafterend\nitskip
}%

\edef\ag{\hbox{à}}

\def\cc{\hbox{ç}}

\def\ea{\hbox{é}}

\edef\Ea{\hbox{É}}

\edef\uD{\hbox{ü}}

\let\scr\mathscr
\let\bbf\mathbb

\def\abs#1{{\lvert#1\rvert}}

\let\ee\varepsilon

\mathchardef\DLT="7001

\let\l\lambda

\let\W\Omega

\let\G\Gammait

\let\dd\delta
\let\f\phi
\let\ff\varphi

\def\bbF{{\bbf F}}

\def\bbG{{\bbf G}}
\def\C{{\scr C}}

\def\bbP{{\bbf{P}}}

\def\DD{{\scr D}}

\def\DB{{\bbf D}}

\def\bDDc{{\skew0\bar\DD}^{\lower3pt\hbox{$\scriptstyle\complement$}}}

\def\qq{{\scr q}}

\def\R{{\bbf R}}

\def\XXX{{\cal X}}

\def\AA{{\cal A}}

\def\C{{\scr C}}

\def\LLL{{\cal L}}
\def\M{{\cal M}}

\def\PP{{\scr P}}
\def\PPP{{\cal P}}

\def\OO{{\scr O}}
\def\EE{{\scr E}}
\def\EEE{{\cal E}}
\def\T{{{\bbf T}}}
\def\TT{{\scr T}}
\def\R{{\bbf R}}
\def\Rp{{\R_{+}}}
\def\Rpp{{\R_{++}}}

\def\SSS{{\cal S}}

\def\TTT{{\scr T}}

\def\sii{{\tts\raise1pt\hbox{$\scriptstyle\mathord *$}\tts}}

\let\a\alpha
\let\b\beta

\let\h\eta
\let\s\sigma

\let\q\theta

\let\qq\vartheta

\def\d{\mskip1.75mu{\rm d}\mskip.75mu}

\def\dm{{\rm d}\mskip.75mu}

\def\emph#1{\textbf{\textit{#1}}}

\def\uvec{{\overset{\lower1.25pt\hbox{$\scriptstyle\rightharpoonup$}}{1}}}
\def\zvec{{\overset{\lower1.25pt\hbox{$\scriptstyle\rightharpoonup$}}{0}}}
\def\tts{\kern .065em\ignorespaces }

\def\OX{{{}^{\scriptscriptstyle \OO}\kern-1ptX}}
\def\PX{{{}^{\scriptscriptstyle \PP}\kern-1ptX}}

\def\ts{\kern .1em\ignorespaces }

\def\rlbkt{\mathclose[}
\def\llbkt{\mathopen[}
\def\lrbkt{\mathopen]}

\def\lbkt{\mathopen]}

\def\`#1{{\accent"00 #1}}
\def\'#1{{\accent"01 #1}}
\def\v#1{{\accent"07 #1}} \let\^^_=\v
\def\u#1{{\accent"08 #1}} \let\^^S=\u
\def\=#1{{\accent"09 #1}}
\def\^#1{{\accent"02 #1}} \let\^^D=\^
\def\.#1{{\accent"0A #1}}
\def\~#1{{\accent"03 #1}}
\def\"#1{{\accent"04 #1}}

\def\df{\buildrel \scriptscriptstyle {\textrm{def}}\over=}

\let\le\leq
\let\ge\geq
\let\smc\sc

\def\trn{{\kern-1pt\intercal}}

\def\stoch #1 #2 #3 {{\smint{#2}{\d #3}}}
\def\stochi #1 #2 #3 {{\int\nolimits_0^{#1}\!\!#2\d #3}}

\let\phd\placehold
\let\smint\smallint

\chardef\og="13
\chardef\fg="14

\def\si{{\ts\raise2pt\hbox{$\scriptscriptstyle\bullet$}\ts}} 
\def\ssi{{\ts\raise1.5pt\hbox{$\scriptscriptstyle\bulletS$}\ts}} 
\def\RWarrow{{|\kern-4.25pt\Rrightarrow}}

\def\d{\mskip1.75mu{\mrm d}\mskip.75mu}

\def\dm{{\mrm d}\mskip.75mu}

\def\rnd{\ts\raise1pt\hbox{$\scriptstyle\odot$}\ts} 
\let\mrm\rm

\let\timesorig\times
\def\times{\tts\raise.5pt\hbox{$\scriptstyle\mathord\timesorig$}\tts}

\def\hide#1{\iffalse#1\fi}

\makeatletter
\newtoks\inifil \inifil={1fil}
\def\display#1{\null\,\vcenter{\openup\jot\m@th
  \ialign{&\hfil\strut\everymath={\displaystyle{}}##
&\hfil\everymath={\displaystyle{}}##\hfil
&\everymath={\displaystyle{}}##\hfil
      \crcr#1\crcr}}\,}

\def\mdisplay#1{\null\,\vcenter{\openup\jot\m@th
  \ialign{&\hfil$\displaystyle{{}##{}}$
&\hfil$\displaystyle{{}##{}}$\hfil
&$\displaystyle{{}##{}}$\hfil
      \crcr#1\crcr}}\,}

\def\alignno#1{\displ@y \tabskip0pt
  \halign to\displaywidth{
  \kern\displaywidth\llap{$\@lign##$}\tabskip-\displaywidth plus1fil
  &&\hfil$\@lign\displaystyle{{}##{}}$\tabskip\z@skip
    &$\@lign\displaystyle{{}##{}}$\hfil\tabskip=0pt plus1fil
    	\crcr
    #1\crcr}}

\def\mdisplayno#1{\displ@y \tabskip0pt
  \halign to\displaywidth{
  \kern\displaywidth\llap{$\@lign##$}\tabskip-\displaywidth plus\the\inifil
  &&\hfil$\@lign\displaystyle{{}##{}}$\tabskip\z@skip
   &\hfil$\@lign\displaystyle{{}##{}}$\hfil\tabskip\z@skip
    &$\@lign\displaystyle{{}##{}}$\hfil\tabskip=0pt plus1fil
    	\crcr
    #1\crcr}}

\def\mldisplayno#1{\displ@y \tabskip0pt
  \halign to\displaywidth{
  \rlap{$\@lign##$}\tabskip0pt plus\the\inifil
  &&\hfil$\@lign\displaystyle{{}##{}}$\tabskip\z@skip
    &\hfil$\@lign\displaystyle{{}##{}}$\hfil\tabskip\z@skip
    &$\@lign\displaystyle{{}##{}}$\hfil\tabskip=0pt plus1fil
    	\crcr
    #1\crcr}}

\def\align#1{\null\,\vcenter{\openup\jot\m@th
  \ialign{&\strut\hfil$\displaystyle{{}##{}}$&$\displaystyle{{}##{}}$\hfil
      \crcr#1\crcr}}\,}

\def\alignno#1{\displ@y \tabskip0pt
  \halign to\displaywidth{
  \kern\displaywidth\llap{$\@lign##$}\tabskip-\displaywidth plus1fil
  &&\hfil$\@lign\displaystyle{{}##{}}$\tabskip\z@skip
    &$\@lign\displaystyle{{}##{}}$\hfil\tabskip=0pt plus1fil
    	\crcr
    #1\crcr}}

\def\displayno#1{\displ@y \tabskip0pt
  \halign to\displaywidth{
  \kern\displaywidth\llap{$\@lign##$}\tabskip-\displaywidth plus1fil
  &&\hfil\@lign\everymath={\displaystyle{}}##\tabskip\z@skip
   &\hfil\@lign\everymath={\displaystyle{}}##\hfil\tabskip\z@skip
    &\@lign\everymath={\displaystyle{}}##\hfil\tabskip=0pt plus1fil
    	\crcr
    #1\crcr}}

\def\overset#1#2{{\mathop{\kern\z@#2}\limits^{#1}}}
\def\underset#1#2{{\mathop{\kern\z@#2}\limits_{#1}}}

\makeatother

\input mac-2.tex
\ReadWriteBib{5}
\gdef\bibshape#1#2{\vskip\bibskip\par\hangindent\bibindent\hangafter1\noindent\ignorespaces
\hbox to\bibindent{\hfill}\ignorespaces
\llap{\relax\hypertarget{#1}{\bibmark.}\hbox to \bibnoskip{\hfill}}\ignorespaces#2\ignorespaces}
\input mac-3.tex
\input ushyphex.tex
\newif\iftoshow
\def\toshow#1{\iftoshow #1\else\medskip\relax\fi}
\toshowtrue 

\def\cint{{\lbkt 0,\bar c\,]}}

\def\pdg{{{}^\dag}\kern-1.5pt}
\def\pddg{{{}^\ddag}\kern-1.5pt}
\def\pst{{{}^{\raise0.7pt\hbox{$\scriptstyle*$}}}\kern-1.5pt}
\def\csd{{F}}%
\let\T\Thetait
\def\pfc{{\TT}} 
\def\csdp{\pdg\csd_{y}}%
\def\csds{\pst\csd_{y}}%
\def\gp{{\pi}}
\let\Lmp\Phiit%
\let\DB\bbF
\def\Up{{\partial U}}

\def\qedsymb{{$\bulletS$}}   
\def\qed{\ifmmode\nobreak\hbox{\quad\qedsymb}\else\nobreak\relax\nobreak\hbox{\quad\qedsymb}\fi}

\def\chptr{paper}
 
\def\aut#1{{\smc #1}.}
\def\btitle#1{{\it  #1\/}.}
\def\jtitle#1{{\rm #1.}} 
\def\anno#1{({\rm #1}).} 
\def\book#1{{\rm #1}}
\def\journal#1 #2 #3{{\it #1\/}~{\bf #2} #3}
\def\journalnt#1 #2 #3{{\it #1\/} {\bf #2} {#3}}
\numberwithin{equation}{section}
\theoremstyle{plain}  
\title{The Time-Interlaced Self-Consistent Master System\\
of
Heterogeneous-Agent Models
}
\author{{}Andrew Lyasoff\ts\thanks{{}email: mathema@lyasoff.net, Git repository:
\href{https://github.com/AndrewLyasoff}{https://github.com/AndrewLyasoff}}%
}
\date{{}\today} 
\usepackage{titlesec}
\titleformat{\section}[hang]{\center\bf\large}{\llap{\bf\large\thesection.\hbox to0.5em{\hfill}}}{0.0em}{}
\titleformat{\figure}[hang]{\normalsize\rmshape}{\textsc{\thefigure}}{0.5em}{}
\titlespacing*{\section}{0pt}{1.25ex plus 0.35ex minus .125ex}{1.0ex  plus 0.35ex minus 0.125ex}
\usepackage{titling}
\usepackage{fancyhdr}
\fancypagestyle{FirstPage}{
\fancyhf{}
\fancyfoot{}
\fancyfoot[C]{\thepage} 
\fancyfoot[L]{} 
}
\begin{document}

\maketitle
{\small
\vskip-5cm
\begin{flushright}
--- L'essentiel est invisible pour les yeux, répéta le

petit prince, afin de se souvenir.
\footnote{{\smc de Saint-Exupéry, Antoine}.~(1943).~{Le Petit Prince, Ch.~XXI}. New York, NY: Reynal \& Hitchcock.}
\end{flushright}
}

\bigskip
\allowdisplaybreaks
\thispagestyle{FirstPage}
\def\citex[#1]#2{[\cited{#2}, #1]}
\begin{abstract}\normalsize\noindent 
\noindent%
\noindent%
\small{}%
\noindent 
\textbf{Abstract:}\enspace%
It is shown that the structure of general equilibrium incomplete
market models is intrinsically self-consistent and time-interlaced, with 
mean field interactions that are only implicit and also endogenous.
Novel mathematical tools that can handle such structures and do not rely on the
representative agent point of view are developed.
The study was
prompted by the surprising discovery that the common strategy for
resolving the classical Aiyagari-Bewley-Huggett
model fails to achieve its objective in a widely cited benchmark
study. 
In addition to providing a numerically verifiable solution to such models,
the scope of the approximate
aggregation conjecture
of Krusell and Smith (still an open problem in macroeconomics)
is clarified.
New features of 
Krusell-Smith's model are uncovered and novel computational technique, which does not
involve simulation, is developed. 
\medskip

\noindent%
\textsc{Keywords:} Markov chains, transport problems, mean field games, 
general equilibrium, incomplete markets, heterogeneous agent mo\-dels, numerical methods
\end{abstract}

\section{Background and Introduction}
\label{sec:Intro}\setcounter{paragraph}{0}

\parskip=0.0pt

\noindent
To motivate what follows in this \chptr,
consider  the familiar savings problem described in Sec.~18.2 in 
the landmark text~\cite{LjunSar00}.
Its~least involved version is Huggett's pure credit economy, first proposed in~\cite{Hug93}.
In it the agents have exogenous endowments that
follow statistically identical but independent dis\-crete-time Markov chains with state space $\SSS$ and transition
matrix~$\PPP$. The elements of $\SSS$ have the meaning of work-hours per period and
the agents trade a single riskless asset
that is in net supply of zero.
The individual asset holdings are restricted to a finite uniform grid $\AA$ over an interval
$[u,v]\subset\R$, the choice of which is ad hoc.
The classical approach to producing an equilibrium
comes down to postulating infinite time horizon and calculating,
for a given (and fixed) interest rate $r$, the aggregate demand for the traded security.
The idea is to vary the choice of $r$ until the aggregate demand becomes null.
The common implementation of this program (see~\cite{LjunSar00})
boils  down to calculating (with fixed $r$) the time-invariant (long-run) optimal policy function
$g_\infty \colon{\AA}\times\SSS\mapsto {\AA}$
for a generic household
\footnote{{}The optimal policy function maps the pair of previous capital holdings and present
employment state into present capital holdings.}
and the associated long run
distribution of agents, $\l_\infty(\cdot,\cdot)$, treated as a distribution of unit mass
over the finite space $\AA\times\SSS$. This distribution obtains~-- see 
\citex[Sec.~18.2.1]{LjunSar00}~-- by iterating to convergence as $t\to\infty$,
starting from the uniform distribution
$\l_0(\cdot,\cdot)$, the equation
{\abovedisplayskip=9pt plus 1pt minus 1pt\belowdisplayskip=9pt plus 1pt minus 1pt\belowdisplayshortskip=3pt plus 0.5pt minus 0.5pt
\begin{equation}\label{lmb-iter}
\l_{t+1}(a',s')=\sum\nolimits_{a\in{\AA},\,s\in\SSS,\,g_\infty(a,s) = a'}\l_{t}(a,s){\PPP}(s,s')\,,
\quad a'\in{\AA}\,,\ s'\in\SSS\,,
\end{equation}
}%
which is a replica of  \citex[(18.2.4)]{LjunSar00}.
This relation still holds~-- see  \citex[Sec.~18.2.2]{LjunSar00}~-- 
if $\l_t(\cdot,\cdot)$ is re-interpreted as the probability distribution at time $t$ of the state of a
generic household that follows the optimal policy $a'=g_\infty(a,s)$, which policy obtains in the
obvious way from iterating to convergence as $t\to\infty$ the Bellman equation
\begin{subequations}\label{Bellman}
{\abovedisplayskip=9pt plus 1pt minus 1pt\belowdisplayskip=9pt plus 1pt minus 1pt\belowdisplayshortskip=3pt plus 0.5pt minus 0.5pt
\begin{equation}\label{Bellman-a}
V_t(a,s)=\max\nolimits_{\,c,\,a'}
\Bigl(U(c)+\b\sum\nolimits_{s'\in\SSS}{\PPP}(s,s')V_{t+1}(a',s')\Bigr)\,,\quad
(a,s)\in\AA\times\SSS\,,
\end{equation}
}%
where the maximization is subject to the constraints ($a\in\AA$ and $s\in\SSS$ are given)
{\abovedisplayskip=9pt plus 1pt minus 1pt\belowdisplayskip=9pt plus 1pt minus 1pt\belowdisplayshortskip=3pt plus 0.5pt minus 0.5pt
\begin{equation}\label{Bellman-b}
c+a'=(1+r) a + w s \,,\ \ c\in\Rpp\,,\ \ a'\in{\AA}\,,
\end{equation}
}%
\end{subequations}%
the discount factors $\b>0$ and the wage $w>0$ are given, and so is also the risk aversion
parameter $R$ in $U(c)\df c^{1-R}/(1-R)$. The parameter~$r$ is then varied until the following
identity holds (within an acceptable numerical accuracy)
{\abovedisplayskip=7pt plus 1pt minus 1pt\belowdisplayskip=7pt plus 1pt minus 1pt\belowdisplayshortskip=3pt plus 0.5pt minus 0.5pt
\begin{equation}\label{mclr}
\sum\nolimits_{a\tts\in\tts\AA\tts,\,s\tts\in\tts\SSS} g_\infty(a,s)\l_\infty(a,s)=0\,.
\end{equation}
}%
Note that because of the dual meaning of the distribution $\l_\infty(\cdot,\cdot)$ the left side
can be interpreted as the long-run expected demand of a
representative household.   
The strategy just described is illustrated in Fig.~\ref{fg1}.
{\captionsetup{belowskip=-10pt}
\captionsetup{aboveskip=0pt}
\begin{figure}[!htbp]
\centering 
\begin{subfigure}{.47\textwidth}  
  \centering
\leavevmode\raise0.85cm\hbox{\rotatebox{90}{\tiny expected representative demand}}%
\ %
\toshow{\includegraphics[width=7.1cm]{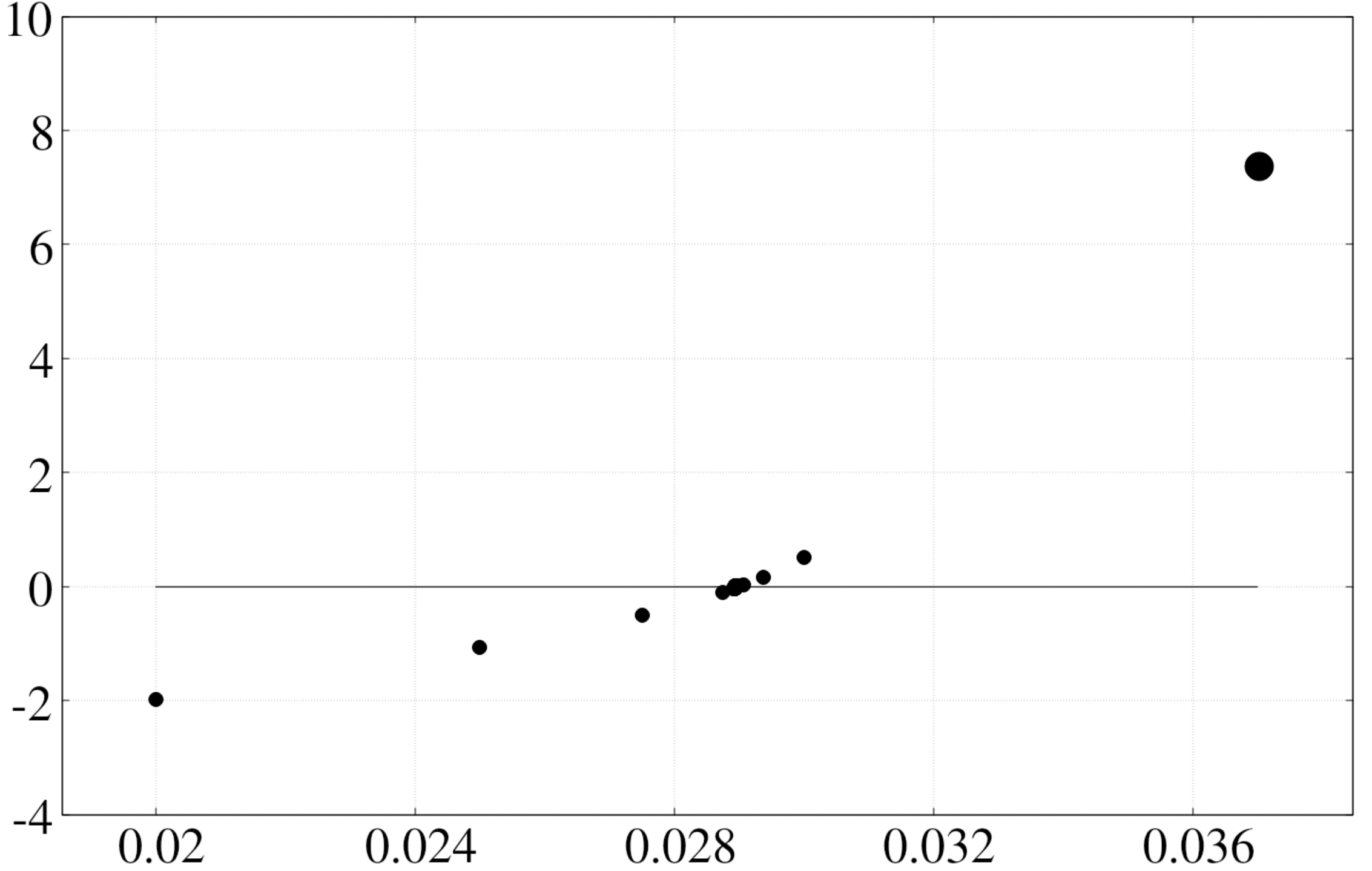}}%

\leavevmode\smash{\hbox to0.3cm{\hfill} \raise6pt\hbox{\tiny interest rate}}

\end{subfigure}\quad
\begin{subfigure}{.47\textwidth}  
  \centering 
\leavevmode\raise0.85cm\hbox{\rotatebox{90}{\tiny expected representative demand}}%
\ %
\toshow{\includegraphics[width=7.1cm]{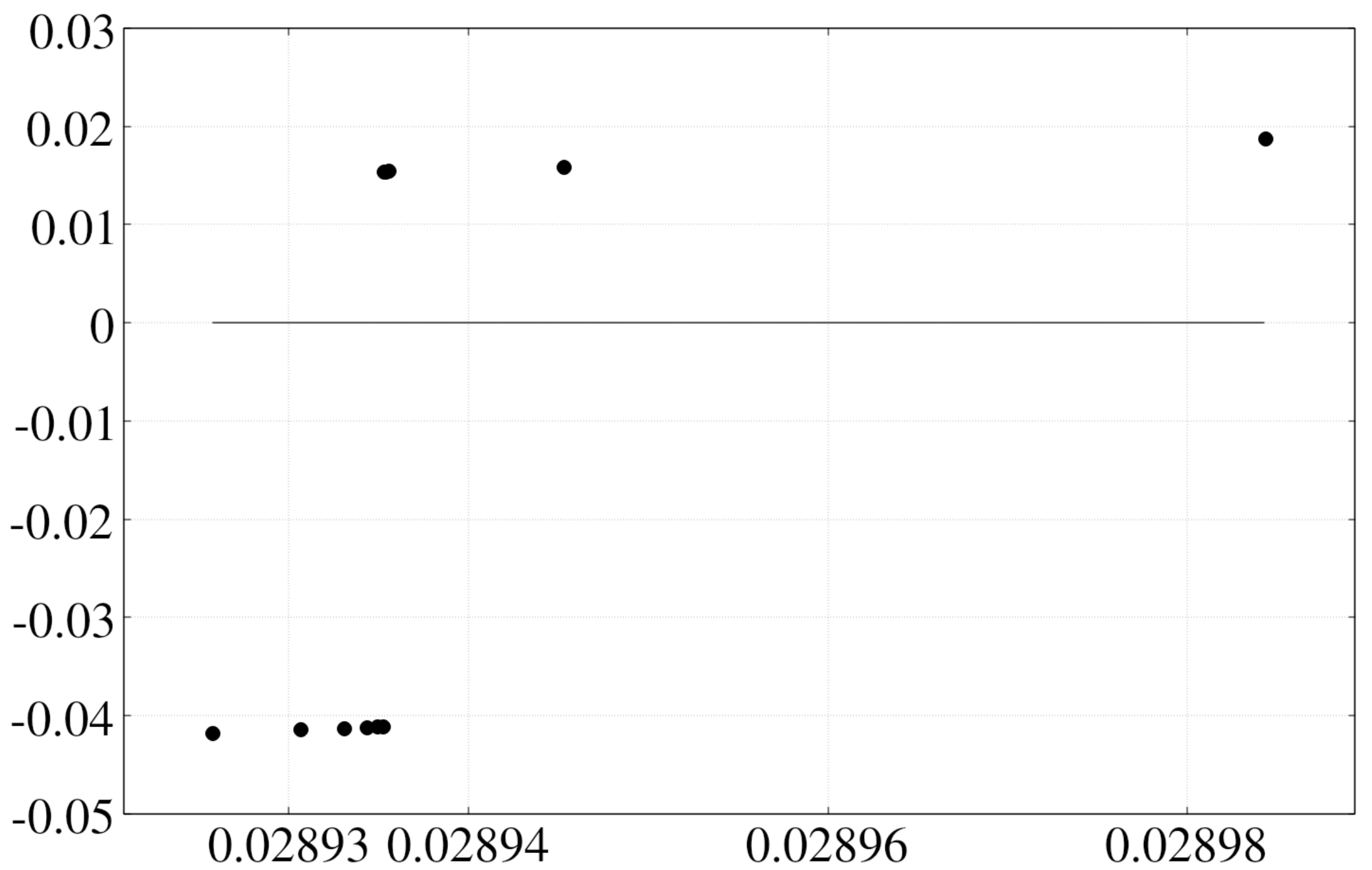}}   

\leavevmode\smash{\hbox to0.3cm{\hfill} \raise6pt\hbox{\tiny interest rate}} 

\end{subfigure} 
\caption{Illustration of the strategy for calculating the equilibrium rate in a pure credit
economy with  asset holdings constrained to a grid $\AA$ of 200 equally spaced points (the right plot is
a microscopic view of a portion of the left).} 
\label{fg1} 
\end{figure} }%
The~parameter values ($\b>\nobreak 0$, $w>\nobreak 0$, $\SSS\in\R^7$, $\PPP\in\R^{7\otimes7}$)
and the ad hoc range
$[u,v]$ are borrowed
from the first specification from  \citex[Sec.~18.7]{LjunSar00} and so is also the program
for computing the expected long-run representative demand for a given interest rate~$r$.
%
\footnote{{}The~computer code (in Julia) with which the plots in Figures~\ref{fg1}
and \ref{fg2} 
are generated emulates
the \hbox{MATLAB} program that accompanies \cite{LjunSar00}, except in the 
following step: 
the iterations are terminated  after the first simultaneous repetition
of both the policy function and the value function (within a prescribed threshold), not just after
the first repetition of the policy function, as in the original code.
This modification 
is necessary because, due to the discretization
of the state space, the 
value function can still improve after the first repetition of the policy function,
if the former has not yet converged to its time-invariant state.
The main reason for translating the original code into
Julia is the ability of the latter to handle very
large grid sizes with only marginal effect on performance.%
}
%
The~left plot shows the expected demands corresponding
to 20 different choices for the rate.
The~first three rates are chosen arbitrarily and every consecutive rate is the arithmetic average of
the latest rate that yields positive expected demand and the latest rate that yields negative expected
demand (a straightforward implementation of the classical bisection method).
The~right plot shows the expected demands in the last $10$ trials.
While the convergence of the inte\-rest rate is of order $10^{-8}$ (the distance between the last
two rates), there
appears to be a lower bound on how close to $0$ the expected demand can get. 
Interestingly, if the same experiment is repeated on a substantially 
more refined grid $\AA$ over the same domain of asset holdings,
then the discontinuity in the expected demand as a function of the inte\-rest
becomes much more pronounced~-- see Fig.~\ref{fg2}.
{\captionsetup{belowskip=-10pt}
\captionsetup{aboveskip=0pt}
\begin{figure}[!htbp]
\centering 
\begin{subfigure}{.47\textwidth} 
  \centering
\leavevmode\raise0.85cm\hbox{\rotatebox{90}{\tiny expected representative demand}}%
\ %
\toshow{\includegraphics[width=7.1cm]{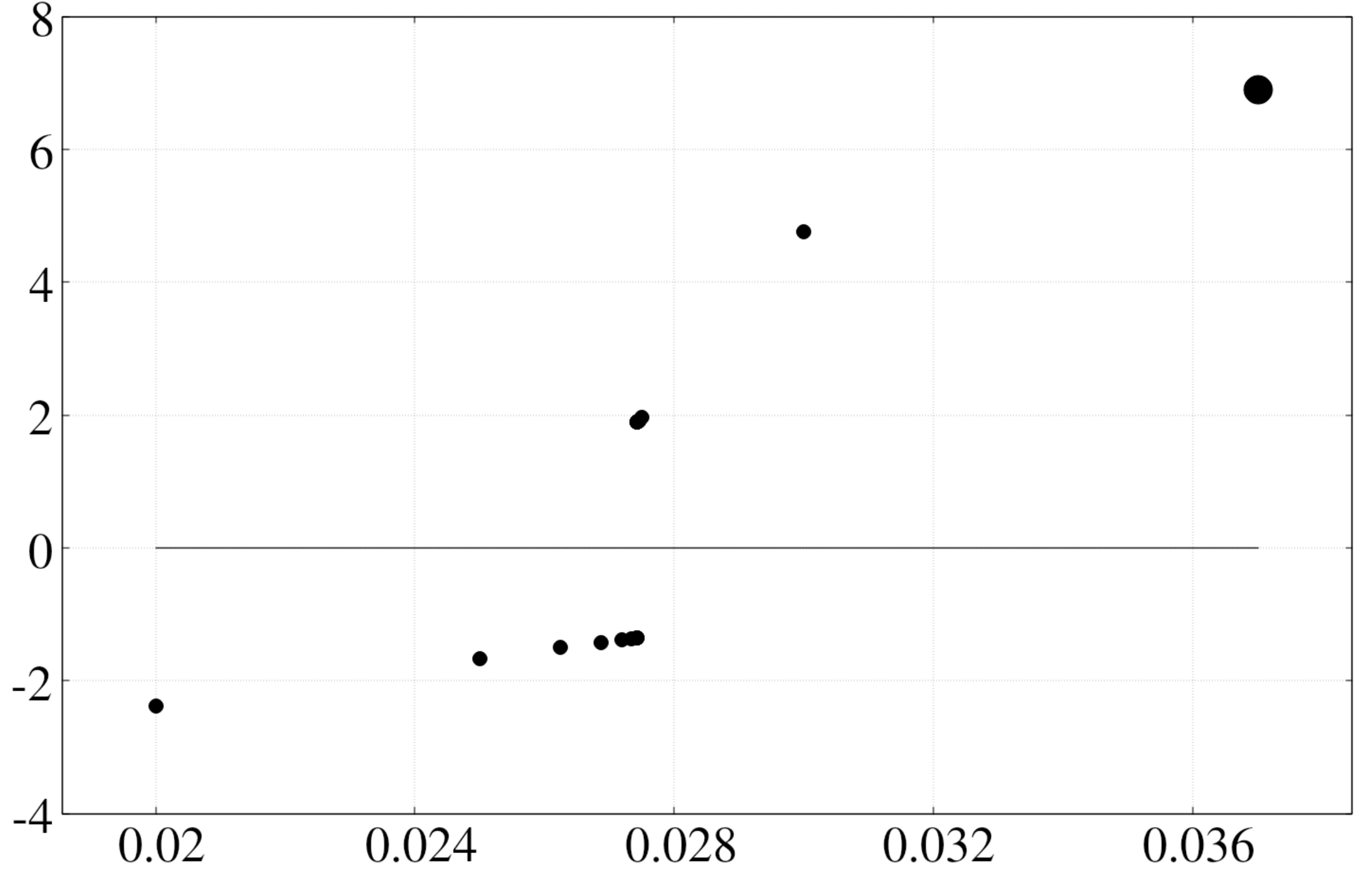}}%

\leavevmode\smash{\hbox to0.3cm{\hfill} \raise6pt\hbox{\tiny interest rate}}

\end{subfigure}\quad
\begin{subfigure}{.47\textwidth}  
  \centering
\leavevmode\raise0.85cm\hbox{\rotatebox{90}{\tiny expected representative demand}}%
\ %
\toshow{\includegraphics[width=7.1cm]{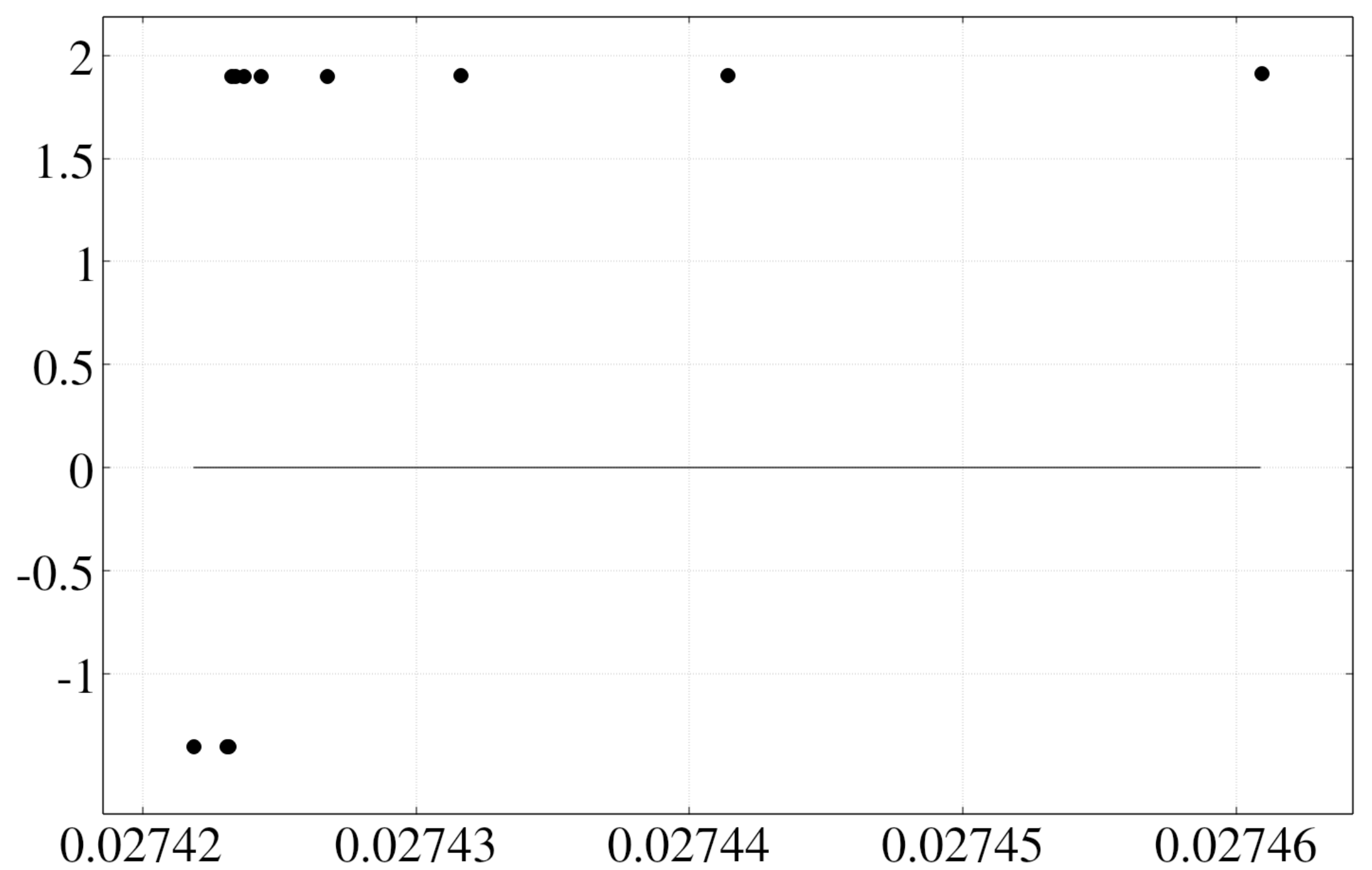}}%

\leavevmode\smash{\hbox to0.3cm{\hfill} \raise6pt\hbox{\tiny interest rate}} 

\end{subfigure}
\caption{Illustration of the strategy for calculating the equilibrium rate in a pure credit
economy with asset holdings constrained to a grid $\AA$ of $2\mathord,000$ equally spaced points
(the right plot is 
a microscopic view of a portion of the left).}
\label{fg2}
\end{figure} }%
The~culprit for this phenomenon is illustrated in Fig.~\ref{fg3}:
the last two in the list of 20 trial rates differ by less than $10^{-7}$ but
the corresponding stationary distributions (obtained by iterating \eqref{lmb-iter} from the uniform
distribution) are very different and so are the respective expected demands,
which differ by more than $3.25$~-- see the right plot in Fig.~\ref{fg2}.
{\captionsetup{belowskip=-9pt} 
\captionsetup{aboveskip=0pt}
\begin{figure}[!htbp]
\centering 
\begin{subfigure}{.47\textwidth} 
  \centering
\leavevmode\raise1.2cm\hbox{\rotatebox{90}{\tiny population distribution}}%
\ %
\toshow{\includegraphics[width=7.1cm]{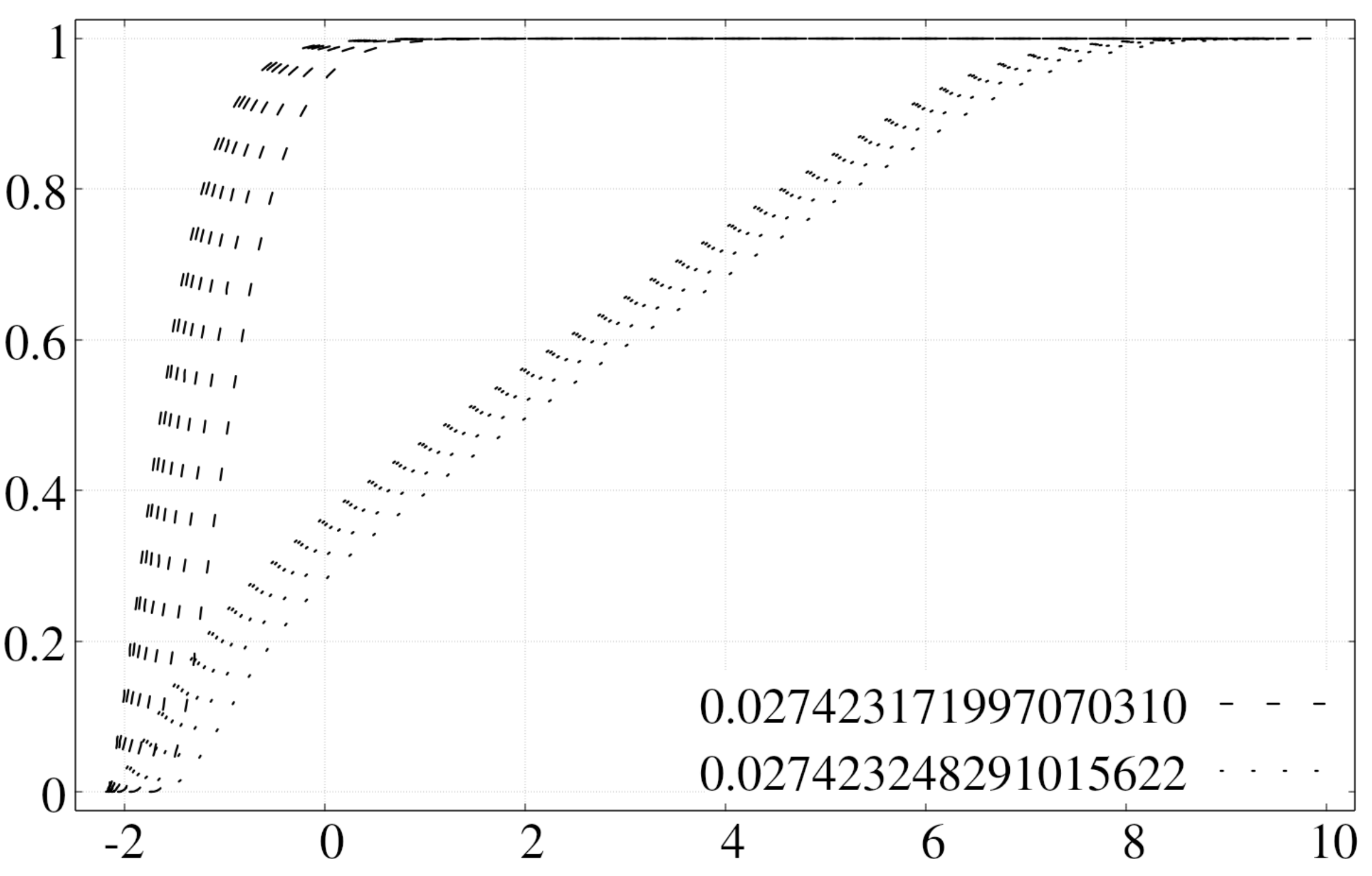}}%

\leavevmode\smash{\hbox to0.3cm{\hfill} \raise6pt\hbox{\tiny asset holdings ($2\mathord,000$ grid points)}}

\end{subfigure}\quad
\begin{subfigure}{.47\textwidth}  
  \centering
\leavevmode\raise1.2cm\hbox{\rotatebox{90}{\tiny population distribution}}%
\ %
\toshow{\includegraphics[width=7.1cm]{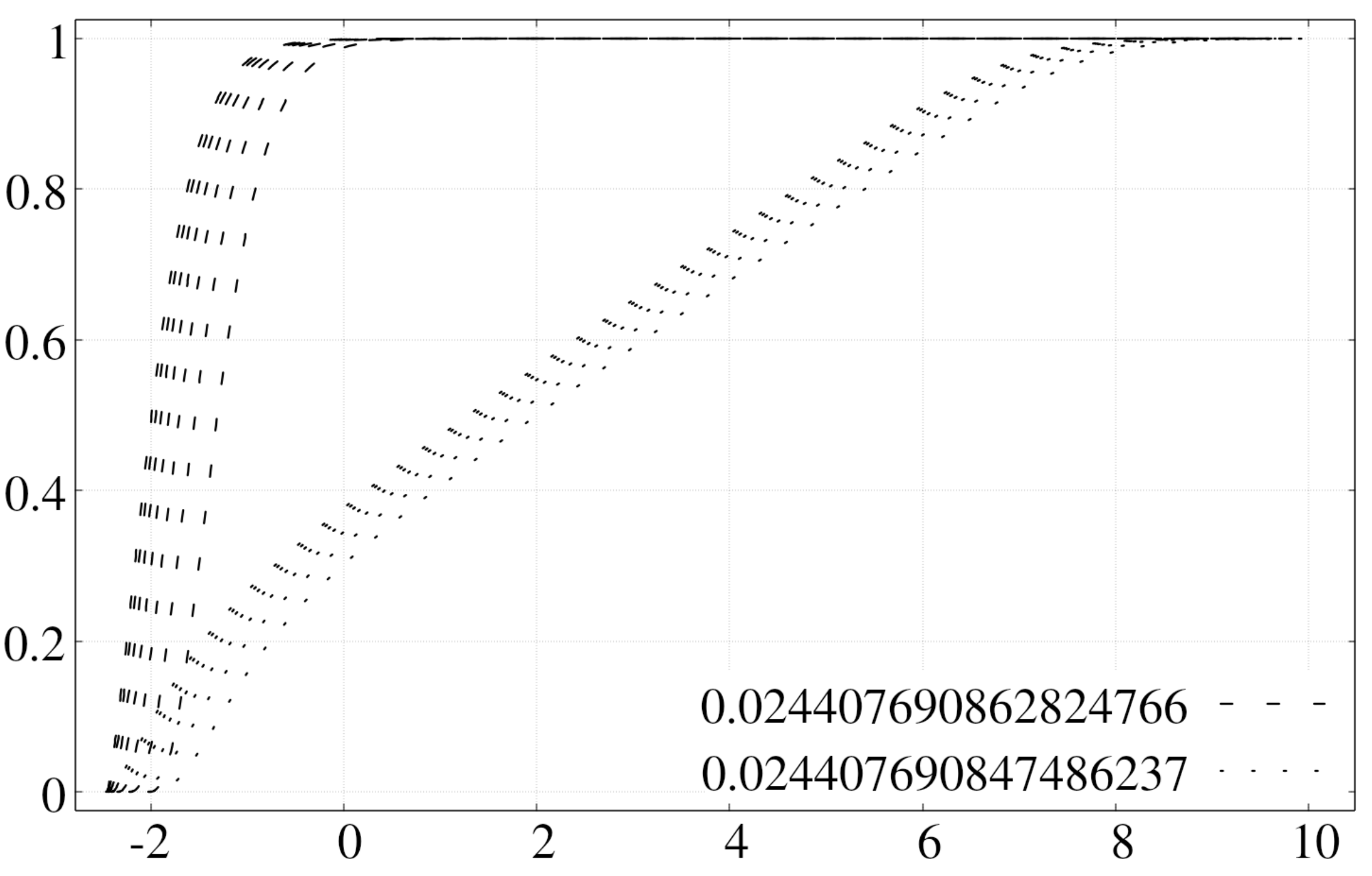}}%

\leavevmode\smash{\hbox to0.3cm{\hfill} \raise6pt\hbox{\tiny asset holdings ($3\mathord,000$ grid points)}} 

\end{subfigure}
\caption{The stationary distributions in each of the seven employment categories calculated over
$2\mathord,000$ grid points (left) and $3\mathord,000$ grid points (right), for two interest rates
that differ by less than 
$10^{-7}$ (left) and $10^{-10}$ (right).%
}
\label{fg3}
\end{figure} }%
Pushing the CPU to
$3\mathord,000$ grid points~-- see the right plot in Fig.~\ref{fg3}~--
does not remove the discontinuity in the distribution. The rate at which the jump occurs
only moves slightly 
to the left as the density of the grid increases, but the gap in the expected demand
remains larger 
than $3$ even with $4\mathord,000$ grid points, with neither the left nor the right
limit being 
close enough to zero.

In sum, when applied to the particular example borrowed here from \cite{LjunSar00}
the common strategy (see~\nocite{LjunSar00}{ibid.})
fails to identify~-- within an acceptable numerical tolerance~--
the equilibrium rate, despite the temptation to accept as ``almost equilibrium''
the rate suggested by
Fig.~\ref{fg1} (it will be shown below that the true equilibrium rate
is considerably bigger).
\footnote{Here the state space of the underlying optimization problem
is the continuum and
the uniform grid $\AA$ is an approximation of that state space.
Most basic intuition
demands that, as the density of the uniform grid increases, the associated equilibrium rate
must converge to some hypothetical value that represents
the true rate in the underlying model. In the case described here not only
that the equilibrium rate does not converge, it actually becomes indeterminate~--
see~Fig.~\ref{fg2}. This makes the ``almost solution'' from Fig.~\ref{fg1} difficult
to accept~-- even if one is willing to accept a deviation from the market clearing of
order $10^{-2}$, together with the unrealistic
confinement of all private choices to a rather
coarse discrete grid.}
Because the purpose of this \chptr\ is to develop a new methodology
that allows for computing the equilibrium in the same concrete setting verifiably,
i.e., without the problems described above,
it would be instructive for what follows  to identify
the reasons for the phenomenon 
that Figures~\ref{fg2}~\&~\ref{fg3} reveal.
As~is well known, 
the failure of the stationary distribution of a Markov chain 
to depend continuously on a parameter when the transition matrix depends on that
parameter continuously, which is what Fig.~\ref{fg3} illustrates,
implies multiplicity of the stationary distribution for certain values of
the parameter.
\footnote{It is easy to show that if $I\subseteq\R$ is an interval,
the transition matrix of a
particular Markov chain depends continuously on $\l\in I$ and,
furthermore, admits a unique
stationary distribution for every $\l\in I$, then that stationary
distribution is also a continuous
function $\l\in I$.
Hence, the stationary distribution can be discontinuous only if
its uniqueness fails
for certain values of $\l\in I$, which is quite intuitive.}
Theorem~2 in \cite{Hug93} provides conditions under which such
phenomena do not occur in the case of
two idiosyncratic states. While formulating these conditions
(which essentially boil down to
certain monotonicity in the transition probabilities)
for any finite number of idiosyncratic states is
straightforward, they become less natural~-- and thus
difficult to impose generically~-- 
in the presence of more than two idiosyncratic states.
The transition probabilities in the example considered here do
not satisfy such conditions and, for this reason, the
discontinuity in Fig.~\ref{fg3} is not a surprise.
In~particular, the example shows that, in general, the transition mechanism encoded
into equation~\eqref{lmb-iter} cannot be expected to have a unique fixed point, i.e.,
the Markov chain followed by the optimal state of the representative
household (under the time invariant policy) may
have infinitely many stationary distributions. Clearly, in equilibrium (if one exists
with constant risk-free rate and constant distribution of the population) the long-run
distribution of households must belong to the collection of stationary distributions,
i.e., fixed points, for~\eqref{lmb-iter}, but if this collection is not a singleton
then there would be no obvious way to identify a solution to~\eqref{lmb-iter} that
also satisfies the market clearing~\eqref{mclr}. To~put it another way, iterating to
convergence~\eqref{lmb-iter} from an arbitrary initial distribution cannot be expected
to produce a distribution that is compatible with the notion of equilibrium. In~what
follows we shall identify  two very different fixed points for~\eqref{lmb-iter}, both
of which correspond to the equilibrium rate~$r$ obtained with the method developed in
the present \chptr. While one of these two stationary distributions  yields expected
demand that is very close to~$0$, the other one, produced by
iterating~\eqref{lmb-iter} from the uniform distribution as above, yields expected
demand that is very far from~$0$~-- see the large dot in Fig.~\ref{fg2}. One is then
led to conclude that, as it stands, the classical framework described above is
incomplete, in the sense that a solution to  \eqref{lmb-iter} and \eqref{Bellman} that
also satisfies \eqref{mclr} cannot be identified generically within that
framework. To~see why this should not come as a surprise, notice that if the economy
is to converge to its equilibrium state, then the interest, the individual optimal
policies, and the population distribution will all adjust toward their steady-state
regimes simultaneously. One must then note that iterating~\eqref{lmb-iter} with fixed
interest~$r$ and fixed optimal policy $g_\infty(\cdot,\cdot)$  has the effect that the
economy is assumed to be in some form of partial equilibrium, where some endogenous
quantities have already found their equilibrium values, while the population
distribution  is yet to do so through the dynamics of~\eqref{lmb-iter}. In particular,
the representative household is in steady state (in terms of its optimal policy) but
the cross-sectional distribution of the population is not. There is no intuition to
suggest that the economy ever enters such a  partially equilibrated state, which is to
say, there is no reason to suppose
that the progression of the population distribution toward its time-invariant
configuration must be governed by~\eqref{lmb-iter}~-- even in situations where
that configuration happens to be a fixed point~of~\eqref{lmb-iter}. It is important to
also recognize that a modeling framework built exclusively around the optimization
problem attached to  a single representative household cannot account for the
price-agreement among a large population of
households. 
This price-agreement~-- see \eqref{ze5} below~--
is an important component of the notion of general equilibrium and, as we are about to see,
plays a crucial rôle in its calculation.

The observations from the last paragraph suggest very strongly that one needs to
develop
an alternative law of motion, i.e., alternative to equation \eqref{lmb-iter},  for the
distribution of households over the collection of private states. Moreover,
this law of motion must incorporate the interdependence
(in both space and time) between the
dynamics of prices, optimal private choices, and
population distribution before any of
these quantities has attained its steady state regime.
This, seemingly innocuous, task involves several crucial steps, with 
consequences that are far-reaching and in some sense radical.
In particular, it becomes necessary to develop the notion of economic equilibrium with
finite time horizon~$T$ (of any length), and
only then investigate the joint limit of all endogenous variables
as~$T\to\infty$. This step imposes certain dependencies
across time and across primal and dual variables that
cannot be resolved with existing mathematical tools~-- see \ref{iterlaced-rem} below.
Most important, the law of motion of the population distribution
may not be possible to interpret
as the low of motion of the probability distribution of the state of a single
representative household. Indeed, noting requires 
the transport of the population from one period to the next to be given as a
transition in the probability distribution of a single private state (however
defined). Insisting on such an interpretation amounts to a constraint that
is not intrinsic to the model, and therefore acts as a restriction on the
available solutions~-- and, ultimately, may lead to an outcome similar
to the one illustrated in Fig.~\ref{fg3}. 
In practical terms, this means that the transport of the population
must be understood and treated in a way that is broadly similar
to the way G.~Monge understood the transport of one pile of sand into another
pile of sand,
setting aside the time evolution of the probability distribution of any one
state (however defined). 
All these steps are made precise and assembled into a
computable program in Sec.~\ref{sec:gen-model} below, together with
an explanation for
the name ``time-interlaced backward induction''~-- see \ref{iterlaced-rem} and
\ref{gei-rem} for details.

Ultimately, the method developed in the present \chptr\  allows one to construct
a numerically verifiable solution
to Huggett's example introduced above 
and returns an equilibrium rate of approximately $0.037$
with market clearing of order $10^{-6}$. This rate  differs 
significantly from the one suggested by the left plot in Fig.~\ref{fg1}, which is around
$0.029$.
The plots in Fig.~\ref{fg4} illustrate the nature of the proposed new approach.
The~left plot shows the equilibrium (produced with the new method)
long-run entering and exiting cross-sectional distributions of households in different employment
categories over asset holdings.
We stress that these two sets of distributions need not be identical, though both obtain
from the same long run distribution over consumption~-- see Sec.~\ref{sec:IOU}.
{\captionsetup{belowskip=-10pt}
\captionsetup{aboveskip=0pt}
\begin{figure}[!htbp]
\centering 
\begin{subfigure}{.47\textwidth} 
  \centering
\leavevmode\raise0.65cm\hbox{\rotatebox{90}{\tiny distribution within emplpyment group}}%
\ %
\toshow{\includegraphics[width=7.1cm]{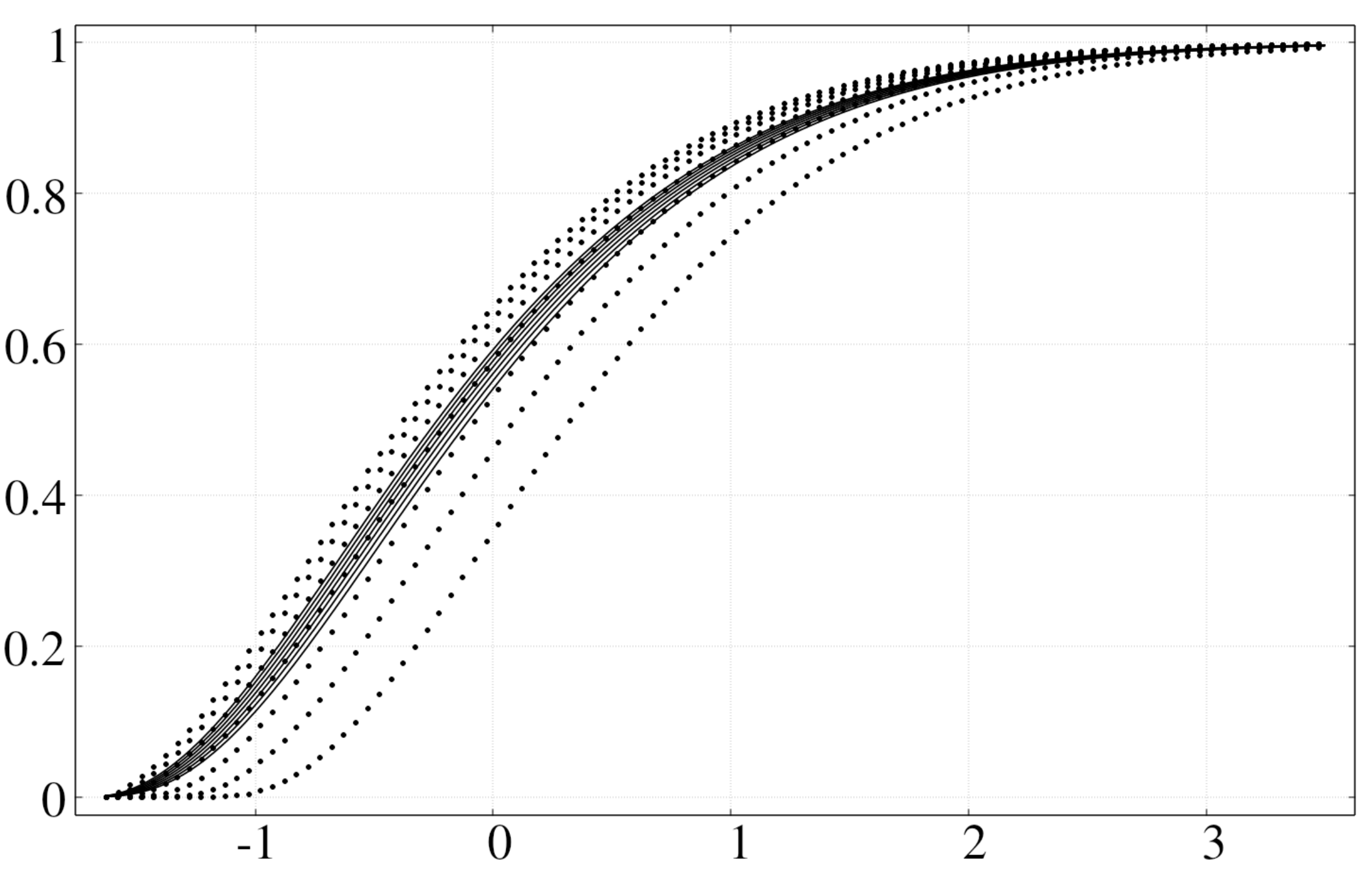}}%

\leavevmode\smash{\hbox to0.3cm{\hfill} \raise6pt\hbox{\tiny asset holdings}}

\end{subfigure}\quad
\begin{subfigure}{.47\textwidth}  
  \centering
\leavevmode\raise0.65cm\hbox{\rotatebox{90}{\tiny distribution within employment group}}%
\ %
\toshow{\includegraphics[width=7.1cm]{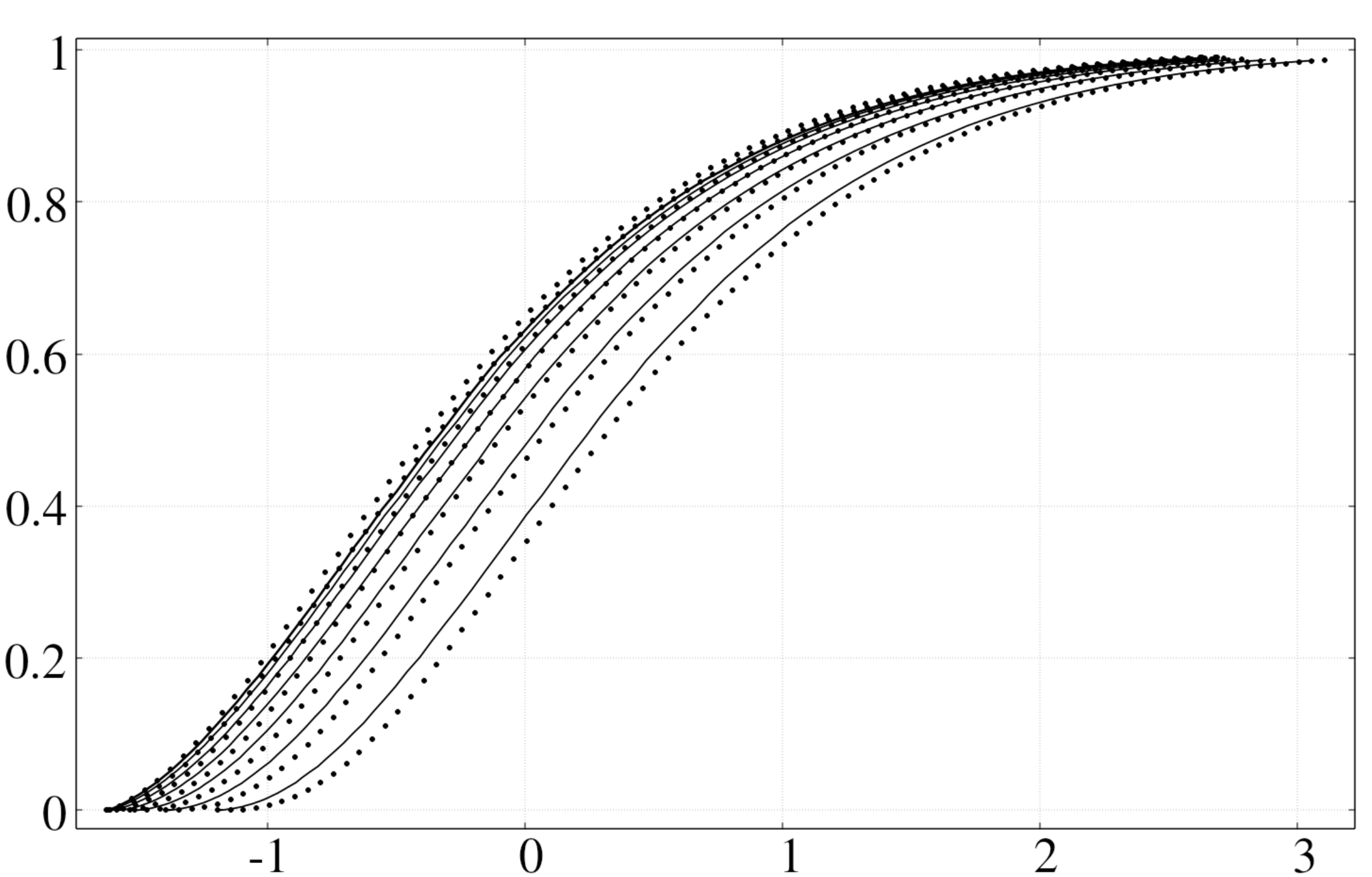}}%

\leavevmode\smash{\hbox to0.3cm{\hfill} \raise6pt\hbox{\tiny asset holdings}} 

\end{subfigure}
\caption{Left plot: entering (solid lines) and exiting (dotted lines)
equilibrium distributions of households in every employment category
over asset holdings produced with the new method.
Right plot: the population distribution produced
with the classical (see above)
method over $2\mathord,000$ grid points (solid lines)  when the iterations are initiated 
with the equilibrium rate of $0.037$ and with the exiting distribution (dotted lines) obtained with
the new method (replicated from the left plot).}  
\label{fg4} 
\end{figure} }%
It~is instructive to note that
if the classical program (see \eqref{lmb-iter} and \eqref{Bellman})
leading to Fig.~\ref{fg2} is initiated with the equilibrium rate
obtained with the new method, i.e., $r\approx0.037$, and
with the discretized (over $2\mathord,000$ grid points) version
of the exiting distribution from the left plot in Fig.~\ref{fg4} (instead of the uniform one), then
it~re\-turns expected demand of order~$10^{-3}$, together with the distribution shown in solid lines
on the right plot.
This illustrates the multiplicity of the stationary distribution that was noted earlier
\footnote{{}Note that the existence of multiple stationary distributions does not
amount to an existence 
of multiple equilibria, since only one of these distributions is found to clear the market.}
:
the large dot in Fig.~\ref{fg2} corresponds to the same rate $r\approx 0.037$,
but the expected demand of~$\mathord\approx 6.901$
is only possible if the distribution obtained by
iterating~\eqref{lmb-iter} from the uniform one is vastly
different from the distribution shown in solid lines on the right plot in Fig.~\ref{fg4},
which yields expected demand of $\mathord\approx10^{-3}$.
\footnote{{}We see that the classical approach described above is capable of locating
the equilibrium rate after all, 
but with the caveat that 
one needs to know how to locate the correct distribution (out of infinitely many)
with which to initiate the iterations of \eqref{lmb-iter}.}
The small difference between that distribution and the one in dotted lines
(obtained with the new method) on the same plot
is due to the fact that  the distribution $\l_\infty(\cdot,\cdot)$
(see \eqref{lmb-iter} and \eqref{Bellman})
is defined over pairs of an employment state attached to the 
present period and exiting wealth (taken from the finite grid) attached to the previous period,
i.e., the private state is exiting in terms of wealth but entering in
terms of employment.
\footnote{{}Such an approach is useful only in the absence of aggregate shocks.}
In contrast, the distribution obtained with the new method, which does not involve~\eqref{lmb-iter}, 
is over pairs of employment and exiting wealth attached to the same period.
It is interesting to note that the distributions shown in Fig.~\ref{fg4}~-- whether
produced with the new method 
developed later in this \chptr, or with the classical method described above~-- have no
points of accumulation.
We stress that the general method with which the
left plot in Fig.~\ref{fg4} was produced does 
not require the borrowing limit to be fixed in the outset and does not involve
boundary conditions at that limit,
i.e., the borrowing limit is endogenized. 

Perhaps the most interesting application of the new mathematical framework proposed
in this \chptr\  is to models with production that is subjected to common
(for all agents) productivity shocks.
This is the scenario where, generically, no time-invariant distribution of the
population exists, not even conditioned to the realized productivity state.
The only time-invariance that one may hope for is for the
population distribution, treated as a stochastic process, to be Markov in random environment, with time-invariant
transition (transport) mechanism.
The general strategy, adopted throughout most of
the literature on Krusell-Smith's model,
\footnote{{}\label{ftl1}%
There is an extensive body of research~-- see vol.~34 (2010) of Journal of Economic
Dynamics \& Control and the references therein~-- concerned with  the robustness and the
accuracy of the algorithm used in the benchmark case study of \cite{KS98}.
In particular, the paper by Den Haan in the same volume discusses the weaknesses (referred  to as ``fatal
flaws'') of the $R^2$ and
the standard regression error tests, as used in \cite{KS98},
to measure compliance with the relations defining the equilibrium.
Nevertheless, it will be shown in Sec.~\ref{sec:KS} below that
the results obtained by the simulation technique proposed in
\cite{KS98}, though narrower in scope, are reasonably accurate~-- at least with model
parameters chosen as in \nocite{KS98}{ibid.}} 
including  \cite{KS98},
is to reduce the cross-sectional distribution to a ﬁnite list of moments and then 
describe~-- somehow~-- the way in which the next period's cross-sectional moments
depend on the current moments and aggregate productivity state, which then determines
the individual policies. In most cases, including in \cite{KS98}, this law of motion is studied 
only in the long run, and is deciphered 
from the simulated behavior of a large population of households over a large number of periods.
\footnote{{}In Krusell and Smith's benchmark case study \cite{KS98}
the law of motion of the first moment in the long run is
extracted by way of least square fitting from the simulated behavior of 5,000 households over 11,000 time periods. }
This is suboptimal because 
both the individual and the collective behavior (hence, the equilibrium 
itself) depend only on the distribution of the
population, and full information about the position of each and every agent is massively 
superfluous. 
Another drawback from this approach is the insistence on~a ``sufficiently large''
time horizon and 
the persistence of i.i.d.\ prediction errors, which lack a clear economic interpretation.
One is also faced with the need to postulate a particular type of dependence
(log-linear in the benchmark study 
of \cite{KS98}) in the outset.

In addition to resolving the problem illustrated in Fig.~\ref{fg2} and providing a numerically
verifiable solution to the classical  Aiyagari-Bewley-Huggett model, the present \chptr\  makes the
following 
contributions to the study of models with shared risk, in particular to 
Krusell-Smith's model with production and
aggregate risk associated with shocks in the productivity factor.
First, it identifies analytically~-- not
empirically, on a case-by-case basis~-- general conditions under which the approximate aggregation
hypothesis holds (approximately).
Second, it is shown
that even if all aggregate variables and individual choices are forced
to depend on the population distribution only through its mean,
the exact form of the law of motion of the full population distribution can still be
identified; in particular, the population distribution cannot be chosen arbitrarily,
subject to the 
only requirement for its mean to follow a prescribed set of dynamics.
This feature reveals fluctuations in the disparity across the population that are
substantially larger than the fluctuations in the productivity shocks~-- see Fig.~\ref{fgKS5}
below~-- and cannot be captured by a model that is confined to the population mean alone
(somehow small fluctuations in the productivity shocks lead to much larger fluctuations in the
disparity%
\footnote{To the best of this author's knowledge such phenomena have not been documented before.}%
).  
Third, it is shown that movements of the population distribution
take place in the random environment of the transition in
the productivity state~-- not in the random environment of the productivity state
itself. 
To~put it another way, the transport from the present period to the
next depends on the productivity states in both, present and future, time periods.
Lastly, the general methodology proposed in the present \chptr\  is meaningful for
any, large or 
small, 
time horizon; in particular, models with infinite time horizon are merely limits of
models with 
finite time horizon. This feature is important for two main reasons.
The first one is that even in the classical examples borrowed
here from \cite{LjunSar00} and \cite{KS98} all time-invariant features
are attained after at least several hundred periods, whereas no real economy can
remain unchanged for that long. The second reason is that although postulating
infinite time horizon 
simplifies the matters enormously, it also makes invisible important
connections across time that may be needed 
in order to identify the equilibrium~-- see above.

Several important warnings and disclaimers are now in order. The computational
strategy developed in 
the present \chptr\  seeks to endogenize internally quantities that, traditionally,
have not been endogenized~-- 
at least not internally.
\footnote{{}An example of external endogenization would be a program that calculates
demands with 
given (as if exogenously specified) prices, and then varies the prices until market clearing
is attained in the long run, i.e., the prices are kept unchanged for many periods.
Traditionally, the borrowing constraint has been specified only exogenously.}
In addition, the strategy is deeply rooted in a special time-reordering of the endogenous
variables, which does not comply with the common Markovian structure
and has no precedent in the literature other than the paper~\cite{DL12}.
The benefits from this new approach notwithstanding (see above), one drawback is that
existence  
(nothing to say about uniqueness) of the equilibria is impossible to establish generically~--
at least not with
tools that are currently available. The main reason is that most of the dependencies
that the program 
seeks to resolve are implicit, in which case the use of the classical fixed-point
type argument~-- as in, 
say, the default reference \cite{DGMM94}~-- becomes very difficult.
In~particular, all variations of the general program outlined
in Section~\ref{sec:gen-model} involve several layers of iterations the convergence
of which is not 
guaranteed. Although endless loops do not occur in the examples included in
this \chptr\  and the 
convergence is quite fast, any computer code that implements the program
must limit the number of iterations in order to prevent potential endless loops.
We~stress, however, that the program tests for accuracy and convergence at every
step, and as long as it
completes, the result is always a numerically verifiable equilibrium. Another
drawback is that a 
continuous-time analog of the model described below is not currently available for two main
reasons: the time-interlaced structure of the model,
which is a departure from the classical Markovian setup,
and the effect of the random environment, which is
given by the 
transition in the productivity state~-- not by the productivity state alone.
In particular, the transport  operators
from~\ref{main-q} below do not appear to have an easily identifiable analog in any
known continuous-time framework.

The \chptr\  is organized as follows. Sec.~\ref{sec:gen-model} describes the limit of a
generic
heterogeneous agent model with finite time horizon and finite number of households
and outlines a metaprogram for identifying an equilibrium. Sec.~\ref{sec:IOU}
specializes that metaprogram
to the case of an economy with infinite time horizon and no aggregate risk. 
It is shown there that, the plots in Fig.~\ref{fg2} notwithstanding, the new
strategy can locate an equilibrium in the same benchmark study.
Sec.~\ref{sec:KS} implements the metaprogram from Sec.~\ref{sec:gen-model} in
the context of the benchmark economy of Krusell and Smith \cite{KS98},
compares the results, and draws new insights.



\section{General Equilibrium and the Time-Interlaced Master System}%
\label{sec:gen-model}\setcounter{paragraph}{0}
 
\noindent
The main goal in this section is to formulate precisely, and put together in the form
of a computable program, all the steps in the proposed new technique outlined in
Sec.~\ref{sec:Intro}~-- see \ref{main-proc} below.
In particular, the law of motion of the cross-sectional distribution
of all agents 
over the range of private states is derived without any reliance on the concept
of representative agent,
i.e., without any connection with the transitions in the probability distribution of
a single observable quantity~--
see \ref{main-q} below and the discussion in Sec.~\ref{sec:Intro}.
The main difficulty to overcome comes from the fact that
the notion of general equilibrium
imposes certain intrinsic connections
between the endogenous variables that are time-interlaced, 
in the sense 
that the search for certain
endogenous variables attached to one period
in time can only be done simultaneously with the search
for other endogenous variables attached to the next period.
\footnote{On some very general grounds,
the time-interlaced structure of the general equilibrium is quite
intuitive (recall that all prices are endogenous): the prices today depend on the
demand for securities today, which depends on the returns tomorrow, which returns are
affected by the prices today, which are affected by the demand today $\ldots$}
Another difficulty is that, in the parlance of MFG, the coupling function, i.e., the
effect of the population distribution on the private states and decisions, is
endogenous and can be determined only implicitly while solving for the equilibrium. 
As~a~re\-sult, one is faced with an impossibly
large system composed of all first-order and market
clearing conditions attached to all agents, all time periods, and all
states (aggregate and idiosyncratic).
Because of its time-interlaced structure and because of
the implicit nature of the coupling,
this system cannot be resolved with common methods
borrowed from the domain of optimal control and MFG. 
Just as an example, 
re-arranging all conditions that define the equilibrium
into a pair of coupled recursive programs,
one moving backward and one moving forward,
similarly to a coupled MFG system,
does not appear to be possible. 
A~workaround was proposed in the paper
\cite{DL12}, which shows that without any reliance on the representative
agent framework, the giant system of market clearing and private first order
conditions across time and across the population of agents can be broken into smaller
systems, 
which can then be chained into a computable backward induction program. 
This program parallels the familiar backward induction in dynamic
programming, except that the system to be solved for at every step involves
some endogenous quantities attached to
period~$t$ and other endogenous quantities attached to period~$t+1$. 
One major drawback from the method developed in \cite{DL12} is that
it assumes a finite number of agents and  
requires tracking the
individual state of every agent, which is
practical only if the number of agents is very small (usually, just 2 in most
workable examples). 
Most of what follows in this section is essentially a revision and extension
of the method developed
in  \cite{DL12} with the goal of: introducing
the distribution of all agents over the collection
of private states as an endogenous variable
\footnote{That the cross-sectional distribution of the population is a
``sufficient statistic,''
in that it makes the full information about the state of every single agent superfluous,
has been know to economists at least since the
publications \cite{HHK74} and \cite{Hil74}.},
removing the need to track the individual state of every agent, and 
pushing the number of agents
is to $\infty$. 
Thus, the main challenges in front of us are:
(a)~passing to the limit as the number of agents increases to~$\infty$;
(b)~working with an endogenous variable that belongs to an infinite dimensional space;
(c)~dealing with the intrinsic time-interlaced structure of the model;
(d)~developing the law of motion of the population distribution without any reliance
on the concept of representative household.

The reader must be forewarned that, just as in the work \cite{DL12},
in the present \chptr\ %
consumption is used as a state variable instead of wealth, so that the population
distribution is 
treated as a distribution of unit mass over states of employment and levels of
consumption~-- not 
over states of employment and levels of wealth, as is more common in the
literature.
\footnote{That consumption carries all the necessary information about
wealth, together with the benefits from using consumption as a state variable
(instead of wealth), has been known to economists
for some time~-- see \cite{Hall78} and \cite{DL12}.}
The reasons for this choice essentially boil down to the benefits from identifying (through a
particular homeomorphism) consumption as both state and costate variable. Another
reason is that in every given time period there is only one consumption level to
attach to every household, whereas the household's wealth at the end of the period is
generally different from 
that at the beginning. One must also note that there is an obvious lower bound on
consumption, namely~$0$, and this lower bound never binds. In contrast, the lower
bound on wealth is a priori unknown.
The formal description of the model is next. 

The time parameter $t$ is restricted to the finite set
$\{0,1,\ldots,T\}$ and the total number of households (alias: agents), $N$, is
assumed, for now, to be finite.
Economic output is generated in every period and is expressed in units
of a single numéraire good, which 
can be either consumed, or turned into productive capital 
except during period $T$.
Every household extracts utility from consuming the numéraire good.
All households share the same
impatience parameter $\b>0$ and the same time-separable utility from 
intertemporal consumption given by the mapping $U\colon\R\mapsto \R$,
which is twice continuously differentiable in $\Rpp$ with $\partial U>0$, $\partial^2 U<0$,
and is such that
$\lim_{c\searrow 0}\partial U(c)=+\infty$, $\lim_{c\searrow 0}U(c)=-\infty$, and
$U(c)=-\infty$ for $c\le 0$.  
Economic output is generated by two inputs: the net labor supplied during the period
the output is delivered
and the net productive capital installed during the previous period. Two investment
instruments are available to all households: capital stock
and locally risk-free private lending instrument (alias: IOU), which is in net supply
of zero.
\footnote{{}The numéraire good, which is also the currency, cannot be stored from one
period to 
the next, but entitlements to it can be carried from one period to the next by means
of financial 
contracts. The assumption that the private lending instrument is in zero net supply is
imposed here 
merely for simplicity.}
The collection of all (idiosyncratic) private  employment states 
is $\EEE\subset\Rp$ and the collection of all
productivity states is $\XXX\subset\Rpp\tts$.
These sets have finite cardinalities denoted $\abs{\EEE}$ and~$\abs{\XXX}$, both
assumed to be at 
least~$2$.
The~elements of $\XXX$ have the meaning of total factor productivity (TFP)
and the elements of $\EEE$ have the meaning of  physical units of labor.
The productivity state, which is shared by all households, follows an irreducible
Markov chain in 
the state space $\XXX$
with transition probability matrix~$Q$ (of size $\abs{\XXX}$-by-$\abs{\XXX}$)
that has a unique set of steady-state probabilities $(\psi(x),\, x\in\nobreak\XXX)$.
The transitions in the individual employment states, which are independent from one another
when conditioned to a particular transition in the productivity state $x \to y$,
are governed by the transition probability
matrices~$P_{x,y}\in\R^{\abs{\EEE}\otimes\abs{\EEE}}$, 
$x,y\in\XXX$. All elements of the
matrices $Q$ and $P_{x,y}$ are assumed strictly positive. The pair consisting of 
the shared productivity state and the employment state
of a particular household follows a Markov chain on the state-space $\XXX\times\EEE$
with transition 
from $(x,u)$ to $( y,v)$ occurring with probability  $Q(x, y)P_{x, y}(u,v)$.

Households that are in the same state of employment would have identical consumption levels only if
they start the period with identical asset holdings,
\footnote{{}Both assets, private lending and capital, are assumed to be perfectly liquid, in which
case the composition of asset holdings is irrelevant.}
in which case their
investment decisions would be identical as well (see \ref{thm1} below for an explanation).
To put it another way,
households that are in the same state of employment and choose the same consumption
level $c\in\Rpp$, which quantity represents physical units of the numéraire good, are indistinguishable.
For this reason, in what follows consumption will be used as a state variable instead of wealth. 
Thus, the mathematical metaphor for the collective
state of the population is the distribution of unit mass over the product space
$\EEE\times\Rpp$ (the space of household characteristics, i.e., employment and
consumption levels).
To~be able to work with such objects,
we now introduce the space~$\bbP(\EEE)$ of strictly positive
unit-mass (a.k.a. probability) measures over $\EEE$,
the space $\DB$ of all (cumulative)
càdlàg distribution functions over~$\Rpp$, and the collection~$\DB^\EEE$ of all assignments
$\csd\colon \EEE\mapsto \DB$. An~element~$\csd\in\DB^\EEE$ can be
identified as a finite list of distribution functions $\csd\equiv(\csd^u\in\DB)_{u\tts\in\tts\EEE}$, in which
one cumulative distribution function over~$\Rpp$ is assigned to every employment category $u\in\EEE$.
Any probability measure on $\EEE\times\Rpp$ can be disintegrated into the form $\pi(\dm u)\d
F^u(c)$ and treated as a pair $(\pi,\csd)\in\bbP(\EEE)\times\DB^\EEE$.
When representing the distribution of~$N$
households in the state space $\EEE\times\Rp$, a measure of this form would have at most~$N$
atoms every one of which has a mass that is an integer multiple of $1/N$, so
that $\pi(u)N$ gives the total number of households who happen to be in employment state~$u$ and
$\pi(u)\csd(c)N$ gives the total number of households who happen to be in employment state~$u$ and
choose consumption level that is not strictly larger than $c\in\Rpp$.
If the private states of $N$ agents are distributed over the space
$\EEE\times\Rpp$ with law $\pi(\dm u)\d F^u(c)$ then the average employment level across the population
can be cast as
{\abovedisplayskip=8pt plus 3.5pt minus 3.5pt\belowdisplayskip=8pt plus 3.5pt minus 3pt\belowdisplayshortskip=8pt plus 3.5pt minus 3.5pt
\[
L(\gp) 
= \sum\nolimits_{u\ts\in\ts\EEE}u\,\gp(u)  = \gp\ts\EEE\,,
\]
}%
where $\gp$ and $\EEE$ are treated as vector row and vector column respectively.
In particular, the aggregate amount of installed labor is given by $L(\pi)N=\gp\ts\EEE\ts N$.
Similarly, if $\qq_u(c)$
denotes the capital invested by every household who happens to be in state
$(u,c)\in\EEE\times\Rpp$, then the average private capital
invested across the population is
{\abovedisplayskip=8pt plus 3.5pt minus 3.5pt\belowdisplayskip=8pt plus 3.5pt minus 3pt\belowdisplayshortskip=8pt plus 3.5pt minus 3.5pt
\[
K(\gp,\csd) 
= \sum\nolimits_{u\ts\in\ts\EEE}\,\gp(u)\int_\Rpp\qq_u(c)\d\csd^u(c)\,,
\]
}%
and the aggregate installed capital is $K(\gp,\csd) N$.

\begin{nit}{Aggregate level vs. population average}
It is very common in the literature to integrate the level of a particular private
variable against 
a probability measure that represents the distribution of the population and to
declare that the integral 
gives the aggregate level of that same variable across the entire population~--
see \cite{KS98} as just one example.
One must note, however, that 
such integrals only represent the weighted average level across the population~-- not the
aggregate level. As~the number of agents increases to $\infty$ the aggregate level can remain
finite only if the private levels become negligible and the private levels can remain
non-negligible only if the aggregate level is allowed to explode.
There are two common scenarios in which ignoring the difference between the average
level across 
the population and the aggregate level is innocuous. The~first one is when
the aggregate level must be adjusted to $0$, as in equation \eqref{mclr} above. The
second one is 
when a Cobb-Douglas production function is postulated.\qed
\end{nit}

As a next step, we postulate the usual ``competitive firm'' with two factors of production,
capital and labor, and with production technology 
given by  a Cobb-Douglas constant return to scale production function with
capital share parameter $0<\a<1$. 
Thus, the rates of return on capital and labor, realized during the future period, can be treated as
functions of the average privately installed capital~$K$ during the present period.
These functions must depend on the future productivity state $y\in\XXX$ and the future distribution
over states of employment $\varpi\in\bbP(\EEE)$.
In~equilibrium factor prices maximize the firm's profits, so that
the rates of return on capital and labor are given (as functions of the average $K$)
by
\footnote{Note that the average capital $K$ belongs to the present period while the pair
$(y,\varpi)$ belongs to the future (i.e., the next) period, during which both returns are realized.}
{\abovedisplayskip=7pt plus 1.5pt minus 5pt\belowdisplayskip=7pt plus 1.5pt minus 5pt\belowdisplayshortskip=3pt plus 1.5pt minus 6pt
\begin{equation}\label{returns}
\begin{gathered}
\Rpp\ni K \leadsto \rho_ {y,\varpi}(K)\df  y\times\a\times
\Bigl({K N\over L(\varpi) N}\Bigr)^{\a-1}\equiv y\times\a\times \Bigl({K\over L(\varpi)}\Bigr)^{\a-1}\,,\\
\Rpp\ni K \leadsto \ee_ {y,\varpi}(K)\df
y\times(1-\a)\times \Bigl({K \over L(\varpi N)}\Bigr)^{\a}
\equiv y\times(1-\a)\times \Bigl({K \over L(\varpi)} \Bigr)^{\a}\,.
\end{gathered}
\end{equation}
}%
We see that with this special choice of the production function replacing the aggregate capital
$K N$ and the aggregate labor $L(\varpi) N$ with the respective population averages
$K$ and $L(\varpi)$ would not matter. As the returns in \eqref{returns} depend only on the
population averages, they are perfectly meaningful for any population distribution (expressed as a
probability measure on $\EEE\times\Rpp$) that may or may not correspond to a finite population of agents. 
Installed productive capital is assumed to depreciate at constant rate $\dd>0$ and,
in order to generate paychecks at time $t=0$, we postulate the fictitious quantity
$K_{-1}$, which has the meaning of a primordial average endowment with capital
that is shared equally among all households.
\footnote{{}There is no need for the households to be identical before time $0$.
This assumption is imposed only for the sake of simplicity.} 

Because economic agents are concerned only with returns, they are
concerned exclusively with the distribution over the space $\EEE\times\Rpp$ on which the averages
depend.
We postulate now that the number of agents is infinite and that their distribution over
$\EEE\times\Rpp$ can be any measure of the form $\pi(\dm u)\d F^u(c)$
for some (any) choice of the pair $(\pi,\csd)\in\bbP(\EEE)\times\DB^\EEE$.
Thus, the aggregate state of the economy during any given period is understood to be
a triplet of the form $(x,\pi,\csd)$, for some
choice of a productivity state $x\in\XXX$ and population
distribution $(\pi,\csd)\in\bbP(\EEE)\times\DB^\EEE$.
If the dependence on the time period needs to be emphasized in the
notation we shall write $(x_t,\pi_t,F_t)$ instead of $(x,\pi,F)$ and shall use similar
conventions for all other quantities that we may introduce, with the understanding that the
subscript~$t$ may be omitted, if the association with a specific time period is irrelevant.
It is important to recognize that the first two elements of the triplet $(x,\pi,\csd)$ have
dynamics that are fully exogenous and unrelated to the individual choices across the population,
i.e., have no relation to the domain of economics. 
For this reason, we call the pair $(x,\pi)\in\XXX\times\bbP(\EEE)$ exogenous aggregate state, or
simply exogenous state. The third component of the aggregate state $(x,\pi,F)$, i.e.,
$F\in\DB^\EEE$, we call endogenous aggregate state, or simply endogenous state.

The assumption that the population of agents is infinite has important implications which
we now address.
The major simplification that takes place in the limit as $N\to\infty$ is in the following.
Assuming that the present period productivity state $x\in\XXX$ transitions in the
next period to 
productivity state $y\in\nobreak\XXX$, every household presently in employment state
$u\in\EEE$ 
would sample its future employment state from the collection~$\EEE$, independently from all other households,
according to the distribution law over~$\EEE$ given by the vector~$P_{x, y}(u,\cdot)\in\R^{\abs{\EEE}}$.
By Glivenko-Cantelli's theorem, as the number of households in state $u$ increases
to~$\infty$, 
the proportion of all households presently in
employment state~$u$ who transition to employment state~$v$ must converge to $P_{x,y}(u,v)$.
Once the number of households is postulated to be infinite, one can set aside the
assumption that the 
shocks in employment are independent (and hence forfeit the reliance on Glivenko-Cantelli's
theorem)
and simply postulate that the collection of households in state
$u$ who transition to state $v$ can be weighted against the population of households
in state $u$, 
with relative weight given by $P_{x,y}(u,v)$~-- this is all that matters in the model
and this is what will be assumed from now on.

\begin{nit}{Infinite collections of agents}\label{issues}%
The technical problems associated with measuring and comparing infinite collections
of agents are well 
known.
\footnote{See the extensive discussion and references
in \cite{Mal72}, \cite{FG85}, \cite{Judd85}, \cite{Sun06}, \cite{DS12}.}
First, one cannot distribute a finite mass uniformly across a countably infinite collection of
agents.  
It is possible to distribute a finite mass uniformly over a continuum, but then the
only sets that can be compared would be the elements of a particular $\s$-field. This is
inadequate because if all agents sample their employment state independently, there
would be no 
reason for the set of agents who fall into a particular employment category to belong
to that 
special \hbox{$\s$-field}. The~workaround that we use below has two main
aspects. The~first one is dispensing 
with the notion of ``uniform distribution of weights'' and simply postulating the 
relative weights suggested 
by Glivenko-Cantelli's theorem. The second one is dispensing with the need for a
universal $\s$-field 
specified in the outset, and taking advantage of the fact that relative weights can be
assigned in a  
consistent fashion to the elements of finite partitions 
and sub-partitions (however defined) as they come along period by period.\qed
\end{nit}

\begin{nit}{Sets of agents and their relative weights}\label{weighing}%
Let $\W$ stand for the collection of all economic agents. Let $A_u\subset\W$ be the
set of all agents 
who happen to be in state $u\in\EEE$ during a given period~$t$. Then $(A_u,\,u\in\EEE)$ is a
finite partition of $\W$. We make no assumptions about the structure of the sets $\W$
and $A_u$ other than  
insisting that $A_u$ is not a finite set for any $u\in\EEE$. We do assume, however, that
for every $u\in\EEE$ the collection of agents $A_u$ can be weighted against the
collection $\W$ and 
the associated list of relative weights is given by some $\pi\in\bbP(\EEE)$. Let
$B_{u,v}$ be the 
subset of $A_u$ consisting of all agents who transition to state $v\in\EEE$ in period
$t+1$, so  
that $A_u=\cup_{v\tts\in\tts\EEE}B_{u,v}$.  We make no assumptions about the
structure of the sets 
$B_{u,v}$ other than insisting that none of these sets is finite. Furthermore, we
suppose that when the 
productivity states in periods $t$ and $t+1$, respectively $x$ and $y$, are known,
then every set $B_{u,v}$ can be weighted against $A_u$ and its relative weight is given by
$P_{x,y}(u,v)$.  In particular, $B_{u,v}$ can be weighted against $\W$ with relative weight
$\pi(u)P_{x,y}(u,v)$.
Next, given any $u,v\in\EEE$ and any $c\in\Rpp$, let $E_{u,v}(c)$ stand
for the subset of $B_{u,v}$ consisting of all agents whose consumption level during
period~$t$ does not 
exceed~$c$ (strictly). We again suppose that $E_{u,v}(c)$ can be weighted against
$B_{u,v}$ with 
relative weight that is independent from $v$ and given by $\csd^u(c)$ for some choice of
$\csd\in\DB^\EEE$.
\footnote{During period $t$ all agents in state $u$ face the same uncertain future in terms of
employment and make investment-consumption decisions before their future employment state is
realized. Therefore, the individual choices across any ``sufficiently representative'' subset of $B_u$
must be statistically indistinguishable from those in any other ``sufficiently representative''
subset.
The insistence that the relative weight of  $E_{u,v}(c)$ does not depend on $v$ is a mathematical
metaphor for the claim  
that $B_{u,v}$ is a ``sufficiently representative'' subset of $A_u$ for any $v$,
but we stress that the independence of the relative weight of $E_{u,v}(c)$
from the future employment state $v$ is an assumption that we impose, not a conclusion that we
arrive at.}
Hence, the relative weight of
$E_{u,v}(c)$ against~$A_u$ is $P_{x,y}(u,v)\csd^u(c)$ and against~$\W$ it is
$\pi(u)P_{x,y}(u,v)\csd^u(c)$, but these relative weights become available only if the present and
the future productivity states, $x$ and $y$, are known.
Finally, any finite union of sets of the form  $E_{u,v}(c)$ for various choices of $u,v\in\EEE$
and $c\in\Rpp$, or of the form $B_{u,v}$ for various choices of $u,v\in\EEE$,
has a well defined relative weight against~$\W$.
\footnote{This is because any such finite union can be written in a unique way as the finite union
of disjoint sets that were already assigned relative weights.}
In particular, the collection of
agents in state $u$ whose consumption level does not exceed~$c$ can be weighted against~$A_u$ with
relative weight given by $\sum_{v\tts\in\tts\EEE} P_{x,y}(u,v)\csd^u(c)=\csd^u(c)$, and therefore
also weighted
against~$\W$ with relative weight $\gp(u)\csd^u(c)$.
Similarly, given any $v\in\EEE$, the collection of agents in state $v$ during period $t+1$
is noting but the union of disjoint sets $\cup_{u\tts\in\tts\EEE}B_{u,v}$, so that the
period-$(t+1)$ distribution over states of employment is given by
{\abovedisplayskip=8pt plus 1.5pt minus 1.55pt\belowdisplayskip=8pt plus 1.5pt minus 1.55pt\belowdisplayshortskip=5pt plus 1.5pt minus 3pt
\[
\varpi(v)=\sum\nolimits_{u\ts\in\ts\EEE}\pi(u)\, P_{x, y}(u,v)\quad 
\text{for all}\ \ v \in\EEE\,,
\]}%
which we may abbreviate as
$
\varpi=\pi\, P_{x, y}\,,
$
treating $\pi$ and $\varpi$ as vector rows.\qed
\end{nit}

The last relation illustrates one of the key advantages of working with an infinite population of
agents: the period-$(t+1)$ distribution over employment is fully determined by the period-$t$
distribution over employment, in conjunction with the productivity states in both periods.
Nevertheless, without further restrictions on the model
the aggregate exogenous state $(x,\pi)\in\XXX\times\bbP_{++}(\EEE)$ would not be constrained
to a finite set and this complicates enormously all practical aspects of the model.
It~turns out to be possible to reduce, at the expense of certain restriction on the
transition 
probabilities $P_{x,y}(u,v)$, the range of the exogenous state to the finite 
collection~$\XXX$, as is explained next. 

\begin{nit}{Simplifying assumption and remark}\label{n1}%
In the benchmark economy studied in the landmark paper \cite{KS98}  the
con\-di\-ti\-onal transition matrices
$P_{x, y}$ are chosen in such a way that it becomes possible to attach a unique distribution,
$\gp_x\in\nobreak\bbP(\EEE)$, to every productivity state $x\in\XXX$ so that
$\gp_ y =\gp_xP_{x, y}$ for all choices of $x, y\in\XXX$ (the future distribution over
employment depends only on the future productivity state).
With this special choice for the matrices $P_{x, y}$ the steady state regime of the
population in 
terms of employment is such that the distribution of households over states of employment
fluctuates randomly but in perfect sync with the productivity state, so that when the
productivity state is $x\in\XXX$ the distribution of households over states 
of employment is exactly $\gp_x\in\nobreak\bbP(\EEE)$.
\footnote{{}This feature comes from the fact that
the Markov chain on $\XXX\times\EEE$ with transition matrix $Q(x, y)P_{x, y}(u, v)$ admits a
steady-state distribution in which state $(x,u)$ occurs with probability $\psi(x)\gp_x(u)$.}
This is a vast simplification, because in such a regime the distribution of
households over states of employment 
fluctuates through the finite collection $\{\gp_x\colon x\in\XXX\}\subset \bbP(\EEE)$, as opposed
to fluctuating through the infinite collection~$\bbP(\EEE)$.
In~what follows we shall suppose that this simplification is in force
and shall assume without further notice that productivity and employment are fluctuating according
to the steady state regime just described (even at time $t=0$). In~addition, we shall
exclude the scenario where productivity or employment
can get absorbed in a single state. The population distribution in every productivity
state can now 
be given as an element of $\DB^\EEE$, since the distribution over $\EEE$ is fixed by
the productivity state.
Thus, 
an ``aggregate state of the economy'' will be understood to mean a pair
of the form $(x,\csd)\in\XXX\times\DB^\EEE$, consisting of a productivity state $x$
and a list of 
distribution functions $(\csd^u\in\DB,\,u\in\EEE)$. The exogenous state is then just the
productivity state $x\in\XXX$. For the sake of simplicity we set
$L(x)=L(\pi_x)$, $\rho_y(K)=\rho_ {y,\pi_y}(K)$, and $\ee_ {y}(K)=\ee_ {y,\pi_y}(K)$.\qed
\end{nit}

Because of the assumption postulated in \ref{n1},
the movement of the population distribution over the space $\EEE\times\Rpp$
can now be treated as a movement in the space $\DB^\EEE$ which
is affected by
the random environment given by the transition in the productivity state~-- the only
exogenous aggregate variable faced by all agents. Describing these movements is one
of the key 
aspects of the model and is the task we turn to next.

\begin{nit}{Time-dependent transport}\label{et}%
The transport of the population distribution from period~$t$ to period $t+1$ must be
allowed to depend 
on the period-$t$ productivity state $x\in\XXX$ and on the period-$(t+1)$ productivity state
$y\in\XXX$.
\footnote{{}\label{ftn82148}%
This structure is consistent with the way in which Krusell and Smith cast
their general  
model in \citex[II-B]{KS98}, in which the updating rule
is written as $\G'=H(\G,z,z')\tts$. However, the computational strategy described in
\citex[II-C]{KS98} assumes an updating rule for 
the mean of the form ${\bar k\tts}'=h(\bar k, z)$. This later form persists   
throughout most
of the literature on Krusell-Smith's model (see, just as an
example, \citex[Sec.~18.15.2]{LjunSar00}). 
We will see in \ref{main-q} below (see also \ref{n2} and \ref{T-inf})
that, some special cases notwithstanding, neither
the dependence on the present productivity state $x$ nor the dependence on the future
productivity state $y$ can be ignored.} 
We express this transport as a collection of mappings
$\T_{t,x}^y\colon\DB^\EEE\mapsto\nobreak \DB^\EEE$, $x,y\in\XXX$.
Rational private decisions about consumption and investment during period $t$, given
the period-$t$ 
productivity state $x\in\XXX$, are only possible if an assumption about the period-$t$
population 
distribution $F\in\DB^{\EEE}$ and about the collection of transport mappings
$\{\T_{t,x}^y\colon y\in\nobreak \XXX\}$ is made.
At~the same time, the realized population distributions in periods $t$ and $t+1$,
respectively $F_t^*$ and $F_{t+1}^*$, are the result of a multitude of individual
choices made during periods 
$t$ and $t+1$, with the understanding that $F_{t+1}^*$ depends on the realized
period-$(t+1)$ 
productivity state \smash{$y^*\in\XXX$}, which affects the consumption choices during
period~$t+1$. 
Equilibrium considerations (see \ref{def-equil} below for a precise definition)
require that the assumed distribution and its
transport coincide 
with the realized ones, i.e, $F_t^*=F$ and \smash{$F_{t+1}^*=\T_{t,x}^{y^*}(F)$} in
every possible   
realization of the future
productivity state $y^*\in\XXX$.
This requirement, which we call ``self-consistency of the transport,''
is the main challenge in the study of incomplete-market models with a large number
of heterogeneous agents and addressing it beyond the realm of what is commonly known as
``stationary recursive equilibria''
is the main goal in what~follows in this \chptr.
\footnote{The self-consistency of the transport is very similar in nature to the
self-consistency of the mean field in mean field theory, whence the borrowed
term.}\qed%
\end{nit}

The features outlined in~\ref{et} have two important aspects: (a)~during period $t$
the households do not 
observe $F_t^*$ (the realized population distribution after all private decisions are made)
but assume that $F_t^*=F$ for some percieved $F\in\DB^\EEE$, and (b) all private
choices during period $t$ take into account the shared prediction for the
period-$(t+1)$ state of the 
population, expressed as the list $\{\T_{t,x}^y(F)\colon y\in\XXX\}$.

\begin{nit}{Parallel with classical transport problems}%
There is an obvious parallel bet\-ween the transport
(for fixed $t$) introduced in \ref{et} and the transport of mass arising 
in the classical Monge-Kantorovich problem~--
see \cite{Vil03}, \cite{Vil09}, 
for example.  
There is also a crucial difference:
there is no single surplus function 
whose average is to be
optimized and the target measure 
is endogenous (see below).
This makes the problem more challenging and also more interesting: the transport is determined
from optimizing over a large number of individual objectives rather than a single global one.
Because of this shift in the paradigm, most of the tools for solving the
Monge-Kantorovich problem 
developed in the course of the last two
centuries will not be possible to utilize in the present context~-- at least not
directly. Nevertheless, 
some of the general ideas can still be mimicked and put to use. The method developed
below still 
relies on the idea of randomization introduced by Kantorovich, but not by way of
coupling of two 
probability measures. It also relies on the idea of Monge coupling, but not by way of pure
assignment~-- see \ref{rnd-Monge} below.\qed 
\end{nit}

The next step is to introduce the field of individual savings problems and the field
of their duals.
All households observe
the history of the realized productivity state, together with the
history of their own individual employment states and asset holdings, 
share the same belief about the transition probabilities that govern the employment and
productivity shocks (encrypted in the matrices~$P$ and~$Q$), 
share the same belief about the initial (time~$0$) population distribution, and, finally,
share the same belief about the entire collection of
transport mappings
\footnote{{}Allowing the transport to be time dependent is one of the key differences between
the approach adopted here and previous works~-- see \cite{KS98}.}
{\abovedisplayskip=5pt plus 2pt minus 2pt\belowdisplayskip=8pt plus 2pt minus 2pt\belowdisplayshortskip=5pt plus 1.5pt minus 1.5pt
$$
\T\df \bigl\{ \T_{t,x}^y \colon 0\le t<\nobreak T,\, x,y\in\XXX \bigr\}\,.
$$
}%
Hence, during any given period all households have identical beliefs about
the present population distribution (over the range of consumption and levels
of employment), once the history of the productivity state until
that period is revealed. As a result,
given the stream of present and future transport mappings,
private decisions about savings and consumption depend on the current employment state,
the asset holdings at the beginning of the period (entering wealth),
the current productivity state, and the perceived population
distribution during the current period.
\footnote{{}The individual savings problems are influenced by the history of the productivity shocks only
through their dependence on the current population distribution.}
In particular, all global (shared) endogenous variables, namely the average installed capital and
risk-free rate, depend on the time period $0\le t<T$ and
on the aggregate state of the economy $(x,F)\in\XXX\times\DB^\EEE$ during that period
(again: for a given stream of present and future transport mappings).

Suppose that during period $0\le t<T$ the economy happens to be in state $(x,F)\in\XXX\times\DB^\EEE$, the
average installed capital is $K=K_t(x,F)>0$, and the (one period) risk-free rate is $r=r_t(x,F)>-1$.
Consider a generic household that enters employment state $u\in\EEE$ 
with its personal wealth measuring~$w$ units of the numéraire good,
which quantity aggregates the wage received during period $t$, the return on capital invested in the
previous period,
and the holdings of private loans carried from the previous period.
Treating the entering wealth $w $
as a given resource and taking $K$ and $r$ as given (the household is a price taker),
the household must determine its consumption level $c$,
its investment $\q$ in the private lending instrument (IOU),
and its investment $\qq$ in productive capital
\footnote{All three quantities $c$, $\q$ and $\qq$ are understood to represent physical
units of the numéraire good.}
by maximizing over $(c,\q,\qq)\in\R^3$ and $(W_{y,\tts v}\in\R)_{ y\tts\in\tts\XXX, v\tts\in\tts \EEE}$ the
objective
\begin{subequations}\label{ze1212} 
{\abovedisplayskip=8pt plus 2pt minus 2pt\belowdisplayskip=8pt plus 2pt minus 2pt\belowdisplayshortskip=5pt plus 1.5pt minus 1.5pt
\begin{equation}\label{ze1}
\begin{aligned}
J_t\Bigl(c ,\,&(W_{y,\tts v} )_{ y\tts\in\tts\XXX, v\tts\in\tts \EEE}\Bigr)\\
&\df U(c )
+\b\sum\nolimits_{ y\ts\in\ts\XXX,\,v\ts\in\ts\EEE}
V_{t+1,\tts y,\tts \T_{t,\tts x}^y(F),\tts v}\bigl(W_{y,\tts v} \bigr)\, Q(x, y)P_{x, y}(u, v)\,,
\end{aligned}
\end{equation}}%
subject to
{\abovedisplayskip=8pt plus 2pt minus 2pt\belowdisplayskip=8pt plus 2pt minus 2pt\belowdisplayshortskip=5pt plus 1.5pt minus 1.5pt
\begin{equation}\label{ze2}
\begin{gathered}
W_{y,\tts v}=\bigl(1+r\bigr)\q + \bigl(\rho_ y(K)+ 1-\dd\bigr)\qq + 
\ee_ y(K)v\,,\  \  y\in\XXX\,,\  v \in\EEE\,,\\
\llap{and\qquad\qquad}c+\q+\qq=w\,,
\end{gathered}%
\end{equation}%
}%
\end{subequations}%
with the understanding that  $V_{t,\tts x,\tts F,\tts u}(w )$ is the constrained maximum attained in
\eqref{ze1} for every $0\le t<\nobreak T$ and
$V_{T,\tts y,\tts \T_{T-1,\tts x}^y(F),\tts v}\bigl(W_{y,\tts v} \bigr)=U\bigl(W_{y,\tts v} \bigr)$,
i.e., during the last period the household can only consume (note that $U(c )=-\infty$ if $c\le 0$ and
$U\bigl(W_{y,\tts v} \bigr)=-\infty$ if $W_{y,\tts v}\le 0$ by the very definition of
$U\phd$). 
We~stress that no~borrowing constraints are imposed extraneously in the individual
savings problems. The borrowing 
limits arise 
endogenously from the notion of equilibrium~-- see~\ref{def-equil} below.
\footnote{\label{ftn-no-b}%
To~put it another way, the agents formulate their private savings problems
under the assumption 
that~$K$ and~$r$ are chosen (and given to them) so that, after all private instances of
\eqref{ze1212} have been solved, all lenders lend what they
perceive as optimal to lend, all borrowers borrow what they perceive as optimal to borrow, all
private budgets are balanced at all times (present and future),
and the market for every security clears.
Any such arrangement removes the need for extraneously imposed limits on borrowing. 
One~may consider the infimum over the private demands so obtained as being the
``borrowing limit,'' but such 
a quantity becomes meaningful only after all individual savings problems have been
solved and does  
not constrain any agent.}
The range of the value function is the interval
$\llbkt -\infty,\infty\rlbkt$, with the value of $(-\infty)$ attained only if, for
the given entering wealth, 
there is no policy that can fund
strictly positive consumption in all possible future aggregate and idiosyncratic states.
By convention, the derivatives of any function will be treated as undefined on any
domain in which 
its value is $(-\infty)$ and the appearance of derivatives implies that the argument
belongs to a 
domain in which the function is finite;
as an example, the appearance of any of the symbols $\partial U(c)$ and $\partial^2 U(c)$ 
implies~$c>0$.

\begin{nit}{A field of optimization problems}\label{1350}%
All relations in \eqref{ze1212} are understood to represent
a field of optimization problems over the range of
employment states $u$ and entering wealth $w$.
Since the agents are distinguished only by their employment and wealth, or,
equivalently, by employment and consumption (see \ref{thm1} below),
the relations in \eqref{ze1212}
represent a field of optimization problems over the population of agents.
In particular, $(c,\q,\qq)\in\R^3$ is a vector field over the collection of agents
and so is $(\q,\qq)\in\R^2$. 
We will see below how this field gives rise to a field of first order conditions over
the range of employment and consumption, i.e., over the population of agents.
Since the variables that make the agents distinct (employment and wealth, or
employment and consumption) vary from one
period to the next, the sequence of systems \eqref{ze1212} obtained for $t=T-1,\ldots,0$
does not give rise to a ``Bellman equation'' that can be attached to a single
(possibly hypothetical) agent.
The paths of the individual optimal states (generated by the individual Bellman
equations) never enter the model developed in this
section and no labeling set for the collection of agents is ever used.\qed 
\end{nit}
\nitskip

Suppose next that all value functions $V_{t+1,\tts y,\tts \T_{t,\tts x}^y(F),\tts v}\phd$
in \eqref{ze1} are strictly concave and in $\C^2$ on the domain in which they are finite,
in which case the objective function in \eqref{ze1} is also strictly concave and in $\C^2$
in the domain where it is finite.
As a result, if $V_{t,\tts x,\tts F,\tts u}(w )$ happens to be finite, then the first order Lagrange, also known as
Karush-Kuhn-Tucker (KKT), conditions attached to the problem \eqref{ze1212} must hold at any 
$(c,\q,\qq)\in \R^3$ that happens to be a solution.
Furthermore, these conditions are also sufficient: if  the
KKT conditions attached to \eqref{ze1212}
hold at some $(c,\q,\qq)\in\R^3$, at which the objective in \eqref{ze1} is finite,
\footnote{{}This is the only situation in which the KKT conditions are meaningful.}
then $(c,\q,\qq)$ is the (unique)
solution to~\eqref{ze1212}. 
The KKT conditions are instrumental in what follows and are introduced next.
The idea again is to use those conditions as a tool for deciphering the channels through which
population distribution gets transported  from one period to the next.
To this end, we now restate the first set of constraints in \eqref{ze2} in the form
(meant to place the resource $w$ in the right side of every constraint) 
{\abovedisplayskip=5pt plus 1pt minus 1pt\belowdisplayskip=5pt plus 1pt minus 1pt\belowdisplayshortskip=3pt plus 0.5pt minus 0.5pt
$$   
W_{y,v}-\bigl(1+r\bigr)\q-\bigl(\rho_ y(K)+ 1-\dd\bigr)\qq -
\ee_{ y}(K)v + c + \q + \qq = w\,,\ \  y\in\XXX\,,\  v \in\EEE\,.
$$}%
The costate variable (Lagrange multiplier) attached to each of these constraints we deliberately cast
in the factor form
{\abovedisplayskip=5pt plus 1pt minus 1pt\belowdisplayskip=5pt plus 1pt minus 1pt\belowdisplayshortskip=3pt plus 0.5pt minus 0.5pt
$$
\l_{y,\tts v}=\Lmp_{y,\tts v} \times {\b }\ts Q(x, y)P_{x, y}(u,v)\,,\ \  y\in\XXX\,,\  v \in\EEE\,,
$$}%
i.e., we will be working with $\Lmp_{y,\tts v}$ instead of the true costate variable $\l_{y,\tts v}$,
and  the costate variable attached to the last condition in \eqref{ze2} we denote by $\ff $.
The Lagrange dual of the optimization problem in \eqref{ze1212} can be stated as
{\abovedisplayskip=5pt plus 1pt minus 1pt\belowdisplayskip=5pt plus 1pt minus 1pt\belowdisplayshortskip=3pt plus 0.5pt minus 0.5pt
$$
\mathop{\text{minimize}}_{\ff,\tts (\Lmp_{y,\tts v})_{ y\tts\in\tts\XXX,\, v\in\EEE}}
\biggl(\,\mathop{\text{maximize}}_{c,\tts \q,\tts \qq,\tts (W_{y,\tts v})_{ y\in\XXX,\tts v\in\EEE}}
\LLL\Bigl(c,\q,\qq,(W_{y,\tts v})_{ y\tts \in\tts \XXX,\, v\tts \in\tts \EEE},
\ff,(\Lmp_{y,\tts v})_{ y\tts \in\tts \XXX,\, v\tts \in\tts \EEE}\Bigr)
\biggr)\,,
$$}%
where
{\abovedisplayskip=5pt plus 1pt minus 1pt\belowdisplayskip=5pt plus 1pt minus 1pt\belowdisplayshortskip=3pt plus 0.5pt minus 0.5pt 
\begin{align*}
\LLL\Bigl(c,\q,&\qq,(W_{y,\tts v})_{ y\tts \in\tts \XXX,\, v\tts \in\tts \EEE},
\ff,(\Lmp_{y,\tts v})_{ y\tts \in\tts \XXX, \, v\tts \in\tts \EEE}\Bigr)\\
&\qquad\df J_t\Bigl(c ,\,(W_{y,\tts v} )_{ y\tts\in\tts\XXX,\, v\tts\in\tts\EEE}\Bigr) 
+ {\ff }\bigl(w-c-\q-\qq\bigr)\\
&\qquad\qquad +
{\b }\sum\nolimits_{ y\ts\in\ts\XXX,\,v\ts\in\ts\EEE}\Lmp_{y,\tts v}\Bigl(w
-W_{y,\tts v}+\bigl(1+r\bigr)\q+\bigl(\rho_y(K)+ 1-\dd\bigr)\qq\\
&\hbox to4.5cm{\hfill}+\ee_{ y}(K)\ts v-c - \q -\qq \Bigr) Q(x, y)P_{x, y}(u,v)\,.
\end{align*}
}%
With the substitution
{\abovedisplayskip=5pt plus 1.5pt minus 1.5pt\belowdisplayskip=5pt plus 1.5pt minus 1.5pt\belowdisplayshortskip=6pt plus 1.5pt minus 1.5pt
\begin{equation}\label{bsde}
\f\df \varphi + {\b}\sum\nolimits_{ y\ts\in\ts\XXX,\,v\ts\in\ts\EEE}\Lmp_{y,\tts v}Q(x, y)P_{x, y}(u,v)\,,
\end{equation}}%
equating to $0$ the derivatives of the Lagrangian
relative to $W_{y,\tts v}$, $c$, $\q$, and $\qq$ gives, respectively,
{\abovedisplayskip=5pt plus 1.5pt minus 1.5pt\belowdisplayskip=7pt plus 1.5pt minus 1.5pt\belowdisplayshortskip=6pt plus 1.5pt minus 1.5pt
\begin{equation}\label{ze4}
\begin{gathered}
\Lmp_{y,\tts v}=\partial V_{t+1,\tts y,\tts \T_{t,\tts x}^y(F),\tts v}\bigl(W_{y,\tts v} \bigr)\,,\quad
\f={\Up}(c )\,,\\
\f=\bigl(1+r\bigr)\b\sum\nolimits_{ y\ts\in\ts\XXX,\,v\ts\in\ts\EEE}\Lmp_{y,\tts v}\,Q(x, y)P_{x, y}(u,v)\,,\\
\f=\b\sum\nolimits_{ y\ts\in\ts\XXX,\,v\ts\in\ts\EEE}\Lmp_{y,\tts v}\,\bigl(\rho_{ y}(K)
+1-\dd\bigr) Q(x, y)P_{x, y}(u,v)\,,
\end{gathered}
\end{equation}}%
and, after a straightforward application of the envelope theorem,
{\abovedisplayskip=5pt plus 1.5pt minus 5pt
\belowdisplayskip=5pt plus 1.5pt minus 1.5pt
\belowdisplayshortskip=6pt plus 1.5pt minus 1.5pt
\begin{equation}\label{ze4a1}
\partial V_{t,\tts x,\tts F,\tts u} (w )=\varphi +{\b}\sum\nolimits_{ y\tts\in\tts\XXX,\,v\tts\in\tts\EEE}
\Lmp_{y,\tts v}\,Q(x, y)P_{x, y}(u,v)=\f={\Up}(c )\,.
\end{equation}}%
If~the system \eqref{ze2}\ts\&\ts\eqref{ze4}  
can be solved for $c$, $q$, $\qq$, $W_{y,v}$, $\f$, and
$\Lmp_{y,v}$, then any such solution must depend on the time period $t$, the aggregate state
$(x,F)$, and the employment state~$u$. If~this dependence must be emphasized in the notation,
we shall embellish the symbols $c$, $q$ and $\qq$ with subscripts in the obvious way
(recall that in the above $K$ and $r$ stand for $K_t(x,F)$ and $r_t(x,F)$).
The entire system \eqref{ze2}\ts\&\ts\eqref{ze4}   depends 
on the future distributions $\T_{t,x}^y(F)$, $y\in\nobreak\XXX$, i.e., depends on the assumed
structure of the transport mappings~$\T_{t,x}^y\phd$, $y\in\nobreak\XXX$~-- not just on
the assumed statistical behavior of the aggregate and idiosyncratic shocks.
The~system depends also on the entering wealth (resource)
$w$, and all these dependencies will be incorporated into the notation when needed
(and suppressed when understood from the context, for the sake of simplicity). 

The next step is to restate the KKT conditions in a more useful form. First, \eqref{ze2}
and \eqref{ze4} reduce to the following system of three equations for the unknowns $c$, $\q$
and~$\qq$, i.e., for $c_{t,\tts x,\tts \csd,\tts u}$, $\q_{t,\tts x,\tts \csd,\tts u}$ and 
$\qq_{t,\tts x,\tts \csd,\tts u}\,$:
{\abovedisplayskip=5pt plus 1.5pt minus 1.5pt\belowdisplayskip=7pt plus 1.5pt minus 1.5pt\belowdisplayshortskip=6pt plus 1.5pt minus 1.5pt
\begin{equation}\label{z2-no-lm}
\begin{gathered}
c +\q +\qq -w =0\,,\\
{\Up}(c ) -  \sum\nolimits_{ y\ts\in\ts\XXX,\,v\ts\in\ts\EEE}
(1+r)\,\b\,\partial V_{t+1,\tts y,\tts \T_{t,\tts x}^y(F),\tts v}\bigl(W_{y,\tts v} \bigr)\\
\hbox to 7cm{\hfill}\times\, Q(x, y)P_{x,y}(u,v)=0\,,\\
{\Up}(c )- \sum\nolimits_{ y\ts\in\ts\XXX,\,v\ts\in\ts\EEE}
\bigl(\rho_y(K)+1-\dd\bigr)\,\b\,
\partial V_{t+1,\tts y,\tts \T_{t,\tts x}^y(F),\tts v}\bigl(W_{y,\tts v} \bigr)\\
\hbox to 7cm{\hfill}\times\, Q(x, y)P_{x,y}(u,v) =0\,,
\end{gathered}
\end{equation}
}%
where the expressions $W_{y,\tts v}$ are as defined in \eqref{ze2}.
Next, observe that with \eqref{ze4a1} applied to period $t+1$ the first equation in \eqref{ze4}
states:
{\abovedisplayskip=5pt plus 1.5pt minus 5pt
\belowdisplayskip=5pt plus 1.5pt minus 1.5pt
\belowdisplayshortskip=3pt plus 1.5pt minus 1.5pt
\[
\Lmp_{y,\tts v}=\partial V_{t+1,\tts y,\tts \T_{t,x}^y(F),\tts v}\bigl(W_{y,\tts v} \bigr)
={\Up}(c_{t+1,\tts y,\tts \T_{t,x}^y(F),\tts v} )\,.
\]}%
It is now easy to remove the costate variables from the KKT conditions
by casting the last two equations in \eqref{ze4} in
the familiar form of ``kernel conditions:''
{\abovedisplayskip=10pt plus 1.5pt minus 2pt\belowdisplayskip=10pt plus 1.5pt minus 1.5pt\belowdisplayshortskip=8pt plus 1.5pt minus 1.5pt
\begin{equation}\label{ze5} 
\begin{gathered}
1=\bigl(1+r\bigr)\,
\b\sum\nolimits_{ y\tts\in\tts\XXX,\,v\ts\in\ts\EEE}{{\Up}(c_{t+1,\tts y,\tts \T_{t,x}^y(F),\tts v} )
\over {\Up}(c_{t,\tts x,\tts F,\tts u} )} Q(x, y)P_{x, y}(u,v)\,,\\
1=\b\sum\nolimits_{ y\tts\in\tts\XXX,\,v\tts\in\tts\EEE}{{\Up}(c_{t+1,\tts y,\tts \T_{t,x}^y(F),\tts v} )
\over {\Up}(c_{t,\tts x,\tts F,\tts u} )}\bigl(\rho_{ y}(K)+1-\dd\bigr) Q(x, y)P_{x, y}(u,v)\,.
\end{gathered} 
\end{equation}}%
The meaning of these conditions is that in equilibrium all agents agree on the returns that
the two traded securities generate (the right sides in \eqref{ze5} must be identical across all agents).
In~all concrete implementation presented later in the \chptr\  the utility function
$U\phd$ is chosen to be isoelastic, in which case
{\abovedisplayskip=8pt plus 1.5pt minus 5pt\belowdisplayskip=8pt plus 1.5pt minus 1.5pt\belowdisplayshortskip=6pt plus 1.5pt minus 1.5pt
$$
{{\Up}(c_{t+1,\tts y,\tts \T_{t,x}^y(F),\tts v} )\over {\Up}(c_{t,\tts x,\tts F,\tts u} )}
={\Up}\Bigl({c_{t+1,\tts y,\tts \T_{t,x}^y(F),\tts v}\over c_{t,\tts x,\tts F,\tts u}}\Bigr)\,.
$$}%
For the sake of simplicity of the notation, we shall assume this form from now on,
and note that, 
simplicity aside, it also makes the model invariant to any re-scaling the consumption
variable.  

\begin{nit}{An useful homeomorphism}%
Clearly,
$\Up\colon\lrbkt0,\infty\rlbkt\mapsto\lrbkt0,\infty\rlbkt$
provides a homeomorphism between
$c_{t,\tts x,\tts F,\tts u} $ and $\f$ and between $c_{t+1,\tts y,\tts \T_{t,x}^y(F),\tts v} $ and
$\Lmp_{y,v}$. In particular, the system \eqref{ze5} does not really exclude the costate
variables, as it retains their homeomorphic copies.
Thus, consumption plays multiple rôles: it~is a household
descriptor, state variable, control parameter, and, up to a homeomorphism, a costate variable.\qed
\end{nit}

All three equations in \eqref{z2-no-lm}~--
recall that $V_{t+1,\tts y,\tts \T_{t,\tts x}^y(F),\tts v}\phd$
is assumed strictly concave and in $\C^2$ wherever it is finite~-- 
define the vector $(c,\q,\qq)$ as an implicit
$\C^1$-function of the entering wealth $w$ and the last two equations in \eqref{z2-no-lm} define
the portfolio vector $(\q,\qq)$ as an implicit $\C^1$-function of the consumption level $c\in\Rpp$. This
last feature is instrumental for what follows, as it allows for the use of consumption as a state
variable (instead of wealth). The next theorem makes these statements precise
(recall that $(c,\q,\qq)\in\R^3$ and $(\q,\qq)\in\R^2$ are understood as vector fields
over the collection of agents~-- see \ref{1350}).

\begin{nit}{Theorem}\label{thm1}%
If  $w\in\R$ is such that $V_{t,x,F,u}(w )$ is finite and
the system \eqref{z2-no-lm} admits a solution $(c,\q,\qq)$,
then \eqref{z2-no-lm} admits a unique solution 
for every entering wealth from some open neighborhood of $w$ and that solution is a
$\C^1$-function  
of the entering wealth with $\partial c>0$.  
Moreover, $V_{t,x,F,u}\phd$ is a $\C^1$-mapping, and hence also a $\C^2$-mapping, with 
$\partial V_{t,x,F,u}\phd > 0$ and $\partial^2 V_{t,x,F,u}\phd<0$ in some
neighborhood of~$w $. 
In~addition,  if the system composed of the
last two equations in \eqref{z2-no-lm} admits a solution $(\q,\qq)$ for some
fixed $c\in\Rpp$, then that system admits a unique solution  for every consumption level
in some open neighborhood of~$c$ and that solution is 
a $\C^1$-function in the neighborhood of~$c$. All results continue to hold if one of
the traded 
assets is removed from the model, i.e., the households invest only in the private lending
instrument, or only in productive capital.\qed
\end{nit}

The proof of \ref{thm1} is given in Appendix~\ref{sec:A2}.
One consequence from this theorem is that the value function is a strictly concave
$\C^2$-function 
on any domain on which it happens to be finite, and this property holds in all time
periods and for all 
realizations of the aggregate and idiosyncratic states. 
The theorem also shows that there is a one-to-one correspondence between optimal consumption
and entering wealth;
specifically, consumption is a strictly increasing $\C^1$-function
of entering wealth, and so, entering wealth is a strictly increasing $\C^1$-function
of consumption.
Since in every period the
households differ only in their entering wealth and state of employment, they can be
distinguished 
just as well by their state of employment and consumption level~-- a~strategy that we
have adopted 
already. We stress that although households are identified as elements of
$\EEE\times\Rpp$, this 
set is not a true labeling set, in the sense that an element $(u,c)\in\EEE\times\Rpp$
identifies a 
collection of households that are indistinguishable as economic agents, rather than a single
physical household. Nevertheless, any vector-valued function of the pair $(u,c)$ can
be understood as a vector field over the collection of households.

Another crucially important consequence from the last theorem is that re\-sol\-ving
the individual sa\-vings problem comes down to assigning a consumption level $c\in\nobreak \Rpp$ to any period
$t<T$ and any realized aggregate and idiosyncratic state in such a way that the system composed of the
last two equations in \eqref{z2-no-lm} admits a solution $(\q(c),\qq(c))\in\R^2$ and this solution is
such that $c+\q(c)+\qq(c)$ exactly matches the entering wealth~$w$. The existence of such an assignment
ensures both: the value function remains finite (consumption is always strictly positive) and the
KKT conditions hold. As the system \eqref{z2-no-lm} depends on the time period $t$, the aggregate
state $(x,F)$, and the private state $u$, then so does also the solution $(\q(c),\qq(c))$. In~most
cases this dependence needs to be emphasized in the notation and
we shall often write $\q_{t,\tts x,\tts F,\tts u}(c)$ and $\qq_{t,\tts x,\tts F,\tts u}(c)$.
Of course, these objects depend
also on the selection of average installed capital, risk-free rate, and population transport mappings. 

The matter to address next is the compatibility of the optimal private
allocations. Let us suppose that, with a given choice for the initial population 
distribution, for the collection of transport mappings, and for the assignment of average 
installed capital and interest to every period and every aggregate state of the economy, all households
are able to solve their private savings problems with a unique (individual)
optimal allocation along every
realized path of the productivity state and the private employment state.
In general, such allocations have no reason to be consistent in that: (a)~exercising all optimal
private policies may generate transport mappings that are different from the given
(and assumed by 
all agents) ones, (b)~the
averages of all private capital investments may differ from the given
(and assumed by all agents) ones, and (c)~the given
interest rates may generate
aggregate demands for borrowing and lending that do not match.
``General equilibrium'' is any
arrangement in which no mismatch of type (a), or (b),
or (c) in any 
period and in any state occurs. The precise definition is given next
(note that, as it stands, this definition  is perfectly meaningful
if the population of agents is finite). 

\begin{nit}{Global general equilibrium}\label{def-equil}%
In the context of the economy introduced above, global general equilibrium, or simply
equilibrium, is given by:
\vskip 3pt plus1.0pt

(1)~an initial population distribution $F_0\in\DB^\EEE$ and a collection of transport mappings
{\abovedisplayskip=5pt plus 4.5pt minus 1.5pt\belowdisplayskip=5pt plus 4.5pt minus 1.5pt\belowdisplayshortskip=3pt plus 4.5pt minus 1.5pt
$$
\T_{t,x}^y\colon \DB^\EEE \mapsto \DB^\EE\,,\quad x,y\in\XXX\,,\ \ 0\le t<T\,;
$$
}%

(2)~a collection of mappings $K_t\colon\XXX\times\DB^\EEE \mapsto\Rpp$ and
$r_t\colon\XXX\times\DB^\EEE \mapsto\lrbkt -1,\infty\,\rlbkt $, $0\le t<T$,
\vskip 2pt plus 0.5pt

\noindent
all chosen so that, for any realized path of the productivity state
$(x_t\in\XXX)_{0\tts\le\tts t\tts\le \tts T}$, the realized population distribution in period~$t$
is exactly $F_0$ if $t=0$ and exactly
{\abovedisplayskip=8pt plus 1.5pt minus 4.5pt\belowdisplayskip=8pt plus 4.5pt minus 1.5pt\belowdisplayshortskip=3pt plus 4.5pt minus 1.5pt
$$
F_t\df \bigl(\T_{t-1,x_{t-1}}^{x_t}\circ \T_{t-2,x_{t-2}}^{x_{t-1}}\circ \cdots\circ \T_{0,x_{0}}^{x_{1}}\bigr)(F_0)
\quad\text{if \ $0<t\le T$}\,,
$$
}%
the realized average installed capital is exactly $K_t(x_t,F_t)$, and the realized
average demand for the risk free private lending instrument is
exactly~$0$ in all periods $0\le \nobreak t<\nobreak T$,
provided that all households choose their savings policies by way of solving for their private KKT
conditions with population distribution $F_t$, with installed capital $K_t(x_t,F_t)$, with interest
$r_t(x_t,F_t)$, and with a collection of transport mappings
$\T_{t,x_t}^y$, $y\in\XXX$, and provided  this choice results
into a strictly positive consummation for every household at all times
and in all aggregate and private states.\qed
\end{nit}

The main step in the calculation of the global general equilibrium is to establish
the connections 
across time between the population distributions
${(\csd_{t})}_{0\tts\le\tts t\tts\le\tts T}$ that all private KKT conditions and
the notion of equilibrium dictate. This is the task we turn to next.

\begin{nit}{State transition assignments}\label{rnd-Monge}%
Households that are in the same state of emp\-loyment, choose the same consumption level, and
experience the same shock in employment, would choose the same consumption level during the next
period, too. Let $\pfc_{t,x,F}^{ y,v}(u,c)$ denote the period $t+1$ consumption level
of any household that happens to be of type $(u,c)$ during period $t$, when the economy is in
state $(x,F)$, provided that during period $t+1$ the household faces
transition to employment state $v\in\EE$ and
the productivity state transitions to $y\in\XXX$.
The assignment
{\abovedisplayskip=8pt plus 1.5pt minus 1.5pt\belowdisplayskip=8pt plus 1.5pt minus 1.5pt\belowdisplayshortskip=5pt plus 1.5pt minus 1.5pt
\begin{equation*}\label{rnd-Monge-a}
\EEE\times\Rpp \ni (u,c) \leadsto (v,\pfc_{t,x,F}^{ y,v}(u,c))\in \EEE\times\Rpp
\end{equation*}
}%
is analogous to the Monge assignment in the classical transportation of mass problem,
except that it is not pure, in that it depends on the
idiosyncratic shock in employment, which differs across the population of households that are of
type $(u,c)$. In~addition to being random in that sense, this assignment takes place in the random environment
determined by the transition in productivity from $x$ to $y$~-- and we stress the dependence on
both $x$ and $y$.
In~what follows the mappings $\pfc_{t,x,F}^{ y,v}(\cdot,\cdot)$
are referred to as state transition assignments, or simply as assignments.
Accordingly, the mappings 
$\pfc_{t,x,F}^{ y,v}(u,\cdot)$ are referred to as conditional assignments, or
employment-specific 
assignments. The rôle  the assignments $\pfc_{t,x,F}^{ y,v}(\cdot,\cdot)$ 
play in the model developed here
may seem similar to that of the familiar flux in fluid mechanics,
but there is also a fundamental difference (the very reason for using the term ``transitions'' instead of
``flux'' or ``flow''),
in  that
$(v,\pfc_{t,x,F}^{ y,v}(u,c))$ is not the next ``location'' of ``particle'' $(u,c)$; rather,
$(v,\pfc_{t,x,F}^{ y,v}(u,c))$ is the type that certain particles of type $(u,c)$
turn into when 
time changes from~$t$ to $t+1$.
Furthermore, knowledge about the transition assignments is not sufficient to restore the
trajectory in the state space followed by a given household. Indeed, even if it is
known that a 
particular household is of type $(u,c)$, one must know the next period employment $v$ of
that same household in
order to identify its next period type as $(v,\pfc_{t,x,F}^{ y,v}(u,c))$.
We~stress that the individual
trajectories followed by the states of the households, whether in the range of wealth or
consumption, never enter the model developed in this \chptr.
While the general strategy in the domain of fluid mechanics
and MFG is to first derive the individual trajectories and then derive the flow of
probabilities 
along those, the strategy adopted here is to go directly to the distribution transfer
and ignore 
the individual paths altogether. Naturally, the state transitions affect the
transport of the population (treated as a probability measure on $\EEE\times\Rpp$)~--
see \ref{main-q} 
below~-- but this transport is very different in nature from the flow of
probabilities along a given family of 
trajectories (recall the Lagrangian formulation of MFG in [\cited{BB00},\cited{BCS17}] and
the relaxed MFG equilibrium in~\cite{CanCap18}).%
\qed
\end{nit}

Basic intuition
suggests that the mappings
$\pfc_{t,x,F}^{ y,v}(u,\cdot)$ must be increasing,
\footnote{{}An agent who consumes at least as much as another agent during the present period will
consume at least as much during the next period as well, if both agents experience the same shock in their
employment status.}
and we shall
seek equilibria in which these mappings are also continuous. 
Consequently, all functions
$\pfc_{t,x,F}^{ y,v}(u,\cdot)$ have inverses given~by
{\abovedisplayskip=5pt plus 1.5pt minus 1.5pt\belowdisplayskip=5pt plus 1.5pt minus 1.5pt\belowdisplayshortskip=3pt plus 1.5pt minus 1.5pt
\begin{equation}\label{inversion}
\Rpp\ni \a \leadsto \hat\pfc_{t,x,F}^{ y,v}(u,\a)\df\inf\{c\in\Rp\colon \pfc_{t,x,F}^{ y,v}(u,c)> \a\}\,.
\end{equation}}%
The postulated features of $\pfc_{t,x,\csd}^{ y,v}(u,\cdot)$ guarantee that
{\abovedisplayskip=5pt plus 1.5pt minus 1.5pt\belowdisplayskip=5pt plus 1.5pt minus 1.5pt\belowdisplayshortskip=3pt plus 1.5pt minus 1.5pt
\[
c\le\hat\TT_{t,x,F}^{ y,v}(u,\a)\quad\text{and}\quad
\pfc_{t,x,F}^{ y,v}(u,c)\le \a
\]
}%
are equivalent relations.
Intuitively, the state transition assignments $\pfc_{t,x,F}^{ y,v}(\cdot,\cdot)$ govern the transport of the
population from period $t$ to $t+1$ and the next proposition makes this feature precise.

\begin{nit}{Proposition}\label{main-q}%
The transport mappings $\T_{t,x}^y\colon\DB^\EEE\mapsto\DB^\EEE$
obtain from the state transition assignments introduced above according to the rule
{\abovedisplayskip=8pt plus 2pt minus 2pt\belowdisplayskip=8pt plus 2pt minus 2pt\belowdisplayshortskip=3pt plus 1.5pt minus 1.5pt
\begin{equation*}\tag{${\text{d}}_t$}\label{zze666}    
\begin{multlined}
\T_{t,x}^y(F)^{v}(\a) = \sum\nolimits_{\,u\ts\in\ts\EEE} 
{\gp_x(u) P_{x, y}(u, v)\over \gp_ y( v)} 
\csd^{u}(\,\hat\pfc_{t,x,F}^{ y,v}(u,\a))\,,\\
\a\in\Rpp\,,\ v\in\EEE\,,\ F\in\DB^\EEE
\end{multlined} 
\end{equation*}}%
for all $t<T$ and all $x,y\in\XXX$.\qed
\end{nit}

Most of what follows in this chapter stands on the last result, which is justified next:
%
%
\begin{nit}{Proof of \ref{main-q}}%
Adopt the terminology, the notation, and the results from \ref{weighing}.
Let $v\in\EEE$ be fixed. Then $\gp_ y( v)\T_{t,x}^y(F)^{v}(\a)$ is the relative (with respect to
the entire population) weight of the collection of agents who happen to be in employment state~$v$
during period $t+1$ and happen to choose consumption level (during that same period) that is not
strictly larger than $\a\in\Rpp$. Consider the collection $B_{u,\tts v}$ of agents who happen to
be in employment state~$u$ during period~$t$ and transition to state $v$ during period~$t+1$.
The period-$(t+1)$ consumption level of an agent from the set $B_{u,\tts v}$ would not exceed~$\a$
only if and only if the period-$t$ consumption level,~$c$, of that same agent is such that 
$\pfc_{t,x,\csd}^{ y,v}(u,c)\le \a$,
which property is the same as $c\le\hat\pfc_{t,x,\csd}^{ y,v}(u,\a)$, i.e., the agent must belong
to the set $E_{u,v}\bigl(\hat\pfc_{t,x,\csd}^{ y,v}(u,\a)\bigr)$, which has relative weight
(against the entire population) of $\gp_x(u) P_{x, y}(u, v)\csd^{u}(\,\hat\pfc_{t,x,F}^{
y,v}(u,\a))$.
Observing that  the finite union of disjoint set
$\cup_{u\tts\in\tts\EEE}E_{u,v}\bigl(\hat\pfc_{t,x,\csd}^{ y,v}(u,\a)\bigr)$
is nothing but the collection of agents who happen to be in employment state $v$
during period $t+1$ and choose consumption level that is not strictly larger than $\a\in\Rpp$ completes
the proof.\qed
\end{nit}

\begin{nit}{Remark}\label{n2}%
The~rôle that equation (\ref{zze666}) plays in the present study is analogous to the rôle
of the master  
equation in MFG, or the rôle
of the Kolmogorov forward equation in the classical approach to
heterogeneous models (see \eqref{lmb-iter}, for example).
However, its
structure and intrinsic nature differ from either of these two tech\-ni\-ques in at least
these aspects:
The first one is that the transport is driven by the state transition assignments,
which, though similar to a flux, are not really a flux.
The structure of these assignments 
is deri\-ved  
below and reflects the time-interlaced structure noted earlier.
As we are about to see in \ref{self-dep}, the
assignments in the right side 
of \eqref{zze666} must depend on the left side (the transport is self-consistent).
In~addi\-tion, the transport
encoded into \eqref{zze666} acts in the random environment of the transition in the
productivity 
state, not just in the random environment of the productivity state alone.
Indeed, the right side of \eqref{zze666} depends on both the present and future
states $x$ and 
$y$~-- not on $x$ alone, or on $y$ alone. 
Moreover, the very rule that
transports the distribution~$\csd$ depends on~$\csd$ (the symbol $\csd$ appears twice
in the right 
side). 
This feature may appear reminiscent to a Kolmogorov forward
equation with coefficients that depend (explicitly) on the
distribution that the equation drives, but we stress that, notwithstanding the
self-consistency of 
the transport, the dependence on the second appearance of $\csd$ in the right
side of \eqref{zze666} is only implicit and comes from solving (simultaneously) all
private KKT 
conditions, together with the collective market clearing requirement.
Most important, (\ref{zze666}) is only meaningful in conjunction
with the other first order and market clearing conditions,
which, in turn, are only meaningful in conjunction with \eqref{zze666}~--
see \ref{cross-sys} below. This simultaneity is unavoidable and is
the main difficulty to overcome.\qed
\end{nit}


In order to address the time-interlaced structure noted earlier in a more
practical fashion, 
we must find a way to mimic the approach proposed in \cite{DL12}. The idea is to
break the large  
system across all periods and all aggregate and idiosyncratic states into smaller ones
that can be chained into a computable backward induction program. 

\begin{nit}{The local time-interlaced master system}\label{cross-sys}%
Given any period $0\le t<T$ and any aggregate state $(x,\csd)\in\XXX\times\DB^\EEE$
and idiosyncratic (employment) state $u\in\EEE$ associated with that period,
define the system (parameterized by $x$, $\csd$ and $u$): 
{\abovedisplayskip=8pt plus 1.5pt minus 1.5pt\belowdisplayskip=8pt plus 1.5pt minus 1.5pt\belowdisplayshortskip=8pt plus 1.5pt minus 1.5pt
\begin{equation*}\tag{$\text{n}_t$}\label{zze5a} 
\begin{gathered}
1=\bigl(1+r_{t}(x,\csd)\bigr)
\b\sum\nolimits_{y\ts\in\ts\XXX,\,v\ts\in\ts\EEE}{{\Up}\bigl(\pfc_{t,x,\csd}^{y,v}(u,c)/c\bigr)}
\,Q(x, y)P_{x, y}(u,v)\,,\\
1=\b\sum\nolimits_{ y\ts\in\ts\XXX,\,v\ts\in\ts\EEE}{{\Up}\bigl(\pfc_{t,x,\csd}^{y,v}(u,c)/c\bigr)}
\bigl(\rho_{ y}(K_{t}(x,\csd))+1-\dd\bigr) \\
\hbox to 6cm{\hfill}\times\,Q(x, y)P_{x, y}(u, v)\,,
\end{gathered}
\end{equation*}
}%
{\abovedisplayskip=8pt plus 1.5pt minus 1.5pt\belowdisplayskip=8pt plus 1.5pt minus 1.5pt\belowdisplayshortskip=8pt plus 1.5pt minus 1.5pt
\begin{equation*}\tag{$\text{e}_{t+1}$}\label{zze5xb} 
\begin{gathered}
\begin{multlined}
(1+r_{t}(x,\csd)){\q_{t,x,\csd,u}(c)} + 
\bigl(\rho_ y(K_{t}(x,\csd))+1-\dd\bigr){\qq_{t,x,\csd,u}(c)}\qquad\qquad\\
+\ee_ y(K_{t}(x,\csd))v
={\pfc_{t,x,\csd}^{ y,v}(u,c)}
+ {\q_{t+1, y,\T_{t,x}^y(\csd),v}\bigl(\pfc_{t,x,\csd}^{ y,v}(u,c)\bigr)}\\
\hbox to 1.5cm{\hfill}+ {\qq_{t+1, y,\T_{t,x}^y(\csd),v}\bigl(\pfc_{t,x,\csd}^{ y,v}(u,c)\bigr)}\,,
\quad \text{for all }\   y\in\XXX\,,\ v \in\EEE\,,
\end{multlined} 
\end{gathered}
\end{equation*}
}%
{\abovedisplayskip=8pt plus 1.5pt minus 1.5pt\belowdisplayskip=8pt plus 1.5pt minus 1.5pt\belowdisplayshortskip=8pt plus 1.5pt minus 1.5pt
\begin{equation*}\tag{$\text{m}_{t}$}\label{ze9}
\begin{gathered}
\sum\nolimits_{u\ts\in\ts\EEE}\gp_x(u)\int_0^\infty{\q_{t,x,\csd,u}(c)}\d \csd^u(c)=0\,,\\
\sum\nolimits_{u\ts\in\ts\EEE}\gp_x(u)\int_0^\infty{\qq_{t,x,\csd,u}(c)}\d \csd^u(c) =
K_{t}(x,\csd)\,,
\end{gathered}
\end{equation*}
}%
in which \eqref{zze5a}  and \eqref{zze5xb} are understood as identities between functions of
$c\in\Rpp$ and the distributions $\T_{t,x}^y(\csd)\in\DB^\EEE$, $y\in\XXX$,
are given by (a replica from \ref{main-q})
{\abovedisplayskip=8pt plus 1.5pt minus 1.5pt\belowdisplayskip=8pt plus 1.5pt minus 1.5pt\belowdisplayshortskip=8pt plus 1.5pt minus 1.5pt
\begin{equation*}\label{zze667}\tag{${\text{d}}_t$}
\begin{multlined}
\T_{t,x}^y(F)^{v}(\a)\\
= \sum\nolimits_{\,u\ts\in\ts\EEE} 
{\gp_x(u) P_{x, y}(u, v)\over \gp_ y( v)} 
\csd^{u}(\,\hat\pfc_{t,x,F}^{ y,v}(u,\a))\,,\ \ \a\in\Rpp\,,\ v\in\EEE\,.
\end{multlined} 
\end{equation*}
}%
Let $\M_t$, for $0\le t<T$, stand for the collection of
equations $\bigl\{\text{\eqref{zze5a}, \eqref{zze5xb}, \eqref{ze9}, \eqref{zze667}}\bigr\}$.
We call $\M_t$ the local time-interlaced master system,
or simply the local master system.
Solving for the general equilibrium comes down to solving the global master system
$\{\M_t\colon 0\le t<T\}\cup\text{\eqref{zze5ab}}$, where \eqref{zze5ab} is the collection of all
period~$t=0$ balanced budget conditions, which is described next.
Since at time $t=0$ all households share the same entering wealth, 
households that  happen to be in employment state $u\in\EEE$ are identical and thus choose the same
consumption level $\bar c_{u}$.  In~particular, the period $t=0$ population distribution
$\csd_0\in\DB^\EEE$ has the form $(\csd_0)^u(c)=0$ for $c< \bar c_{u}$ and
$(\csd_0)^u(c)=1$ for $c\ge \bar c_{u}$. Thus, there are $\abs{\EEE}$ balanced budget
equations attached to period $t=0$, namely, 
{\abovedisplayskip=5pt plus 1.5pt minus 1.5pt\belowdisplayskip=5pt plus 1.5pt minus 1.5pt\belowdisplayshortskip=3pt plus 1.5pt minus 1.5pt
\begin{equation*}\tag{$\text{e}_0$}\label{zze5ab}
\begin{multlined}
\bar c_{u} + {\q_{0,x,\csd_0,u}(\bar c_{u})} + {\qq_{0,x,\csd_0,u}(\bar c_{u})}
= \bigl(\rho_x(K_{-1}) + 1-\dd\bigr)K_{-1} + \ee_x(K_{-1})u\\
\text{for all }\ u\in\EEE\,,
\end{multlined}
\end{equation*}
}%
which are to be solved for the (same number of) unknowns $\bar c_u$, $u\in\EEE$.

The reason for organizing all conditions that define the equilibrium in such a way that the
master system $\M_t$ includes some equations associated with period $t$ and other
equations associated 
with period $t+1$ is to make it possible to seek a solution by solving, sequentially, the systems
$\M_{T-1}$, $\ldots$, $\M_0$, \eqref{zze5ab}. This process requires organizing and
connecting accordingly 
the unknowns that are being solved for, while keeping in mind that $\M_t$ is a system
parameterized 
by $x\in\XXX$, $F\in\DB^\EEE$, and $u\in\EEE$. 
To be a bit more precise, when solving  $\M_t$ the collection of functions 
$\q_{t+1, y,\tilde\csd,v}\phd$ and
$\qq_{t+1,y,\tilde\csd,v}\phd$, for all possible choices of $y\in\XXX$,
$\tilde\csd\in\DB^\EEE$ and $v\in\EEE$, are 
assumed given, while the unknowns are the functions (since $\M_t$ is parameterized
by~$x$, $F$, 
and~$u$, so are also the unknowns)
{\abovedisplayskip=5pt plus 1.5pt minus 1.5pt\belowdisplayskip=5pt plus 1.5pt minus 1.5pt\belowdisplayshortskip=3pt plus 1.5pt minus 1.5pt
$$
\q_{t,x,\csd,u}\phd\,, \ \ \qq_{t,x,\csd,u}\phd\ \ \text{and}\ \   
\pfc_{t,x,\csd}^{y,v}(u,\cdot) \quad \text{for all } y\in\XXX\,, \  v\in\EEE\,.
$$
}%
We stress that $\q_{t,x,\csd,u}\phd$ and  $\qq_{t,x,\csd,u}\phd$ map period-$t$
consumption into 
period-$t$ asset holdings, while $\pfc_{t,x,\csd}^{y,v}(u,\cdot)$ map period-$t$
consumption into 
period $t+1$ consumption, i.e., some of the unknowns are associated with  period $t$,
while other 
unknowns are associated with period $t+1$.
This feature, together with the fact that $\M_t$ involves equations associated
with two consecutive time periods, is compressed into the term
``time-interlaced.''\qed
\end{nit}

\begin{nit}{Time-interlaced vs.\ forward-backward systems and MFG}\label{iterlaced-rem}
The main reason for introducing the 
term ``time-interlaced'' is to avoid
using the term ``forward-backward,'' which may be confusing,
as this latter term is already taken and commonly associated with 
a coupled pair of one forward and one backward equation
(as in FBSDE, for example).
It~is important to note that in the present setup splitting the global system
$\{\M_{T-1},\,\ldots,\,\M_0\}$ into two recursive programs, one moving forward and
one moving backward, does not appear to be possible.
In what follows
we shall not seek such an arrangement as a way to solve the global system.
Instead, a new recursive program will be developed from scratch and organized in such
a way that  
it moves in only one direction and is thus computable~-- see \ref{main-proc} below.
The main challenge is that while moving in only one direction, the program must still
incorporate the time-interlaced structure of the systems $\M_t$, $0\le t <T$.
The algorithm so obtained is one of the main innovations in the \chptr.
 
One must also note that, unlike in the standard MFG setup,
here the coupling function, i.e., the interaction between the private choices and the
cross-sectional distribution of the population (i.e., the ``mean field''),
is only implicit and is also endogenized. Indeed, the dependence of the unknowns
$\q_{t,x,\csd,u}\phd$, $\qq_{t,x,\csd,u}\phd$ and  
$\pfc_{t,x,\csd}^{y,v}(u,\cdot)$ on the distribution $\csd$ comes only through
solving the entire master system $\M_t$, which is affected by $\csd$ only through the
market clearing condition \eqref{ze9}, some components of which depend on $\csd$ only
through solving the entire system $\M_t$.
This implicit structure presents a substantial computational challenge, and is one of
the key differences between the approach developed in this \chptr\ and the one based
on mean field theory, which requires the coupling function to be fixed in the
outset.\qed%
\end{nit}

\begin{nit}{On the notion of equilibrium}\label{gei-rem}
It is important to recognize that the local master system $\M_t$ obtains from the
notion of general equilibrium (see \ref{def-equil}),
not from the notion of Nash
equilibrium. It has been known at least since the seminal work of Auman \cite{Aum64}
that the  notion  
of Nash-equilibrium is not meaningful in the context of general equilibrium
incomplete market 
models with finite number of agents. This is because in such models no agent can
change their 
asset allocation without affecting the prices (and hence the agent's own optimization
problem) and without forcing other agents to change their allocations as
well.
\footnote{As was already noted, the definition of general equilibrium
in \ref{def-equil} is perfectly meaningful with any, finite or infinite, collection of
agents.} 
The (classical, by now) remedy
proposed in 
ibid.\ is to introduce a continuum of agents, in which case all agents can be
considered negligible. 
This approach still leaves open the question of how infinitely many negligible capital
allocations 
aggregate into a quantity that is both non-negligible and finite, or how infinitely
negligible agents are distributed over the space of private states. 
In this section the
passage  to the limit as the number of agents increases to $\infty$ was taken solely
for the purpose of computational simplicity~-- not because the model would be
meaningless otherwise. Specifically, the transport equation becomes much more
involved if the number of agents is finite, since one can no longer resort on the
Glivenko-Cantelli's theorem when transcribing the transformations of the relative
weights of the employment
categories across time.
To wit, if $10$ agents flip a coin independently, for any $n\in\{0,\ldots,10\}$
there would be a nontrivial probability
that the percentage of agents who get $H$ is
$n\times 10\%$, but if
$10^{100}$ agents flip a coin the percentage of those who get~$H$ could be assumed to be
$50\%$ with probability~1. This is the only aspect of the model where the
stipulation that the agents are ``infinitely many'' happens to be
relevant and, indeed, very useful.\qed   
\end{nit}

The recursive program for solving the  system
$\{\M_t\colon 0\le t<T\}\cup\text{\eqref{zze5ab}}$
(see \ref{cross-sys}) now suggests itself. 
At every step (associated with period $t$) the program must compute the demand
functions
{\abovedisplayskip=5pt plus 1.5pt minus 1.5pt\belowdisplayskip=5pt plus 1.5pt minus 1.5pt\belowdisplayshortskip=3pt plus 1.5pt minus 1.5pt
$$  
\Rpp\ni c \leadsto \q_{t,x,\csd,u}(c)\,,\qq_{t,x,\csd,u}(c)\in\R
\quad \text{for all }\ x\in\XXX\,,\ \csd\in\DB^\EEE\,,\ u\in\EEE\,,
$$
}%
and the transition assignments
{\abovedisplayskip=5pt plus 1.5pt minus 1.5pt\belowdisplayskip=5pt plus 1.5pt minus 1.5pt\belowdisplayshortskip=3pt plus 1.5pt minus 1.5pt
$$
\EEE\times\Rpp\ni (u,c) \leadsto \pfc_{t, x, \csd}^{y, v}(u,c)\in\Rpp
\quad \text{for all }\ x,y\in\XXX\,,\ \csd\in\DB^\EEE\,,\ v\in\EEE\,,
$$
}%
while taking the demand functions
{\abovedisplayskip=5pt plus 1.5pt minus 1.5pt\belowdisplayskip=5pt plus 1.5pt minus 1.5pt\belowdisplayshortskip=3pt plus 1.5pt minus 1.5pt
$$
\Rpp\ni \tilde c \leadsto \q_{t+1,\tts y,\tts \tilde\csd,\tts v}(\tilde c)\,,\ \qq_{t+1,\tts y,\tts \tilde\csd,\tts v}(\tilde c)\,,
\quad \text{for all }\ y\in\XXX\,,\ \tilde\csd\in\DB^\EEE\,,\ v\in\EEE\,,
$$
}%
as given, i.e., already computed during the previous step (associated with period $t+1$) if $t<T-1$
or taken to be $0$ if $t=T-1$.
What complicates this plan is the following salient feature.

\begin{nit}{Self-consistent transport}\label{self-dep}%
Since the transition assignments $\pfc_{t, x, \csd}^{y, v}(\cdot,\cdot)$ must
obey \eqref{zze5xb}, 
they must depend (through the period-$(t+\nobreak 1)$ portfolios in the right side)
not only on the \hbox{period-$t$} distribution $\csd$,
but also on its transport, $\T_{t,x}^y(\csd)$, to period~$t+\nobreak 1$. In
particular, the right side of the 
transport equation \eqref{zze667} depends on the left side; that is to say,
the mechanism that transports the population 
(i.e., transfers its distribution)
from one period to the next must depend on the result from
the transport (and the identity in \eqref{zze667} must hold).
This feature is very similar to the property ``self-consistent'' as is commonly used
in reference to a
mean field, but notice that in the present setting it implies consistency across
time, which affects the solution method (see below).\qed
\end{nit}

The general program for constructing a general equilibrium is the following.

\begin{nit}{Time-interlaced backward induction}\label{main-proc}%
{\it Initial Backward Step:} Set $t=T-1$ and for every $x\in\XXX$ do:

{\parskip=0pt\nobreak
For every choice of the distribution (state variable)
$\csd\in\DB^\EEE$ do:

{ (1)}~Make an ansatz choice for the values $K_{t}(x,\csd)$ and $r_{t}(x,\csd)$. Go to (2).

{ (2)}~For every 
$(u,c)\in\EEE\times\Rpp$ solve (\ref{zze5a}-\ref{zze5xb}) with
$\q_{t+1, y, \T_{t,x}^y(\csd), v}\equiv 0$ and  
$\qq_{t+1, y, \T_{t,x}^y(\csd), v}\equiv\nobreak 0$
(total of $\abs{\XXX}\times\abs{\EEE}+2$ equations) for the (same number of)
unknowns:
{\abovedisplayskip=5pt plus 1.5pt minus 1.5pt\belowdisplayskip=5pt plus 1.5pt minus 1.5pt\belowdisplayshortskip=3pt plus 1.5pt minus 1.5pt
$$
\{\pfc_{t, x, \csd}^{ y,v}(u,c)\colon  y\in\XXX,\,v \in\EEE\}\,,\quad
\q_{t, x, \csd, u}(c)\,, \quad 
\qq_{t, x, \csd, u}(c)\,.
$$
}%

{ (3)}~Test the market clearing conditions (\ref{ze9}).
If at least one of these conditions fails by more than some prescribed threshold,
go back to (2) with appropriately revised values for $K_{t}(x,\csd)$ and $r_{t}(x,\csd)$; otherwise stop
and record (i.e., accept) the most recently computed scalars $K_{t}(x,\csd)$ and $r_{t}(x,\csd)$ and functions
$\q_{t,\tts x,\tts \csd,\tts u}(\cdot)$, $\qq_{t,\tts x,\tts \csd,\tts u}(\cdot)$ and
$\pfc_{t,\tts x,\tts \csd}^{ y,v}(u,\cdot)$, for all $y\in\XXX$ and $v\in\EEE$. Proceed to
the next step.
}

{\it Generic Backward Step:}  If $t-1<0$, go to the final backward step below; else set $t=t-1$ and
for every $x\in\XXX$ do:

For every choice of the distribution (state variable) $\csd\in\DB^\EEE$ do:

{(1)}~Set $\csdp =\csd$ for every $y\in\XXX$
(the next period distribution, i.e., state variable, is initially guessed to be the
same as the one in the present period, irrespective of the realized future productivity state~$y$).

{ (2)}~Set $K_{t}(x,\csd)=K_{t+1}(x,\csd)$ and $r_{t}(x,\csd)=r_{t+1}(x,\csd)$ (initial guess taken
from the previous iteration).

{ (3)}~For every fixed
$(u,c)\in\EEE\times\Rpp$ solve (\ref{zze5a}-\ref{zze5xb})
with $\T_{t,x}^y(\csd)$ replaced by $\csdp$ (total of $\abs{\XXX}\times\abs{\EEE}+2$
equations) 
for the (same number of) unknowns
$$
\{\pfc_{t,x,\csd}^{ y,v}(u,c)\colon  y\in\XXX,\,v \in\EEE\}\,,\ %
\q_{t,x,\csd,u}(c)\,\ \text{and}\ \qq_{t,x,\csd,u}(c)\,. 
$$

Go to (4).

{ (4)}~Test the market clearing conditions (see (\ref{ze9}))
{\abovedisplayskip=7pt plus 1.5pt minus 1.5pt\belowdisplayskip=7pt plus 1.5pt minus 1.5pt\belowdisplayshortskip=8pt plus 1.5pt minus 1.5pt
\begin{equation*}
\begin{gathered}
\sum\nolimits_{u\ts\in\ts\EEE}\gp_x(u)\int_0^\infty{\q_{t,x,\csd,u}(c)}\d \csd^u(c)=0\\
\text{and}\quad 
\sum\nolimits_{u\ts\in\ts\EEE}\gp_x(u)\int_0^\infty{\qq_{t,x,\csd,u}(c)}\d \csd^u(c)
= K_{t}(x,\csd)\,.
\end{gathered}
\end{equation*}}%
If at least one of these conditions fails by more than some prescribed threshold, go back to (3)
with appropriately revised values for $K_{t}(x,\csd)$ and $r_{t}(x,\csd)$; otherwise, proceed to (5).

{ (5)}~With the computed functions $\pfc_{t,x,\csd}^{ y,v}(u,\cdot)$,
which now depend on the choice of $\csdp$, $y\in\XXX$, compute the distributions
$\csds\in\DB^\EEE$, $y\in\XXX$, as (see \eqref{zze667})
{\abovedisplayskip=5pt plus 1.5pt minus 1.5pt\belowdisplayskip=5pt plus 1.5pt minus 1.5pt\belowdisplayshortskip=8pt plus 1.5pt minus 1.5pt
\[
(\pst\csd_{y})^v(\a) = \sum\nolimits_{\,v\ts\in\ts\EEE} 
{\gp_x(u) P_{x, y}(u, v)\over \gp_ y( v)} 
\csd^{u}(\hat\pfc_{t,x,\csd}^{ y,v}(u,\a))\,,\quad
\a\in\Rpp\,,\ \ v \in\EEE\,.
\]}%
If the largest Kolmogorov-Smirnov distance between $(\pst\csd_{y})^v\phd$ and 
$(\pdg\csd_{y})^v \phd$, for the various choices of $y\in\XXX$ and $v \in\EEE$,
is not acceptably close to~$0$, set $\csdp=\csds$ and go back
to (3) without changing $K_{t}(x,\csd)$ and $r_{t}(x,\csd)$; otherwise, stop
and record (i.e., accept) the most recently obtained scalars $K_{t}(x,\csd)$ and $r_{t}(x,\csd)$ and functions
$\q_{t,x,\csd,u}\phd$, $\qq_{t,x,\csd,u}\phd$ and
$\pfc_{t,x,\csd}^{ y,v}(u,\cdot)$, $y\in\XXX$, $v\in\EEE$. Go~to the beginning of the generic
backward step.

{\it Final Backward Step:}~For every $x\in\XXX$ do:

For every $u\in\EEE$ determine the period $t=0$
consumption level $\bar c_u$ for all households in employment state~$u$ 
(all households in employment category $u$ are identical in period $t=0$ and consume the same
amount) by solving the following system of $\abs{\EEE}$ equations (see (\ref{zze5ab}))
{\abovedisplayskip=7pt plus 1.5pt minus 1.5pt\belowdisplayskip=7pt plus 1.5pt minus 1.5pt\belowdisplayshortskip=8pt plus 1.5pt minus 1.5pt
\[
\bar c_u + {\q_{0,x,u,{\csd_{0}}}(\bar c_u)} + {\qq_{0,x,u,{\csd_{0}}}(\bar c_u)} = \bigl(\rho_x(K_{-1})
+1-\dd\bigr)K_{-1} + u\,\ee_x(K_{-1})\,,\quad u\in\EEE\,,
\]}%
in which $\csd_{0}\in\DB^\EEE$
is given~by $(\csd_{0})^u(c)=1$ if $c\ge \bar c_u$ and $(\csd_{0})^u(c)=0$ if
$c< \bar c_u$, \ $u\in\EEE\,$.

{\it Initial Forward Step:}~In period $t=0$ the initial productivity state $x\in\XXX$ is revealed and so
is also the (idiosyncratic) employment state of every household.
As all households in employment category $u\in\EEE$ have the same income in period $t=0$ and are faced
with the same uncertain future, they are identical and adopt the same
consumption plan $\bar c_u$, calculated during the final backward step.
Define the period $t=0$ population distribution
$\csd_{0}\in\DB^\EEE$ as the corresponding list of Heaviside step functions ($\abs{\EEE}$ in number).
As all quantities
$K_{0}(x,\csd)$,  $r_{0}(x,\csd)$, $\q_{0,x,\csd,u}(c)$ and $\qq_{0,x,\csd,u}(c)$ 
have been precomputed for every $\csd\in\DB^\EEE$ and $c\in\Rpp$, the period $t=0$ average productive
capital $K_{0}(x,\csd_{0})$ is available, and so is also the period $t=0$ exiting portfolio,
$\{\q_{0,x,\csd_{0},u}(\bar c_u),\qq_{0,x,\csd_{0,x},u}(\bar c_u)\}$, for all
households in employment state $u\in\EEE$.

{\it Generic Forward Step:} The economy exits period $(t-1)$ from productivity state $x\in\XXX$
with 
population distribution $\csd_{t-1}$ and in period $t$ enters a new
productivity state $ y\in\XXX$.
As~all $(t-\nobreak1)$-to-$t$  transition assignments
$(u,c) \leadsto \pfc_{t-1,x,\tilde\csd}^{ y,v}(u,c)$ are available from the backward steps for all
\smash{$\tilde\csd\in\DB^\EEE$}, the period $t$ consumption levels of all
households become known:
a period $(t-1)$ household of type $(u,c)\in\EEE\times\Rpp$ 
that changes employment from $u$ to $v$ becomes, during period~$t$, household of type 
$(v,\tilde c)$ with $\tilde c = \pfc_{t-1,x,\csd_{t-1}}^{y,v}(u,c)$.
The period $t$  population distribution is then given~by
{\abovedisplayskip=4pt plus 1.0pt minus 1.5pt\belowdisplayskip=4pt plus 1.0pt minus 1.5pt\belowdisplayshortskip=4pt plus 1.0pt minus 1.5pt
\begin{equation*}
\begin{multlined} 
(\csd_{t})^v(\a) = \sum\nolimits_{\,u\,\in\,\EEE} 
{\gp_x(u) P_{x, y}(u, v)\over \gp_ y( v)}\,
(\csd_{t-1})^u(\hat\pfc_{t-1,x,\csd_{t-1}}^{ y,v}(u,\a))\,,\\
\text{for all }\  
\a\in\Rpp\,,\ \ v \in\EEE\,.
\end{multlined} 
\end{equation*}}%
As all quantities $K_{t}(y,\csd)$, $r_{t}(y,\csd)$, $\q_{t, y,\csd,v}(\tilde c)$
and $\qq_{t, y,\csd,v}(\tilde c)$, assumed to be $0$ if $t=T$,
have been precomputed during the backward steps for every period\hbox{-$t$}
population distribution  $\csd\in\DB^\EEE$ and all individual consumption levels $\tilde c$,
they are meaningful with $\csd=\csd_{t}$. The period~$t$ average installed
productive capital is  $K_{t}(y,\csd_{t})$, the agreed upon interest is $r_{t}(y,\csd_{t})$,
and the period-$t$ exiting portfolio of any household of type $(v,\tilde c)$ is
$\{\q_{t, y,\csd_{t},v}(\tilde c),\,\qq_{t, y,\csd_{t},v}(\tilde c)\}$.\qed
\end{nit}

\begin{nit}{Endless loops warning and disclaimer}\label{endless-loop}%
There are no theoretical results to guarantee that the iterations between steps (3) and (4) and (3)
and (5) converge, or to guarantee that step~(3) in the generic backward step is always feasible,
in that a numerical solution to the system exists generically.\qed
\end{nit}

\begin{nit}{Parallel with self-consistent mean fields}\label{MFG-rem}%
The iterations between steps (2) and (5) in the generic backward step are meant
to ensure that, in every possible realization of the future productivity state,
the result from the transport of the population distribution coincides with the
one assumed by the transporting mechanism~-- see \ref{self-dep} above.
Indeed, before being solved for during stage (3)
in the generic backward step, the system (\ref{zze5a}-\ref{zze5xb}) is made to
depend on the guessed period-$(t+1)$
distributions $\csdp$, $ y\in\XXX$, which guesses are being iterated until they become
consistent with the structure of (\ref{zze667}).
These iterations are very similar in nature
to the way in which mean fields are iterated until they become self-consistent,
though the objects that are iterated and the connections among them are different. 
We again stress that the transport 
equation (\ref{zze667}) is meaningful only in conjunction with the associated budget,
kernel, and market clearing conditions~-- not as a stand-alone equation. 
Note also that this
adjustment (coordination) is local in time, in that the program does not
move to the next period going backward (which is the previous period in real time)
until the correct transport from the current period is established~-- recall that the transport is
time dependent and may become time invariant only in the limit.\qed
\end{nit}

The metaprogram described in \ref{main-proc} differs from other similar procedures in
a number of 
key aspects.  
Some were already noted in \ref{self-dep} and \ref{MFG-rem} above.
Another key aspect is that all backward steps involve the simultaneous computation
of future consumption and present demand, i.e., at every iteration the programs
solves for variables attached to two different periods from equations attached to
different periods as well, whence
the qualifier ``time-interlaced.''
Generally, such a program would be difficult to implement in concrete
models
mainly due to the lack of an adequate computing technology for representing
general (nonlinear) functions on the space of distributions.
This is  a common problem in all heterogeneous agent models, since
the population distribution is inevitably a state variable.
Nevertheless, there are important special cases where the program outlined
in \ref{main-proc} can still be 
carried out. One is the absence of aggregate shocks, in which case a stationary
distribution of the 
population is available and it becomes possible to organize the program so that
it keeps track of just one distribution~-- see Sec.~\ref{sec:IOU} below.
Another one is the possibility to
approximate the population distribution with the vector of its conditional mean
values specific to the various 
employment groups, or even just with the unconditional mean value across the entire
population, which then removes the impossibility of having to work
on an infinite dimensional
state space~-- see Sec.~\ref{sec:KS} below.

\section{Models With Infinite Time Horizon and No Aggregate Risk}
\label{sec:IOU}\setcounter{paragraph}{0}
\def\inc{{A}}

\noindent 
In this section we revisit the benchmark Huggett economy borrowed from \cite{LjunSar00}  and
already reviewed in Sec.~\ref{sec:Intro}.
The enormous simplification that comes
from the removal of the
production function and the common shocks is that time-invariant distribution of the
population 
and time-invariant value for the interest rate become available. In~the search for these two
objects, we now
restate the global system
{\abovedisplayskip=5pt plus 1.5pt minus 1.5pt\belowdisplayskip=3.5pt plus 1.0pt minus 1.0pt\belowdisplayshortskip=5pt plus 1.5pt minus 1.5pt
\[
\{\M_t\colon 0\le t\nobreak <T\}\cup\text{\eqref{zze5ab}}
\]}%
(see \ref{cross-sys}) 
with all wages $\ee_y(K_{t}(x,F))$ replaced by a
fixed value $\ee$, 
with all quantities $\qq$ set to~$0$ (no capital investment takes place),
with the second equations in \eqref{zze5a} and \eqref{ze9} removed, with all instances of $x$ and $y$
as sub/super-scripts removed, and with all transition probabilities $Q(x,y)$ set to 1.
The transition probability matrix~$P$, which governs the idiosyncratic
transitions in every individual employment state, admits a unique list of steady state
probabilities $\gp=(\gp(u)>0,\,u\in\nobreak\EEE)$, which we treat as a vector-row with 
$\gp\,P=\gp$ and with $\sum_{u\ts\in\ts\EEE}\gp(u)=1$. Assuming that all independent private Markov
chains have reached steady-state, the
average income in the cross-section of the population is fixed at
{\abovedisplayskip=5pt plus 1.5pt minus 1.5pt\belowdisplayskip=7pt plus 1.5pt minus 1.5pt\belowdisplayshortskip=5pt plus 1.5pt minus 1.5pt
\[
\inc=\ee\sum\nolimits_{u\ts\in\ts\EEE}{u} \, \gp(u)\,.%
\]}%
As we seek time-invariant equilibrium, we drop the subscript ``$t$'' throughout.
For technical reasons, instead of seeking the equilibrium interest $r$, we seek
the spot price, $B=\inc/(1+r)$, of a risk-free bond with face value $\inc$.
The relative risk aversion for all agents is the constant $R\ge1$.

\begin{nit}{Time-invariant (recursive) equilibrium}\label{time-inv}%
It consists of:

{(1)}~a fixed scalar $B\in\R$;

{(2)}~a collection of continuous and non-decreasing functions
$\q_u\colon\Rpp\to\R$, $u\in\EEE$;

{(3)}~a collection of continuous and non-decreasing functions
$\pfc^v(u,\cdot)\colon\Rpp\to\Rpp$, $u,v\in\EEE$ with inverses $\hat\pfc^v(u,\cdot)$;

{(4)}~a collection of cumulative distribution functions $\csd^u\in\bbF$, $u\in\EEE$,

\noindent
all chosen so that the following four conditions (kernel, balanced budget, market clearing, and
self-consistent transport)
are satisfied:
{\abovedisplayskip=5pt plus 1.5pt minus 1.5pt\belowdisplayskip=5pt plus 1.5pt minus 1.5pt\belowdisplayshortskip=5pt plus 1.5pt minus 1.5pt
\begin{gather}%
B=\b\,\sum\nolimits_{\,v\ts\in\ts\EEE} {\inc}\, \Bigl({c\over \pfc^v(u,c)}\Bigr)^{R} {P}(u,v)
\quad\text{for all}\  c\in\Rpp\ \text{and all}\ u\in\EEE\,;
\tag{$\text{n}$}\label{focfoc-a}\\ 
{\q_u(c)}{\inc}+{\ee\ts v}
={\pfc^v(u,c)}+{\q_v(\pfc^v(u,c))}B\quad\text{for all}\   c\in\Rpp\ \text{and all}\ u,v\in\EEE\,;
\tag{$\text{e}$}\label{focfoc-b}\\
\sum\nolimits_{\,u\ts\in\ts\EEE}\gp(u)\int_0^\infty{\q_u(c)}\d \csd^u (c) = 0\,;
\tag{$\text{m}$}\label{focfoc-c}\\
\begin{multlined}
\csd^v(c) = \sum\nolimits_{u\ts\in\ts\EEE} 
{\gp(u){P}(u,v)\over \gp(v)}\, 
\csd^u\bigl(\hat \pfc^v(u,c)\bigr)\\
\hbox to4.5cm{\hfill}\text{for all}\  c\in\Rpp\ \text{and all}\ v\in\EEE\,.\qed
\tag{$\text{d}$}\label{focfoc-d}
\end{multlined}
\end{gather}}%
\end{nit}

As the balanced budget constraints in \ref{time-inv}-(\ref{focfoc-b}) obtain in
the limit with $t\to\infty$ in \ref{cross-sys}-(\ref{zze5xb}), those constraints give rise to the following 
iteration program (convention: the token~$\pst$ marks new values and the token~$\pdg$ marks previously
computed values)
{\abovedisplayskip=5pt plus 1.5pt minus 1.5pt\belowdisplayskip=5pt plus 1.5pt minus 1.5pt\belowdisplayshortskip=5pt plus 1.5pt minus 1.5pt
\begin{equation}\label{4iter-q}
\pst{\q_u(c)}\,{\inc}+{\ee\,v} 
={\pst\pfc^v(u,c)} + \pdg\q_v(\pst\pfc^v(u,c))B\,,\quad c\in\Rpp\,,\ u,v\in\EEE\,.
\end{equation}}%
Since the functions
{\abovedisplayskip=5pt plus 1.5pt minus 1.5pt\belowdisplayskip=5pt plus 1.5pt minus 1.5pt\belowdisplayshortskip=5pt plus 1.5pt minus 1.5pt
\[
\Rpp\ni \pfc \leadsto  \pdg{{H}}_v(\pfc)\df \pfc +\pdg\q_v(\pfc)\,B\,,\quad v\in\EEE\,,
\]}%
are strictly increasing and continuous, they can be inverted in the usual way.
Letting $\pdg{\hat{H}}_v\phd$ denote the inverse of $\pdg{{H}}_v\phd$ gives 
$\pst\pfc^v(u,c)=\pdg{\hat{H}}_v\bigl(\pst\q_u(c)\,A+{\ee\ts v}\bigr)$~-- notice that
$\pst\pfc^v(u,\cdot)$ is constructed from both $\pdg\q_v\phd$ and $\pst\q_u\phd$.
As a result,
the functions $\pst\q_u\phd$, $u\in\EEE$, obtain implicitly from the relations
{\abovedisplayskip=5pt plus 1.5pt minus 1.5pt\belowdisplayskip=5pt plus 1.5pt minus 1.5pt\belowdisplayshortskip=5pt plus 1.5pt minus 1.5pt
\begin{equation}\label{reduced-k}
B=\beta\,\inc\,\sum\nolimits_{v\ts\in\ts\EEE}\Bigl({c\over 
 \pdg{\hat{H}}_v\bigl(\pst\q_u(c)\,A+{\ee\ts v}\bigr)}\Bigr)^{R} {P}(u,v)\,,
\quad c\in\Rpp\,,\ u\in\EEE\,,
\end{equation}}%
or, which amounts to the same but is easier, $c$ can be
written as an explicit function of $\pst\q_u(c)$.
\footnote{{}In terms of computer code, if the connection $c=f(\q)$ can be expressed as a cubic
spline, then $\q$ can be written as a function of $c$ (again as a cubic spline) by merely swapping
the lists of abscissas and the ordinates in the routine that produces the spline objects. Thus,
writing $\q$ as a function of $c$ is no different from writing $c$ as a function of $\q$~-- as long
as the dependence is monotone and smooth.}
Furthermore, \ref{time-inv}-(\ref{focfoc-d}) gives rise to the iteration program
{\abovedisplayskip=7pt plus 1.5pt minus 1.5pt\belowdisplayskip=7pt plus 1.5pt minus 1.5pt\belowdisplayshortskip=5pt plus 1.5pt minus 1.5pt
\begin{equation}\label{4iter-F}
\pst\csd^v(\cdot) \df \sum\nolimits_{u\ts\in\ts\EEE} 
{\gp(u){P}(u,v)\over \gp(v)}\, 
\pdg\csd^u\bigl(\pst\hat \pfc^v(u,\cdot)\bigr)\,,\quad v\in\EEE\,,
\end{equation}}%
In the present context, the general time-interlaced backward induction 
program from \ref{main-proc} reduces to the following.
\nitskip

\begin{nit}{Time-interlaced backward induction}\label{IOU-proc}%
{\it Step 0:} Make an ansatz choice for the collection of portfolio mappings $\pdg\q_v(\cdot)$, $v\in\EEE$.
Then make an ansatz choice for the spot price~$B$ (these two choices are independent).
Go to Step~1. 

{\it Step 1:} Set $\pdg{{H}}_v\phd = \phd +\pdg\q_v\phd\,B$
and  compute the inverse $\pdg{\hat{H}}_v\phd$ for every $v\in\EEE$.

{\it Step 2:} For every choice of $u\in\EEE$ and certain choices of $c\in\Rpp$, 
solve \eqref{reduced-k} (there is one equation for every $u\in\EEE$ and every $c\in\Rpp$) 
for the unknowns $\pst\q_u(c)$ and set
{\abovedisplayskip=5pt plus 1.0pt minus 1.5pt\belowdisplayskip=5pt plus 1.5pt minus 1.5pt\belowdisplayshortskip=5pt plus 1.5pt minus 1.5pt
$$\pst\pfc^v(u,c)
=\pdg{\hat{H}}_v\bigl(\pst\q_u(c)\,\inc+{\ee\,v}\bigr)\quad\text{ for every $u,v\in\EEE$}\,.
$$}%
Find the smallest $c\in\Rpp$, denoted $\bar
c$, with the  property $\bar c\ge \pst\pfc^v(u,\bar c)$ for all
$u,v\in\EEE$.
\footnote{{}This step is meant to endogenize the upper bound on consumption.}
Construct a uniform (equidistant) finite grid, denoted $\bbG_\cint$, on the
interval $\cint$. Go to the next step.

{\it Step 3:}  For every $u\in\EEE$ and every grid-point
$c\in\bbG_\cint$,
solve for $\pst\q_u(c)$ from \eqref{reduced-k} and set
$\pst\pfc^v(u,c)=\pdg{\hat{H}}_v\bigl(\pst\q_u(c)\,\inc + {\ee\,v}\bigr)$  for all
$v\in\EEE$.
\footnote{Note that $\pst\pfc^v(u,\cdot)$ is fully determined by $\pdg\q_u\phd$ and $\pst\q_u\phd$.}
By interpolating the respective values define the functions $\pst\q_u(\cdot)$ and
$\pst\pfc^v(u,\cdot)$, $v\in\EEE$, as cubic splines over the grid $\bbG_\cint$ in the obvious way.
Define
uniform interpolation grids over the ranges of the functions  $\pst\pfc^v(u,\cdot)$, compute 
the inverse values at those grid-points and, finally, 
define the inverse functions $\pst\hat \pfc^v(u,\cdot)$ as the cubic
splines obtained by interpolating the inverse values over the respective grids. Go to the next
step.

{\it Step 4:} 
If the family of distribution functions $\pdg\csd^u$, $u\in\EEE$, 
has not been updated before (this is the first visit to Step~4), 
define $\pdg\csd^u$ to be the distribution function associated with the uniform probability
measure  on $\cint$ for every $u\in\EEE$. Otherwise, do nothing and go to the next step.

{\it Step 5:}
Calculate
{\abovedisplayskip=2pt plus 1.0pt minus 1.5pt\belowdisplayskip=5pt plus 1.5pt minus 1.5pt\belowdisplayshortskip=5pt plus 1.5pt minus 1.5pt
\begin{equation}\tag{a}\label{step-4-dist}
\pst\csd^v(c) \df \sum\nolimits_{u\ts\in\ts\EEE} 
{\gp(u){P}(u,v)\over \gp(v)}\, 
\pdg\csd^u\bigl(\pst\hat\pfc^v(u,c)\bigr)
\end{equation}}%
for every  $c\in\bbG_\cint$ and every $v\in\EEE$ and construct the
distribution functions $\pst\csd^v(\cdot)$, $v\in\EEE$, as cubic splines over 
the grid $\bbG_\cint$ in the obvious way. Compute the error term
{\abovedisplayskip=5pt plus 1.5pt minus 1.5pt\belowdisplayskip=5pt plus 1.5pt minus 1.5pt\belowdisplayshortskip=5pt plus 1.5pt minus 1.5pt
\[
\max\nolimits_{\tts v\ts\in\ts\EEE,\,c\ts\in\ts \bbG_\cint}
|\pst\csd^{v} (c)- \pdg\csd^{v}(c)|\,.
\]}%
If this error term exceeds some  prescribed
threshold, set $\pdg\csd^v(\cdot)=\pst\csd^v(\cdot)$, 
$v\in\EEE$, go back to the beginning of this step and
repeat.
\footnote{These iterations are determined by the choice of $\pdg\q_u\phd$ and $\pst\q_u\phd$.}
Otherwise, set $\pdg\csd^v(\cdot)=\pst\csd^v(\cdot)$, $v\in\EEE$, and
go to the next step.

{\it Step 6:} Test the market clearing
{\abovedisplayskip=5pt plus 1.5pt minus 1.5pt\belowdisplayskip=5pt plus 1.5pt minus 1.5pt\belowdisplayshortskip=5pt plus 1.5pt minus 1.5pt
\begin{equation}\tag{b}\label{step-5-mktc}
\sum\nolimits_{u\ts\in\ts\EEE}\gp(u)\int_0^{\bar c}{\pst\q_u(c)}\d\,\pst\csd^u(c) = 0\,.
\end{equation}}%
If this identity fails by more
than some  prescribed threshold, discard the splines
$\pst\q_u(\cdot)$, $u\in\nobreak\EEE$, while still
keeping $\pdg\q_u(\cdot)$ on
record,  modify the most recent choice for the spot price $B$
accordingly, and go back to Step~1. Otherwise go to the next step.

{\it Step 7:} If this is the first visit to Step 7, set
$\pdg\q_u(\cdot)=\pst\q_u(\cdot)$
and go to Step~1 with the most recently updated value for the spot price~$B$. 
Otherwise, compute the error terms
{\abovedisplayskip=5pt plus 1.5pt minus 1.5pt\belowdisplayskip=5pt plus 1.5pt minus 1.5pt\belowdisplayshortskip=5pt plus 1.5pt minus 1.5pt
\begin{equation}\tag{c}\label{conv-test}
\begin{gathered}
\max\nolimits_{u\ts\in\ts\EEE,\,c\in \bbG_\cint}
|\pst\q_u(c)-\pdg\q_u(c)|\\
\llap{\text{and}\quad}
\max\nolimits_{u,v\in\EEE,\,c\in \bbG_\cint}
|\pst\pfc^v(u,c)-\pdg \pfc^v(u,c)|\,.
\end{gathered}
\end{equation}}%
If at least one of these terms exceeds some prescribed threshold, set
$\pdg\q_u(\cdot)=\pst\q_u(\cdot)$ and go to Step~1 with the most recently updated value for the
spot price~$B$. 
Otherwise stop. Declare that the
equilibrium is given by the most recently updated spot price $B$,
portfolio mappings $\pst\q_u(\cdot)$, $u\in\EEE$, 
state transitions 
$\pst\pfc^v(u,\cdot)$, $u,v\in\EEE$, and family of distribution
functions $\pst\csd^u(\cdot)$, $u\in\EEE$.\qed
\end{nit}

\begin{nit}{Abridged version of \ref{IOU-proc}}\label{abridged}%
Make an ansatz choice for $(B,\pdg\q)$ and record it.
Gi\-ven $(B,\pdg\q)$, produce $\pst\q$. 
Then produce $\pst{\csd}$ as a fixed point of the transport determined by~$\pdg\q$
and~$\pst\q$. If market clearing with~$\pst\q$ and~$\pst{\csd}$ fails, forget
$\pst\q$ and $\pst{\csd}$, change the value of~$B$, and repeat with the modified $B$ and
with $\pdg\q$. If the market clears, record~$\pst\q$,  the latest $B$, and the latest
transition assignments.
If~this is the first incidence of market clearing, set $\pdg\q=\pst\q$ and repeat from
the beginning with the latest $B$ and the new $\pdg\q$. If~not, test the uniform distance
between $\pst\q$ and $\pdg\q$ and between the two most recent collections of state transition assignments.
If~this distance is not acceptable, set $\pdg\q=\pst\q$ and repeat from
the beginning with the latest $B$ and the new $\pdg\q$. Otherwise, stop.\qed
\end{nit}

\begin{nit}{Remark}%
It is instructive to note the key
differences between the program in~\nobreak\ref{IOU-proc} and the classical strategy outlined
in Sec.~\ref{sec:Intro}: (a)~The portfolio mappings $\pdg\q_u(\cdot)$, $u\in\EEE$, capture
the investment decisions in the cross-section of all households that share the same state of
employment~-- not the investment decision of one representative household.
(b)~The law of motion in the space of distributions (of unit mass) encrypted in
\ref{IOU-proc}-(\ref{step-4-dist}) is not sought as the law of motion of the probability
distribution of any particular Markovian state. 
(c)~It is the price that gets adjusted to the portfolio mappings, and then new portfolio mappings
are obtained with the new price, i.e., the adjustments in portfolio mappings and prices alternate.
(d)~The search for a fixed point of the transport equation in Step~5, for every instance of
$\pst\q_u(\cdot)$ and $\pdg\q_u(\cdot)$, removes the need to write the endogenous variables as
functions on the space of distributions, in that there is always a unique
distribution associated with every instance of $\pst\q_u(\cdot)$ and $\pdg\q_u(\cdot)$.
Thus, instead attaching values of the endogenous variables to every instance of the population
distribution, the program associates such values only with a single distribution, namely, the one
that remains invariant (under the most recent instance of portfolio mappings)~-- an
enormous simplification, possible only if stationary distribution of the population exists and
one is concerned with the infinite time-horizon case alone.
\qed
\end{nit}

\begin{nit}{Remark}%
Step~1 in \ref{IOU-proc} is nothing but the search for the
endogenous upper bound on consumption, which then translates into an upper bound on
investment, 
since the functions $c \leadsto \pst\q_u(c)$,
$u\in\EEE$, are increasing.
The lower bound on the investment (i.e., the borrowing limit)
is then $\min_u\lim_{c\to0}\q_u(c)B$.
We stress that both bounds are determined
endogenously throughout the iterations.
In the benchmark economy discussed here these bounds
are never reached and the cross-sectional distribution of the population has no mass at
them.\qed
\end{nit}
\nitskip

What follows next is a brief summary of the concrete results from implementing the
time-interlaced backward induction
program from \ref{IOU-proc} in the context of the benchmark Huggett economy
borrowed from \cite{LjunSar00}  and
already introduced in Sec.~\ref{sec:Intro}. 
All model parameters are taken from the first specification in 
\citex[Sec.~18.7]{LjunSar00}. 
The initial ansatz choice for the portfolio functions is $\pdg\q_v(c)=40c-8$ for all
$v\in\EEE$ and 
for the bond price the initial choice is $B=\inc$, corresponding to zero interest as
an initial guess.  
The convergence (the largest amount
in \hbox{\ref{IOU-proc}-(\ref{conv-test})}) is $9.08447\times 10^{-5}$ after 235
iterations, which 
the program completes on a single core for about 127 minutes, and 
returns equilibrium interest rate of $0.03702$ 
and market clearing (the left side of~\ref{IOU-proc}-(\ref{step-5-mktc})) of
$-1.73878\times10^{-6}$.
The distribution of households in every one of the $7$ employment categories over the consumption space
is shown in Fig.~\ref{fg4X} 
{%
\begin{figure}[!htbp]  
\centering
\begin{subfigure}{.5\textwidth}
  \centering
\leavevmode\raise0.55cm\hbox{\rotatebox{90}{\tiny population distribution in states $u\in\EEE$}}%
\ %
\toshow{\includegraphics[width=7.1cm]{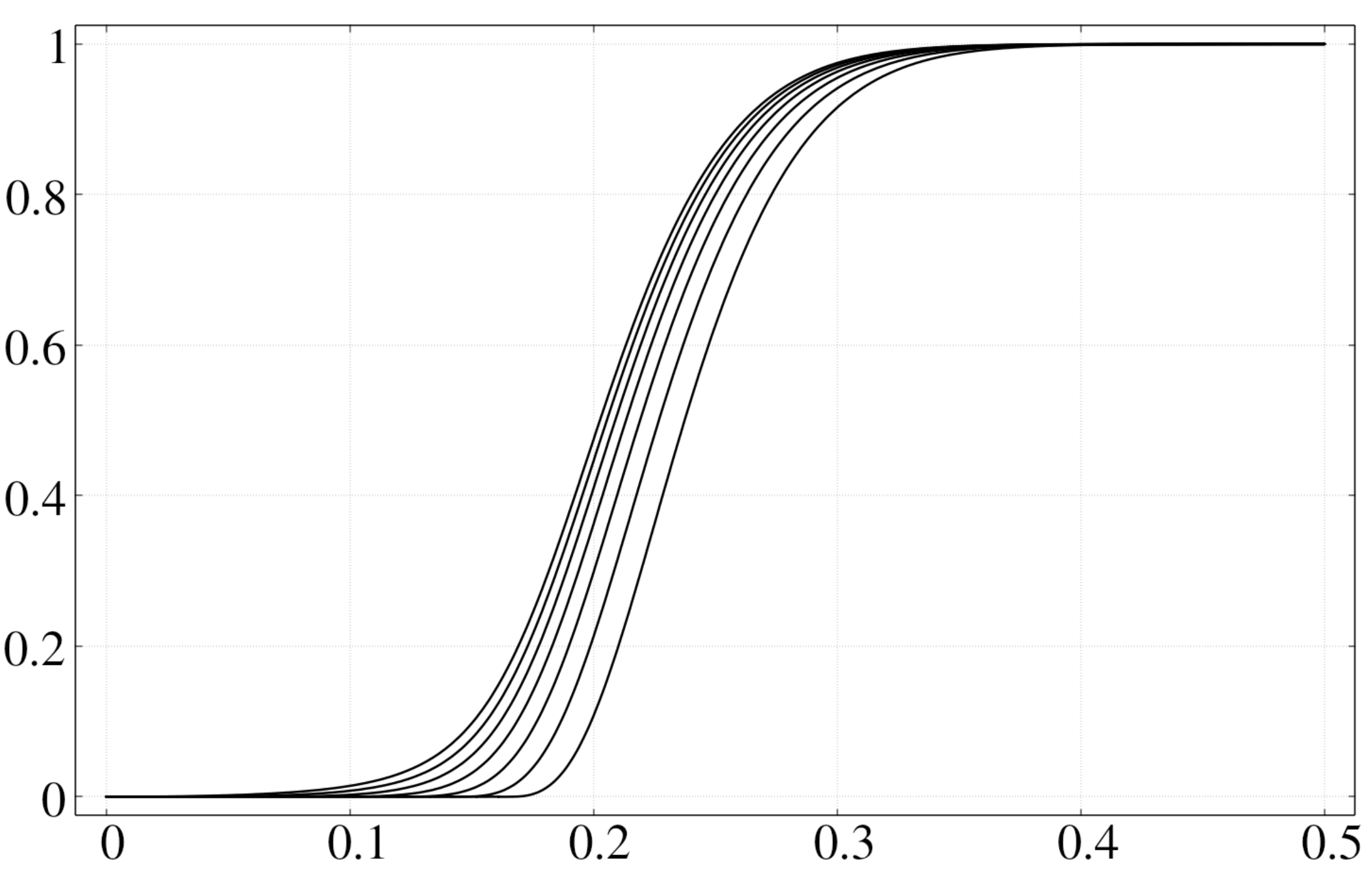}}

\leavevmode\smash{\raise6pt\hbox{\tiny consumption}}
\end{subfigure}%
\begin{subfigure}{.5\textwidth}
  \centering
\leavevmode\raise0.9cm\hbox{\rotatebox{90}{\tiny population density in states $u\in\EEE$}}%
\ %
\toshow{\includegraphics[width=7.1cm]{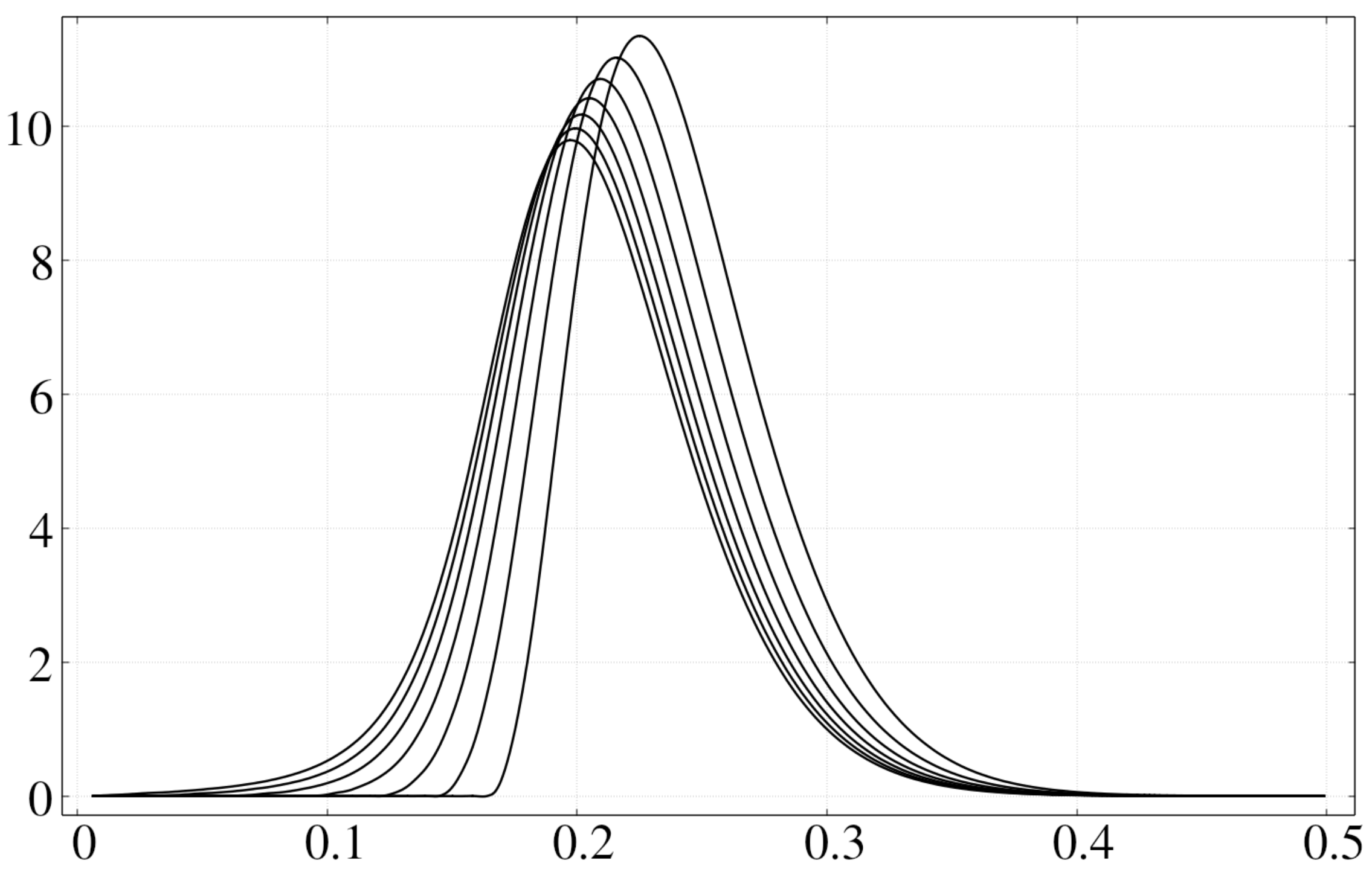}}  

\leavevmode\smash{\raise6pt\hbox{\tiny consumption}}
\end{subfigure}
\caption{The distribution (cumulative left, density right) of households over the range of consumption.}
\label{fg4X} 
\end{figure}}%
and the plots in Fig.~\ref{fg8X} show the investment level in the cross-section of the population, i.e.,
the mappings $c \leadsto \q_u(c)\times B$, $u\in\nobreak\EEE\,$.
%
%
{%
\begin{figure}[!htbp]  
\centering
\begin{subfigure}{.5\textwidth}
  \centering
\leavevmode\raise1.2cm\hbox{\rotatebox{90}{\tiny investment in states $u\in\EEE$}}%
\ %
\toshow{\includegraphics[width=7.1cm]{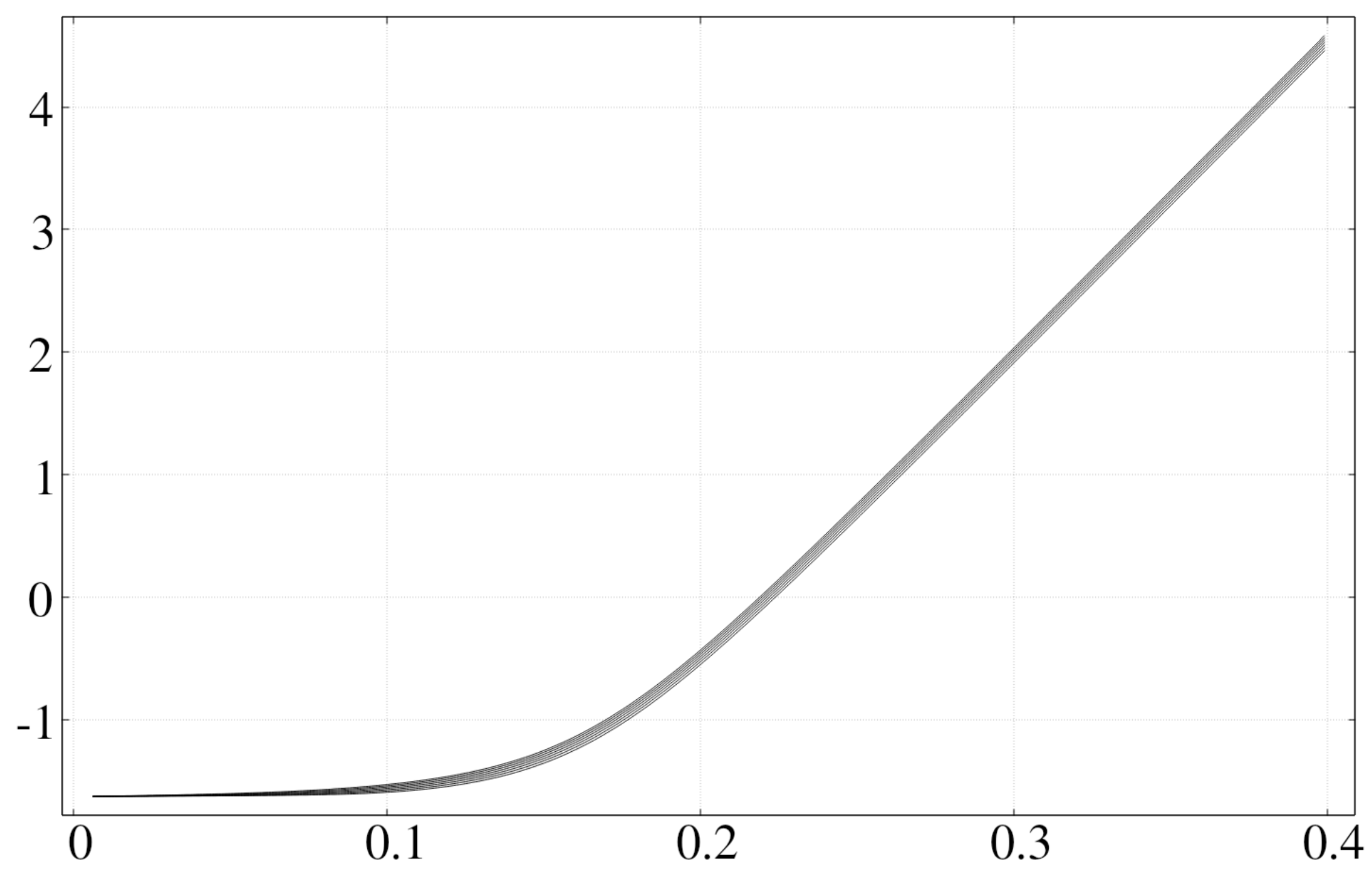}}

\leavevmode\smash{\raise6pt\hbox{\tiny consumption}}
\end{subfigure}%
\begin{subfigure}{.5\textwidth}
  \centering
\leavevmode\raise1.2cm\hbox{\rotatebox{90}{\tiny investment in states $u\in\EEE$}}%
\ %
\toshow{\includegraphics[width=7.1cm]{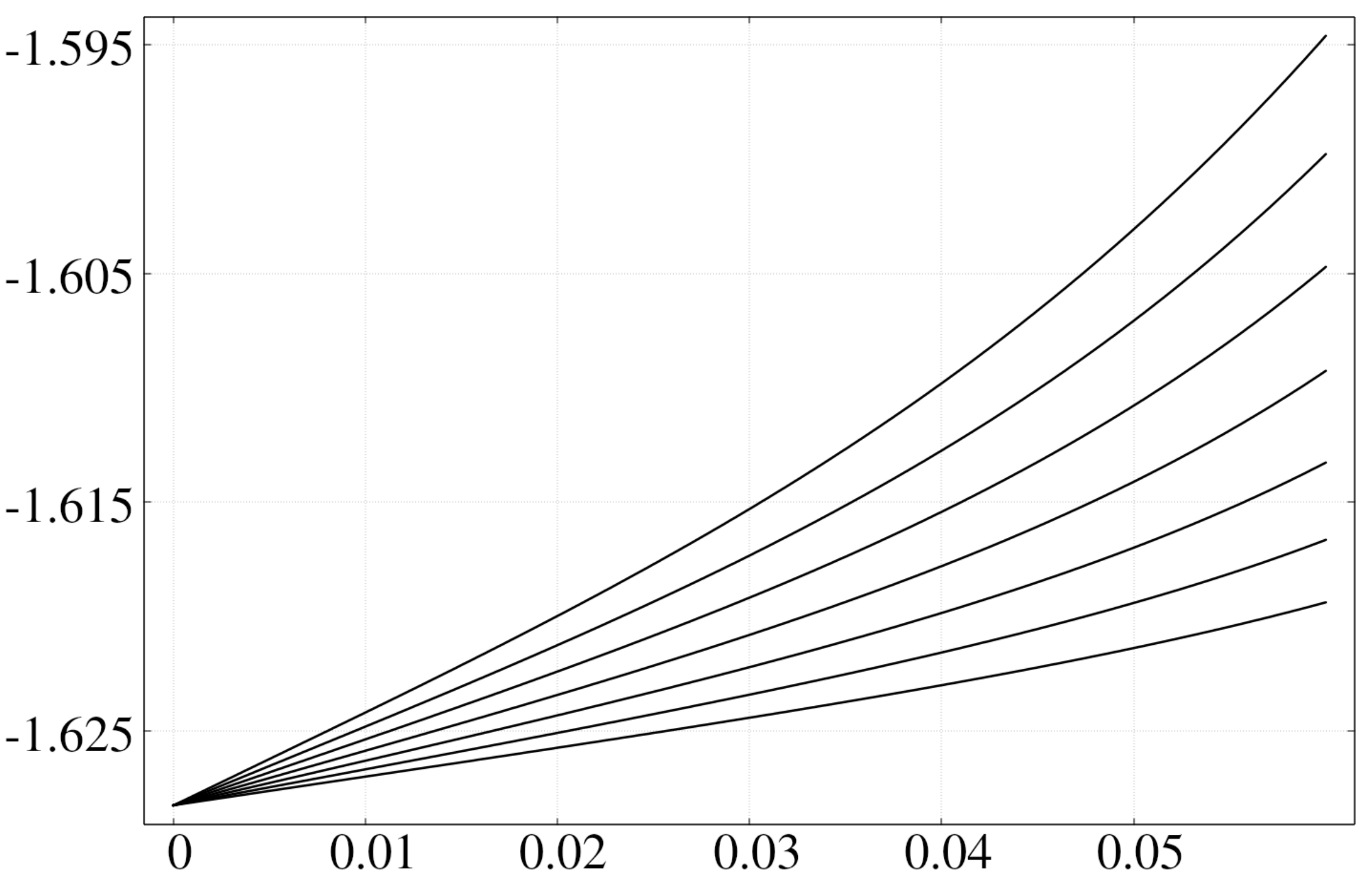}} 

\leavevmode\smash{\raise6pt\hbox{\tiny consumption}} 
\end{subfigure}
\caption{Investment in the bond as a function of consumption  shown on two different scales.}
\label{fg8X}
\end{figure}}%
The left limit in these graphs (the endogenous borrowing limit)
is around $-1.62826$ and is remarkably close to Aiyagari's
natural borrowing limit (in this case, the same as the ``ad hoc'' limit~-- see \cite{LjunSar00} and
\cite{Aiya94}),
which is around $-1.62726$. The endogenous upper bound on investment~-- see Step~2
in \ref{IOU-proc}~-- is around $17.93751$, 
but we see from Figures~\ref{fg4} and~\ref{fg5X} that most of the population is amassed over a much narrower
range.
Since the mappings $c\leadsto \q_u(c)\ts B$ and $c\leadsto c+\q_u(c)\ts B-\ee\ts u$ are both
strictly increasing and continuous, the equilibrium distribution over consumption from Fig.~\ref{fg4X} is easy to
transform into entering and exiting distribution of households over the asset space~--
this is how the left plot in Fig.~\ref{fg4} was produced. The left plot in
Fig.~\ref{fg5X} gives a detailed view of the left plot in Fig.~\ref{fg4} near the borrowing
limit.
%
{%
\begin{figure}[!htbp]  
\centering
\begin{subfigure}{.5\textwidth}
  \centering
  \leavevmode\raise0.55cm\hbox{\rotatebox{90}{\tiny population distribution in states $u\in\EEE$}}%
\ %
\toshow{\includegraphics[width=7.1cm]{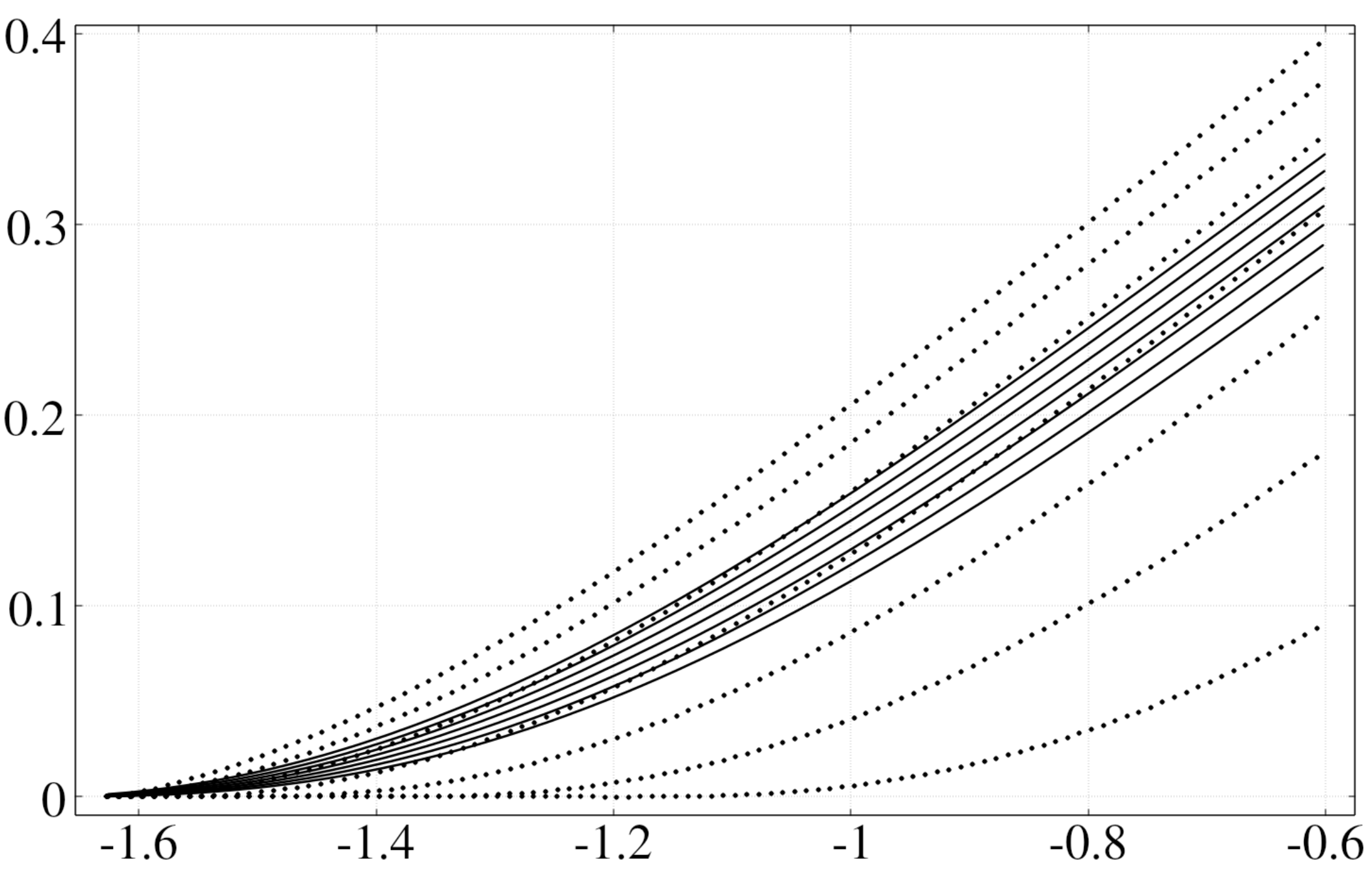}}

\leavevmode\smash{\raise6pt\hbox{\tiny asset holdings}}
\end{subfigure}%
\begin{subfigure}{.5\textwidth}
  \centering
  \leavevmode\raise0.9cm\hbox{\rotatebox{90}{\tiny population density in states $u\in\EEE$}}%
\ %
\toshow{\includegraphics[width=7.1cm]{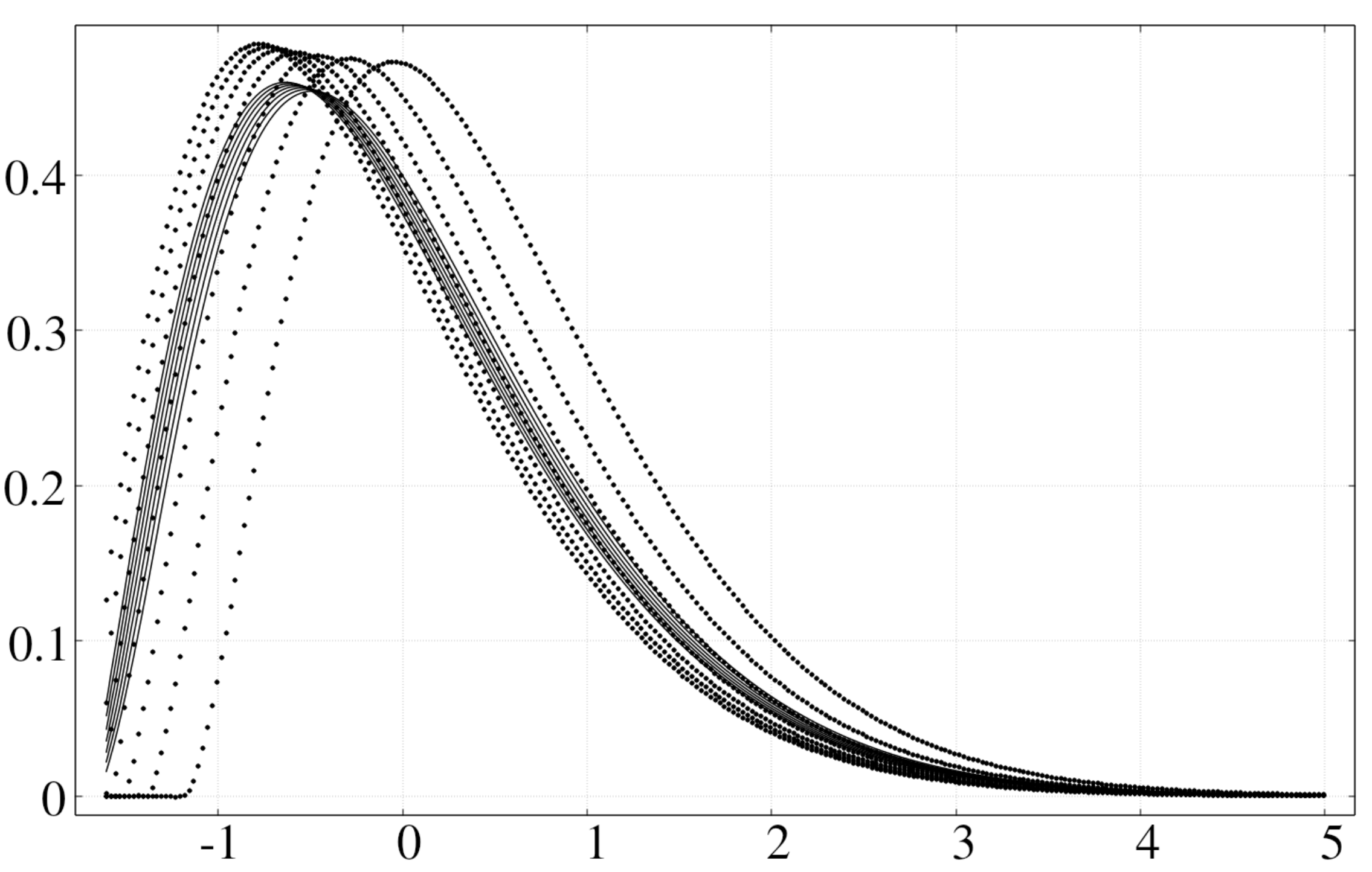}} 

\leavevmode\smash{\raise6pt\hbox{\tiny asset holdings}}
\end{subfigure}
\caption{The entering (solid lines) and exiting (dotted lines) distribution of households over
asset holdings.}
\label{fg5X}
\end{figure}}%
The right plot
in Fig.~\ref{fg5X} is simply the density version of the left plot in Fig.~\ref{fg4}. 
The graphs of the conditional transition assignments  $c \leadsto \pfc^v(u,c)$, $v\in\EEE$, for the lowest
and the highest employment category $u\in\EEE$ are shown on the left plot in Fig.~\ref{fg8Y}. 
%
%
{\captionsetup{belowskip=-10pt}
\begin{figure}[!htbp]  
\centering
\begin{subfigure}{.5\textwidth}
  \centering
\leavevmode\raise0.75cm\hbox{\rotatebox{90}{\tiny future consumption in states $v\in\EEE$}}%
\ %
\toshow{\includegraphics[width=7.1cm]{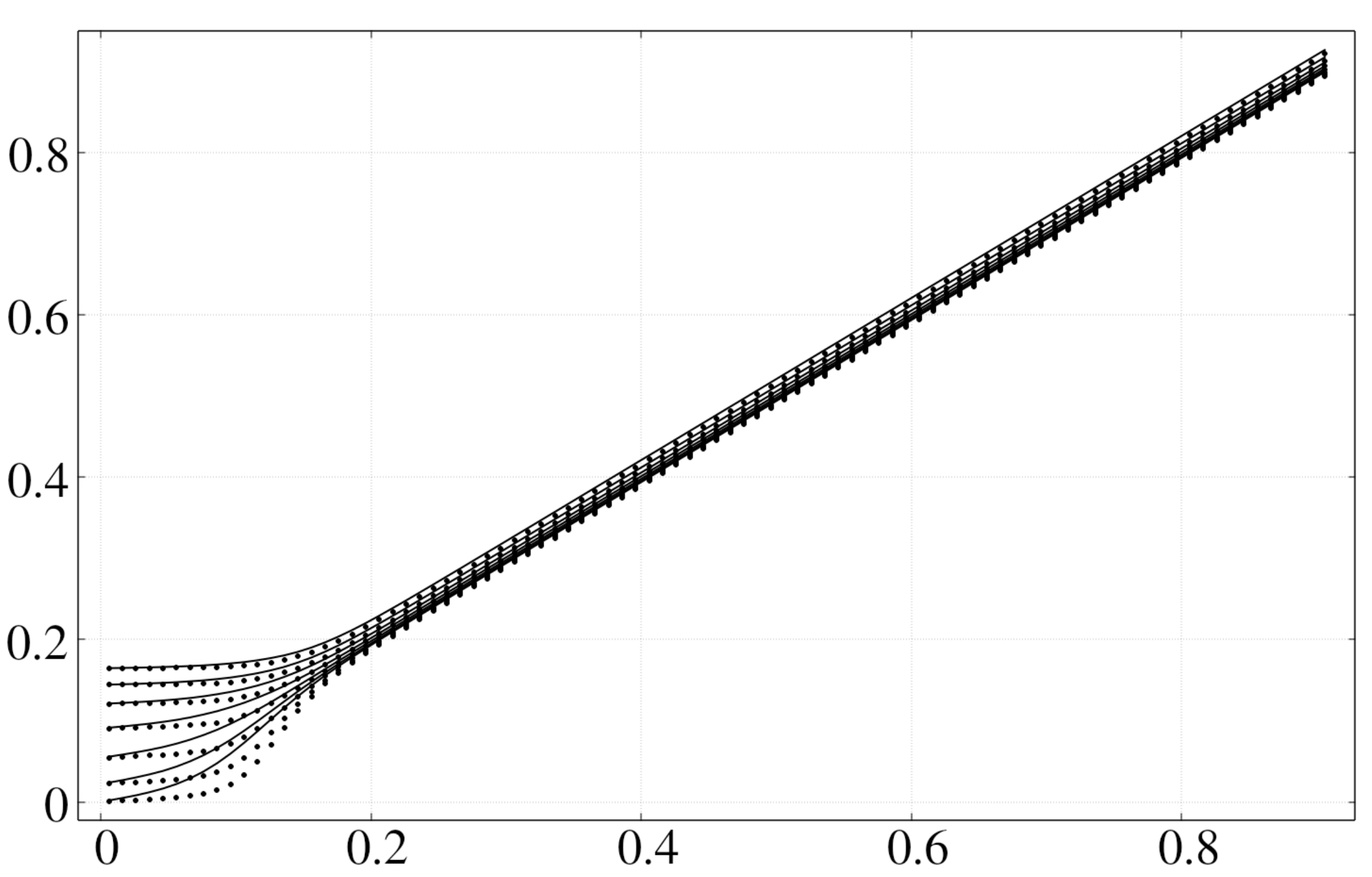}}

\leavevmode\smash{\raise6pt\hbox{\tiny consumption in state $u=\EEE_1$ (solid) and in state $u=\EEE_7$ (dotted)}}
\end{subfigure}%
\begin{subfigure}{.5\textwidth}
  \centering
\leavevmode\raise1.25cm\hbox{\rotatebox{90}{\tiny consumption in states $v\in\EEE$}}%
\ %
\toshow{\includegraphics[width=7.1cm]{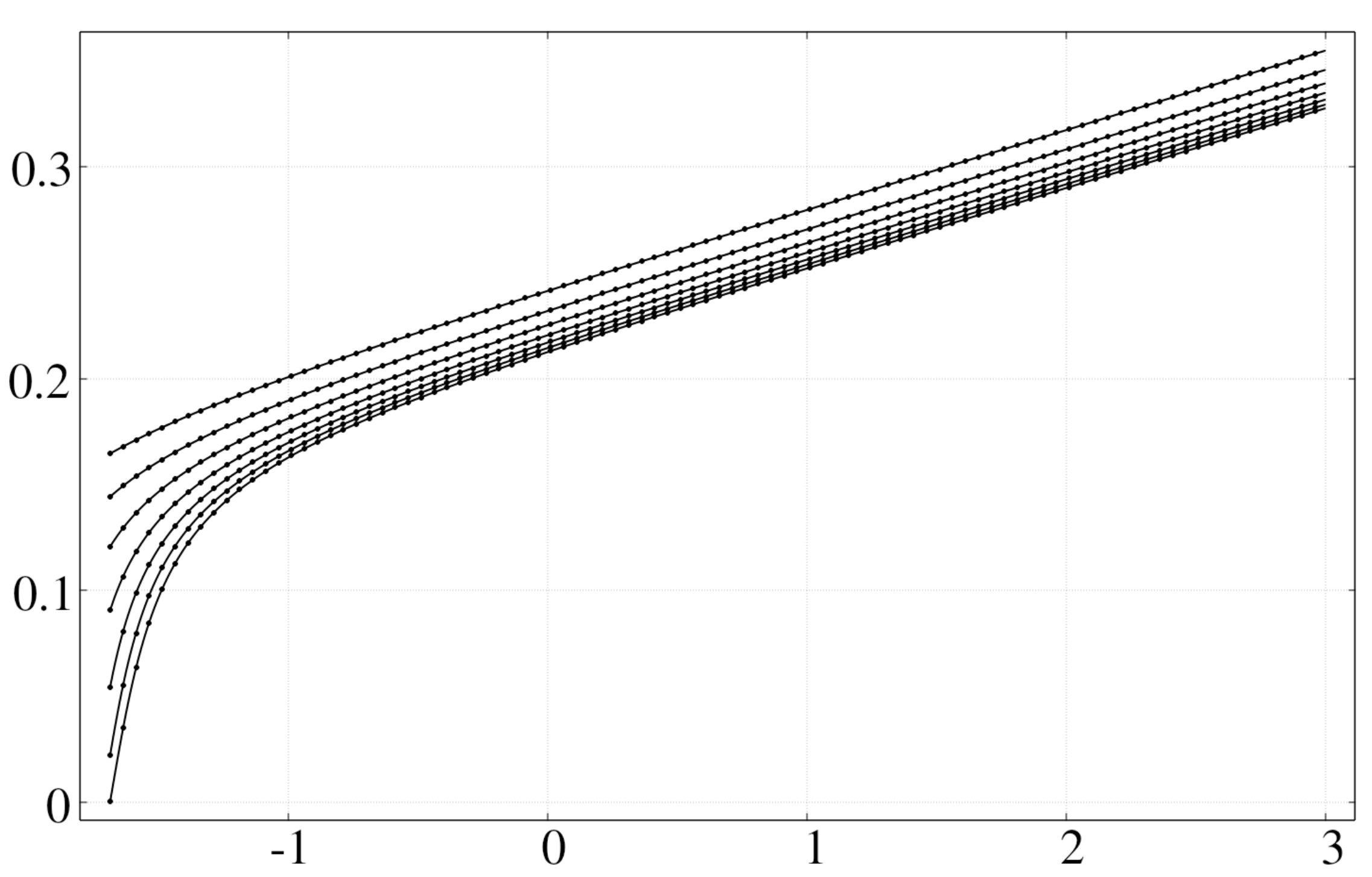}} 

\leavevmode\smash{\raise6pt\hbox{\tiny entering wealth from state $u=\EEE_1$ (solid) and from $u=\EEE_7$ (dotted)}} 
\end{subfigure}
\caption{Future consumption as a function of present consumption (left) and entering future wealth (right).} 
\label{fg8Y}
\end{figure}}%
The right plot in Fig.~\ref{fg8Y} provides an important verification of the program developed in
this section: on the one hand the employment specific transition assignments depend on both the exiting and the
entering employment states, while on the other hand future consumption must depend only on the future
employment state and the entering wealth in that state, irrespective of what employment state that
wealth is carried from. Since any household of type $(u,c)\in\EEE\times\Rpp$ enters its future state with
assets $a=\a_u(c)\df \q_u(c)\times A$, letting $\hat\a_u\phd$ denote the inverse of the assignments
$c \leadsto \a_u(c)$, this means that $\TTT^v(u,\hat\a_u(a))$ must depend on~$v$ and~$a$ but not
on~$u$. Such a connection was never imposed in the system that produced the equilibrium, but the
right plot in Fig.~\ref{fg8Y} shows that it nevertheless holds~-- as it should.
Finally, the left plot in Fig.~\ref{fg9Z} shows consumption as a function of total wealth
(i.e., consumption plus investment)
and the right plot shows the marginal propensity to consume (simply
the gradient of the splines generating the left plot). 
%
%
{\captionsetup{belowskip=-10pt}
\captionsetup{aboveskip=10pt}
\begin{figure}[!htbp]  
\centering
\begin{subfigure}{.5\textwidth}
  \centering
\leavevmode\raise0.75cm\hbox{\rotatebox{90}{\tiny consumption in states $u\in\EEE$}}%
\ %
\toshow{\includegraphics[width=7.1cm]{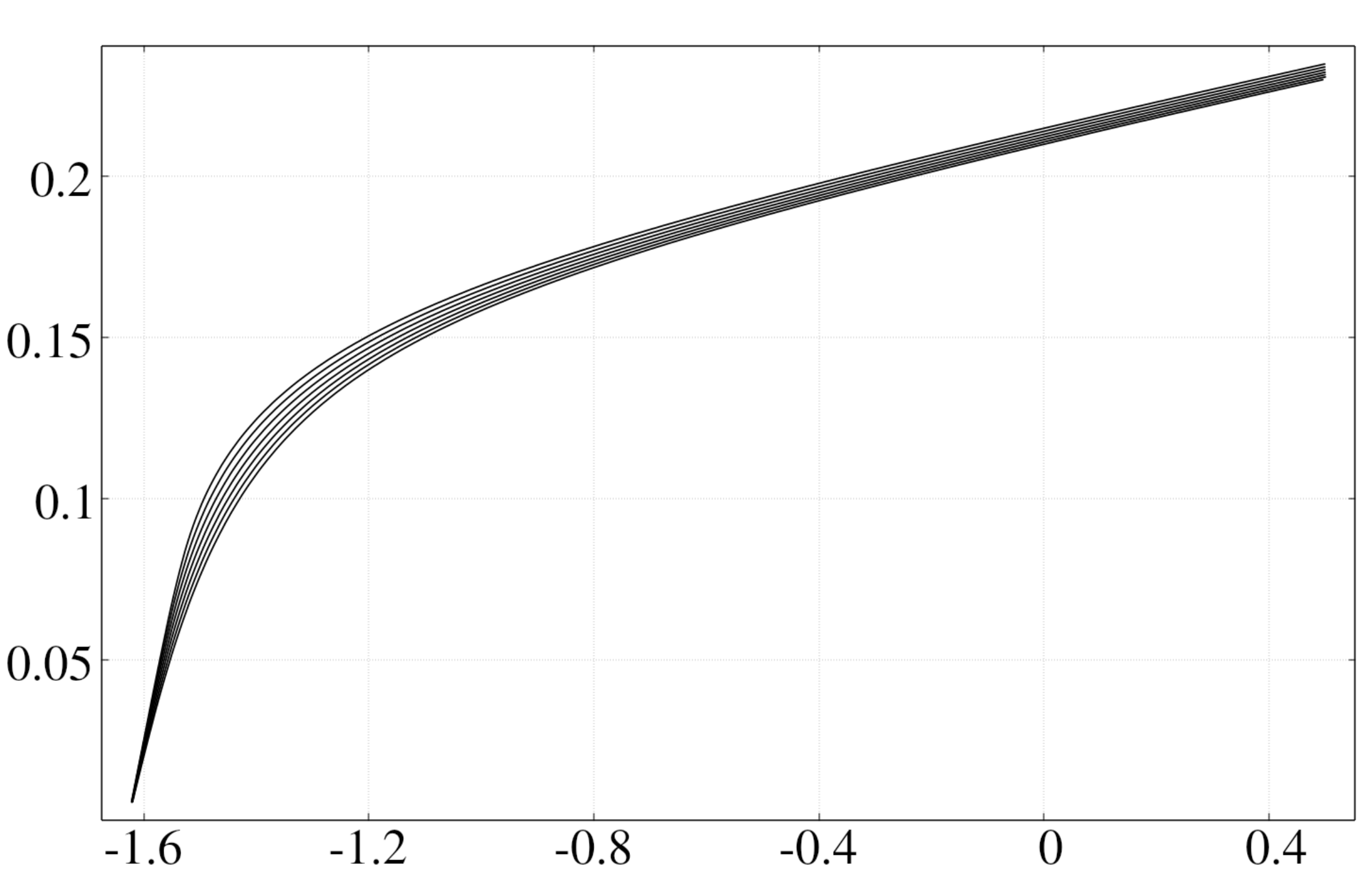}}

\leavevmode\smash{\raise6pt\hbox{\tiny \qquad total wealth (entering plus income)}}
\end{subfigure}%
\begin{subfigure}{.5\textwidth}
  \centering
\leavevmode\raise0.15cm\hbox{\rotatebox{90}{\tiny marginal propensity to consume in states $u\in\EEE$}}%
\ %
\toshow{\includegraphics[width=7.1cm]{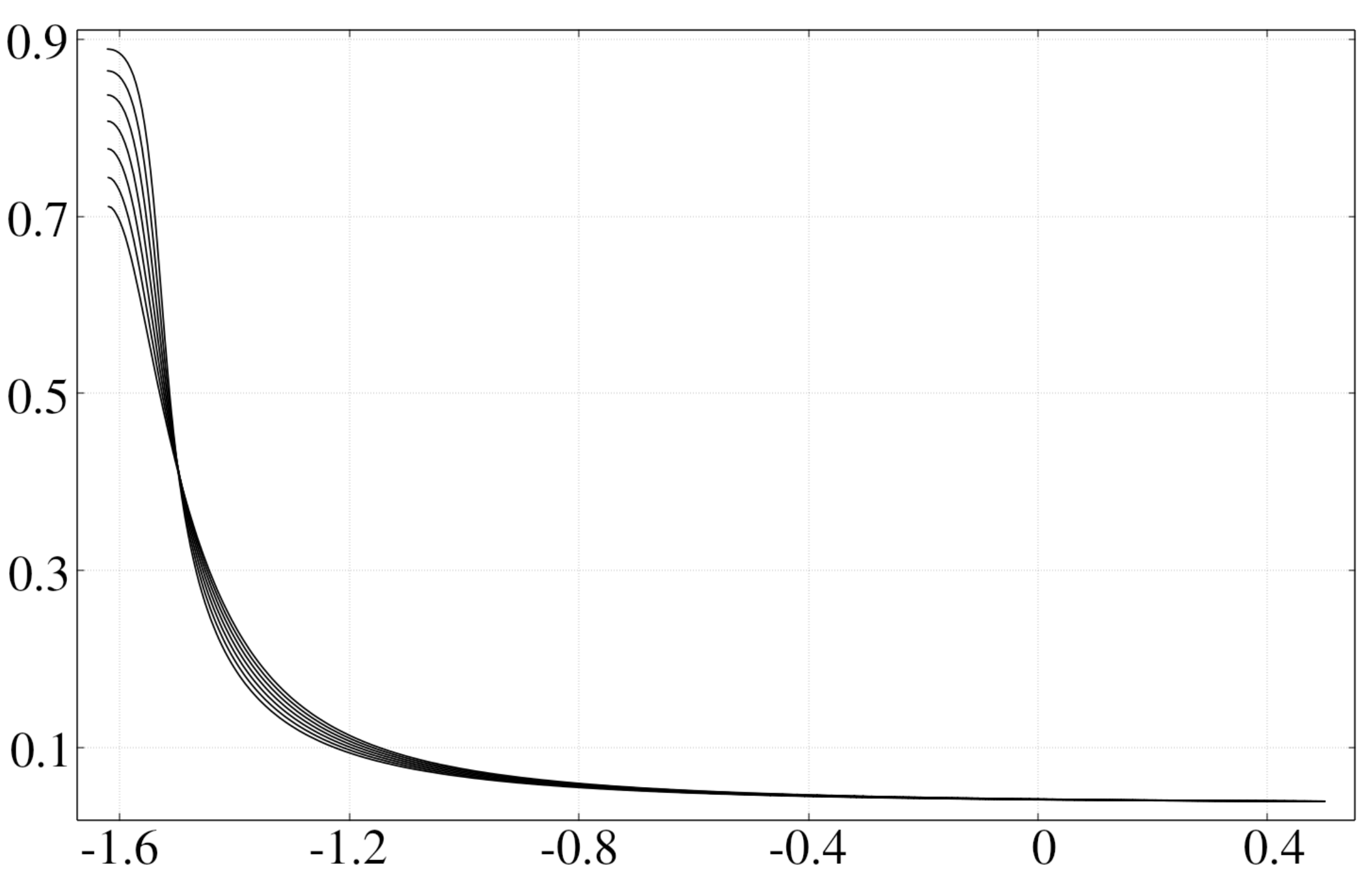}} 

\leavevmode\smash{\raise6pt\hbox{\tiny \qquad total wealth (entering plus income)}} 
\end{subfigure}
\caption{Consumption as a function of total wealth (entering plus income) and its gradient.} 
\label{fg9Z}
\end{figure}}%
%
\begin{nit}{Implications}%
Fig.~\ref{fg8X} shows that the employment state has only a marginal effect on the dependence
between investment and consumption (all households face the same stream of employment shocks in the
long run and the effect of temporary differences in employment is too
small relative to the entire stream of anticipated future shocks).
In addition, the right plot shows that the graphs on the left are not flat near~$0$, with slopes
ranging between  $0.12502$ and $0.40615$.
Fig.~\ref{fg4X} and Fig.~\ref{fg5X} show that the distribution of households (both entering and
exiting) over asset holdings is
substantially more dispersed and skewed toward the wealthy than it is over consumption.
The standard deviation over consumption ranges between  
$0.0458$ and $0.03827$, over entering wealth it ranges between
$0.97125$ and $0.97857$, and ranges between  $0.92818$ and $0.93861$ over exiting wealth. For the
skewness these numbers are  $0.11558$ $\sim$ $0.84976$ for consumption,  $0.96578$ $\sim$ $0.93788$
for entering wealth, and  $1.00653$ $\sim$ $0.9316$ for exiting wealth.
In contrast to the continuous-time model discussed in \cite{AHLLM17},
which involves exogenously imposed
boundary conditions and is based on different model parameters,
it is clear from Fig.~\ref{fg5X} and Fig.~\ref{fg9Z} that here the cross-sectional
distribution of the 
population does not accumulate at the borrowing limit and the marginal propensity to
consume does 
not explode at that limit.
\footnote{These features cannot be claimed to hold universally.}
%
More details and illustrations are included with the Julia code that accompanies the \chptr.\qed
\end{nit} 



\section{Approximate Equilibrium in Models with Aggregate Risk}\setcounter{paragraph}{0}
\label{sec:KS}

\noindent
The main objective in this section is to specialize the general model developed in
Sec.~\ref{sec:gen-model} to the 
setup of the widely cited paper \cite{KS98},
in which the economy is endowed with production technology and 
the households invest only in productive capital, with no risk-free private lending available.
The main premise in the paper \cite{KS98} is what the authors call ``collapse of the
state space,'' or 
``approximate aggregation,'' i.e., the idea that ``in~equilibrium, all aggregate variables~--
consumption, the capital stock, and relative prices~-- can be almost perfectly described
as a function of two simple statistics: the mean of the wealth distribution and the aggregate
productivity shock.'' Simply put: ``only the mean matters.''
The insight offered in \cite{KS98}  is that ``utility costs from
fluctuations in consumption are quite small and that self-insurance with only one
asset is quite 
effective''~-- a phenomenon that  the authors link to the permanent income hypothesis
from the paper \cite{Bew77}. 
The present section is a follow-up to this insight, but with a number of deflections and
clarifications, which are spelled out next.
In~what follows the approximate aggregation hypothesis is placed in the context of a broader framework,
conditions under which this result  
holds (as an approximation) are identified, new methodology that allows one to
quantify the notions 
of ``approximate'' and ``almost perfectly'' is developed, and a more refined
aggregation strategy is 
outlined.  
It~is to be noted that in the present context ``the mean of the population
distribution'' is understood as the mean 
of the distribution over consumption~-- not wealth.
Since this distribution enters the model only through the market 
clearing condition, 
the claim ``only the mean matters'' boils down to the claim that, as functions of
consumption, the portfolio 
mappings are approximately affine with employment-invariant slopes.
\footnote{{}Stroock \protect\cite{Str23} provides a particularly
elegant formal justification.} 
Hence, quantifying ``approximate'' in ``approximate aggregation''
comes down to quantifying the
error from replacing the portfolio mappings
with affine functions and identifying conditions under
which the slopes of those functions are employment invariant. 
In~addition, closed-form expressions for the slopes are derived and
the time-interlaced backward induction program from \ref{main-proc} 
is adapted to this special structure. One consequence from this new
approach is that one no longer 
needs to postulate infinite time horizon and resort on simulation;
instead, the program
boils down to solving sequentially linear systems~-- for as
many periods as needed.

The only-the-mean-matters point of view has one obvious limitation: it fails to
capture the dynamics of the relative disparity among households~-- an
important macroeconomic
characteristic. The idea that ``the mean of the wealth distribution and the aggregate
productivity shock'' are ``two simple statistics'' that describe
``almost perfectly'' all aggregate 
variables may lead one to conclude that, in the study of general equilibrium,
there is no further inference to be made about the
distribution of all households over the range of wealth~-- or,  
equivalently, over the range of consumption~-- other than its mean.
The approach developed in this \chptr\  shows that this is not the
case: even if the state transition
assignments in the right side of 
\ref{main-q}-\eqref{zze666} were to depend only on the mean value of
the distribution $\csd$, this relation 
would still provide the exact law of motion of the full population
distribution. For~the sake of  
simplicity, in what follows we do not pursue this law of motion in its
full generality, but nevertheless 
derive the exact law 
of motion of the vector of conditional population means attached to
every employment group 
(in effect, reducing the population distribution to a vector of mean
values, rather than a single 
mean value)~-- see \eqref{no-fp-2-0} below.
Most important, the dynamics of this vector
reveal substantial fluctuations in the wealth-inequality across the population
that are not possible to capture if the model remains
confined to the total population mean alone~-- this feature is
illustrated in Fig.~\ref{fgKS5} below.  

There is one subtle~-- or not so subtle, depending on the point of
view~-- aspect of the claim ``only the 
mean matters,'' which is commonly swept under the proverbial ``rug,'' although the
work \cite{KS98} is almost explicit about it: the individual
investment-consumption decisions during the present period depend not
only on the present mean 
and present productivity state, but also on the transport of the present mean into the future,
conditioned on the realized future productivity state of the economy. To put it another way, the
claim ``only the mean matters'' is in fact the claim that
``only the mean and its transport matter,'' which then implies~-- as
we have seen before~-- that the transport must be self-consistent.
\footnote{The search for parameters in the log-linear  AR(1) predictor  that maximize the goodness
of fit in Krusell-Smith's method can be seen as a way to bring the transport as close as possible to
being self-consistent~-- in the long run.}
Such features are  less apparent in models with
infinite time horizon~-- see \cite{KS98}, for example~-- but this is not the approach adopted here:
we insist that infinite time horizon models are to be understood only as limits of models with
finite time horizon, which must be developed first.
The reason for this (admittedly, not very
common) point of view transcends the obvious need for methodological coherence: it~will be shown below that
even in the classical model, borrowed here from \cite{KS98},
with only~2 employment and~2
productivity states, the economy typically needs to run for 
hundreds of periods before it can achieve its time-invariant regime,
whereas no real-world economy can exist unchanged for that long~-- whence the need for a methodology that does
not resort on time invariance. In~the illustrations that follow in this section the infinite time
horizon scenario is pursued mainly for the purpose of benchmarking and aims to demonstrate that the
finite time horizon methodology developed in this \chptr\  is consistent with previously known
results.

Consistent with the general model introduced in
Sec.~\ref{sec:gen-model}, we set
$\q_{t,x,\csd,u}(c)\equiv\nobreak 0$ (the agents do not invest in a risk-free asset),
ignore the 
first kernel condition in \ref{cross-sys}-(\ref{zze5a}), and ignore the first market
clearing condition in 
\ref{cross-sys}-(\ref{ze9}). The cross-sectional distribution of the population
enters the model 
only through the market clearing condition and also through the transport
equation \ref{main-q}-\eqref{zze666}. 
To develop a better grasp of its rôle, we now turn to the second market clearing condition
in \ref{cross-sys}-(\ref{ze9}) 
and introduce the vector of conditional (to employment) mean values, namely
\footnote{{}The obvious dependence between the symbols $A$ and $\csd$ will be suppressed in the notation
for simplicity.}
%
%
{\abovedisplayskip=5pt plus 1.5pt minus 1.5pt\belowdisplayskip=8pt plus 1.5pt minus 1.5pt\belowdisplayshortskip=6pt plus 1.5pt minus 1.5pt
\[
A\df (A^{u},u\in\EEE)\in\R^{\abs{\EEE}}\,,\quad \text{where}\quad 
A^{u}\df \int_0^\infty c \d\csd^{u}(c)\,,\quad \csd\in\DB^\EEE\,. 
\]}%
If the cross-sectional distribution $\csd$ affects the model only through the associated vector
$A$, then  the left side of \ref{cross-sys}-\eqref{ze9} must be 
a function of that vector alone. In~that case, if all portfolio 
mappings $\qq_{t,x,F,u}\phd$ happen to be continuous, by Stroock's
argument \cite{Str23} 
the left side can only be of the form
{\abovedisplayskip=7pt plus 1.5pt minus 1.5pt\belowdisplayskip=8pt plus 1.5pt minus 1.5pt\belowdisplayshortskip=7pt plus 1.0pt minus 1.5pt
\[
\R^{\abs{\EEE}}\ni A \leadsto  \sum\nolimits_{u\tts\in\tts\EEE}\pi_x(u)\qq_{t,x,\csd,u}(A^u)
\]
}
and, in addition, all portfolio mappings $\qq_{t,x,F,u}\phd$, $u\in\EEE$,
must be affine functions. It is clear from the above expression that the
population distribution affects the market clearing only through its total (unconditional to employment) mean value
$A^* = \sum_{u\tts\in\tts\EEE}\pi_x(u) A^u$ precisely when the slopes of the affine
functions $\qq_{t,x,F,u}\phd$ do not depend on the employment state $u\in\EEE$.
It~will be shown below that such an
arrangement is indeed possible~-- as a reasonably good approximation~-- if all
consumption levels across the population are sufficiently large.

In general, the households' demand for capital cannot be an affine function of  the  consumption
level~$c$, since the latter affects the kernel condition in \ref{cross-sys}-(\ref{zze5a})
in nonlinear fashion.
In~order to uncover the way in which this kernel condition affects the structure of the portfolio
mappings and identify conditions under which the population distribution $\csd\in\DB^\EEE$ affects,
at least as an approximation, 
the model only through the vector of its employment specific  mean values $A\in\R^{\abs{\EEE}}$~--
or, as a special case, only through the total population mean
$A^* = \sum_{u\tts\in\tts\EEE}\pi_x(u) A^u$~-- 
let us suppose, contrary
to fact, that all portfolio and  
future consumption mappings have the affine structure
(the dependence on $\csd\in\DB^\EEE$ is now collapsed to dependence on the associated vector of
conditional means $A\in\R^\abs{\EEE}$)
{\abovedisplayskip=8pt plus 2.5pt minus 1.5pt\belowdisplayskip=8pt plus 2.5pt minus 1.5pt\belowdisplayshortskip=8pt plus 1.0pt minus 1.5pt
\begin{equation}\label{lin-subst}
\begin{gathered}
\qq_{t,\tts x,\tts A,\tts u}(c) =
a_{t,\tts x,\tts  A,\tts u}+b_{t,\tts x,\tts A,\tts u}\times c\\
\text{and}\quad \pfc_{t,\tts x,\tts A}^{ y , v }(u,c) = g_{t,\tts x,\tts  A,\tts u}^{ y , v }
+ h_{t,\tts x,\tts  A,\tts u}^{ y , v }\times c\,,\quad c\in\Rpp\,,
\end{gathered}
\end{equation}}%
for some yet to be determined coefficients
{\abovedisplayskip=8pt plus 2.5pt minus 1.5pt\belowdisplayskip=8pt plus 2.5pt minus 1.5pt\belowdisplayshortskip=8pt plus 1.0pt minus 1.5pt
\begin{equation}\label{4coef}
a_{t,\tts x,\tts A,\tts u}\,, \ \ b_{t,\tts x,\tts A,\tts u}\,, \ \ 
g_{t,\tts x,\tts  A,\tts u}^{ y , v }\,, \ \ \text{and} \ \ h_{t,\tts x,\tts  A,\tts u}^{ y , v }\,.
\end{equation}
}%
With the choice just made and
with risk aversion parameter of $R=1$
(same as in the benchmark case study of \cite{KS98}), the second kernel
condition in \ref{cross-sys}-(\ref{zze5a}) can be cast as
{\abovedisplayskip=8pt plus 1.5pt minus 1.5pt\belowdisplayskip=8pt plus 1.5pt minus 1.5pt\belowdisplayshortskip=8pt plus 1.5pt minus 1.5pt
\begin{equation}\label{only-q-bn2-2}
\begin{multlined}
1 = \b\,\sum_{ y \tts \in\tts\XXX, v\tts\in\tts\EEE}
{ 1 \over g_{t,\tts x,\tts A,\tts u}^{ y,\tts v}/c  + h_{t,\tts x,\tts
A,\tts u}^{ y ,\tts v}}
\times\Bigl(\rho_ y \bigl(K_{t}(x, A)\bigr) + 1-\dd \Bigr)\\
\times  Q(x, y )\,  P_{x,\tts y }( u ,v)\,.
\end{multlined}
\end{equation}}%
With the substitution \eqref{lin-subst} in mind, the balance equation 
in \ref{cross-sys}-(\ref{zze5xb}) becomes (the transport of the distribution $\csd$ is now transport of
the associated vector $A$~-- see \eqref{no-fp-2-0} below)
{\abovedisplayskip=5pt plus 1.5pt minus 1.5pt\belowdisplayskip=5pt plus 1.5pt minus 1.5pt\belowdisplayshortskip=5pt plus 1.5pt minus 1.5pt
\begin{equation}\label{e10-bn3}
\begin{aligned}
&({ {a_{t,\tts x,\tts A,\tts u}}}+{ {b_{t,\tts x,\tts A,\tts u}}}\times c)\times
\Bigl(\rho_ y \bigl({ {K_{t}(x, A)}}\bigr) +1-\dd\Bigr) 
+ v\times \ee_ y \bigl({ {K_{t}(x, A)}}\bigr)\hbox to1.5cm{\hfil}\\
&\hbox to0.5cm{\hfil} = \bigl({ {g_{t,\tts x,\tts A,\tts u}^{ y ,v}}} + { {h_{t,\tts x,\tts A,\tts u}^{ y ,v}}}\times c\bigr)\\
&\hbox to1.5cm{\hfil}
+ \Bigl({{a_{t+1,\tts y,\tts\T_{t,x}^y( A),\tts v}}}
+{{b_{t+1,\tts y,\tts \T_{t,x}^y( A),\tts v}}}\times\bigr({ {g_{t,\tts x,\tts A,\tts u}^{ y ,v}}}
+ { {h_{t,\tts x,\tts A,\tts u}^{ y ,v}}}\times c\bigr)\Bigr)\,.
\end{aligned}
\end{equation}}%
Just as before, the strategy is to take the future portfolio mappings encrypted in the pairs
{\abovedisplayskip=5pt plus 1.5pt minus 1.5pt\belowdisplayskip=5pt
plus 1.5pt minus 1.5pt\belowdisplayshortskip=5pt plus 1.5pt minus 1.5pt
$$
\bigl(a_{t+1,\tts y ,\tts \T_{t,x}^y( A),\tts v},b_{t+1,\tts y ,\tts  \T_{t,x}^y( A),\tts v}\bigr)\,,\quad
y\in\XXX\,,\ \ v\in\EEE\,,
$$
}%
as
given and treat \eqref{e10-bn3} as a system for the unknowns (present portfolios and future
consumption mappings) $a_{t,\tts x,\tts A,\tts u}$, $b_{t,\tts x,\tts A,\tts u}$,
$g_{t,\tts x,\tts A,\tts u}^{ y ,v}$, and $h_{t,\tts x,\tts A,\tts u}^{ y ,v}$, $x, y \in\XXX$,
$u, v\in\EEE$.
Thus, for every fixed  period-$t$ state of the economy $(x,A)\in\XXX\times\R^{\abs{\EEE}}$,
there is a total of $2\abs{\EEE}+2\abs{\EEE}^2\abs{\XXX}$ unknowns to solve for,
provided that an ansatz choice for the average capital $K_{t}(x, A)$ is somehow
made (the market clearing will be addressed later).
The next step is to
extract the same number of equations from \eqref{only-q-bn2-2} and \eqref{e10-bn3}. This task is
non-trivial since \eqref{only-q-bn2-2} and \eqref{e10-bn3}
are, in fact, systems of infinitely many equations~-- one for every
$c\in\Rpp$.  
Treated as an identity between two polynomials of
degree~$1$ over the variable $c\in\Rpp$, \eqref{e10-bn3} can be split into two systems:
\begin{subequations}\label{lin-split}
{\abovedisplayskip=5pt plus 1.5pt minus 1.5pt\belowdisplayskip=5pt plus 1.5pt minus 1.5pt\belowdisplayshortskip=5pt plus 1.5pt minus 1.5pt
\begin{gather}
\begin{aligned}\label{lin-split-a}
&\hbox to0.5cm{\hfill}{ {b_{t,\tts x,\tts A,\tts u}}}\times
\Bigl(\rho_ y \bigl({ {K_{t}(x, A)}}\bigr) +1-\dd\Bigr) 
= { {h_{t,\tts x,\tts A,\tts u}^{ y ,v}}}
+ {{b_{t+1,\tts y ,\tts \T_{t,x}^y(A),\tts v}}}\times{ {h_{t,\tts x,\tts A,\tts u}^{ y ,v}}}\\
&\hbox to5.0cm{\hfill}\text{for all }\ \  y \in\XXX\,,\  u, v\in\EEE\,.
\end{aligned}%
\nobreak\\
\noalign{\rlap{\smash{and}}}
\begin{aligned}\label{lin-split-b}
&{ {a_{t,\tts x,\tts A,\tts u}}}\times  
\Bigl(\rho_ y \bigl({ {K_{t}(x, A)}}\bigr) +1-\dd\Bigr) 
+v \times \ee_ y \bigl({ {K_{t}(x, A)}}\bigr)\\
&\hbox to3.0cm{\hfil}= {{a_{t+1,\tts y ,\tts \T_{t,x}^y( A),\tts v}}}+\bigl(1+{{b_{t+1,\tts y
,\tts \T_{t,x}^y( A),\tts v}}}\bigr)\times { {g_{t,\tts x,\tts A,\tts u}^{ y ,v}}}\\
&\hbox to6.5cm{\hfil}\quad\text{for all }\ \  y \in\XXX\,,\  u, v\in\EEE\,,
\end{aligned}
\end{gather}}%
\end{subequations}
Each system in \eqref{lin-split} provides $\abs{\EEE}^2\abs{\XXX}$ equations and solving both
guarantees that \eqref{e10-bn3} holds exactly for every $c\in\Rpp$.
Thus, $2\abs{\EEE}$ additional
equations, that can only come from \eqref{only-q-bn2-2}, are needed.
Unfortunately, as written (with the stipulated affine structure in mind), it is not
possible to enforce the kernel condition \eqref{only-q-bn2-2} exactly for every $c\in\Rpp$, and this
is the main reason why the affine structure imposed in \eqref{lin-subst}~-- that is to say, the
stipulation that ``only the population mean matters''~-- can hold only as an approximation.
Hence, estimating the accuracy of the approximate aggregation point of
view comes down to estimating the
deviation of the right side in \eqref{only-q-bn2-2} from the constant $1$~-- see
\ref{mod-accu} below.
One approximation of \eqref{only-q-bn2-2} that immediately comes to mind, and
provides exactly $2\abs{\EEE}$ additional equations, is replacing the right side with
its first-order 
Taylor expansion over the variable ${1\over c}$. As we are about to see, the easiest
such expansion 
is around ${1\over c}=0$. This is quite intuitive: in a neighborhood of $c=\infty$
the right side is 
nearly invariant under the choice of $c$, so that the first order Taylor
approximation should be 
quite accurate.  Coincidentally, this choice will turn out to be consistent with the
approximate aggregation hypothesis~-- see below. For every fixed
$(x,A)\in\XXX\times\R^{\abs{\EEE}}$, the first-order Taylor expansion around ${1\over c}=0$
transforms \eqref{only-q-bn2-2} into the following two systems:
\begin{subequations}\label{Taylor}
{\abovedisplayskip=6pt plus 1.5pt minus 1.5pt\belowdisplayskip=6pt plus 1.5pt minus 1.5pt\belowdisplayshortskip=6pt plus 1.5pt minus 1.5pt
\begin{gather}\label{Taylor-a}
\begin{multlined}
1 = \b\,\sum_{ y \tts \in\tts\XXX, v\tts\in\tts\EEE}
{ 1 \over h_{t,\tts x,\tts A,\tts u}^{ y ,\tts v}}\times
\Bigl(\rho_ y \bigl(K_{t}(x, A)\bigr) + 1-\dd \Bigr)\\
\hbox to5cm{\hfill}\times  Q(x, y )\,  P_{x,\tts y }( u ,v)\,,\quad u\in\EEE\,,
\end{multlined}\\
\noalign{\rlap{{and}}}
\label{Taylor-b}
\begin{multlined}
0 = \sum_{ y \tts \in\tts\XXX, v\tts\in\tts\EEE}
{ 1 \over (h_{t,\tts x,\tts A,\tts u}^{ y ,\tts v})^2}
\times g_{t,\tts x,\tts A,\tts u}^{ y,\tts v} \times
\Bigl(\rho_ y \bigl(K_{t}(x, A)\bigr) + 1-\dd \Bigr)\\
\hbox to5cm{\hfill} \times  Q(x, y )\,  P_{x,\tts y }( u ,v)\,,\quad u\in\EEE\,.
\end{multlined}
\end{gather}
}%
\end{subequations}
Keeping in mind that
${a_{t+1,\tts y ,\tts \T_{t,x}^y( A),\tts v}}$ and ${b_{t+1,\tts y ,\tts \T_{t,x}^y( A),\tts v}}$
are treated as given,
\eqref{lin-split-a} and \eqref{Taylor-a} provide a closed system of 
$\abs{\EEE}+\abs{\EEE}^2\abs{\XXX}$ equations for the same number of
unknowns, namely
{\abovedisplayskip=5pt plus 1.5pt minus 1.5pt\belowdisplayskip=5pt plus 1.5pt minus 1.5pt\belowdisplayshortskip=5pt plus 1.0pt minus 1.5pt
\[
{ {b_{t,\tts x,\tts A,\tts u}}}\,, \ { {h_{t,\tts x,\tts A,\tts u}^{ y ,v}}}\,,\quad
 u, v\in\EEE\,,\  y \in\XXX\,.
\]}%
This system simplifies substantially, once it is observed that the dependence of the unknowns on
the employment states $u,v\in\EEE$ can be suppressed~-- coincidentally,
this is precisely the arrangement that one needs in order to proclaim that the state variable
$A\in\R^{\EEE}$ can be collapsed to the total population mean $A^*
= \sum_{u\tts\in\tts\EEE}\pi_x(u) A^u$ alone.
Indeed, with $h_{t,\tts x,\tts A,\tts u}^{ y ,v}\equiv h_{t,\tts x,\tts A}^{ y}$ the
system \eqref{Taylor-a} collapses to the single equation
\begin{subequations}\label{Taylor-x}
{\abovedisplayskip=5pt plus 1.5pt minus 1.5pt\belowdisplayskip=5pt plus 1.5pt minus 1.5pt\belowdisplayshortskip=5pt plus 1.5pt minus 1.5pt
\begin{equation}\label{Taylor-x-a}
1=\b\,\sum\nolimits_{ y \tts \in\tts\XXX}
{ 1 \over h_{t,\tts x,\tts A}^{ y}}\times
\Bigl(\rho_ y \bigl(K_{t}(x, A)\bigr) + 1-\dd \Bigr)
\times  Q(x, y )
\end{equation}
}%
and with $b_{t,\tts x,\tts A,\tts u}\equiv b_{t,\tts x,\tts A}$ \eqref{lin-split-a}
collapses to the system of only $\abs{\XXX}$ equations
{\abovedisplayskip=5pt plus 1.5pt minus 1.5pt\belowdisplayskip=5pt plus 1.5pt minus 1.5pt\belowdisplayshortskip=5pt plus 1.5pt minus 1.5pt
\begin{equation}\label{Taylor-x-b}
{ {b_{t,\tts x,\tts A}}}\times
\Bigl(\rho_ y \bigl({ {K_{t}(x, A)}}\bigr) +1-\dd\Bigr) 
= { {h_{t,\tts x,\tts A}^{ y }}}
+ {{b_{t+1,\tts y ,\tts \T_{t,x}^y(A)}}}
\times{ {h_{t,\tts x,\tts A}^{ y}}}\,,\quad y\in\XXX\,,
\end{equation}
}%
\end{subequations}
for the total of $1+\abs{\XXX}$ unknowns, namely $b_{t,\tts x,\tts A}$ and $h_{t,\tts x,\tts A}^{ y}$,
$y\in\XXX$ (recall that the state $(x,A)$ is fixed). Remarkably, the system  \eqref{Taylor-x}
admits a closed-form solution~-- see \ref{closed-form} below. In~any
case, with a closed-form
solution or without, once the unknowns $b_{t,\tts x,\tts A,\tts u}\equiv b_{t,\tts x,\tts A}$ and
$h_{t,\tts x,\tts A,\tts u}^{ y,\tts v}\equiv h_{t,\tts x,\tts A}^{ y}$,
$y\in\XXX$  are solved for from \eqref{Taylor-x},
the system composed of \eqref{lin-split-b} and \eqref{Taylor-b} would provide a linear system
of $\abs{\EEE}+\abs{\EEE}^2\abs{\XXX}$ equations for the same number of
unknowns, namely
$
{ {a_{t,\tts x,\tts A,\tts u}}}\,, \ { {g_{t,\tts x,\tts A,\tts u}^{ y ,v}}}\,,$
$ u, v\in\EEE\,,\  y \in\XXX\,.$
If~a solution for the unknowns
{\abovedisplayskip=8pt plus 2.5pt minus 1.5pt\belowdisplayskip=8pt plus 2.5pt minus 1.5pt\belowdisplayshortskip=5pt plus 1.5pt minus 1.5pt
\begin{equation}\label{unkn2}
a_{t,\tts x,\tts A,\tts u}\,, \ b_{t,\tts x,\tts A}\,, \ g_{t,\tts x,\tts A,\tts u}^{ y ,v}\,,
 \ h_{t,\tts x,\tts  A}^{ y}\,,\quad  u, v\in\EEE\,,\  y \in\XXX\,,
\end{equation}
}%
can indeed be found (for a fixed state $(x,A)$) as described,
the ansatz choice for ${ {K_{t}(x, A)}}$ can then be tested
with  the market clearing condition 
{\abovedisplayskip=8pt plus 1.5pt minus 1.5pt\belowdisplayskip=8pt plus 1.5pt minus 1.5pt\belowdisplayshortskip=6pt plus 1.5pt minus 1.5pt
\begin{equation}\label{mclr-lin-a}
\sum\nolimits_{u\ts\in\ts\EEE}\gp_x(u)\bigl(a_{t,\tts x,\tts  A,\tts u}
+b_{t,\tts x,\tts A} A^u\bigr)={ {K_{t}(x, A)}}\,.
\end{equation}
}%
If the test fails, then the value for ${ {K_{t}(x, A)}}$ will need to be adjusted
accordingly and the procedure will need to be repeated until the last relation becomes
numerically acceptable. 
Due to the affine structure imposed on the conditional transition assignments, the transport encrypted in
\ref{main-q}-(\ref{zze666}) can now be stated as transport of the vector of employment means $A$
in the form
\footnote{{}The change of variables formula gives:
$\displaystyle \int \a \d \csd^u(\,\hat\pfc_{t,\tts x,\tts A}^{ y,\tts v}(u,\a))
= \int \pfc_{t,\tts x,\tts A}^{ y ,v}(u,\tts\a) \d \csd^{u}(\a)\,.$}
{\abovedisplayskip=8pt plus 1.5pt minus 1.5pt\belowdisplayskip=8pt plus 1.5pt minus 1.5pt\belowdisplayshortskip=5pt plus 1.5pt minus 1.5pt
\begin{equation}\label{no-fp-2-0}
\begin{multlined}
\T_{t,x}^y( A)^v
= \sum\nolimits_{u\tts\in\tts\EEE}
{\gp_x(u)  P_{x, y }( u,v)\over \gp_ y (v)}\,\Bigl(g_{t,\tts x,\tts A,\tts u}^{ y ,v} + 
h_{t,\tts x,\tts  A,\tts u}^{y,v}\times  A^u\Bigr)\\
\text{for all }\ x, y \in\XXX\,,\  v\in\EEE\,,
\end{multlined}
\end{equation}}%
where the left side is understood to be the mean of the distribution $\T_{t,x}^y(\csd)^v\in\DB$.
Hence, in this reduced (due to the affine structure)
model the transport operator $\T_{t,x}^ y $ introduced in \ref{et} now maps
$(\R_{++})^{\abs{\EEE}}$ into $(\R_{++})^{\abs{\EEE}}$, as opposed to mapping $\bbF^\EEE$
into~$\bbF^\EEE$. We again stress that the transport in \eqref{no-fp-2-0} is meaningful only in 
conjunction with the system composed of \eqref{lin-split}, \eqref{Taylor} and \eqref{mclr-lin-a}~-- and the
system composed of \eqref{lin-split}, \eqref{Taylor} and \eqref{mclr-lin-a} depends on the transport
in \eqref{no-fp-2-0}~-- so that the system attached to period~$t$ is composed of all 4 relations
\eqref{lin-split}, \eqref{Taylor}, \eqref{mclr-lin-a} and \eqref{no-fp-2-0}.

It is important to recognize that, as long as the slopes
$b_{t,\tts x,\tts A,\tts u}\equiv b_{t,\tts x,\tts A}$ and
$h_{t,\tts x,\tts A,\tts u}^{ y,\tts v}\equiv h_{t,\tts x,\tts A}^{ y}$ can be chosen to be
invariant to the choice of the employment states $u,v\in\EEE$, then \eqref{mclr-lin-a} can be cast as
{\abovedisplayskip=8pt plus 1.5pt minus 1.5pt\belowdisplayskip=8pt plus 1.5pt minus 1.5pt\belowdisplayshortskip=8pt plus 1.5pt minus 1.5pt
\begin{equation}\label{mclr-lin-b}
b_{t,\tts x,\tts A}\times A^*+\sum\nolimits_{u\ts\in\ts\EEE}\gp_x(u) a_{t,\tts x,\tts  A,\tts u} = {{K_{t}(x, A)}}
\end{equation}
}%
where $A^* = \sum_{u\tts\in\tts\EEE}\pi_x(u) A^u$ is the total population mean,  
and \eqref{no-fp-2-0} can be stated as
{\abovedisplayskip=8pt plus 3.5pt minus 1.5pt\belowdisplayskip=8pt plus 3.5pt minus 1.5pt\belowdisplayshortskip=7pt plus 3.5pt minus 1.5pt
\begin{equation}\label{no-fp-x}
\begin{multlined}
\T_{t,x}^y( A^*)\df\sum\nolimits_{v\tts\in\tts\EEE}\gp_y(v)
\times\T_{t,x}^y( A)^v\\
\hbox to3cm{\hfill}= h_{t,\tts x,\tts A}^y\times  A^*
+ \sum\nolimits_{u,\tts v\tts \in\tts \EEE}
{\gp_x(u)\times  P_{x, y }( u,v)}\times g_{t,\tts x,\tts A,\tts u}^{ y ,v}
\end{multlined}
\end{equation}
}%
for every $x,y\in\XXX$. 

\begin{nit}{Collapse of the state space revisited}\label{collapse}%
The main corollary from \eqref{mclr-lin-b} and \eqref{no-fp-x} is that, 
if the slopes $b_{t,\tts x,\tts A,\tts u}$ and $h_{t,\tts x,\tts A,\tts u}^{ y,\tts v}$ can be
chosen to be
invariant to the choice of the employment states $u,v\in\EEE$, then the dependence on the state variable
$A\in\R^\abs{\EEE}$ collapses to dependence only on the total population mean
$A^*= \sum_{u\tts\in\tts\EEE}\pi_x(u) A^u$, i.e., the period-$t$ state of the economy can be expressed
as $(x,A^*)\in\XXX\times\Rpp$, the unknowns in \eqref{unkn2} can be written as
{\abovedisplayskip=8pt plus 2.5pt minus 2.5pt\belowdisplayskip=8pt plus 2.5pt minus 2.5pt\belowdisplayshortskip=5pt plus 1.5pt minus 1.5pt
\[
a_{t,\tts x,\tts A^*,\tts u}\,, \ b_{t,\tts x,\tts A^*}\,, \ g_{t,\tts x,\tts A^*,\tts u}^{ y ,\tts v}\,,
 \ h_{t,\tts x,\tts  A^*}^{ y }\,,\quad  u, v\in\EEE\,,\  y \in\XXX\,,
\]
}%
and, consequently, the average installed capital $K_t(x,A)$ can be cast as $K_t(x,A^*)$.
As a result, all variables that define the equilibrium can be written in terms of the total
population mean~$A^*$.
In particular, in both \eqref{mclr-lin-b} and \eqref{no-fp-x} the symbol $A$ can be
replaced everywhere 
with~$A^*$. 
While the idea of a ``collapsed state space'' is nothing new~-- see~\cite{KS98}, for
example~-- the present 
analysis allows one to identify the main source of this feature: it comes from 
the first order Taylor approximation of the kernel
condition around ${1\over c}=0$, i.e., $c=\infty$, and from the resulting decoupling
of the system  
composed of \eqref{lin-split} and \eqref{Taylor} into two sub-systems, one of which
is self-contained.  
This observation makes it possible to quantify the error introduced by collapsing the
state space~--  
see~\ref{mod-accu} below. In addition, as we are about to see, it becomes possible to
develop a new 
computational strategy, which does not involve simulation and is meaningful for any,
small or large, 
time-horizon; in particular, it provides closed-form analytic expressions for the slopes
$b_{t,\tts x,\tts A^*}$ and $h_{t,\tts x,\tts A^*}^{ y}$, and, most important,
provides the exact form 
of the law of motion of the total population mean~$A^*$~-- see~\eqref{no-fp-x}. 
Another interesting consequence from the approach developed in this section is that,
even as the 
state variable $A\in\R^\abs{\EEE}$ collapses to the scalar $A^*\in\Rpp$, it is still
possible to 
describe exactly the dynamics of the vector of employment-specific mean
values~$A$. Indeed, for that purpose one 
merely needs to replace in the right side of~\eqref{no-fp-2-0} the coefficients
$g_{t,\tts x,\tts A,\tts u}^{ y ,\tts v}$ and $h_{t,\tts x,\tts  A,\tts u}^{ y ,\tts v}$ with,
respectively, 
$g_{t,\tts x,\tts A^*,\tts u}^{ y ,\tts v}$ and $h_{t,\tts x,\tts  A^*}^{ y}$.
The main message here is that the collapse of the state space does not imply that the
law of motion 
of the population distribution could be chosen arbitrarily, as long as its mean
complies with the 
law of motion of the mean; in fact, the law of motion of the population distribution
is fixed, once 
the law of motion of the mean is fixed.\qed
\end{nit}

\begin{nit}{A more realistic endogenous state variable}\label{beyond}%
The choice of the abscissa ${1\over c}=0$ for the Taylor expansion in the kernel condition
implies that equilibrium is sought exclusively
for relatively large consumption levels. Altho\-ugh this approach provides reasonably satisfactory
results in the examples given later in this section, it still leaves something to be desired. Another~--
perhaps somewhat more intuitive, depending on the point of view~-- approach is to develop
first-order Taylor expansion in the right side of \eqref{only-q-bn2-2} around
the abscissa ${1\over c}={1\over A^u}$. The idea is to ensure that the kernel condition is exact for
households that consume exactly at the mean of their employment group and is asymptotically exact
for households with consumption levels that are not very far from their group mean. One drawback
from this approach is that the decoupling of the system as in \eqref{lin-split}
and \eqref{Taylor} is no longer possible. In particular, the endogenous variable that
captures the state of the population will have to be taken to be
the entire vector of group-specific mean values
$A=(A^u,\,u\in\EEE)$ and can no longer be collapsed to the total mean value
$A^*= \sum_{u\tts\in\tts\EEE}\pi_x(u) A^u$ alone. In addition, one would be forced to seek a
numerical solution to a nonlinear system with twice as many equations (because the closed-form
solution for half of them will no longer be available). We are not going to pursue this
program for two main reasons. First, methodologically such a program would differ from the one carried out below
only in the increased computational complexity. Second, our primary objective here is
to benchmark the 
methodology developed in this \chptr\  to widely cited methods and results, most of
which are based on the 
infinite time horizon and what Krusell and Smith \cite{KS98} call ``approximate
aggregation'' point 
of view.

One possible arrangement that would still allow one to use the total population mean
$A^*$ as an 
endogenous state variable is to consider Taylor expansion around the abscissa
${1\over c}={1\over 
A^*}$. One major objection to this approach is that it ignores the variations in the
group-specific 
mean values, which are illustrated in Fig.~\ref{fgKS5} below.\qed
\end{nit}

The model reduction brought by the affine structure introduced in \eqref{lin-subst}
leads to some useful closed-form expressions for the slopes
$b_{t,\tts x,\tts A^*,\tts u}\equiv b_{t,\tts x,\tts A^*}$
$h_{t,\tts x,\tts  A^*,\tts u}^{y,v}\equiv h_{t,\tts x,\tts  A^*}^{y}$ which are developed next. 
As was noted earlier, we seek a solution for the slopes in the portfolios and the future
consumption assignments that do not depend on the state of employment, i.e., rely on Taylor's expansion
in the kernel condition around the abscissa ${1\over c}=0$. This allows us to
use \eqref{lin-split}, \eqref{Taylor}, \eqref{mclr-lin-b} and \eqref{no-fp-x} with the symbol $A$
replaced everywhere with $A^*$ and with the dependence of the slopes on the state of employment suppressed.
For the sake of better readability, we restate (for a fixed aggregate state
$(x,A^*)\in\XXX\times\Rpp$) these relations as
%
%
%
{\abovedisplayskip=5pt plus 1.5pt minus 1.5pt\belowdisplayskip=8pt plus 1.5pt minus 1.5pt\belowdisplayshortskip=8pt plus 1.5pt minus 1.5pt
\begin{equation}\label{eq-b-1}
{ {b_{t,\tts x,\tts A^*}}}\times 
\Bigl(\rho_ y \bigl({ {K_{t}(x, A^*)}}\bigr) +1-\dd\Bigr) 
= { {h_{t,\tts x,\tts A^*}^{ y}}}
+ {{b_{t+1,\tts y ,\tts \T_{t,x}^y(A^*)}}}\times{ {h_{t,\tts x,\tts A^*}^{ y }}}\,,\quad
y \in\XXX\,,
\end{equation}
}%
{\abovedisplayskip=5pt plus 1.5pt minus 1.5pt\belowdisplayskip=8pt plus 1.5pt minus 1.5pt\belowdisplayshortskip=8pt plus 1.5pt minus 1.5pt
\begin{equation}\label{eq-b-0}
\begin{aligned}
&{ {a_{t,\tts x,\tts A^*,\tts u}}}\times  
\Bigl(\rho_ y \bigl({ {K_{t}(x, A^*)}}\bigr) +1-\dd\Bigr) 
+v \times \ee_ y \bigl({ {K_{t}(x, A^*)}}\bigr)\\
&\hbox to2.5cm{\hfil}= {{a_{t+1,\tts y ,\tts \T_{t,x}^y( A^*),\tts v}}}+\bigl(1+{{b_{t+1,\tts y
,\tts \T_{t,x}^y( A^*)}}}\bigr)\times { {g_{t,\tts x,\tts A^*,\tts u}^{ y ,v}}}\,,\\
&\hbox to7.5cm{\hfill}y \in\XXX\,,\  u, v\in\EEE\,,
\end{aligned}
\end{equation}
}%
{\abovedisplayskip=5pt plus 1.5pt minus 1.5pt\belowdisplayskip=8pt plus 1.5pt minus 1.5pt\belowdisplayshortskip=8pt plus 1.5pt minus 1.5pt
\begin{equation}\label{eq-n-1}
1 = \b\,\sum\nolimits_{ y \tts \in\tts\XXX}
{ 1 \over h_{t,\tts x,\tts A^*}^{ y }}\times
\Bigl(\rho_ y \bigl(K_{t}(x, A^*)\bigr) + 1-\dd \Bigr)
\times  Q(x, y )\,,\quad y\in\XXX\,,
\end{equation}
}%
{\abovedisplayskip=5pt plus 1.5pt minus 1.5pt\belowdisplayskip=8pt plus 1.5pt minus 1.5pt\belowdisplayshortskip=8pt plus 1.5pt minus 1.5pt
\begin{equation}\label{eq-n-0}
\begin{multlined}
0 = \sum_{ y \tts \in\tts\XXX, v\tts\in\tts\EEE}
{ -\b \over (h_{t,\tts x,\tts A^*}^{ y })^2}
\times g_{t,\tts x,\tts A^*,\tts u}^{ y,\tts v} \times
\Bigl(\rho_ y \bigl(K_{t}(x, A^*)\bigr) + 1-\dd \Bigr)\\
\times  Q(x, y )\,  P_{x,\tts y }( u ,v)\,,\quad  u\in\EEE\,,
\end{multlined}
\end{equation}
}%
{\abovedisplayskip=5pt plus 1.5pt minus 1.5pt\belowdisplayskip=8pt plus 1.5pt minus 1.5pt\belowdisplayshortskip=8pt plus 1.5pt minus 1.5pt
\begin{equation}\label{eq-mclr}
b_{t,\tts x,\tts A^*}\times A^*+\sum\nolimits_{u\ts\in\ts\EEE}\gp_x(u) a_{t,\tts x,\tts  A^*,\tts u}
= {{K_{t}(x, A^*)}}\,,
\end{equation}
}%
and       
{\abovedisplayskip=5pt plus 1.5pt minus 1.5pt\belowdisplayskip=5pt plus 1.5pt minus 1.5pt\belowdisplayshortskip=8pt plus 1.5pt minus 1.5pt
\begin{equation}\label{eq-T-star}
\T_{t,x}^y( A^*)
= h_{t,\tts x,\tts A^*}^y\times  A^* + \sum\nolimits_{u,\tts v\tts \in\tts \EEE}
{\gp_x(u)\times  P_{x, y }( u,v)}\times g_{t,\tts x,\tts A^*,\tts u}^{ y ,\tts v}\,,\quad y\in\XXX\,.
\end{equation}
}%
\leavevmode\vbox to 0pt{\vfil}

\begin{nit}{Closed form solution for the slopes}\label{closed-form}%
As no investment takes place in period $T$, with  $t=T-1$ one must have
$a_{t+1,\tts y,\tts \T_{t,x}^y( A^*),\tts v}=0$ and $b_{t+1,\tts y,\tts \T_{t,x}^y( A^*)}=0$,
so that \eqref{eq-b-1} becomes
{\abovedisplayskip=5pt plus 1.5pt minus 1.5pt\belowdisplayskip=5pt plus 1.5pt minus 1.5pt\belowdisplayshortskip=5pt plus 1.5pt minus 1.5pt
\[
h_{t,\tts x,\tts  A^*}^{y} = b_{t,\tts x,\tts A^*} \times
\Bigl(\rho_ y \bigl({ {K_{t}(x, A^*)}}\bigr) +1-\dd\Bigr)\,,\quad  x,\,y \in\XXX\,,\  u,\, v\in\EEE\,.
\]}%
Consequently, \eqref{eq-n-1} gives 
{\abovedisplayskip=8pt plus 1.5pt minus 1.5pt\belowdisplayskip=8pt plus 1.5pt minus 1.5pt\belowdisplayshortskip=6pt plus 1.5pt minus 1.5pt 
\[
\begin{multlined}
1 = { \b \over { {b_{t,\tts x,\tts A^*}}}}\sum\nolimits_{ y \,\in\,\XXX}
Q(x, y )\sum\nolimits_{ v\,\in\,\EEE} P_{x, y }( u ,v)\\
\hbox to5cm{\hfill}={ \b \over { {b_{t,\tts x,\tts A^*}}}}\sum\nolimits_{ y\,\in\,\XXX} Q(x, y )
={ \b \over { {b_{t,\tts x,\tts A^*}}}}
\,,
\end{multlined}
\]}%
so that ${ {b_{T-1,\tts x,\tts A^*}}}=\b$ for every $x\in\XXX$.
Hence, with  $t=T-2$ \eqref{eq-b-1} gives
{\abovedisplayskip=8pt plus 1.5pt minus 1.5pt\belowdisplayskip=8pt plus 1.5pt minus 1.5pt\belowdisplayshortskip=6pt plus 1.5pt minus 1.5pt
\[
h_{t,\tts x,\tts A^*}^{ y } = {b_{t,\tts x,\tts  A^*}\over 1+ \b}\times
\Bigl(\rho_ y \bigl({ {K_{t}(x, A^*)}}\bigr) +1-\dd\Bigr) 
\,,\quad x,\,y \in\XXX\,.
\]}%
Using \eqref{eq-n-1} one more time with $h_{t,\tts x,\tts A^*}^{ y }$ from above we get
{\abovedisplayskip=5pt plus 1.5pt minus 1.5pt\belowdisplayskip=5pt plus 1.5pt minus 1.5pt\belowdisplayshortskip=5pt plus 1.5pt minus 1.5pt
\[
1 = { \b (1+\b)\over b_{t,\tts x,\tts A^*}}
\,,\quad x\in\XXX\,,
\]}%
so that ${ {b_{T-2,\tts x,\tts  A^*}}}=\b+\b^2$.
By induction, for every $n\ge 1$ and  $t=T-n$,
{\abovedisplayskip=5pt plus 1.5pt minus 1.5pt\belowdisplayskip=5pt plus 1.5pt minus 1.5pt\belowdisplayshortskip=5pt plus 1.5pt minus 1.5pt
\begin{equation}\tag{a}\label{explicit-b}
b_{t,\tts x,\tts  A^*}\equiv b_t=\b+\b^2+\cdots+\b^n ={1-\b^{n+1}\over 1-\b}-1
={\b-\b^{T-t+1}\over 1-\b}\,.
\end{equation}}%
As a result, using \eqref{eq-b-1} yet again with $t=T-n$ we get 
{\abovedisplayskip=8pt plus 1.5pt minus 1.5pt\belowdisplayskip=8pt plus 1.5pt minus 1.5pt\belowdisplayshortskip=6pt plus 1.5pt minus 1.5pt
\begin{equation}\tag{b}\label{explicit-h}
\begin{multlined}
{h_{t,\tts x,\tts  A^*}^{ y}} =
{\b+\cdots+\b^n\over 1+\b+\cdots+\b^{n-1}}
\,\Bigl(\rho_ y \bigl({ {K_{t}(x, A^*)}}\bigr) +1-\dd\Bigr)\\
\hbox to5cm{\hfill}=\b\,\Bigl(\rho_ y \bigl({ {K_{t}(x, A^*)}}\bigr) +1-\dd\Bigr)\,.
\end{multlined}
\end{equation}}%
In particular, letting  $n\to\infty$ leads to the following
time-invariant values for the slopes of the portfolios and the employment-specific transition assignments:
{\abovedisplayskip=5pt plus 1.5pt minus 1.5pt\belowdisplayskip=5pt plus 1.5pt minus 1.5pt\belowdisplayshortskip=5pt plus 1.5pt minus 1.5pt
\[
b_\infty={\b\over1-\b}\quad\text{and}\quad
h_{\infty,\tts x,\tts A^*}^{ y } = 
\b\,\Bigl(\rho_ y \bigl({ {K_{\infty}(x, A^*)}}\bigr) +1-\dd\Bigr) \,,
\]}%
provided, of course, that $K_{\infty}(x, A^*)\df \lim_{t\to\infty}K_t(x,A^*)$ exists.\qed
\end{nit}

In~general, 
the intercepts $a_{t,\tts x,\tts A^*,\tts u}$ and $g_{t,\tts x,\tts A^*,\tts u}^{y,\tts v}$
cannot be employment-invariant, and one can only hope that time-invariant versions of the mappings
{\abovedisplayskip=5pt plus 1.5pt minus 1.5pt\belowdisplayskip=5pt plus 1.5pt minus 1.5pt\belowdisplayshortskip=5pt plus 1.5pt minus 1.5pt
\[
(x,u, A^*) \leadsto a_{t,\tts x,\tts A^*,\tts u}\,, \quad
(x,y,u,v, A^*) \leadsto g_{t,\tts x,\tts  A^*,\tts u}^{y,\tts v}\,,\quad\text{and}\quad
(x, A^*) \leadsto K_t(x, A^*)
\]
}%
exist in the limit as $t\to\infty$. 
Because of the implicit structure of the system composed of
\eqref{lin-split}, \eqref{Taylor}, \eqref{mclr-lin-a} and \eqref{no-fp-2-0},
the existence of these limits is
very difficult to establish generically by way of the usual fixed point argument.
Nevertheless, it will be shown below that, at least in the
benchmark example that we borrow here from~\cite{KS98}, the convergence
of the mappings above as $t\to\infty$ is rather easy to establish numerically, i.e.,
by following the general 
iteration strategy from \ref{main-proc} for a sufficiently large number of periods
one arrives at 
successive copies of these mappings that coincide (in uniform distance) within a prescribed
threshold. We stress, however, that the procedure that we are about to outline and implement
does not require time invariance: the program returns equilibrium allocations for any
time horizon, irrespective of 
whether increasing the time horizon results into (numerically confirmed) convergence or not.

\begin{nit}{Remark}\label{T-inf}%
It is clear from \eqref{no-fp-x} that if
$g_{\infty ,\tts x,\tts  A^*,\tts u}^{y,\tts v} = \lim_{t\to\infty}g_{t,\tts x,\tts  A^*,\tts
u}^{y,\tts v}$ were to exist, then time-invariant version of the transport
$\T_{\infty,x}^y( A^*) = \lim_{t\to\infty}\T_{t,x}^y( A^*)$ also exists.
However, the dependence of the transport on both the present productivity state $x$ and on 
the future one $ y $ does not go away.
As a result, the state variable $A^*$ fluctuates in the random environment of
both, present and future, productivity
shocks even when the time horizon is pushed to $\infty$~--
see footnote \ref{ftn82148}, \ref{main-q}, and  \ref{n2}. 
As we are about to see below~-- see Fig.~\ref{fgKS-IK} and \ref{rem85521}~-- 
even in the long run, some very special cases
notwithstanding, one cannot hope to be able to attach
a single population average to every realized productivity state, i.e., the
fluctuations in the population average are much more dispersed than the fluctuations
in the productivity state, even as the latter is the only cause of the former.  
\qed
\end{nit}

\begin{nit}{Krusell-Smith's strategy compared}\label{transport-K}%
In the present setup, the transport of installed ca\-pi\-tal 
is given by the mapping
{\abovedisplayskip=5pt plus 3.5pt minus 1.5pt\belowdisplayskip=5pt plus 3.5pt minus 1.5pt\belowdisplayshortskip=5pt plus 3.5pt minus 1.5pt
\begin{equation}\tag{a}\label{4-2-a}
K_t(x,A^*) \leadsto K_{t+1}(y,\T_{t,x}^y( A^*))\,,
\end{equation}
}%
and since \eqref{eq-mclr} can be solved for $A^*$ against $K_t$, the second
expression above is 
a function of the first,
i.e., if 
written in terms of average installed capital (the approach used in \cite{KS98})
the transport of the population mean
from period $t$ to period $t+1$ can be cast in the form
{\abovedisplayskip=5pt plus 3.5pt minus 1.5pt\belowdisplayskip=5pt plus 3.5pt minus 1.5pt\belowdisplayshortskip=5pt plus 3.5pt minus 1.5pt
\begin{equation}\tag{b}\label{4-2-b}
K \leadsto H_{t,x}^y(K)\,,\quad x,y\in\XXX\,.
\end{equation}
}%
In general, even if time invariant versions of the mappings $H_{t,x}^y\phd$ were to exist,
as was already noted, the
dependence on both the present and future productivity states, $x$ and~$y$, does not go away.
The computational strategy described in \cite{KS98} comes down to approximating the mappings
$H_{\infty,x}^y\phd$ by way of least-square log-linear fit from the simulated
long-run behavior of a 
large population of households, but in the actual implementation the dependence
on~$x$ is ignored, 
i.e., in the concrete example described in \cite{KS98}
there is only one log-linear line for every $y\in\XXX$.
To more indulgent eyes  such a shortcut may appear harmless 
because the choice of model parameters in \cite{KS98} is such
that  all mappings
$H_{\infty,x}^y\phd$, $x,y\in\XXX$, are very close in uniform norm
and are also very close to linear~-- see below.
However, there is no intuition to suggest that such an arrangement persists in
general, and even 
with parameter choices as in \cite{KS98} the mappings $H_{\infty,x}^y\phd$, $x,y\in\XXX$, are
not close enough to be declared numerically identical.

For the purpose of comparison and benchmarking,
the output from the computational
strategy developed in this section can be cast in terms of average capital instead of
consumption as a 
state variable (e.g., one can retrieve \eqref{4-2-b} from \eqref{4-2-a}).
The differences between the method proposed here and the one from \cite{KS98}
can be summarized as follows. First, assuming the affine
structure postulated in \eqref{lin-subst} is in place, 
the mappings $H_{t,x}^y\phd$ can be computed exactly,
i.e, within the accuracy of the numerical solver and the cubic spline interpolation,
for any $t<T$ and any $x,y\in\XXX$~-- without the need to restrict those mappings to
a particular 
type of functional dependence (e.g., log-linear) and without the need to simulate the
individual 
behavior of a large number of households.
The convergence (in uniform norm) $H_{t,x}^y\phd\to H_{\infty,x}^y\phd$ can then be
confirmed numerically. Most important, \ref{mod-accu} below provides a tool for estimating the
error from assuming the affine structure postulated in \eqref{lin-subst}, which is
essentially the 
error from assuming the arrangement known as ``representative agent.''\qed
\end{nit}

Our next step is to reformulate the time-interlaced backward induction program
from \ref{main-proc} in terms of
the setup adopted in the present section and the reduced form postulated
in \eqref{lin-subst}.
For the sake of simplicity the description that follows is written with infinite time
horizon in 
mind, but with the understanding that finite time horizon merely means interrupting
the program before it 
detects convergence.   

\begin{nit}{Time-interlaced backward induction}\label{KS-proc}%
Due to the explicit formulas in
\hbox{\ref{closed-form}-\nobreak\eqref{explicit-b}} and in \ref{closed-form}-\nobreak\eqref{explicit-h},
the only unknowns that need to be computed
are the average installed capital $K_{t}(x, A^*)$ and the intercepts 
$a_{t,\tts x,\tts  A^*,\tts u}$  and $g_{t,\tts x,\tts  A^*,\tts u}^{ y ,\tts v}$,
for all choices of $ u, v\in\nobreak\EEE\,,\ x,  y \in\nobreak\XXX\,,$
and $t<T$. 
As these unknowns depend on the period $t$ average consumption across the population,
the objects we are looking for are functions of $ A^*\in\Rpp$ that are labeled by $t$, $u$, $v$,
$x$ and $y$.
The general program described in \ref{main-proc} comes down to the following steps:

{\it Initial setup:} Make an ansatz choice for the finite interpolation grid
$\bbG\subset\Rpp$, the 
elements of which represent reduced (to the total population mean) cross-sectional
distributions of all households. 

{\it Initial backward step:}  Set $n=1$ and $t=T-n$. For every $x\in\XXX$ do:

For every $ A^*\in\bbG$ perform steps~(1) through~(3) below, then
proceed to~(4):

{(1)} Make an ansatz choice for $K_{t}(x, A^*)>0$ and go to (2).

{(2)} Solve the system composed of \eqref{eq-b-0} with
$a_{t+1,\tts y ,\tts \T_{t,x}^y( A^*),\tts v}\equiv 0$
and \eqref{eq-n-0} (total of $\abs{\EEE}+\abs{\EEE}^2\abs{\XXX}$ equations~--
note that $x$ is fixed) for the unknowns ($\abs{\EEE}+\abs{\EEE}^2\abs{\XXX}$ in number)
$a_{t,\tts x,\tts A^*,\tts u}$ and $g_{t,\tts x,\tts A^*,\tts u}^{y,\tts v}$, $y\in\XXX$, $u,v\in\EEE$. Proceed to (4).

{(3)} Test the market clearing (see \eqref{eq-mclr} and recall that $b_{T-1,\tts x,\tts A^*,\tts u}=\b$)
{\abovedisplayskip=5pt plus 1.5pt minus 1.5pt\belowdisplayskip=5pt plus 1.5pt minus 1.5pt\belowdisplayshortskip=5pt plus 1.5pt minus 1.5pt
\[
K_{t}(x, A^*)=\b A^*+ \sum\nolimits_{u\ts\in\ts\EEE}\gp_x(u) a_{t,\tts x,\tts A^*,\tts u}\,.
\]
}%
If this relation fails by more than a prescribed threshold, set the new value of $K_{t}(x, A^*)$ to the
right side above and go back to (2).

{(4)} Construct spline interpolation objects
\footnote{{}Splines defined over the grid $\bbG$ are treated as functions on $\Rpp$ by way of extrapolation.}
over the grid $\bbG$ from the most recently calculated
values for $K_{t}(x, A^*)$, $a_{t,\tts x,\tts A^*,\tts u}$ and  $g_{t,\tts x,\tts A^*,\tts u}^{ y
,\tts v}$,
$ A^*\in\bbG$, for every $y\in\XXX$ and $u,v\in\EEE$.

{\it Generic backward step:} Set $n=n+1$ and $t=T-n$,
assuming that $ A^* \leadsto a_{t+1,\tts y,\tts A^*,\tts v}$ are already computed functions
(splines with extrapolation)
on $\Rpp$  for every $ y \in\XXX$ and every $ v\in\EEE$. For every $x\in\XXX$ do:

For every $ A^*\in\bbG$ complete steps (1) through (5) below,
then proceed to~(6):

{(1)} Set $\pdg A^*_{y }= A^*$ for every $ y \in\XXX$ (initial guess for the future state of the
population in every future productivity state), then go to (2).

{(2)} Set $K_{t}(x, A^*)=K_{t+1}(x,A^*)$ (initial guess for the average installed capital taken from the
previous iteration), then go to (3).

{(3)} Solve the system \eqref{eq-n-0} \&
\eqref{eq-b-0} with
$a_{t+1,\tts y,\tts \T_{t,x}^y( A^*),\tts v}=a_{t+1,\tts y,\tts \pdg A^*_{y },\tts v}$
 (total of $\abs{\EEE}+\abs{\EEE}^2\abs{\XXX}$ equations~--
note that $x$ is fixed) for the unknowns ($\abs{\EEE}+\abs{\EEE}^2\abs{\XXX}$ in number)
$a_{t,\tts x,\tts A^*,\tts u}$ and $g_{t,\tts x,\tts  A^*,\tts u}^{ y ,\tts v}$, $ y \in\XXX$, $ u, v\in\EEE$. Proceed to (4).

{(4)} Test the market clearing
(see \eqref{eq-mclr} and recall that $b_{T-n,x, A^*,u}=\b+\cdots+\b^n$)
{\abovedisplayskip=5pt plus 1.5pt minus 1.5pt\belowdisplayskip=5pt plus 1.5pt minus 1.5pt\belowdisplayshortskip=5pt plus 1.5pt minus 1.5pt
\[
K_{t}(x, A^*)= (\b+\cdots+\b^n)  A^*
+ \sum\nolimits_{u\ts\in\ts\EEE}\gp_x(u) a_{t,\tts x,\tts  A^*,\tts u}\,.
\]
}%
If this relation fails by more than a prescribed threshold, set the new value of $K_{t}(x, A^*)$ to the
right side and go back to~(3), otherwise proceed to~(5).

{(5)} Compute (see %
\eqref{eq-T-star}%
)
{\abovedisplayskip=5pt plus 2.5pt minus 1.5pt\belowdisplayskip=5pt plus 2.5pt minus 1.5pt\belowdisplayshortskip=5pt plus 2.5pt minus 1.5pt
$$
\pddg A^*
= \b\,\Bigl(\rho_ y \bigl({ {K_{t}(x, A^*)}}\bigr) +1-\dd\Bigr)\times  A^* + \sum\nolimits_{u,\tts v\tts \in\tts \EEE}
{\gp_x(u)\times  P_{x, y }( u,v)}\times g_{t,\tts x,\tts A^*,\tts u}^{ y ,v}\,,
$$
}%
for every $ y \in\XXX$. If the uniform distance
{\abovedisplayskip=5pt plus 1.5pt minus 1.5pt\belowdisplayskip=5pt plus 1.5pt minus 1.5pt\belowdisplayshortskip=5pt plus 1.5pt minus 1.5pt
\[
\max\nolimits_{ y\ts\in\ts\XXX}\abs{\pddg A^*_{y} - \pdg A^*_{y}}
\]}%
exceeds some prescribed threshold (the guessed future averages are not compatible with the
transport),
set $\pdg A^*_{y}=\pddg A^*_{y}$ for all $y\in\XXX$
and go back to (2). 

{(6)} Construct spline interpolation objects over the grid $\bbG$ from the most recently calculated
values for $K_{t}(x, A^*)$, $a_{t,\tts x,\tts A^*,\tts u}$ and  $g_{t,\tts x,\tts A^*,\tts u}^{
y,\tts v}$,
$ A^*\in\bbG$, for every $y\in\XXX$ and $u,v\in\EEE$.

{\it Final backward step\/} (if looking for a time-invariant equilibrium): 
If at least one of the uniform distances
{\abovedisplayskip=5pt plus 1.5pt minus 1.5pt\belowdisplayskip=5pt plus 1.5pt minus 1.5pt\belowdisplayshortskip=5pt plus 1.5pt minus 1.5pt
\[
\begin{gathered}
\max_{x\ts\in\ts\XXX,\, A^*\ts\in\ts\bbG}\abs{K_{t+1}(x, A^*)-K_{t}(x, A^*)}\,,\ %
\max_{x\ts\in\ts\XXX,\,u\ts\in\ts\EEE,\, A^*\ts\in\ts\bbG}\abs{a_{t+1,\tts x,\tts A^*,\tts
u}-a_{t,\tts x,\tts A^*,\tts u}}\,,\\
\max_{x, y \ts\in\ts\XXX,\, u, v\ts\in\ts\EEE,\, A^*\ts\in\ts\bbG}\abs{g_{t+1,\tts x,\tts  A^*,\tts
u}^{ y ,\tts v}-g_{t,\tts x,\tts  A^*,\tts u}^{ y ,\tts v}}
\end{gathered} 
\]}%
exceeds some prescribed threshold, perform another generic backward step.
\footnote{{}The time parameter $t$ may become negative~-- the program moves backward in time for as
many periods as needed to achieve time invariance.
The final backward step would not be necessary if seeking time invariant
transport is not the objective, in which case the program can be terminated at $t=0$.}
Otherwise stop and define
the functions
{\abovedisplayskip=5pt plus 1.5pt minus 1.5pt\belowdisplayskip=5pt plus 1.5pt minus
1.5pt\belowdisplayshortskip=5pt plus 1.5pt minus 1.5pt
$$
 A^* \leadsto K_{\infty}(x, A^*)\,,\quad
 A^* \leadsto a_{\infty,x, A^*,u}\,,\quad
 A^* \leadsto g_{\infty,x, A^*,u}^{ y ,v}\,,\quad x,y\in\XXX\,,\ \ u,v\in\EEE\,,
$$
}%
as the most recently computed
spline objects (with the latest value for $t$).\qed  
\end{nit}

\begin{nit}{Endless loops warning and disclaimer}\label{endless-loop-4}%
There are no theoretical results to guarantee that the iterations between steps (3) and (4) and (3)
and (5) converge, or to guarantee that step (3) is always feasible, i.e., a numerical solution to the system
composed of \eqref{eq-b-0} and \eqref{eq-n-0} always exists, for every possible choice of
the model parameters from some reasonably wide range.\qed   
\end{nit}

\begin{nit}{Remark}\label{MFG-rem-4}%
The iterations between steps (3) and (5) in the generic backward step are meant
to establish the correct connection between future and present
distribution averages, i.e., figuratively speaking, meant to ensure 
that the transport of the mean is self-consistent~-- as it should~be. 
We stress that the 
adjustments that ensure self-consistency are local in time, in that the program does not
move to the next period in the backward induction (which is the previous period in real time)
until the correct (i.e., fully self-consistent)
transport from the current period is established~-- recall that the transport is
time dependent and may become time invariant only in the limit.\qed
\end{nit}

\begin{nit}{Accuracy of the affine approximation}\label{mod-accu}%
Despite the reduction to a model with affine structure, the program outlined
in \ref{KS-proc} solves exactly, meaning, within the nonlinear solver's tolerance of the
infinity norm of the residuals, all equations that define the equilibrium~-- except
for the kernel 
condition \ref{cross-sys}-(\ref{zze5a}), which was replaced by~\eqref{only-q-bn2-2}.
The solution was then sought in such a way that~\eqref{only-q-bn2-2} is
approximately accurate for sufficiently large consumption levels~$c$
(the Krusell-Smith's conjecture is asymptotically accurate as all consumption levels
increase to~$+\infty$).
Hence, quantifying the accuracy of the program comes down to
calculating the aberration in the right side of~\eqref{only-q-bn2-2} from~$1$~-- for
consumption 
levels in some more realistic range. Recall that the economic interpretation of the kernel
condition comes down to the claim that all agents agree on the prices at which all
securities are 
traded. In~the
present setup this feature is tantamount to an agreement about the average of all
private capital 
investments, which, ultimately, boils down to an agreement about future returns and
wages. Thus, an 
aberration from~$1$ in the right side of~\eqref{only-q-bn2-2} has the effect that for
some (not too 
large) consumption levels the present period marginal utility from consumption would
be smaller or larger 
than the expected and discounted future marginal utility from consumption, i.e., agents with
relatively small consumption levels consume less or more than what would be optimal
for them in perfect 
equilibrium. Loosely speaking, the market arrangement favors the objectives of the
big spenders, 
i.e., the wealthy. There is no intuition to suggest that such arrangements occur in
practice, nor is 
there an intuition to suggest that in practice capital markets attain equilibrium
allocation so 
perfectly that the consumption level of every agent is exactly the one provided by
perfect equilibrium. 
The method developed in the present \chptr\  cannot~-- and does not attempt to~--
address such matters. 
Nevertheless, it provides a framework within which the deviation of the
representative agent point 
of view from the theoretical (perfect) equilibrium model can be quantified. 
Indeed, with infinite time horizon in mind, the right side of
\eqref{only-q-bn2-2} can be treated as a function of the collective state 
of the
population~$A^*$ and the consumption level~$c$, i.e., can be cast as
{\abovedisplayskip=7pt plus 1.5pt minus 1.5pt\belowdisplayskip=7pt plus 1.5pt minus 1.5pt\belowdisplayshortskip=5pt plus 1.5pt minus 1.5pt
\begin{equation*}\tag{a}\label{k-check}
\begin{multlined}
R_{\infty,x,u}(A^*,c) \df \sum_{ y\ts \in\ts\XXX,\ts v\ts\in\ts\EEE} \ 
{ \b\,\bigl(\rho_ y \bigl(K_{\infty}(x, A^*)\bigr) + 1-\dd\bigr)
\over g_{\infty,\tts x,\tts A^*,\tts u}^{ y ,v}/c  + \b\,\bigl(\rho_ y \bigl(K_{\infty}(x,
A^*)\bigr) + 1-\dd\bigr)}\\
\times\,
Q(x, y )\,  P_{x, y }( u ,v)\,.
\end{multlined}
\end{equation*}}%
Once the program in \ref{KS-proc} completes,
the value $R_{\infty,x,u}(A^*,c)$ can be calculated for every $c>0$ and for every $ A^*$ in
the long-run range of the population average. There is no obvious choice for the consumption level
$c$, since the only restriction for this variable
is on its distribution across the population.
One possible approach is to calculate the time invariant version (see~\ref{T-inf}) of the transport
in \eqref{no-fp-2-0} and simulate a series of long-run realizations of the vectors $A=(A^u\in\Rpp,
u\in\EEE)$ and the scalars $A^*\df \sum_{u\tts\in\tts \EEE}\pi_x(u)A^u$. The maximum over the
simulated series of the associated quantities $\abs{R_{\infty,x,u}(A^*, A^u)-1}$, $u\in\EEE$, will
then give an estimate of how far from the hypothetical equilibrium is the arrangement in which all
households in the same employment category adopt identical consumption levels. In the illustration
used in this section, which is borrowed from \cite{KS98} (see below), this estimate gives an error of order
$10^{-4}$. Unfortunately, such an estimate can be used only on a case by case basis, i.e,
it does not allow one to conclude that the conjecture
``only the mean and its transport matter'' is reasonably
accurate for a broad class of models. The main difficulty is that
tools to allow for general (not model-specific)
estimates of the long-run range of the conditional averages $A^u$, $u\in\EE$,
are yet to be developed.  

Another,
similar in spirit but more demanding, estimate for the kernel aberration is to compute the
expressions $R_{\infty,x,u}(A^*,\bar c_{x, A^*,u})$ with  $\bar c_{x, A^*,u}$ defined as the investment threshold
for households in employment state $u\in\EEE$,
i.e., the solution to $\qq_{\infty,x, A^*,u}(c)=0$. In our reduced model this is nothing but the intersection of the
line
$c \leadsto a_{\infty,x, A^*,u}+b_{\infty,x, A^*,u}\times c$ with the horizontal axis, i.e., 
{\abovedisplayskip=5pt plus 1.5pt minus 1.5pt\belowdisplayskip=5pt plus 1.5pt minus 1.5pt\belowdisplayshortskip=5pt plus 1.5pt minus 1.5pt
\[
\bar c_{x, A^*,u} = -{a_{\infty,x, A^*,u}\over b_{\infty,x, A^*,u}} = -{(1-\b) a_{\infty,x, A^*,u}\over \b}\,.
\]}%
In the example borrowed here from \cite{KS98} 
this estimate, too, leads to a maximum aberration across the simulated sample  
of order $10^{-4}$.
Appendix \ref{sec:App-B} elaborates on the consumption range in the model with only two employment
states and no risk-free lending. It also shows that in such models borrowing of capital is not
consistent with the notion of equilibrium (in the model studied in \cite{KS98} capital cannot be
borrowed by assumption). We stress, however, that a model with only two employment and two
productivity states is rather narrow in scope.\qed
\end{nit}

To put the methodology developed in the present section to the test,
we now turn to some concrete examples and numerical results.
Since the objective here is to benchmark the new method against those that precede it, in what
follows we focus exclusively on the infinite time horizon case, but stress that the program
developed here is designed to work only with finite time horizon (of any length), and ``infinite
time horizon'' is understood as a ``sufficiently large finite time horizon.'' In all illustrations
below the time horizon is set to $T=1\hbox{,}000$, except for the simulated sample, in which the
time horizon is $T=1\hbox{,}100\hbox{,}000$.
We borrow the general setup and parameter values
from the benchmark economy described in
\cite{KS98}: the list of productivity states is $\XXX=\{1.01,\,0.99\}$
(the economy is either in high state, labeled~``$1$,'' or in low state, labeled ``$2$''), with
transition probability matrix for these states
{\abovedisplayskip=8pt plus 1.5pt minus 1.5pt\belowdisplayskip=8pt plus 1.5pt minus 1.5pt\belowdisplayshortskip=8pt plus 1.5pt minus 1.5pt
\[
 Q=
\begin{bmatrix}
 0.875 & 0.125\\
 0.125 & 0.875
\end{bmatrix}\,,
\]}%
the list of employment states is $\EEE=\{\h,0\}\df \{0.3271,0.0\}$ (the agents can be
either employed or unemployed), with (private)
conditional transition probability matrices for these states
{\abovedisplayskip=5pt plus 1.5pt minus 1.5pt\belowdisplayskip=5pt plus 1.5pt minus 1.5pt\belowdisplayshortskip=5pt plus 1.5pt minus 1.5pt
\begin{gather*}
 P_{1,1}=\begin{bmatrix}
 0.972222 & 0.0277778\\
 0.666667 & 0.333333
 \end{bmatrix}\,,\quad
 P_{1,1}=\begin{bmatrix}
 0.927083 & 0.0729167\\
 0.25   &   0.75
\end{bmatrix}\,,
\end{gather*}
\begin{gather*}
 P_{2,1}=\begin{bmatrix}
 0.983333 & 0.0166667\\
 0.75    &  0.25
  \end{bmatrix}\,,\quad
 P_{2,2}=\begin{bmatrix}
 0.955556 & 0.0444444\\
 0.4     &  0.6
\end{bmatrix}\,,
\end{gather*}}%
which correspond to $\gp_1= [0.96, 0.04]$ and  $\gp_2= [0.9, 0.1]$. The parameter values are $\b=0.99$,
$\dd=0.025$, $\a=0.36$, $R=1$ (risk aversion). The total labor supplied in high state is
$L(\EEE_1)=0.314016$ and in low state it is $L(\EEE_2)= 0.29439$.
The time-invariant solution, i.e., the functions
{\abovedisplayskip=5pt plus 1.5pt minus 1.5pt\belowdisplayskip=5pt plus 1.5pt minus
1.5pt\belowdisplayshortskip=5pt plus 1.5pt minus 1.5pt
$$
 A^* \leadsto K_{\infty}(x, A^*)\,,\quad
 A^* \leadsto a_{\infty,x, A^*,u}\,,\quad
 A^* \leadsto g_{\infty,x, A^*,u}^{ y ,v}\,,\quad x,y\in\XXX\,,\ \ u,v\in\EEE\,,
$$
}%
are constructed as $1$D-splines on a fixed interpolation grid $\bbG\subset \R_{++}$.
The choice of the corresponding domain $[\bbG]$ is ad hoc, but chosen so that it contains the
simulated long-run range of the total population mean~$A^*$.  
The metaprogram in \ref{KS-proc}
was translated into the Julia programming language with parallelization.
With parallelization on $16$ CPUs the program completes $10^3$ iterations for about $3$ minutes
and achieves convergence (in uniform distance over all grid-points between the last two
copies of the respective values) of $4.31651\times 10^{-5}$ across all values
$(a_{\infty,x, A^*,u},\, A^*\in\bbG)$, $7.18267\times 10^{-9}$ across all values
$(g_{\infty,x, A^*,u}^{ y ,v},\, A^*\in\bbG)$, $1.24589\times 10^{-8}$ across all values
$(K_\infty(x, A^*),\, A^*\in\bbG)$,
and $2.42939\times 10^{-10}$ across all transports $(\T_x^{ y,v}( A^*),\, A^*\in\bbG)$. 
In~all iterations the function tolerance in the nonlinear
solver
\footnote{{}The NLsolve package was used despite the fact that in the only-the-mean-matters scenario
the system to solve is linear. This choice was deliberate and meant to make the computer code
usable in other scenarios~-- see~\ref{beyond}.}
was set to $10^{-12}$.
The plots in Fig.~\ref{fgKS10} show the portfolio intercepts $a_{\infty,x, A^*,u}$ as functions of $A^*$ for
all choices of $x\in\XXX$ and $u\in\EEE$.  
{\captionsetup{belowskip=-5pt}
\begin{figure}[!htbp]
\centering
\begin{subfigure}{.5\textwidth}
  \centering
\toshow{\includegraphics[width=7.5cm]{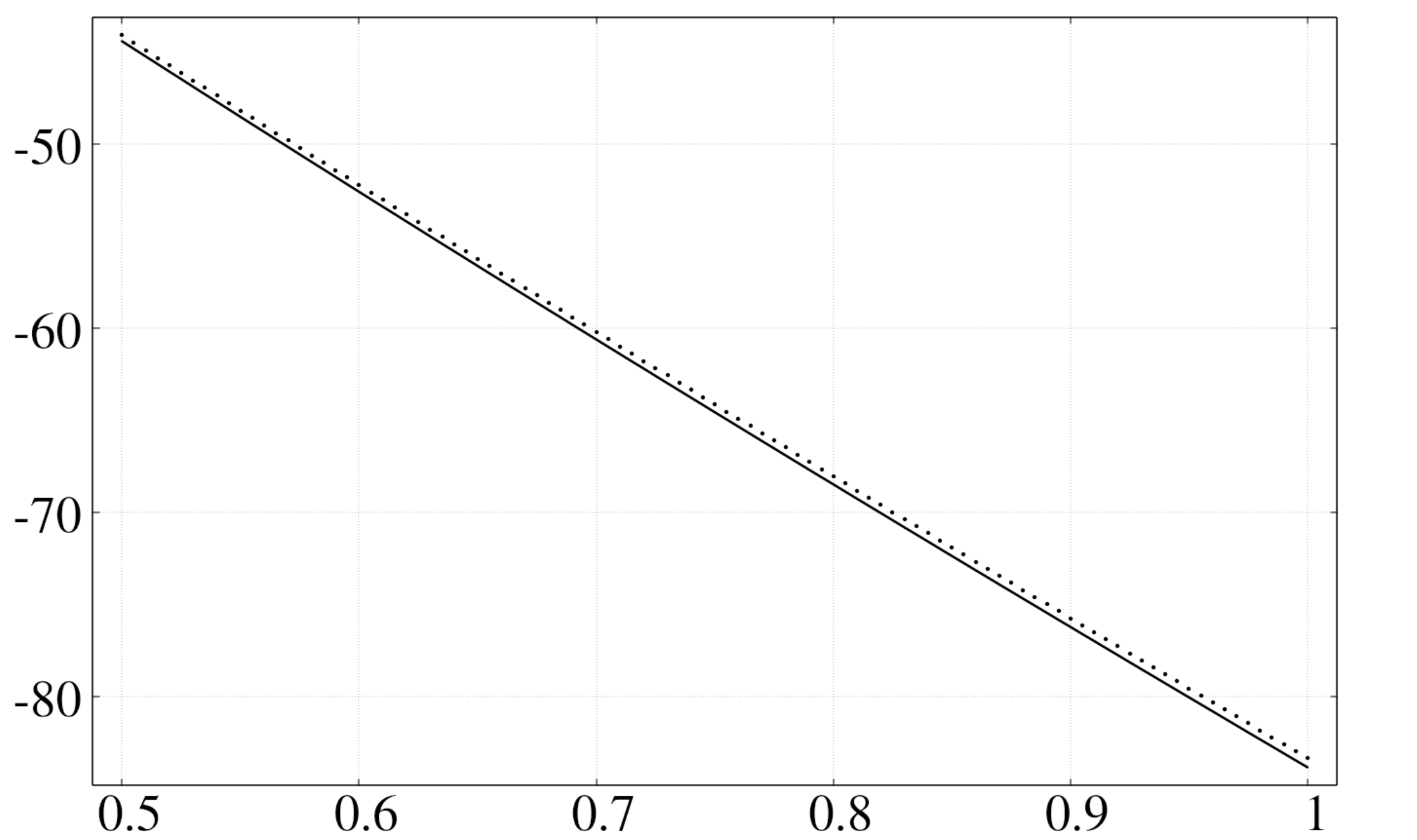}}
\end{subfigure}%
\begin{subfigure}{.5\textwidth}
  \centering 
\toshow{\includegraphics[width=7.5cm]{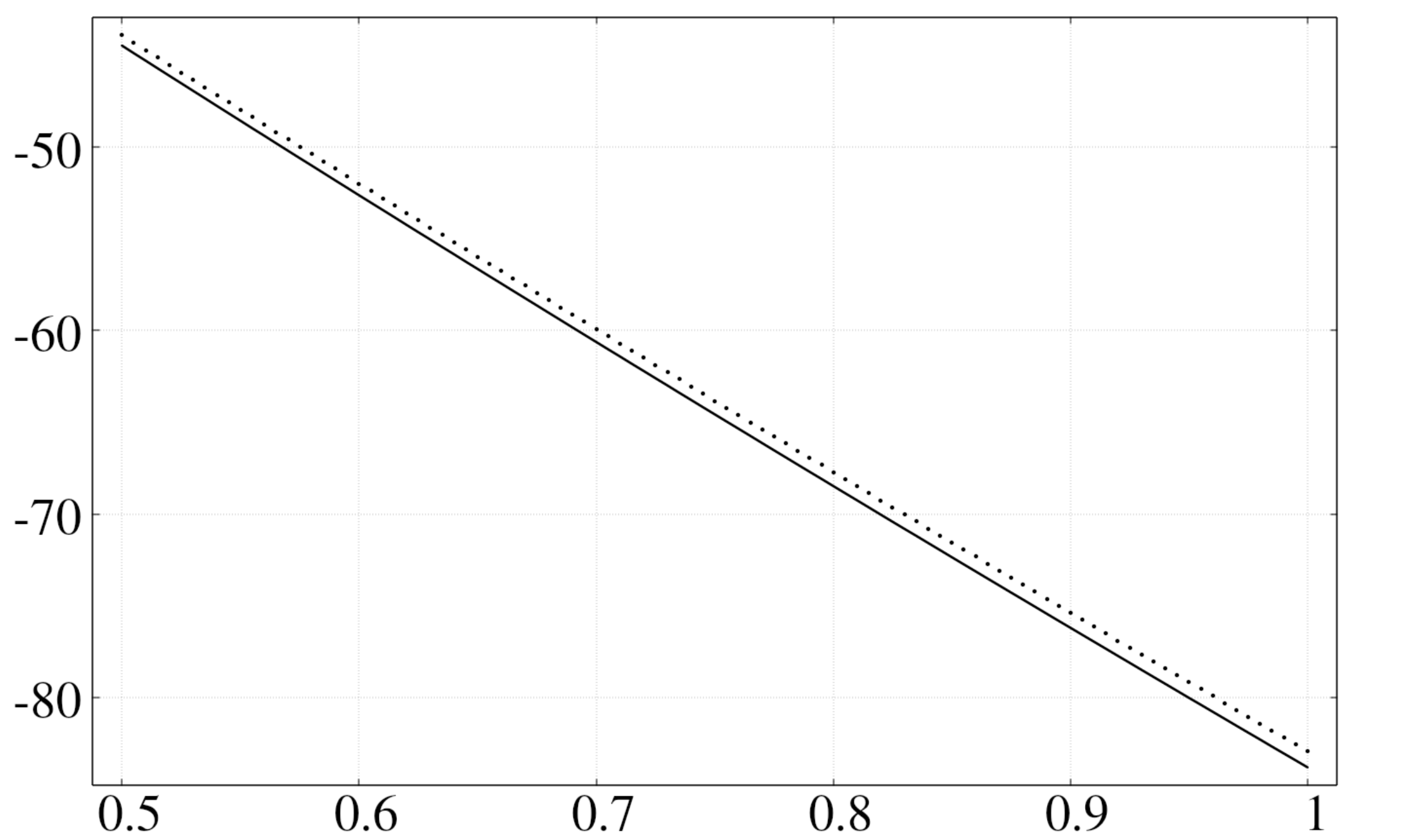}  }      
\end{subfigure}
\caption{Intercepts of the portfolio lines in high state (left plot) and low state (right
plot) for employed (solid lines) and unemployed (doted lines) shown as functions of the total population
mean (over consumption)~$A^*$.}
\label{fgKS10}
\end{figure} }%
The fact that the intercepts decrease as the population average $A^*$ increases
implies that the 
portfolio lines, which give the private demand for capital as a function of the
private consumption 
level, shift downwards as the aggregate consumption level across the population
increases (recall 
that the slopes are fixed). 
The plot in Fig.~\ref{fgKS11} shows the average installed capital $K_{\infty}(x,
A^*)$ as a function of 
$A^*$ for all $x\in\XXX$.
{\captionsetup{belowskip=-5pt}
\begin{figure}[!htbp]
\centering
\toshow{\includegraphics[width=7.5cm]{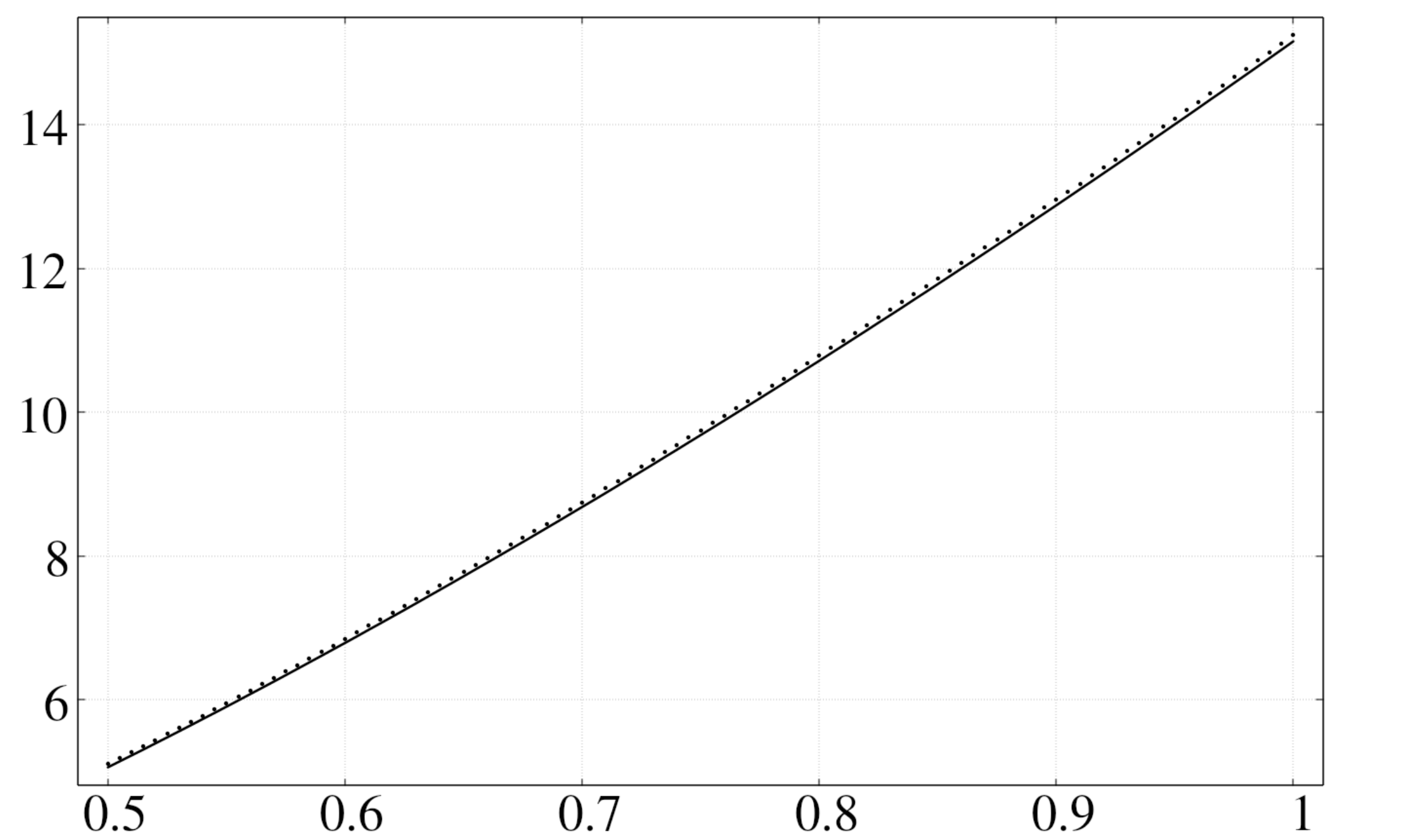}}
\caption{Average installed capital in high state (solid lines)
and low state (doted lines) shown as a function of the population consumption average $A^*$.
The uniform distance between the two lines is around 0.08574.}
\label{fgKS11} 
\end{figure} }%

The transport mappings from \eqref{eq-T-star}~-- the main computation tool in the
approach developed 
in this section~-- are illustrated in Fig.~\ref{fgKS12}.
{\captionsetup{belowskip=-5pt}
\begin{figure}[!htbp]
\centering
\begin{subfigure}{.5\textwidth}
  \centering
\toshow{\includegraphics[width=7.5cm]{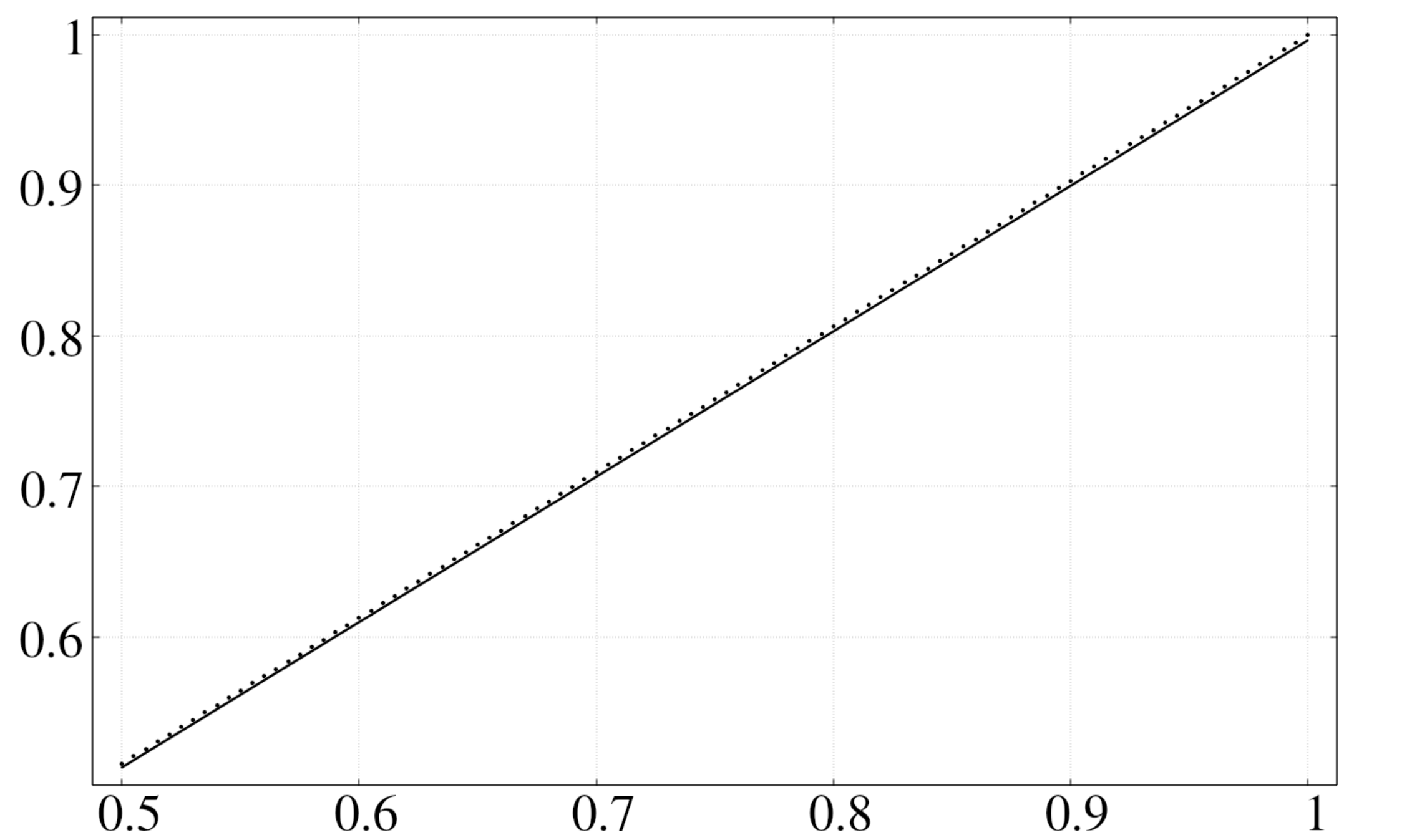}}
\end{subfigure}%
\begin{subfigure}{.5\textwidth}
  \centering 
\toshow{\includegraphics[width=7.5cm]{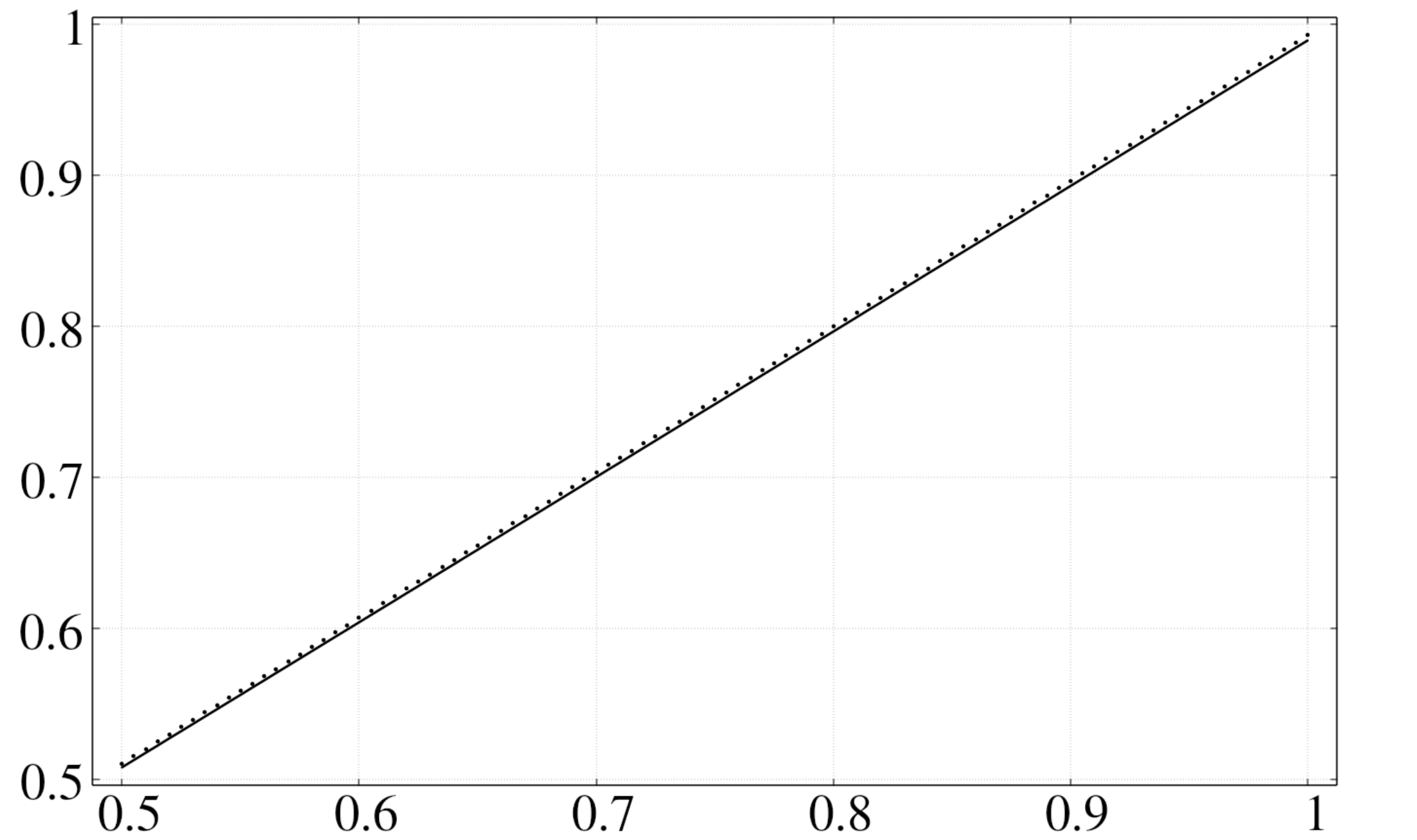}  }      
\end{subfigure}
\caption{The transport of the population consumption mean into high state (left plot) and into
low state (right plot) from high state (solid lines) and from low state (doted lines).
The uniform distance between the solid and the dotted line is around 0.00354
in the left plot and around 0.00353 in the right plot. The
uniform distance between the two solid lines is around 0.00663.}
\label{fgKS12}
\end{figure} }%
It is clear from Fig.~\ref{fgKS12} that, in this particular example,
the dependence
of the transport $\T_{\infty,x}^y$ on the departing aggregate state~$x$ is negligible
(the dotted and the solid lines are very close),
\footnote{Coincidentally~-- or perhaps not so coincidentally~--
in Krusell-Smith's algorithm described in \citex[III]{KS98}
the dependence of the transport on the departing aggregate state is ignored,  i.e., in their
approach there are only two regression lines instead of four.}
and so is also the dependence on the future (arriving) aggregate
state~$y$, although the latter dependence is still some two orders of magnitude bigger
(the solid lines on the left and the right plot are not that close).
While
all four lines in Fig.~\ref{fgKS12} appear to be very close, there is no reason for
this feature 
to persist with other choices for the model parameters,
especially if the numbers of the exogenous aggregate
and idiosyncratic states are increased
\footnote{Just as an example, if the value of the impatience parameter $\b$ is
decreased to $0.96$ 
from its original value of $0.99$, with which Fig.~\ref{fgKS12} is generated, while all other
parameter values are kept unchanged, then the difference between dotted and solid lines in
Fig.~\ref{fgKS12} becomes much more pronounced. This feature is illustrated in the
computer code 
that accompanies this \chptr.}~-- see \ref{T-inf}.

The transport of the population mean in terms of consumption, shown in
Fig.~\ref{fgKS12}, allows 
one to produce the transport of the population mean in terms of average installed capital~--
see \ref{transport-K}~-- which is shown on Fig.~\ref{fgKS13}.  
{\captionsetup{belowskip=-5pt}
\begin{figure}[!htbp]
\centering
\begin{subfigure}{.5\textwidth}
  \centering
\toshow{\includegraphics[width=7.5cm]{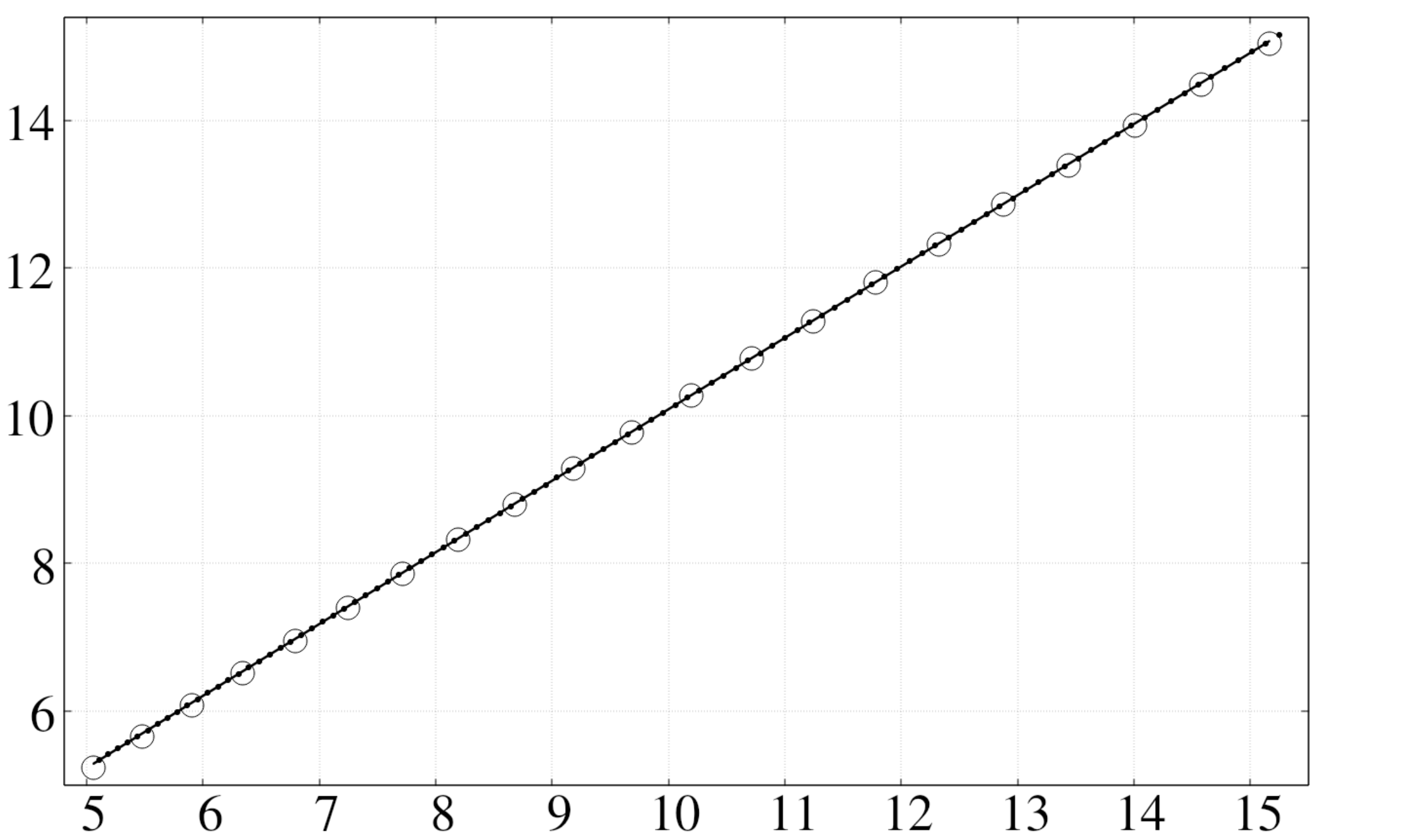}}
\end{subfigure}%
\begin{subfigure}{.5\textwidth}
  \centering 
\toshow{\includegraphics[width=7.5cm]{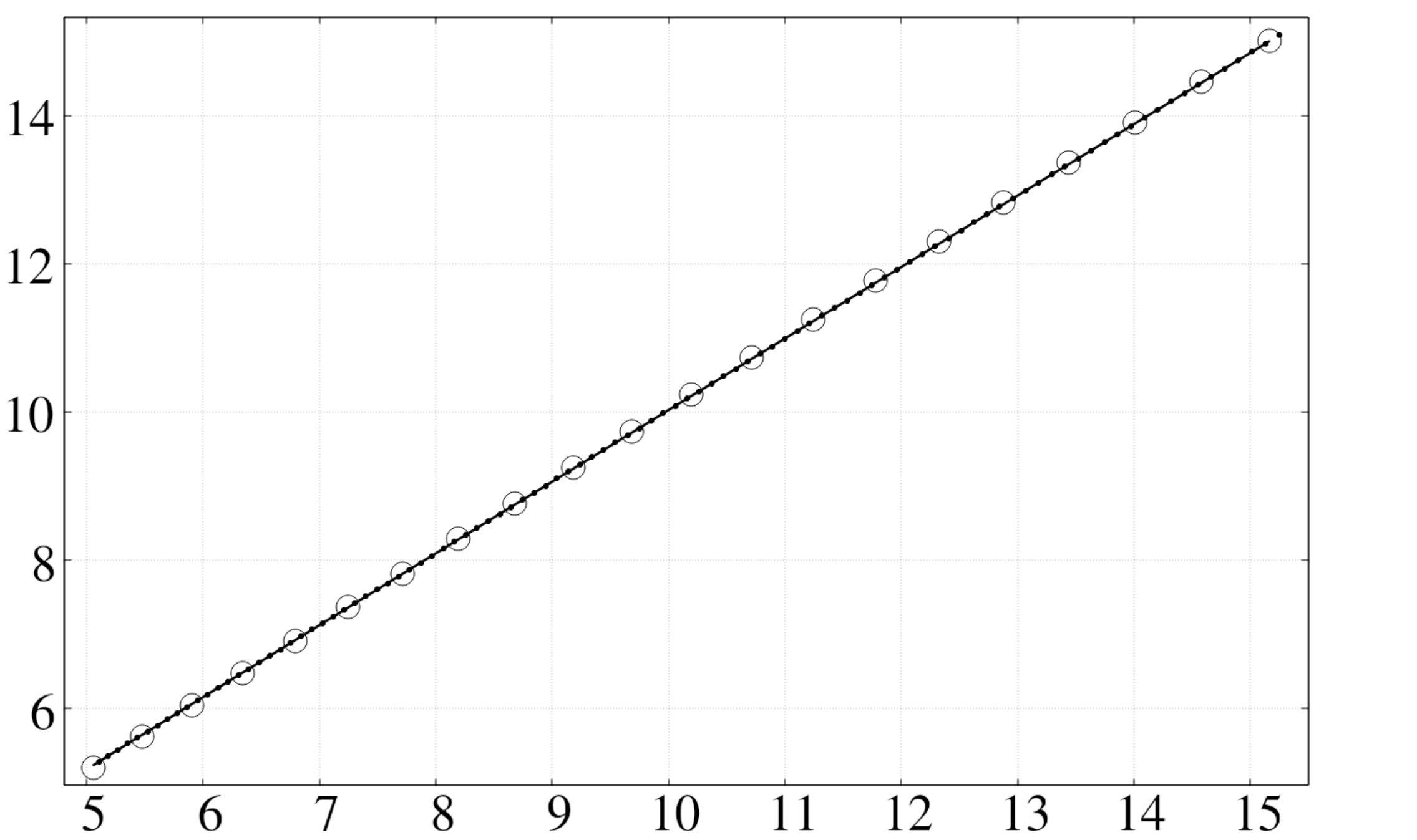}  }      
\end{subfigure}
\caption{The transport of the population capital investment mean into high state (left plot) and into
low state (right plot) from high state (solid lines) and low state (doted lines).
The circles show Krusell-Smith's
 prediction [\protect\cited{KS98}-III] by way of log-linear least square fit from a simulated long-run behavior of
a large population of households. The uniform distance between the solid and the dotted line is around 0.00076
in the left plot and around 0.00083 in the right plot. The largest
disagreement between
Krusell-Smith's prediction and the solid lines is around 0.051 in the left plot and around 0.032 in the
right plot. The uniform distance between the two solid lines is around 0.06952.}
\label{fgKS13}
\end{figure} }%
As can be seen from Fig.~\ref{fgKS13},
this transport is remarkably close to Krusell-Smith's prediction by way of least-square
log-linear fit.
This should not be surprising, given that in this particular example
the transport is nearly affine and the dependence on the
departing aggregate state is negligible.
\footnote{There is no intuition to suggest that the transport mappings shown in Fig.~\ref{fgKS13}
would be nearly affine and nearly identical with any choice of the model parameters whatsoever.
There is therefore no reason to expect that the predictions produced by Krusell-Smith's method
would always be as accurate as they are in the benchmark case illustrated in the paper \cite{KS98}
(and in this section).}

With the transformations shown in Fig.~\ref{fgKS12}
at hand~-- see \eqref{eq-T-star}~-- one can easily simulate the long run behavior of the population
consumption mean 
by merely simulating a time series of the productivity state (not of the private states of a large
population of households) and by applying one of the
four transformations in Figures~\ref{fgKS12} consecutively from some arbitrary initial mean value (one
must test empirically that the choice of the initial value has no effect on the long run
behavior).
With the time series of the productivity state and the population mean at hand,
one can generate the corresponding series of employment specific mean values $(A^\h,A^0)\in\R^2$~--
see \eqref{no-fp-2-0} and the comment in~\ref{collapse}.
Starting with 
$A^\h=0.8$ and $A^0=0.7$ in high productivity state,
the bi-variate series of employment-specific mean values was simulated for~$1.1$ million periods.
The output from the last $10^4$ periods is shown in~Fig.~\ref{fgKS5},
which exhibits a typical law of motion in random environment: with high
probability the productivity state remains unchanged from one period to the next, so that the law of
motion of the population state is deterministic and governed either by
the mapping $\T_{\infty,1}^1\phd$ or by $\T_{\infty,2}^2\phd$ from \eqref{no-fp-2-0}~--
until a change in the productivity state occurs, in which case the population state is transformed either by
$\T_{\infty,1}^2\phd$ or by $\T_{\infty,2}^1\phd$.
\footnote{{}With a slight abuse of the notation we use the same token $\T_{\infty,x}^y\phd$ to denote
the transport of the vector of employment specific mean values from \eqref{no-fp-2-0} and also the
transport of the total population mean from \eqref{eq-T-star}. The precise meaning is given by the context.}
{
\begin{figure}[!htbp]  
\centering
\begin{subfigure}{.5\textwidth}
  \centering
\leavevmode\raise0.8cm\hbox{\rotatebox{90}{\tiny average consumption unemployed}}%
\ %
\toshow{\includegraphics[width=7.1cm]{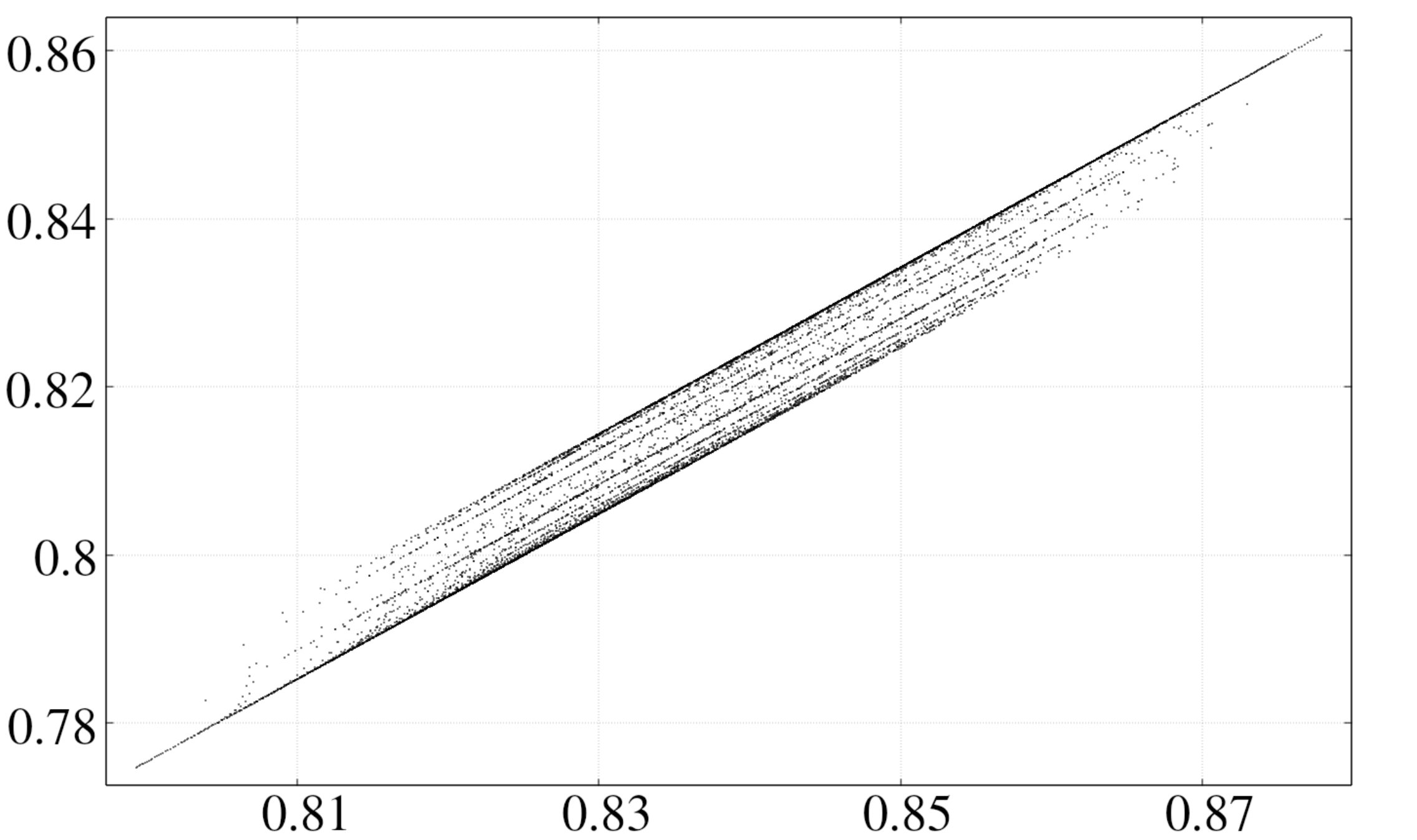}}

\leavevmode\smash{\raise6pt\hbox{\tiny average consumption employed}}
\end{subfigure}%
\begin{subfigure}{.5\textwidth}
  \centering
\leavevmode\raise1cm\hbox{\rotatebox{90}{\tiny average investment unemployed}}%
\ %
\toshow{\includegraphics[width=7.1cm]{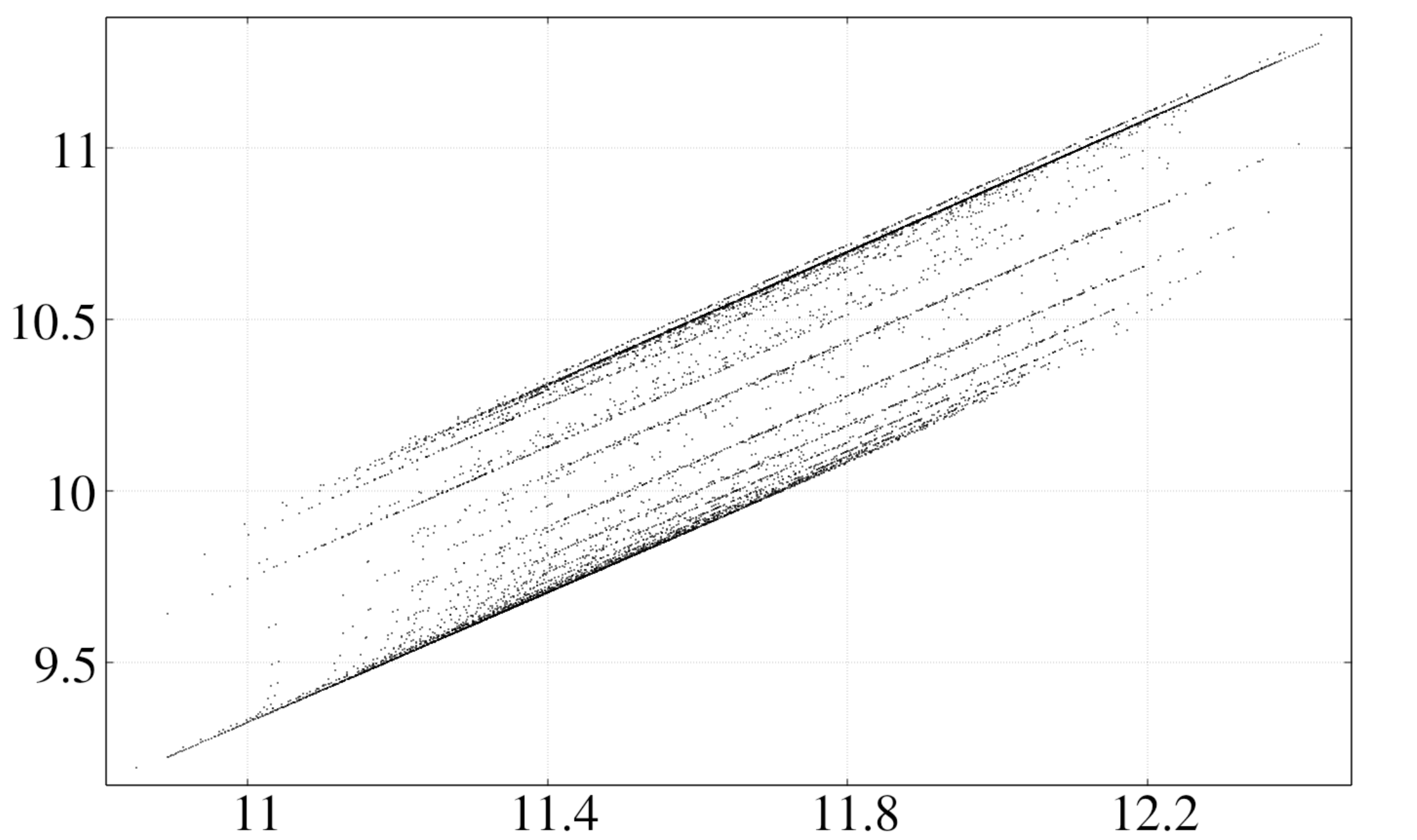}}
 
\leavevmode\smash{\raise6pt\hbox{\tiny average investment employed}} 
\end{subfigure}
\caption{The employment-specific population means in the last 10,000 periods in a simulated
series of $1.1$ million periods.}
\label{fgKS5}
\end{figure} }%
The nearly affine patterns in Fig.~\ref{fgKS5} are easily explained by the fact
that the mappings in \eqref{no-fp-2-0} are very close to affine. As~a result, the
disparity between employed and unemployed, whether in terms of consumption or wealth,
has a nearly affine structure that remains
unchanged for as long as the productivity state remains the
same (whence the straight lines), but that structure changes when the
productivity state flips. 
It is interesting to
note that the data shown in Fig.~\ref{fgKS5} is much more dispersed than the fluctuations
in the random environment (the productivity state) that is causing them, which has
only two values (high and low)~-- a ``ratchet effect'' of a sort.
The larger dispersion in the data presented in the right plot in Fig.~\ref{fgKS5},
in which the state of the population is described in terms of
the employment-specific mean investment level, is quite intuitive: the
households' savings function 
as ``shock absorbers.''
Perhaps the most
important takeaway from these plots is that that they reveal a structure that would not be possible to capture
if the state of the population is collapsed to a single scalar value, whether that value
is the mean consumption level or the mean investment level across the entire
population.
\footnote{Applying Krusell-Smith's approach with higher order moments of the population
distribution, as originally proposed
in \cite{KS98}, would still not allow one to compare the employment specific averages.}
Of course, the simulated employment-specific mean values from the
right plot in Fig.~\ref{fgKS5} can easily be
transformed into total population mean values. These values are shown in Fig.~\ref{fgKS-IK}
and illustrate the long-run range of the average installed capital, which range
is consistent with the results in \citex[III]{KS98}.
{\captionsetup{belowskip=-5pt}
\begin{figure}[!htbp]
\centering
\toshow{\includegraphics[width=7.5cm]{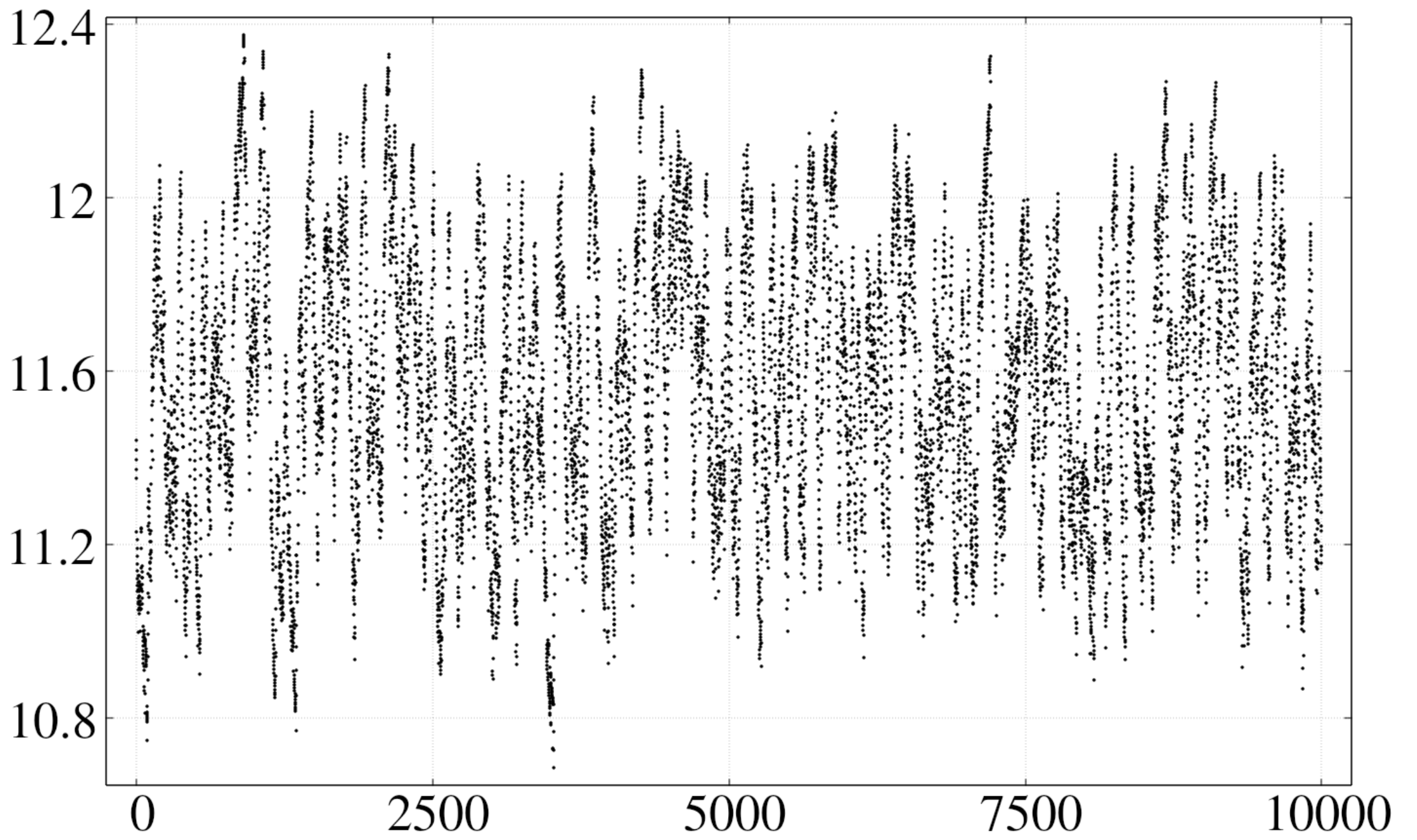}}
\caption{Average installed capital in the last 10,000 periods in a simulated sample of 1.1 million periods.}
\label{fgKS-IK}  
\end{figure} }%

\begin{nit}{Remark}\label{rem85521}%
The plot in Fig.~\ref{fgKS-IK}
reveals fluctuations that are much more dispersed  than the fluctuations in the
exogenous productivity 
state. In particular, the plot shows (by way of an example) that the pair composed of
the aggregate  
productivity state and the capital investment of a ``representative household''
(stipulating that one exists) 
would not have a
stationary long-run distribution, if one is to also stipulate
that the cross-sectional distribution of all households can 
be identified as the probability distribution of a single representative household.
Indeed, if that would be the case, then there would be only one (conditional) population
average to attach to each aggregate state in the long run. Consequently, the plot in
Fig.~\ref{fgKS-IK} would be showing fluctuations between only two points, as there
are only two 
productivity states in this example.\qed 
\end{nit}

Although it is assumed throughout this \chptr\  that the discount (impatience) factor $\b$ is
constant, apart from the desire for greater simplicity,
nothing in the general computational strategy that we have deployed makes such an assumption
necessary. A model with stochastic $\b$ is introduced and discussed extensively
in \cite{KS98}.
While the exploration of such models (and they are many~-- see \cite{KS98})
falls outside the scope of this
\chptr, a feature that can be illustrated here with very little additional effort is to rerun the Julia program
employed in the foregoing with a different value for the impatience parameter, namely with
$\b=0.96$, instead of $\b=0.99$, which was taken from the benchmark case in~\cite{KS98}.
For the sake of brevity, we produce here only the output from the simulation of the bi-variate
state of the population~-- see Fig.~\ref{fgKSlast}.
{
\begin{figure}[!htbp]  
\centering
\begin{subfigure}{.5\textwidth}
  \centering
\leavevmode\raise0.55cm\hbox{\rotatebox{90}{\tiny average consumption unemployed}}%
\ %
\toshow{\includegraphics[width=5.1cm]{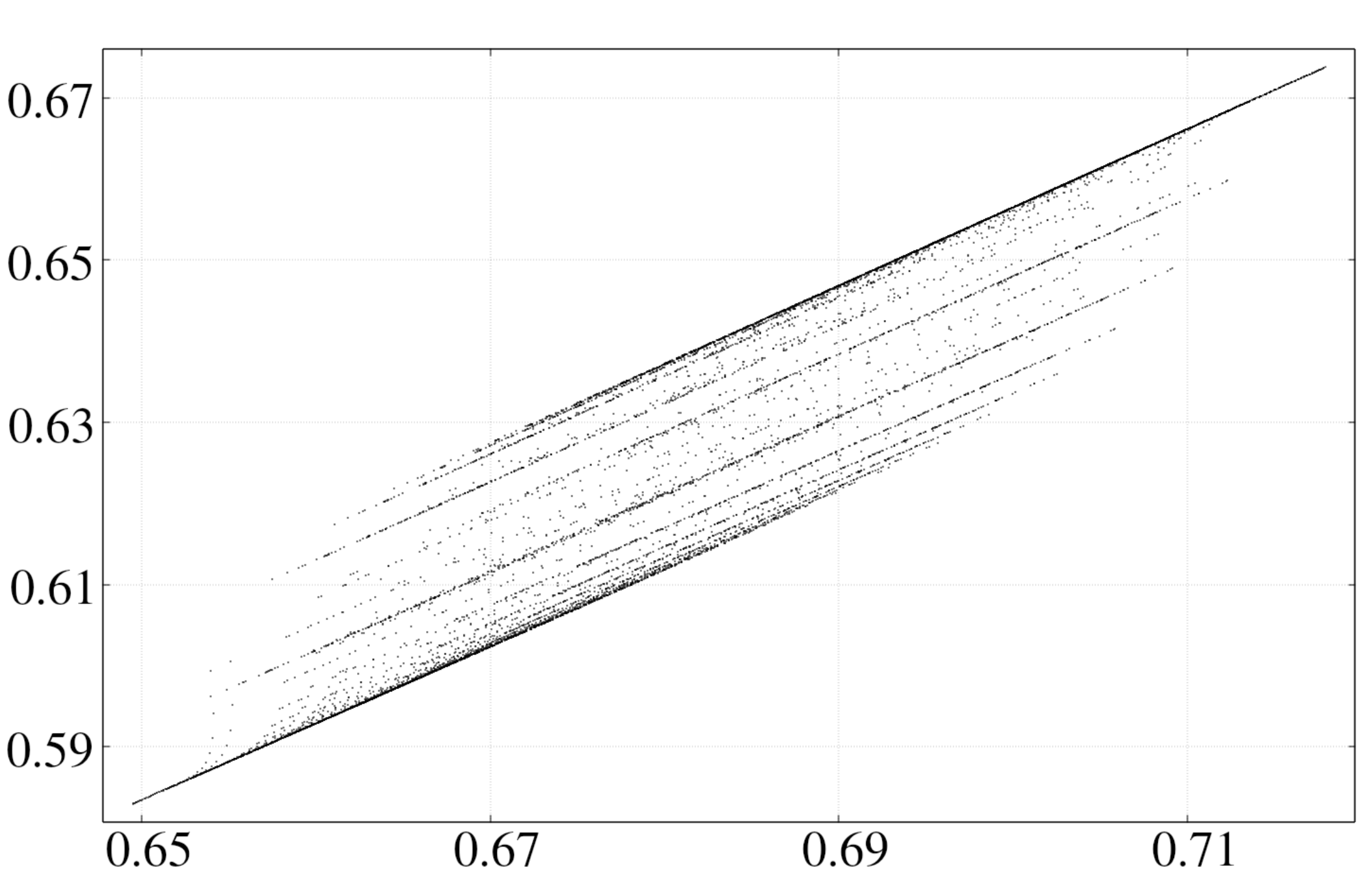}}

\leavevmode\smash{\raise6pt\hbox{\tiny average consumption employed}}
\end{subfigure}%
\begin{subfigure}{.5\textwidth}
  \centering
\leavevmode\raise0.55cm\hbox{\rotatebox{90}{\tiny average investment unemployed}}%
\ %
\toshow{\includegraphics[width=5.1cm]{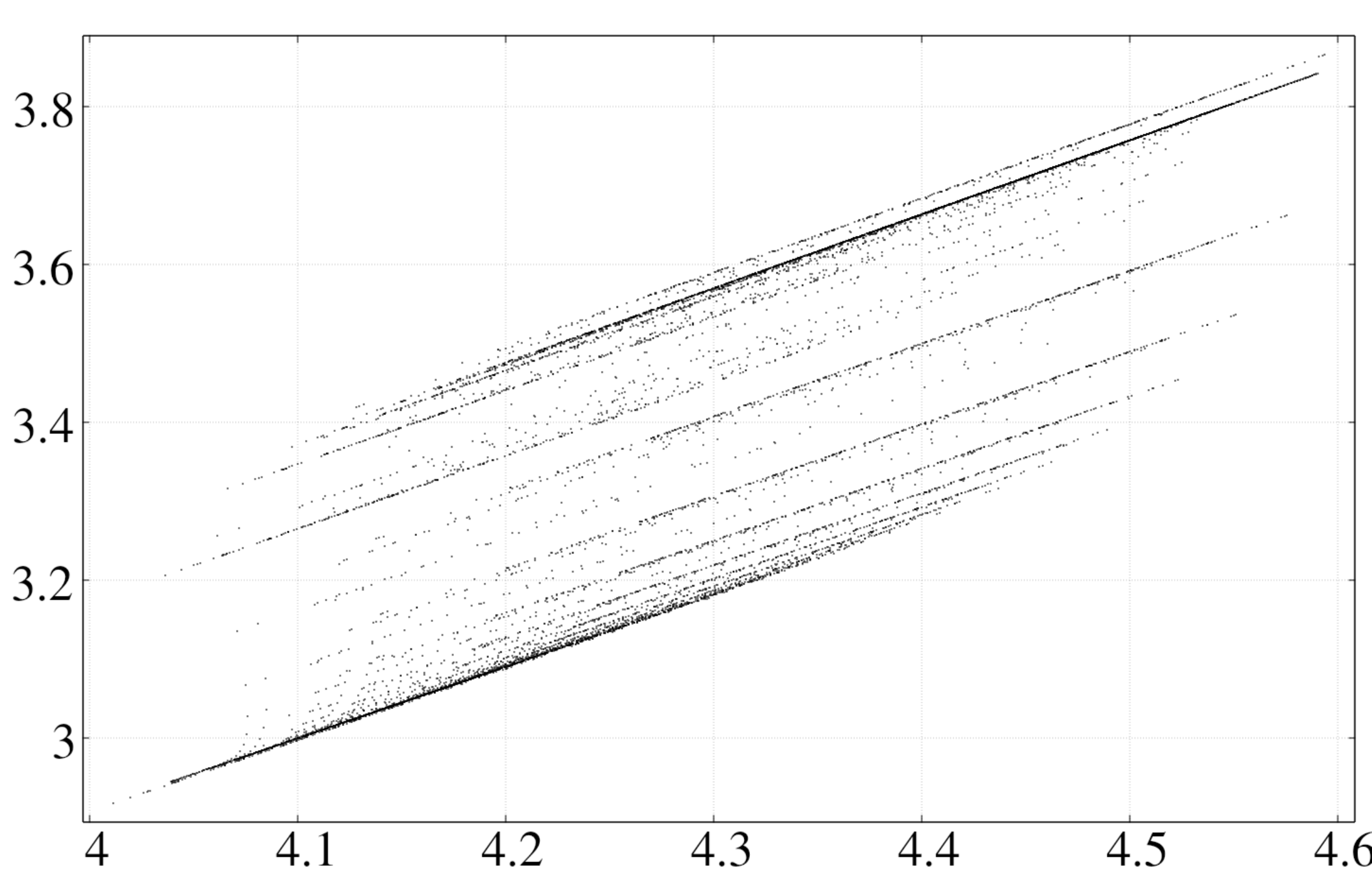}} 

\leavevmode\smash{\raise6pt\hbox{\tiny average investment employed}} 
\end{subfigure}
\caption{The employment specific population means in the last 10,000 periods in a simulated
series of $1.1$ million periods with impatience rate $\b=0.96$ (changed from $\b=0.99$).}
\label{fgKSlast}
\end{figure}
}%
These plots are consistent with the intuition: when the households are more
impatient they invest less, which then leads to a lower output, and ultimately to lower consumption.
What is surprising,
however, is that decreasing the discount factor by only $3\%$ can lead to substantially lower
investment and consumption levels in the long run. It~is also interesting to note that increased impatience leads
to a greater variations in the relative disparity between unemployed and employed~-- in both
consumption and capital investment.

\section{Conclusion}
\label{sec:CONC}\setcounter{paragraph}{0}

\noindent
General equilibrium incomplete market models with a large number of heterogeneous
agents are intrinsically self-consistent (see \ref{et} and \ref{self-dep}) and also
time-interlaced (see \ref{cross-sys}).
In~addition, the mean field interactions are only implicit and, most important,
endogenous (see \ref{iterlaced-rem}). 
Because of this intrinsic structure,
notwithstanding some special cases, the equilibria
in such models do not appear possible to obtain generically with existing tools,
which, as we have seen in Sec.~\ref{sec:Intro}, may fail to produce the equilibrium
even in a very simple setting with a fairly realistic choice for the model parameters. 
The key contributions in this \chptr\ can be summarized as follows:

(a)~Previously unknown features of the population transport are uncovered
(see \ref{main-q} and \ref{n2}).
The key innovation in developing the transport equation \ref{main-q}-(\ref{zze666}) is
in the use of the state transition assignments (see \ref{rnd-Monge}).
The rôle that these assignments play is analogous to the rôle of the flux in the
derivation of the continuity equation in fluid mechanics, but these assignments are
not a flux and the transport equation \ref{main-q}-(\ref{zze666}) does not appear to be
analogous to the continuity equation, or analogous to the master equation in MFG,
mainly because of its time-interlaced and self-consistent structure, in conjunction
with the endogenous nature of the mean field interactions~--
see \ref{rnd-Monge} and \ref{n2} for further details. 

(b)~A new algorithm (see \ref{main-proc}), which yields equilibria that are seemingly
unat\-tain\-able by other means (see Fig.~\ref{fg4} and the discussion in
Sec.~\ref{sec:Intro})
is developed. In~addition, an essentially complete resolution to Krusell-Smith's
conjecture is provided~-- see~\ref{mod-accu}. 
The new algorithm mimics
the usual backward induction program, except that the steps are time-interlaced
in a certain special way (see \ref{cross-sys} and \ref{iterlaced-rem}).
This time-interlaced structure is one of the key innovations in this \chptr.

(c)~The interaction between the private choices and the
cross-sectional distribution of the population (i.e., the ``coupling function'' in
the parlance of MFG) is endogenized. This is important, because specifying the
mean field interactions extraneously amounts to an extraneous restriction on the
equilibrium choices. Another notable departure from the common MFG formulation,
in addition to the time-interlaced reordering of the conditions, is
that the coupling function enters the model only implicitly~--
see~\ref{iterlaced-rem}.

(d)~The new method does not require borrowing constraints to be imposed in the outset,
i.e., the borrowing limit is endogenized~-- see footnote~\ref{ftn-no-b}.
This feature is quite intuitive: the
amount an agent can 
borrow is constrained only by the amount other agents are willing to lend.
Constraints of this type are   
difficult to impose exogenously in the outset,
because they are a consequence of the available wealth and its distribution in the
cross-section of households, 
as well as a consequence of the
collective agreement about the (endogenous) prices, in conjunction with the need for all
private budgets to remain balanced at all times~-- clearly, such an information
becomes available only after the equilibrium arrangement has been identified. 

(e)~The new methodology does not involve the representative agent point of view in
any way and the transport equation (see \ref{main-q}), which governs the time
evolution of the cross-sectional distribution of the population, does not involve the
time evolution of the probability distribution of any observable stochastic  variable
(such as the state of a particular Markov chain, for example).
In fact, the numerical examples presented in the \chptr\ illustrate why
interpreting the cross-sectional distribution of the population as the probability
distribution of a single representative agent may be undesirable even in some simple
scenarios. 
Moreover, the trajectories of the individual states, or even just of a single
state (representative or not), never enter the model (see \ref{1350}), whence the
absence of a Bellman equation in the model.  

The new method can be implemented by way of functional programming
and can produce equilibria that are consistent 
with widely cited case studies, but also equilibria in some critically important
classical models 
that have not been adequately resolved by other means.
It can also replicate
the well documented and widely discussed results of Krusell and Smith
without simulation and without the need to postulate infinite time horizon
(in fact, with a closed-form solution for some important parts of the equilibrium)~--
see \ref{closed-form}, \ref{transport-K},  Fig.~\ref{fgKS13}.
It provides a nearly complete answer to the question whether or not 
only-the-mean-matters point of view, put forward by Krusell and Smith \cite{KS98},
is reasonably accurate and under what restrictions~-- see~\ref{mod-accu}.
Moreover, the new approach
reveals previously unknown
structures and features of Krusell-Smith's model~-- see Fig.~\ref{fgKS5} and
Fig.~\ref{fgKSlast}.   

\section*{References}
\penalty10000%
\begingroup%
\pretolerance=200%
\tolerance=1800%
\ignorespaces%

    \brefitem[Achdou et al$.$ (2014)]{ABLLM14}
    \aut{Achdou, Yves, Francisco J.\ Buera, Jean-Michel Lasry, Pierre-Louis Lions, and Benjamin Moll}
    \anno{2014} 
    \jtitle{Partial differential equations in macroeconomics}
    \journal{Philosohpical Transactions of the Royal Society of London} {A372} 20130397.

    \brefitem[Achdou et al$.$ (2013)]{ACD13}
    \aut{Achdou, Yves, Fabio Camilli, and Italo Capuzzo-Dolcetta}
    \anno{2013} 
    \jtitle{Mean field games: convergence of a finite difference method}
    \journal{Siam Journal on Numerical Analysis} {51} 2585-\nobreak2612.

    \brefitem[Achdou and Capuzzo-Dolcetta (2010)]{AD10}
    \aut{Achdou, Yves, and Italo Capuzzo-Dolcetta}
    \anno{2010} 
    \jtitle{Mean field games: numerical methods}
    \journal{Siam Journal on Numerical Analysis} {48} 1136-\nobreak1162.

         \brefitem[Achdou et al$.$ (2023)]{ACPT23}
         \aut{Achdou, Yves, Guillaume Carlier, Quentin Petit, Daniela Tonon}
	 \anno{2023} 
	 \btitle{A mean field model for the interactions between firms on the markets of their inputs}
         \journal{Mathematics and Financial Economics} (to appear).

    \brefitem[Achdou et al$.$ (2022a)]{AHLLM17}
    \aut{Achdou, Yves, Jiequn Han, Jean-Michel Lasry, Pierre-Louis Lions, and Benjamin Moll}
    \anno{2022} 
    \jtitle{Income and wealth distribution in macroeconomics: a continuous-time approach}
    \journal{Review of Economic Studies} {89} 45-\nobreak86.

         \brefitem[Achdou et al$.$ (2022b)]{ALL22}
         \aut{Achdou, Yves, Jean-Michel Lasry, and Pierre-Louis Lions}
	 \anno{2022} 
	 \btitle{Simulating nu\-me\-ri\-cally the Krusell-Smith model with neural networks} 
\book{\href{https://doi.org/10.48550/arXiv.2211.07698}{DOI: 10.48550 / arXiv.2211.07698}}.

    \brefitem[Achdou and Porretta (2016)]{AP16}
    \aut{Achdou, Yves, and Alessio Porretta}
    \anno{2016} 
    \jtitle{Convergence of a finite difference scheme to weak solutions of the system of partial differential equation arising in mean field games}
    \journal{Siam Journal on Numerical Analysis} {54} 161-\nobreak186.

    \brefitem[Açıkgöz (2018)]{Aci18}
    \aut{Açıkgöz, Ömer T}
    \anno{2018} 
    \jtitle{On the existence and uniqueness of stationary equilibrium in Bewley economies with production}
    \journal{Journal of Economic Theory} {173} 18-\nobreak55.

    \brefitem[Ahn et al$.$ (2018)]{AKMWW18}
    \aut{Ahn, SeHyoun, Greg Kaplan, Benjamin Moll, Thomas Winberry, and Christian Wolf}
    \anno{2018} 
\jtitle{When inequality matters for macro and macro matters for inequality}
\book{In: Eichenbaum, M., Hurst, E., and Parker, J.~A.~(Eds.), {\it NBER Macroeconomics Annual 2017, vol.~32\/}. Chicago, IL: University of Chicago Press, pp.~1-75}.

    \brefitem[Aiyagari (1993)]{Aiya93}
    \aut{Aiyagari, S.~Rao}
    \anno{1993} 
\jtitle{Uninsured idiosyncratic risk and aggregate saving}
\book{\it Working paper 502, Federal Reseve Bank of Minneapolis}.

    \brefitem[Aiyagari (1994)]{Aiya94}
    \aut{Aiyagari, S.~Rao}
    \anno{1994} 
\jtitle{Uninsured idiosyncratic risk and aggregate saving}
\journal{The Quarterly Journal of Economics} 109 659-\nobreak684.

    \brefitem[Aiyagari (1995)]{Aiya95}
    \aut{Aiyagari, S.~Rao}
    \anno{1995} 
    \jtitle{Optimal capital income taxation with incomplete markets, borrowing constraints and constant discounting}
    \journal{Journal of Political Economy} {6} 1158-\nobreak1175.

    \brefitem[Algan et al$.$ (2010)]{AAD10}
    \aut{Algan, Yann, Olivier Allais, and Wouter J.~Den Haan}
    \anno{2010} 
    \jtitle{Solving the incomplete markets model with aggregate uncertainty using parameterized cross-sectional distributions}
    \journal{Journal of Economic Dynamics \& Control} {34} 59-\nobreak68.

    \brefitem[Aumann (1964)]{Aum64}
    \aut{Aumann, Robert J}
    \anno{1964} 
    \jtitle{Markets with a continuum of traders}
    \journal{Econometrica} {32} 39-\nobreak50.

    \brefitem[Aumann (1966)]{Aum66}
    \aut{Aumann, Robert J}
    \anno{1966} 
    \jtitle{Existence of competitive equilibria in markets with a continuum of traders}
    \journal{Econometrica} {34} 1-\nobreak17.

    \brefitem[Bayer and Luetticke (2020)]{BL20}
    \aut{Bayer, Christian, and Ralph Luetticke}
    \anno{2020} 
    \jtitle{Solving discrete time heterogeneous agent models with aggregate risk and many idiosyncratic states by perturbation}
    \journal{Quantitative Economics} {11} 1253-\nobreak1288.

    \brefitem[Benamou and Brenier (2000)]{BB00}
    \aut{Benamou, Jean-David, and Yann Brenier}
    \anno{2000} 
\jtitle{A computational fluid mechanics solution to the Monge-Kantorovich mass transfer problem}
\journal{Numerische Mathematik} {84} 375-\nobreak393.

    \brefitem[Benamou et al$.$ (2017)]{BCS17}
    \aut{Benamou, Jean-David, Guillaume Carlier, and Filippo Santambrogio}
    \anno{2017} 
\jtitle{Variational mean field games}
\book{In: Bellomo, N., Degond, P., and Tadmor, E.~(Eds.), {\it Active Particles, Volume 1\/}. Cham, CH: Springer International, pp.~141-171}.

         \brefitem[Bensoussan et al$.$  (2013)]{BFY13}
         \aut{Bensoussan, Alain, Jens Frehse, and Phillip Yam}
	 \anno{2013} 
	 \btitle{Mean Field Games and Mean Field Type Control Theory} 
         \book{New York, NY: Springer}.

           \brefitem[Bertoin (1996)]{Ber96}
           \aut{Bertoin, Jean}
	   \anno{1996}
           \btitle{Lévy Processes}
           \book{Cambridge, UK: Cambridge University Press}.

    \brefitem[Bewley (1977)]{Bew77}
    \aut{Bewley, Truman F}
    \anno{1977} 
    \jtitle{The permanent income hypothesis: a theoretical formulation}
    \journal{Journal of Economic Theory} {16} 252-\nobreak292.

    \brefitem[Bewley (1980)]{Bew80}
    \aut{Bewley, Truman F}
    \anno{1980} 
\jtitle{The optimum quantity of money}
\book{In: Kareken, J.~H., Wallace, N.\ (Eds.), {\it Models of Monetary Economics\/}. Minneapolis, MI: Federal Reserve Bank of Minneapolis, pp.~169-210}.

    \brefitem[Bewley (1983)]{Bew83}
    \aut{Bewley, Truman F}
    \anno{1983} 
\jtitle{A difficulty with the optimum quantity of money}
\journal{Econometrica} {16} 252-\nobreak292.

    \brefitem[Bewley (1986)]{Bew86}
    \aut{Bewley, Truman F}
    \anno{1986} 
\jtitle{Stationary monetary equilibrium with a continuum of independently fluctuating consumers}
\book{In: Hildenbrand, W., Mas-Colell, A.\ (Eds.), {\it Contributions to Mathematical Economics in Honor of Gerard Debreu\/}. Amsterdam, NL: North Holland, pp.~79-102}.

    \brefitem[Bismut (1973)]{Bis73}
    \aut{Bismut, Jean-Michel}
    \anno{1973} 
\jtitle{Conjugate convex functions in optimal stochastic control}
\journal{Journal of Mathematical Analysis and Applications} {44} 384-\nobreak404.

    \brefitem[Brumm et al$.$ (2017)]{BKK17}
    \aut{Brumm, Johannes, Dominika Kryczka, and Felix Kubler}
    \anno{2017} 
\jtitle{Recursive equilibria in dynamic economies with stochastic production}
\journal{Econometrica} {85} 1467-\nobreak1499.

    \brefitem[Cannarsa and Capuani (2018)]{CanCap18}
    \aut{Cannarsa, Piermarco, and Rossana Capuani}
    \anno{2018} 
    \jtitle{Existence and uniqueness of mean field games with state constraints}
    \book{In: Cardaliaguet, P., Porretta, A., and Salvarani, F.~(Eds.), {\it PDE Models for Multi-Agent Phenomena\/}. Cham, CH: Springer Nature, pp.~49-71}.

    \brefitem[Cannarsa et al$.$ (2021)]{CCC21}
    \aut{Cannarsa, Piermarco, Rossana Capuani, and Pierre Cardaliaguet}
    \anno{2021} 
    \jtitle{Mean field games with state constraints: from mild to pointwise solutions of the PDE system}
    \journal{Calculus of Variations and Partial Differential Equations} {60} 108.

    \brefitem[Cao (2020)]{Cao20}
    \aut{Cao, Dan}
    \anno{2020} 
    \jtitle{Recursive equilibrium in Krusell and Smith (1998)}
    \journal{Journal of Economic Theory} {186} \href{https://doi.org/10.1016/j.jet.2019.104978}{DOI: 10.1016 / j.jet.2019.104978}.

         \brefitem[Cardaliaguet et al$.$  (2019)]{CDLL19}
         \aut{Cardaliaguet, Pierre, Fran{\cc}ois Delarue, Jean-Michele Lasry, and Pierre-Louis Lions}
	 \anno{2019} 
	 \btitle{The Master Equation and the Convergence Problem in Mean Field Games} 
         \book{Princeton, NJ: Princeton University Press}.

         \brefitem[Carmona and Delarue (2014)]{CD14}
         \aut{Carmona, Ren{\ea}, and Fran{\cc}ois Delarue}
	 \anno{2014} 
	 \btitle{The master equation for large population equilibriums} 
         \book{\href{https://doi.org/10.48550/arXiv.1404.4694}{DOI: 10.48550 / arXiv.1404.4694}}.

         \brefitem[Carmona and Delarue (2018a)]{CD18a}
         \aut{Carmona, Ren{\ea}, and Fran{\cc}ois Delarue}
	 \anno{2018a} 
	 \btitle{Probabilistic Theory of Mean Field Games with Applications~I: mean field FBSDEs, control, and games} 
         \book{Cham, CH: Springer}.

         \brefitem[Carmona and Delarue (2018b)]{CD18b}
         \aut{Carmona, Ren{\ea}, and Fran{\cc}ois Delarue}
	 \anno{2018b} 
	 \btitle{Probabilistic Theory of Mean Field Games with Applications~II: mean field games with common noise and master equations} 
         \book{Cham, CH: Springer}.

         \brefitem[Carmona and Delarue (2018)]{CD18}
         \aut{Carmona, Ren{\ea}, and Fran{\cc}ois Delarue}
	 \anno{2018} 
	 \btitle{Probabilistic Theory of Mean Field Games with Applications~I\ts\&\ts II} 
         \book{Cham, CH: Springer}.

         \brefitem[Carmona et al$.$ (2016)]{CDL16}
         \aut{Carmona, Ren{\ea}, Fran{\cc}ois Delarue, and Daniel Lacker}
	 \anno{2016} 
	 \jtitle{Mean Field Games with Common Noise} 
         \journal{Annals of Applied Probability} {44} 3740-\nobreak3803.

    \brefitem[Cass and Shell (1983)]{CS83}
    \aut{Cass, David, and Karl Shell}
    \anno{1983} 
    \jtitle{Do sunspots matter?}
    \journal{Journal of Political Economy} {91} 193-\nobreak227.

    \brefitem[Chassagneux et al$.$ (2014)]{CCD14}
    \aut{Chassagneux, Jean-Fran{\cc}ois, Dan Crisan, and Fran{\cc}ois Delarue}
    \anno{2014} 
    \jtitle{Classical solutions to the master equation for large population equilibria}
    \book{\href{https://doi.org/10.48550/arXiv.1411.3009}{DOI: 10.48550 / arXiv:1411.3009}}.

    \brefitem[Cheridito and Sagredo (2016a)]{ChS16}
    \aut{Cheridito, Patrick, and Juan Sagredo}
    \anno{2016} 
\jtitle{Existence of sequential competitive equilibrium in Krusell-Smith type economies}
\book{\href{https://doi.org/10.2139/ssrn.2794629}{DOI: 10.2139 / ssrn.2794629}}. 

    \brefitem[Cheridito and Sagredo (2016b)]{ChS16b}
    \aut{Cheridito, Patrick, and Juan Sagredo}
    \anno{2016} 
\jtitle{Comment on ``Competitive equilibria of economies with a continuum of consumers and aggregate shocks'' by J. Miao}
\book{\href{https://doi.org/10.2139/ssrn.2815971}{DOI: 10.2139 / ssrn.2815971}}.

    \brefitem[Clarida (1990)]{Cla90}
    \aut{Clarida, Richard H}
    \anno{1990} 
\jtitle{International lending and borrowing in a stochastic, stationary equilibrium}
\journal{International Economic Review} {31} 543-\nobreak558.

    \brefitem[Cociuba et al$.$ (2018)]{CPU18}
    \aut{Cociuba, Simona E., Edward C.~Prescott, and  Alexander Ueberfeldt}
    \anno{2018} 
\jtitle{US hours at work}
\journal{Economics Letters} {169}(C) 87-\nobreak90.

    \brefitem[Cogburn (1980)]{Cog80}
    \aut{Cogburn, Robert}
    \anno{1980} 
    \jtitle{Markov chains in random environment: the case of MArkovian environments}
    \journal{The Annals of Probability} {8} 908-\nobreak916.

    \brefitem[Constantinides (1990)]{Con90}
    \aut{Constantinides, George M}
    \anno{1990} 
\jtitle{Habit formation: a resolution of the equity premium puzzle}
\journal{Journal of Political Economy} {98} 519-\nobreak543.

    \brefitem[Constantinides and Duffie (1996)]{CD96}
    \aut{Constantinides, George M., and Darrell Duffie}
    \anno{1996} 
\jtitle{Asset pricing with heterogeneous consumers}
\journal{Journal of Political Economy} 104 219-\nobreak240.

    \brefitem[Deaton (1991)]{Dea91}
    \aut{Deaton, Angus}
    \anno{1991} 
    \jtitle{Saving and liquidity constraints}
    \journal{Econometrica} {59} 1221-\nobreak1248.

    \brefitem[Delarue et al$.$ (2018)]{DLR18}
    \aut{Delarue, Fran{\cc}ois, Daniel Lacker, and Kavita Ramanan}
    \anno{2018} 
    \jtitle{From the master equation to mean field game limit theory: large deviations and concentration of measure}
    \book{\href{https://doi.org/10.48550/arXiv.1804.08550}{DOI: 10.48550 / arXiv:1804.08550}}. 

           \brefitem[Dellacherie (1972)]{Dell72}
           \aut{Dellacherie, Claude}
	   \anno{1972}
           \btitle{Capacit{\ea}s et Processus Stochastique}
           \book{Berlin, DE: Springer}.

    \brefitem[Dellacherie (1980))]{Dell80}
    \aut{Dellacherie, Claude}
    \anno{1980} 
\jtitle{Un survol de la theorie de l'integrale stochastique}
\journal{Stochastic Processes and Their Applications} {10} 115-\nobreak144.

                 \brefitem[Dellacherie and Meyer (1975)]{DellMey78}
           \aut{Dellacherie, Claude, et Paul-Andr{\ea}~Meyer}
	   \anno{1978}
\btitle{Probabilities and Potential}
\book{Paris, FR: Hermann}.

                 \brefitem[Dellacherie and Meyer (1980)]{DellMey80}
           \aut{Dellacherie, Claude, et Paul-Andr{\ea}~Meyer}
	   \anno{1980}
\btitle{Probabilit{\ea}s et Potentiel, Chapitres V {\ag} VIII: Th{\ea}orie des Martingales}
\book{Paris, FR: Hermann}.

    \brefitem[Den Hann (1996)]{DH96}
    \aut{Den Haan, Wouter J}
    \anno{1996} 
    \jtitle{Heterogeneity, aggregate uncertainty, and the short-term interest rate}
    \journal{Journal of Business \& Economic Statistics} {14} 399-\nobreak411.

    \brefitem[Den Hann (2010a)]{DH10a}
    \aut{Den Haan, Wouter J}
    \anno{2010a} 
    \jtitle{Comparison of solutions to the incomplete markets model with aggregate uncertainty}
    \journal{Journal of Economic Dynamics \& Control} {34} 4-\nobreak27.

    \brefitem[Den Hann (2010b)]{DH10b}
    \aut{Den Haan, Wouter J}
    \anno{2010} 
    \jtitle{Assessing the accuracy of the aggregate law of motion in models with heterogeneous agents}
    \journal{Journal of Economic Dynamics \& Control} {34} 79-\nobreak99.

    \brefitem[Den Hann and Marcet (1990)]{DM90}
    \aut{Den Haan, Wouter J., and Albert Marcet}
    \anno{1990} 
    \jtitle{Solving the stochastic growth model by parameterizing expectations}
    \journal{Journal of Business \& Economic Statistics} {8} 31-\nobreak34.

          \brefitem[Dol{\ea}ans-Dade (1970)]{Dol70}
    \aut{Dol{\ea}ans-Dade, Catherine}
    \anno{1970} 
\jtitle{Quelques applications de la formule de changement de variables pour les semimartingales} 
\journal{Zeitschrift f{\uD}r Wahrscheinlichkeitstheorie und Verwandte Gebiete} 16 181--\nobreak194.

          \brefitem[Dudley (2002)]{Dud02}
	  \aut{Dudley, Richard~M}
	  \anno{2002}
\btitle{Real Analysis and Probability}
\book{Cambridge, UK: Cambridge University Press}.

    \brefitem[Duffie and Epstein (1992a)]{DE92a}
    \aut{Duffie, Darrell, and Larry G.~Epstein}
    \anno{1992a} 
\jtitle{Stochastic differential utility}
\journal{Econometrica} {60} 353-\nobreak394.

    \brefitem[Duffie and Epstein (1992b)]{DE92b}
    \aut{Duffie, Darrell, and Larry G.~Epstein}
    \anno{1992b} 
\jtitle{Asset pricing with stochastic differential utility}
\journal{Review of Financial Studies} {5} 411-\nobreak436.

    \brefitem[Duffie et al$.$ (1994)]{DGMM94}
    \aut{Duffie, Darrell, John Geanakoplos, Andreu Mas-Colell, and Andrew McLennan}
    \anno{1994} 
\jtitle{Stationary Markov equilibria}
\journal{Econometrica} {62} 745-\nobreak781.

    \brefitem[Duffie et al$.$ (2019)]{DQS19}
    \aut{Duffie, Darrell, Lei Qiao, and Yuneng Sun}
    \anno{2019} 
\jtitle{Continuous-time random matching}
\book{\it Working paper}.

    \brefitem[Duffie and Sun (2012)]{DS12}
    \aut{Duffie, Darrell, and Yuneng Sun}
    \anno{2012} 
\jtitle{The exact law of large numbers for independent random matching}
\journal{Journal of Economic Theory} {147} 1105-\nobreak1139.

    \brefitem[Duffie and Zame (1989)]{DZ89}
    \aut{Duffie, Darrell, and William Zame}
    \anno{1989} 
\jtitle{The consumption-based capital asset pricing model}
\journal{Econometrica} {57} 1279-\nobreak1297.

    \brefitem[Dumas (1989)]{Dum89}
    \aut{Dumas, Bernard}
    \anno{1989} 
\jtitle{Two-person dynamic equilibrium in the capital market}
\journal{Review of Financial Studies} {2} 157-\nobreak188.

    \brefitem[Dumas and Lyasoff (2012)]{DL12}
    \aut{Dumas, Bernard, and Andrew Lyasoff}
    \anno{2012} 
\jtitle{Incomplete-market equilibria solved recursively on an even tree}
\journal{Journal of Finance} {67} 1897-\nobreak1941.

    \brefitem[Dumas and Savioz (2022)]{DS22}
    \aut{Dumas, Bernard, and Marcel Savioz}
    \anno{2022} 
\jtitle{A theory of the nominal character of stock securities}
\journal{Review of Finance,} { } forthcoming.

       \brefitem[Emery (1978)]{Eme78}
        \aut{Emery, Michel} 
	\anno{1978} 
        \jtitle{Stabilit{\ea} des solutions des {\ea}quations diff{\ea}rentielles stochastiques application aux int{\ea}grales multiplicatives stochastiques}
        \journal{Zeit\-schrift f{\uD}r Wahr\-schein\-lichkeitstheorie und Verwandte Gebiete} 41 241--\nobreak262.

          \brefitem[Emery (1979)]{Eme79a}
	  \aut{Emery, Michel}
	  \anno{1979}
\jtitle{Une topologie sur l'espace des semimartingales}
\book{In: {\it S{\ea}minaire de Probabilit{\ea}s (Strasbourg) XIII\/}.  Lecture Notes in Mathematics 721. Berlin, DE: Springer, pp.~260--280}.

          \brefitem[Emery (1979)]{Eme79b}
	  \aut{Emery, Michel}
	  \anno{1979}
          \jtitle{{\Ea}quations diff{\ea}rentielles stochastiques lipschitziennes: {\ea}tude de la stabilit{\ea}}
          \book{In: {\it S{\ea}minaire de Probabilit{\ea}s (Strasbourg) XIII\/}.  Lecture Notes in Mathematics 721. Berlin, DE: Springer, pp.~281--293}.

         \brefitem[Emoto and Sunakawa (2021)]{ES21}
         \aut{Emoto, Masakuzi, and Takeki Sunakawa}
	 \anno{2021} 
	 \jtitle{Applying the explicit aggregation algorithm to heterogeneous agent models in continuous time} 
\journal{Economic Letters} {206} 109940.

    \brefitem[Epstein and Zin (1991)]{EZ91}
    \aut{Epstein, Larry G., and Stanley E.~Zin}
    \anno{1991} 
\jtitle{Substitution, risk aversion, and the temporal behavior of consumption and asset returns: an empirical analysis}
\journal{The Journal of Political Economy} {99} 263-\nobreak286.

    \brefitem[Feldman and Gilles (1985)]{FG85}
    \aut{Feldman, Mark and Christian Gilles}
    \anno{1985} 
\jtitle{An expository note on individual risk without aggregate uncertainty}
\journal{Journal of Economic Theory} {35} 26-\nobreak32.

    \brefitem[Feng and Nguyen (2012)]{FN12}
    \aut{Feng, Jin, and Truyen Nguyen}
    \anno{2012} 
    \jtitle{Hamilton-Jacobi equations in space of measures associated with a system of conservation laws}
    \journal{Journal de Mathématiques Pures et Appliquées} {97} 318-\nobreak390.

    \brefitem[Fernández et al$.$ (2020)]{VHN20}
    \aut{Fernández-Villaverde, Jesús, Samuel Hurtado, and Galo Nuño}
    \anno{2020} 
    \jtitle{Financial frictions and the wealth distribution}
    \book{{\it CESifo Working Papers\/} \ No.~8482}. 

    \brefitem[Föllmer (1994)]{Foll94}
    \aut{Föllmer, Hans}
    \anno{1994} 
    \jtitle{Stock price fluctuations as a diffusion in a random environment}
    \journal{Philosohpical Transactions of the Royal Society of London} {A347} 471-\nobreak483.

    \brefitem[Föllmer and Schweizer (1993)]{FS93}
    \aut{Föllmer, Hans, and Martin Schweizer}
    \anno{1993} 
    \jtitle{A microeconomic approach to diffusion models for stock prices}
    \journal{Mathematical Finance} {3} 1-\nobreak23.

           \brefitem[Galichon (2016)]{Gal16}
           \aut{Galichon, Alfred}
	   \anno{2016}
           \btitle{Optimal Transport Methods in Economics}
           \book{Princeton, NJ: Princeton University Press}.

    \brefitem[Geanakoplos et al$.$ (2000)]{GKSS00}
    \aut{Geanakoplos, John, Ioannis Karatzas, Martin Shubik, and William D.\ Sudderth}
    \anno{2000} 
\jtitle{A strategic market game with active bankruptcy}
\journal{Journal of Mathematical Economics} {34} 359-\nobreak396.

         \brefitem[Glynn (2013)]{Glynn13}
         \aut{Glynn, Peter W}
	 \anno{2013} 
	 \btitle{Harris recurrence} 
\book{Stanford University: Stochastic Systems Lecture Notes}.

    \brefitem[Gomes et al$.$ (2010)]{GMS10}
    \aut{Gomes, Diogo A., Joana Mohr, and Rafael Rigão Souza}
    \anno{2010} 
    \jtitle{Discrete time, finite state space mean field games}
    \journal{Journal de Math{\ea}matiques Pures et Apliqu{\ea}es} {93} 308-\nobreak328.

    \brefitem[Greenwald et al$.$ (2019)]{GLL19}
    \aut{Greenwald, Daniel L., Martin Lettau, and Sydney C.~Ludvigson}
    \anno{2019} 
\jtitle{How the wealth was won: factors shares as market fundamentals}
\book{\it NBER Working Paper\/} \ No.~25769. 

    \brefitem[Guéant (2009)]{Gue09}
    \aut{Guéant, Olivier}
    \anno{2009} 
    \jtitle{A reference case for mean field games models}
    \journal{Journal de Mathématiques Pures et Apliquées} {92} 276-\nobreak294.

    \brefitem[Gu{\ea}ant (2009)]{Gue09}
    \aut{Gu{\ea}ant, Olivier}
    \anno{2009} 
    \jtitle{A reference case for mean field games models}
    \journal{Journal de Math{\ea}matiques Pures et Apliqu{\ea}es} {92} 276-\nobreak294.

    \brefitem[Haan (1996)]{Haa96}
    \aut{Den Haan, Wouter J.}
    \anno{1996} 
\jtitle{Heterogeneity, aggregate uncertainty, and the short-term interest rate}
\journal{Journal of Business and Economic Statistics} {14} 399-\nobreak411.

    \brefitem[Haan and Marcet (1990)]{HM90}
    \aut{Den Haan, Wouter J., and Albert Marcet}
    \anno{1990} 
\jtitle{Solving the stochastic growth model by parameterizing expectations}
\journal{Journal of Business and Economic Statistics} {8} 31-\nobreak34.

    \brefitem[Hall (1978)]{Hall78}
    \aut{Hall, Robert E}
    \anno{1978} 
\jtitle{Stochastic implications for the life cycle--permanent income hypothesis: theory and evidence}
\journal{Journal of Political Economy} {86} 971-\nobreak987.

    \brefitem[Han et al$.$ (2017)]{HJE17}
    \aut{Han, Jiequn, Arnulf Jentzen, and Weinan E}
    \anno{2017} 
\jtitle{Deep learning-based numerical methods for high-dimensional parabolic partial differential equations and backward stochastic differential equations}
\journal{Communications in Mathematics and Statistics} {5} 349-\nobreak380.

    \brefitem[Han et al$.$ (2018)]{HJE18}
    \aut{Han, Jiequn, Arnulf Jentzen, and Weinan E}
    \anno{2018} 
\jtitle{Solving high-dimensional partial differential equations using deep learning}
\journal{Proceedings of the National Academy of Sciences of the USA} 115 8505-\nobreak8510.

    \brefitem[Hart et al$.$ (1974)]{HHK74}
    \aut{Hart, Sergiu, Werner Hildenbrand, and Elon Kohlberg}
    \anno{1974} 
\jtitle{On equilibrium allocations as distributions on the commodity space}
\journal{Journal of Mathematical Economics} {1} 159-\nobreak167.

    \brefitem[Heaton (1995)]{Hea95}
    \aut{Heaton, John}
    \anno{1995} 
\jtitle{An empirical investigation of asset pricing with temporally dependent preference specifications}
\journal{Econometrica} {63} 681-\nobreak717.

    \brefitem[Heaton and Lucas (1995)]{HL95}
    \aut{Heaton, John, and Deborah Lucas}
    \anno{1995} 
\jtitle{The importance of investor heterogeneity and financial market imperfections for the behavior of asset prices}
\journal{Carnegie-Rochester Conference Series on Public Policy} {42} 1-\nobreak32.

    \brefitem[Heaton and Lucas (1996)]{HL96}
    \aut{Heaton, John, and Deborah Lucas}
    \anno{1996} 
\jtitle{Evaluating the effects of incomplete markets on risk sharing and asset pricing}
\journal{Journal of Political Economy} {104} 443-\nobreak487.

         \brefitem[Hildenbrand (1974)]{Hil74}
         \aut{Hildenbrand, Werner}
	 \anno{1974} 
	 \btitle{Numerical Methods in Economics} 
\book{Princeton, NJ: Princeton University Press}.

    \brefitem[Holm (2018)]{Holm18}
    \aut{Holm, Martin Blomhoff}
    \anno{2018} 
    \jtitle{Consumption with liquidity constraints: an analytical characterization}
    \journal{Economics Letters} {167} 40-\nobreak42.

    \brefitem[Hopenhayn (1992)]{Hop92}
    \aut{Hopenhayn, Hugo A}
    \anno{1992} 
    \jtitle{Entry, exit, and firm dynamics in long run equilibrium}
    \journal{Econometrica} {32} 39-\nobreak50.

    \brefitem[Huang et al$.$ (2006)]{HMC06}
    \aut{Huang, Minyi, Roland P.~Malham{\ea}, and Peter E.~Caines}
    \anno{2006} 
    \jtitle{Large population stochastic dynamic games: closed-loop McKean-Vlasov systems and the Nash certainty equivalence principle}
    \journal{Communications in Information and Systems} {6} 221-\nobreak252.

    \brefitem[Huggett (1993)]{Hug93}
    \aut{Huggett, Mark}
    \anno{1993} 
\jtitle{The risk free rate in heterogeneous-agent, incomplete-insurance economies}
\journal{Journal of Economic Dynamics and Control} {17} 953-\nobreak969.

    \brefitem[Huggett (1997)]{Hug97}
    \aut{Huggett, Mark}
    \anno{1997} 
\jtitle{The one-sector growth model with idiosyncratic shocks: steady states and dynamics}
\journal{Journal of Monetary Economics} {39} 385-\nobreak403.

    \brefitem[Huggett and Ospina (2001)]{HO01}
    \aut{Huggett, Mark, and Sandra Ospina}
    \anno{1997} 
\jtitle{Aggregate precautionary savings: when is the third derivative irrelevant?}
\journal{Journal of Monetary Economics} {48} 373-\nobreak396.

    \brefitem[İmrohoroğlu (1989)]{Imro89}
    \aut{{\rm İ}mrohoroğlu, Ayşe}
    \anno{1989} 
    \jtitle{Cost of business cycles with indivisibilities and liquidity constraints}
    \journal{Journal of Political Economy} {97} 1364-\nobreak1383.

         \brefitem[Jacod and Shiryaev (2003)]{JacShir87}
           \aut{Jacod, Jean, and Albert N.~Shiryaev}
	   \anno{2003}
\btitle{Limit Theorems for Stochastic Processes}
\book{Berlin, DE: Springer}.

    \brefitem[Jovanovic and Rosenthal (1988)]{JR88}
    \aut{Jovanovic, Boyan, and Robert W.~Rosenthal}
    \anno{1988} 
\jtitle{Anonymous sequential games}
\journal{Journal of Mathematical Economics} {17} 77-\nobreak97.

    \brefitem[Judd (1985)]{Judd85}
    \aut{Judd, Kenneth L}
    \anno{1985} 
\jtitle{The law of large numbers with a continuum of IID random variables}
\journal{Journal of Economic Theory} {35} 19-\nobreak25.

         \brefitem[Judd (1998)]{Judd98}
         \aut{Judd, Kenneth L}
	 \anno{1998} 
	 \btitle{Numerical Methods in Economics} 
\book{Cambridge, MA: MIT Press}.

    \brefitem[Judd et al$.$ (2000)]{JKS00}
    \aut{Judd, Kenneth L., Felix Kubler, and Karl Schmedders}
    \anno{2000} 
\jtitle{Computing equilibria in infinite-horizon finance economies: the case of one asset}
\journal{Journal of Economic Dynamics and Control} {24} 1047-\nobreak1078.

    \brefitem[Judd et al$.$ (2002)]{JKS02}
    \aut{Judd, Kenneth L., Felix Kubler, and Karl Schmedders}
    \anno{2002} 
\jtitle{A Solution Method for Incomplete Asset Markets With Heterogeneous Agents}
\book{\href{https://doi.org/10.2139/ssrn.900165}{DOI: 10.2139 / ssrn.900165}}.

         \brefitem[Kac (1976)]{Kac76}
         \aut{Kac, Mark}
	 \anno{1976} 
	 \btitle{Probability and Related Topics in Physical Sciences} 
\book{Providence, RY: American Mathematical Society}.

    \brefitem[Karatzas et al$.$ (1990)]{KLS90a}
    \aut{Karatzas, Ioannis, John P.~Lehoczky, and Steven E.~Shreve}
    \anno{1990} 
\jtitle{Existence and uniqueness of multi-agent equilibrium in a stochastic, dynamic consumption/investment model}
\journal{Mathematics of Operations Research} {15} 80-\nobreak128.

    \brefitem[Karatzas et al$.$ (1994)]{KSS94}
    \aut{Karatzas, Ioannis, Martin Shubik, and William D.\ Sudderth}
    \anno{1994} 
\jtitle{Construction of stationary Markov equilibria in a strategic market game}
\journal{Mathematics of Operations Research} {19} 975-\nobreak1006.

    \brefitem[Karatzas et al$.$ (1997)]{KSS97}
    \aut{Karatzas, Ioannis, Martin Shubik, and William D.\ Sudderth}
    \anno{1997} 
\jtitle{A strategic market game with secured lending}
\journal{Journal of Mathematical Economics} {28} 207-\nobreak247.

    \brefitem[El Karoui et al$.$ (1997)]{EPQ97}
    \aut{El Karoui, Nicole, Shige G.~Peng, and Marie-Claire Quenez}
    \anno{1997} 
\jtitle{Backward stochastic differential equations in finance }
\journal{Mathematical Finance} {7} 1-\nobreak71.

    \brefitem[Khan and Thomas (2008)]{KT08}
    \aut{Khan, Aubhik, and Julia K.~Thomas}
    \anno{2008} 
\jtitle{Idiosyncratic Shocks and the Role of Nonconvexities in Plant And Aggregate Investment Dynamics}
\journal{Econometrica} {76} 395-\nobreak436.

    \brefitem[Kirby (2018)]{Kir18}
    \aut{Kirby, Robert}
    \anno{2018} 
\jtitle{Bewley-Huggett-Aiyagari models: computation, simulation, and uniqueness of general equilibrium}
\book{\it Macroeconomic Dynamics\/}, forthcoming, published online. 

    \brefitem[Krusell and Smith (1998)]{KS98}
    \aut{Krusell, Per, and Anthony Smith}
    \anno{1998} 
\jtitle{Income and wealth heterogeneity in the macroeconomy}
\journal{Journal of Political Economy} {106} 867-\nobreak896.

    \brefitem[Kubler and Schmedders (2002)]{KS02}
    \aut{Kubler, Felix, and Karl Schmedders}
    \anno{2002} 
\jtitle{Recursive equilibria in economies with incomplete markets}
\journal{Macroeconomic Dynamics} {6} 284-\nobreak306.

    \brefitem[Kunita (2009)]{Kun09}
    \aut{Kunita, Hiroshi}
    \anno{2009} 
\jtitle{Smooth density of canonical stochastic differential equation with jumps}
\journal{Astérisque} {327} 69-\nobreak91.

    \brefitem[Lasry and Lions (2007)]{LaLi07}
    \aut{Lasry, Jean-Michel, and Pierre-Louis Lions}
    \anno{2007} 
\jtitle{Mean field games}
\journal{Japanese Journal of Mathematics} {2} 229-\nobreak260.

          \brefitem[Lepingle and M{\ea}min (1978)]{LepMem78}
    \aut{Lepingle, Dominique, and Jean M{\ea}min}
    \anno{1978} 
\jtitle{Sur l'int{\ea}grabilit{\ea} uniforme des martingales exponentielles} 
\journal{Zeitschrift f{\uD}r Wahrscheinlichkeitstheorie und Verwandte Gebiete} 42 175--\nobreak203.

    \brefitem[Li and Stachurski (2014)]{LS14}
    \aut{Li, Huiyu, and John Stachurski}
    \anno{2014} 
\jtitle{Solving the income fluctuation problem with unbounded rewards }
\journal{Journal of Economic Dynamics and Control} {45} 353-\nobreak365.

    \brefitem[Light (2018)]{Light18}
    \aut{Light, Bar}
    \anno{2018} 
\jtitle{Uniqueness of equilibrium in a Bewley-Aiyagari model}
\book{\it Economic Theory\/}, forthcoming, published online. 

    \brefitem[Lions and Lasry (2007)]{LL07}
    \aut{Lions, Pierre-Louis, and Jean-Michel Lasry}
    \anno{2007} 
\jtitle{Large investor trading impacts on volatility}
\book{In: {\it Paris-Princeton Lectures on Mathematical Finance 2004\/}. Lecture Notes in Mathematics, vol 1919, pp.~173-\nobreak190, Berlin: Springer}.

         \brefitem[Ljungqvist and Sargent (2018)]{LjunSar00}
         \aut{Ljungqvist, Lars, and Thomas J.~Sargent}
	 \anno{2018} 
	 \btitle{Recursive Macroeconomic Theory (4th Ed.)} 
\book{Cambridge, MA: MIT Press}.

    \brefitem[Lucas (1994)]{Luc94}
    \aut{Lucas, Deborah J.}
    \anno{1994} 
\jtitle{Asset pricing with undiversifiable income risk and short sales constraints: deepening the equity premium puzzle}
\journalnt{Journal of Monetary Economics} {34} 325-\nobreak341.

         \brefitem[Lyasoff (2017)]{Lya17}
         \aut{Lyasoff, Andrew}
	 \anno{2017} 
	 \btitle{Stochastic Methods in Asset Pricing} 
\book{Cambridge, MA: MIT Press}.

    \brefitem[Lyasoff (2019)]{Lya19}
    \aut{Lyasoff, Andrew}
    \anno{2019} 
\jtitle{Bewley's incomplete market model revisited}
\book{\it \href{https://drive.google.com/file/d/1VXSzI6ayEqOE_wazXNA7dtS88SKBoDUf/view}{Working Paper}\/}.

    \brefitem[Lyasoff (2019a)]{Lya19aa}
    \aut{Lyasoff, Andrew}
    \anno{2019a} 
\jtitle{Bewley's incomplete market model revisited}
\book{\it Working Paper\/}.

    \brefitem[Lyasoff (2019)]{Lya19b}
    \aut{Lyasoff, Andrew}
    \anno{2019} 
\jtitle{A brief note on incomplete-market equilibria and their connection with mean field games and control}
\book{\it Working Paper\/}.

    \brefitem[Lyasoff (2019b)]{Lya19bb}
    \aut{Lyasoff, Andrew}
    \anno{2019b} 
\jtitle{A brief note on incomplete-market equilibria and their connection with mean field games and control}
\book{\it Working Paper\/}.

    \brefitem[Lyasoff (2019)]{Lya19c}
    \aut{Lyasoff, Andrew}
    \anno{2019} 
\jtitle{A brief note on discrete mean field games and control}
\book{\it Working Paper\/}.

    \brefitem[Lyasoff (2019c)]{Lya19cc}
    \aut{Lyasoff, Andrew}
    \anno{2019c} 
\jtitle{A brief note on discrete mean field games and control}
\book{\it Working Paper\/}.

    \brefitem[Ma et al$.$ (2013)]{MPY94}
    \aut{Ma, Jin, Philip Protter, and Jiongming Yong}
    \anno{1994} 
    \jtitle{Solving forward-backward stochastic differential equations explicitly~-- a fours step scheme}
    \journal{Probability Theory and Related Fields} {98} 339-\nobreak350.

         \brefitem[Ma and Yong (2007)]{MY07}
         \aut{Ma, Jin, and Jiongmin Yong}
	 \anno{2007} 
	 \btitle{Forward-Backward Stochastic Differential Equations and Their Applications} 
\book{Berlin, DE: Springer}.

         \brefitem[Magill and Quinzii (1996)]{MQ96}
         \aut{Magill, Michael, and Martine Quinzii}
	 \anno{1996} 
	 \btitle{Theory of Incomplete Markets} 
\book{Cambridge, MA: MIT Press}.

    \brefitem[Maliar et al$.$ (2021)]{MMW21}
    \aut{Maliar, Lilia, Serguei Maliar, and Pablo Wiant}
    \anno{2021} 
\jtitle{Deep learning for solving dynamic economic models}
\journal{Journal of Monetary Economics} {122} 76-\nobreak101.

    \brefitem[Malinvaud (1972)]{Mal72}
    \aut{Malinvaud, Edmond}
    \anno{1972} 
\jtitle{The allocation of individual risks in large markets}
\journal{Journal of Economic Theory} {4} 312-\nobreak328.

    \brefitem[Marcet and Singleton (1999)]{MS99}
    \aut{Marcet, Albert, and Kenneth J.~ Singleton}
    \anno{1999} 
\jtitle{Equilibrium asset prices and savings of heterogeneous agents in the presence of incomplete markets and portfolio constraints}
\journal{Macroeconomic Dynamics} {3} 243-\nobreak277.

    \brefitem[McKean (1966)]{McK66}
    \aut{McKean, Henry P., Jr}
    \anno{1966} 
\jtitle{A class of Markov processes associated with nonlinear parabolic equations}
\journal{Proceedings of the National Academy of Sciences of the USA} 56 1907-\nobreak1911.

    \brefitem[Mehra and Prescott (1985)]{MP85}
    \aut{Mehra, Rajnish, and Edward C.~Prescott}
    \anno{1985} 
\jtitle{The equity premium: a puzzle}
\journal{Journal of Monetary Economics} {15} 145-\nobreak161.

    \brefitem[Mertens and Judd (2018)]{MJ18}
    \aut{ Mertens, Thomas M., and Kenneth L.~Judd}
    \anno{2018} 
\jtitle{Solving an incomplete markets model with a large cross-section of agents}
\journal{Journal of Economic Dynamics and Control} {91} 349-\nobreak368.

          \brefitem[Me\-yer (1966)]{Mey66}
	  \aut{Meyer, Paul A} 
	  \anno{1966}
	  \btitle{Probability and Potentials} 
\book{Waltham, MA: Blaisdell Publishing}.

    \brefitem[Miao (2002)]{Miao02}
    \aut{Miao, Jianjun}
    \anno{2002} 
\jtitle{Stationary equilibria of economies with a continuum of heterogeneous consumers}
\book{\it Mimeo\/}.

    \brefitem[Miao (2006)]{Miao06}
    \aut{Miao, Jianjun}
    \anno{2006} 
\jtitle{Competitive equilibria of economies with a continuum of consumers and aggregate shocks}
\journal{Journal of Economic Theory} {128} 274-\nobreak298.

         \brefitem[Pakes and McGuire (2001)]{PM01}
         \aut{Pakes, Ariel, and Paul McGuire}
	 \anno{2001} 
	 \jtitle{Stochastic algorithms, symmetric Markov perfect equilibrium, and the `curse' of dimensionality} 
\journal{Econometrica} {69} 1261-\nobreak1281.

    \brefitem[Panageas (2019)]{Pan19}
    \aut{Panageas, Stavros}
    \anno{2019} 
\jtitle{The implications of heterogeneity and inequality for asset pricing}
\book{\href{https://doi.org/10.2139/ssrn.3494558}{DOI: 10.2139 / ssrn.3494558}}. 

    \brefitem[Pardoux and Peng (1990)]{PP90}
    \aut{Pardoux, {\Ea}tienne, and Shige G.~Peng}
    \anno{1990} 
\jtitle{Adapted solution of a backward stochastic differential equation}
\journal{Systems and Control Letters} {14} 55-\nobreak61.

    \brefitem[Peng (1990)]{Peng90}
    \aut{Peng, Shige G}
    \anno{1990} 
\jtitle{A general stochastic maximum principle for optimal control problems}
\journal{SIAM Journal of Control} {28} 966-\nobreak979.

    \brefitem[Pontryagin (1986)]{P86}
    \aut{Pontryagin, Lev Semyonovich}
    \anno{1986} 
\btitle{The Mathematical Theory of Optimal Processes (selected works, Volume 4)}
\book{New York, NY: Gordon and Breach}.

    \brefitem[Pröhl (2021)]{Pro21}
    \aut{Pröhl, Elisabeth}
    \anno{2021} 
\jtitle{Existence and uniqueness of recursive equilibria with aggregate and idiosyncratic risk}
\book{\href{https://doi.org/10.2139/ssrn.3250651}{DOI: 10.2139 / ssrn.3250651}}. 

       \brefitem[Protter (1978)]{Pro78}
        \aut{Protter, Philip} 
	\anno{1978} 
        \jtitle{${\scr H}^p$ Stability of Solutions of Stochastic Differential Equations}
        \journal{Zeitschrift f{\uD}r Wahrscheinlichkeitstheorie und Verwandte Gebiete} 44 337--\nobreak352.

            \brefitem[Reed and Simon (1980)]{RS80}
            \aut{Reed, Michael, and Barry Simon}
	    \anno{1980}
\btitle{Methods of Modern Mathematical Physics I: Functional Analysis}
\book{New York, NY: Academic Press}.

            \brefitem[Revuz and Yor (1999)]{RevYor99}
            \aut{Revuz, Daniel, and Marc Yor}
	    \anno{1999}
\btitle{Continuous Martingales and Brownian Motion}
\book{Berlin, DE: Sprin\-ger-Ver\-lag}.

    \brefitem[Samuelson (1941)]{Sam41}
    \aut{Samuelson, Paul A}
    \anno{1941} 
\jtitle{The Stability of Equilibrium: Comparative Statics and Dynamics}
\journal{Econometrica} {9} 97-\nobreak120.

         \brefitem[Stokey and Lucas (1989)]{SL89}
         \aut{Stokey, Nancy L., and Robert E.\ Lucas Jr}
	 \anno{1989} 
	 \btitle{Recursive Methods in Economic Dynamics} 
\book{Cambridge, MA: Harvard University Press}.

         \brefitem[Stroock (2014)]{Str14}
         \aut{Stroock, Daniel W}
	 \anno{2014} 
	 \btitle{An Introduction to Markov Processes (2d Ed.)} 
\book{Berlin, DE: Sprin\-ger-Ver\-lag}.

         \brefitem[Stroock (2023)]{Str23}
         \aut{Stroock, Daniel W}
	 \anno{2023} 
	 \btitle{Integral functionals of probability that depend only on mean values} 
\book{\href{https://doi.org/10.48550/arXiv.2301.01195}{DOI: 10.48550 / arXiv.2301.01195}}.

    \brefitem[Sun (2006)]{Sun06}
    \aut{Sun, Yuneng}
    \anno{2006} 
\jtitle{The exact law of large numbers via Fubini extension and characterization of insurable risks}
\journal{Journal of Economic Theory} {126} 31-\nobreak69.

    \brefitem[Sun and Zhang (2009)]{SZh09}
    \aut{Sun, Yuneng, and Yongchao Zhang}
    \anno{2009} 
\jtitle{Individual risk and Lebesgue extension without aggregate uncertainty}
\journal{Journal of Economic Theory} {144} 432-\nobreak443.

         \brefitem[Tauchen (1986)]{Tau86}
         \aut{Tauchen, George}
	 \anno{1986} 
	 \jtitle{Finite state Markov chain approximation to univariate and vector autoregressions} 
\journal{Economic Letters} {20} 177-\nobreak181.

    \brefitem[Telmer (1993)]{Tel93}
    \aut{Telmer, Chris I.}
    \anno{1993} 
\jtitle{Asset-pricing puzzles and incomplete markets}
\journal{Journal of Finance} {48} 1803-\nobreak1832.

    \brefitem[Telmer and Zin (2002)]{TZ02}
    \aut{Telmer, Chris I., and Stanley E.~Zin}
    \anno{2002} 
\jtitle{Prices as factors: approximate aggregation with incomplete markets}
\journal{Journal of Economic Dynamics and Control} {26} 1127-\nobreak1157.

         \brefitem[Touzi  (2013)]{Tou13}
         \aut{Touzi, Nizar}
	 \anno{2013} 
	 \btitle{Optimal Stochastic Control, Stochastic Target Problems, and Backward SDE} 
         \book{New York, NY: Springer}.

    \brefitem[Vayanos and Vila (1999)]{VV99}
    \aut{Vayanos, Dimitri, and Jean-Luc Vila}
    \anno{1999} 
\jtitle{Equilibrium interest rate and liquidity premium with transaction costs}
\journal{Economic Theory} {13} 509-\nobreak539.

         \brefitem[Vilani (2003)]{Vil03}
         \aut{Villani, Cédric}
	 \anno{2003} 
	 \btitle{Topics in Optimal Transportation} 
\book{Providence, RI: American Methematical Society}.

         \brefitem[Vilani (2009)]{Vil09}
         \aut{Villani, Cédric}
	 \anno{2009} 
	 \btitle{Optimal Transport: Old and New} 
\book{Berlin, DE: Sprin\-ger-Ver\-lag}.

    \brefitem[Wang (2007)]{Wan07}
    \aut{Wang, Neng}
    \anno{2007} 
    \jtitle{An equilibrium model with wealth distribution}
    \journal{Journal of Monetary Economics} {54} 1882-\nobreak1904.

    \brefitem[Wang (2007)]{Wan07}
    \aut{Wang, Neng}
    \anno{2007} 
    \jtitle{An equilibrium model with wealth distribution}
    \journal{Journal of Monetary Economics} {54} 1882-\nobreak1904.

    \brefitem[Weil (1992)]{Wei92}
    \aut{Weil, Philippe}
    \anno{1992} 
\jtitle{Equilibrium asset prices with undiversifiable labor income risk}
\journal{Journal of Economic Dynamics and Control} {16} 769-\nobreak790.

    \brefitem[Wu and Zhang (2019)]{WZ19}
    \aut{Wu, Cong, and Jianfeng Zhang}
    \anno{2019} 
    \jtitle{Viscosity solutions to parabolic master equations and McKean-Vlasov SDEs with closed-loop controls}
    \journal{Annals of Applied Probability} (to appear).

    \brefitem[Zadeh (1970)]{Zad70}
    \aut{Zadeh, Norman}
    \anno{2016} 
    \jtitle{A Note on the cyclic coordinate ascent method}
    \journal{Management Science} {16} 642-\nobreak644.

    \brefitem[Zeldes (1989)]{Zel89}
    \aut{Zeldes, Stephen P}
    \anno{1989} 
    \jtitle{Optimal consumption with stochastic income: deviations from certainty equivalence}
    \journal{Quarterly Journal of Economics} {104} 275-\nobreak298.

         \brefitem[Zhang (2017)]{Zha17}
         \aut{Zhang, Jianfeng}
	 \anno{2017} 
	 \btitle{Backward Stochastic Differential Equations: from linear to fully nonlinear theory} 
\book{New York, NY: Springer}.

\endgroup
\WriteRefTot

\begin{appendices} 

\setcounter{section}{1}

\titlespacing*{\section}{0pt}{0.0ex plus 0.25ex minus .125ex}{0.55ex  plus 0.125ex minus 0.125ex}

\makeatletter%
\edef\@currentlabel{\Alph{section}}%
\makeatother%
\section*{Appendix \Alph{section}: Proof of Theorem~\ref{thm1}\label{sec:A2}}
\setcounter{paragraph}{0}%

\noindent
This result is a direct application of the implicit function theorem.
The left sides of all three equations
\footnote{{}Recall that $V_{t+1,\tts y,\tts \T_{t,\tts x}^y(F),\tts v}\phd$
is assumed strictly concave and in $\C^2$ wherever it is finite.}
 in \eqref{z2-no-lm} can be treated as a $\R^3$-valued $\C^1$-function, which we write as 
$h(c,\q,\qq,w)$, with the understanding that $W_{y,v}$ substitutes for the right side of the first
equation in \eqref{ze2}. To simplify the notation, set 
{\abovedisplayskip=5pt plus 1.5pt minus 1.5pt\belowdisplayskip=5pt plus 1.5pt minus 1.5pt\belowdisplayshortskip=5pt plus 1.5pt minus 1.5pt
\begin{gather*}
 a_{y,v}\df (1+r)\sqrt{-\b \, \partial^2 V_{t+1,\tts y,\tts \T_{t,\tts x}^y(F),\tts v}\bigl(W_{y,\tts v} \bigr)}\\
\noalign{and}
b_{y,v}\df \bigl(\rho_ y(K)+ 1-\dd\bigr)\sqrt{-\b \, \partial^2 V_{t+1,\tts y,\tts \T_{t,\tts x}^y(F),\tts v}\bigl(W_{y,\tts v} \bigr)}\,,
\end{gather*}
}%
which leads to the following expression%
\footnote{{}By convention, if $a<0$ we write $\sqrt{-a}\sqrt{-a}=-a$.} 
for the Jacobian matrix of the function $h$
{\abovedisplayskip=5pt plus 1.5pt minus 1.5pt\belowdisplayskip=5pt plus 1.5pt minus 1.5pt\belowdisplayshortskip=5pt plus 1.5pt minus 1.5pt
\begin{equation}\label{Jh}
h'(c,\q,\qq,w)=\begin{bmatrix}
1 & 1 & 1 & \,\,\llap{$-1$} \\
\partial^2U(c ) & \sum\nolimits_{y,v} a_{y,v}^2&  \sum\nolimits_{y,v} a_{y,v} b_{y,v} & 0\\
\partial^2U(c ) & \sum\nolimits_{y,v} a_{y,v} b_{y,v} & \sum\nolimits_{y,v} b_{y,v}^2 & 0 
\end{bmatrix}\,.
\end{equation}
}%
Let $h'(c,\q,\qq,w)_1$ denote the $3$-by-$3$ matrix composed of the first three columns in the Jacobian
and let $h'(c,\q,\qq,w)_2$ denote the $3$-by-$1$ matrix composed of the last column.
Hence
{\abovedisplayskip=5pt plus 1.5pt minus 1.5pt\belowdisplayskip=5pt plus 1.5pt minus 1.5pt\belowdisplayshortskip=5pt plus 1.5pt minus 1.5pt
$$
\begin{aligned}
&\bigl|h'(c,\q,\qq,w)_1\bigr| =
\Bigl(\sum\nolimits_{y,v} a_{y,v}^2\Bigr)\Bigl(\sum\nolimits_{y,v}
b_{y,v}^2\Bigr) -\Bigl(\sum\nolimits_{y,v} a_{y,v} b_{y,v}\Bigr)^2\\
&\hbox to0.5cm{\hfill}-\partial^2U(c )\Bigl(\sum\nolimits_{y,v} b_{y,v}^2-\sum\nolimits_{y,v} a_{y,v}b_{y,v}\Bigr)
+\partial^2U(c )\Bigl(\sum\nolimits_{y,v} a_{y,v}b_{y,v}-\sum\nolimits_{y,v} a_{y,v}^2\Bigr)\\
&\hbox to0.5cm{\hfill} =\Bigl(\sum\nolimits_{y,v} a_{y,v}^2\Bigr)\Bigl(\sum\nolimits_{y,v}
b_{y,v}^2\Bigr) -\Bigl(\sum\nolimits_{y,v} a_{y,v} b_{y,v}\Bigr)^2
-\partial^2U(c )\sum\nolimits_{y,v} (a_{y,v}-b_{y,v})^2\,.
\end{aligned}
$$
}%
Since we exclude from the model the degenerate case in which the payoffs from capital investment
are identical in all productivity states, the
determinant above is strictly positive. By the implicit function theorem the equation
$h(c,\q,\qq,w)=(0,0,0)^\trn$ defines 
$(c,\q,\qq)\in\R^3$ as a unique $\C^1$-function in some neighborhood of $w$ with derivative
{\abovedisplayskip=5pt plus 1.5pt minus 1.5pt\belowdisplayskip=5pt plus 1.5pt minus 1.5pt\belowdisplayshortskip=5pt plus 1.5pt minus 1.5pt
$$
\bigl(\partial c,\partial \q,\partial \qq\bigr)
=-h'\bigl(c,\q,\qq,w\bigr)_1^{-1} \, h'\bigl(c,\q,\qq,w\bigr)_2\,,
$$
}%
and since the first entry in the first row of the inverse $h'(c,\q,\qq,w)_1^{-1}$ can be identified
as the strictly positive scalar
{\abovedisplayskip=5pt plus 1.5pt minus 1.5pt\belowdisplayskip=5pt plus 1.5pt minus 1.5pt\belowdisplayshortskip=5pt plus 1.5pt minus 1.5pt
$$
\Bigl(\sum\nolimits_{y,v} a_{y,v}^2\Bigr)\Bigl(\sum\nolimits_{y,v} b_{y,v}^2\Bigr)
-\Bigl(\sum\nolimits_{y,v} a_{y,v} b_{y,v}\Bigr)^2\,,
$$
}%
we see that $\partial c>0$, i.e., the consumption level is a strictly increasing
$\C^1$-function of entering wealth.
Furthermore, the value function of the problem in
\eqref{ze1212} can be cast as
{\abovedisplayskip=5pt plus 1pt minus 1pt\belowdisplayskip=5pt plus 1pt minus 1pt\belowdisplayshortskip=3pt plus 0.5pt minus 0.5pt
\begin{equation*}
\begin{aligned}
&V_{t,x,F,u}(w )\df U(c(w) )\\
&\hbox to0.4cm{\hfill}+\b\sum_{ y\ts\in\ts\XXX,\,v\ts\in\ts\EEE}
V_{t+1,\tts y,\tts \T_{t,\tts x}^y(F),\tts v}\Bigl((1+r)\,\q(w)  + (\rho_ y(K)+ 1-\dd)\,\qq(w)  + 
\ee_ y(K)\, v \Bigr)\\
&\hbox to8cm{\hfill}\times Q(x, y)P_{x, y}(u, v)\,,
\end{aligned}
\end{equation*}}%
and this function is $\C^1$
with respect to the resource $w $ as well. By~the envelope theorem (see \eqref{ze4a1})
{\abovedisplayskip=8pt plus 2pt minus 2pt\belowdisplayskip=8pt plus 2pt minus 2pt
\belowdisplayshortskip=3pt plus 0.5pt minus 0.5pt
$$
\partial V_{t,x,F,u}(w ) =\partial U(c(w) )\,,
$$
}%
with the implication that $\partial V_{t,x,F,u}\phd$
is $\C^1$ and strictly decreasing, since $\partial U\phd$ is strictly decreasing and
$c\phd$ is 
strictly increasing; in particular,
$V_{t,x,F,u}\phd\in\C^2(\R)$ and $\partial^2 V_{t,x,F,u}\phd<\nobreak 0$.

Removing the production technology from the model leads to the removal of the third
row and the 
third column in the Jacobian matrix in \eqref{Jh}. Similarly, removing the private lending
instrument from the model leads to the removal of the second row and the second
column from the 
Jacobian. In either case, the application of the implicit function theorem as above
gives the result.  
 
The second part of the theorem can be established in much the same way. Let
$k(c,\q,\qq)\in\R^2$ 
be the vector composed of the left sides in the last two equations
in \eqref{z2-no-lm}. Then 
$k\in\C^2(\R^3;\R^2)$ with Jacobian matrix  
{\abovedisplayskip=12pt plus 3.5pt minus 2.5pt\belowdisplayskip=12pt plus 3.5pt minus
2.5pt\belowdisplayshortskip=8pt plus 3.5pt minus 1.5pt 
$$
k'(c,\q,\qq)=\begin{bmatrix}
\partial^2U(c ) & \sum\nolimits_{y,v} a_{y,v}^2&  \sum\nolimits_{y,v} a_{y,v} b_{y,v} \\
\partial^2U(c ) & \sum\nolimits_{y,v} a_{y,v} b_{y,v} & \sum\nolimits_{y,v} b_{y,v}^2 
\end{bmatrix}\,.
$$
}%
As the matrix composed of the last two columns in this Jacobian was already shown to
have a strictly 
positive determinant, the implicit function theorem completes the proof.

\stepcounter{section}
\section*{Appendix \Alph{section}: Lower Bounds on Consumption}
\makeatletter%
\edef\@currentlabel{\Alph{section}}%
\makeatother%
\label{sec:App-B}\setcounter{paragraph}{0}

While this result is not used in the \chptr, it is important to note that,
in general, there is no reason why in equilibrium the range of
consumption must expand arbitrarily close to $0$,
i.e., in equilibrium, 
the support of the cross-sectional distribution of agents may exclude a neighborhood of~$0$. %
To~see this, consider the special case where productive capital is the only asset
(i.e., $\q_{t,x,\bar\csd,u}=0$) and $\partial U(c)=1/c$~-- see Sec.~\ref{sec:KS}.
The second kernel condition in \ref{cross-sys}-(\ref{zze5a}) can now be cast as
{\abovedisplayskip=10pt plus 1.5pt minus 1.5pt\belowdisplayskip=10pt plus 1.5pt minus 1.5pt\belowdisplayshortskip=5pt plus 1.5pt minus 1.5pt
\begin{equation}\label{new-kern}   
c=\Bigl(\b\sum\nolimits_{ y\ts\in\ts\XXX,\,v\ts\in\ts\EEE}{1\over\pfc_{t,x,\csd}^{ y,v}(u,c)}
\bigl(\rho_{ y}(K_{t}(x,\csd))+1-\dd\bigr) Q(x, y)P_{x, y}(u,v)\Bigr)^{-1}\,.
\end{equation}
}%
Introducing the strictly increasing functions
{\abovedisplayskip=10pt plus 1.5pt minus 1.5pt\belowdisplayskip=10pt plus 1.5pt minus 1.5pt\belowdisplayshortskip=5pt plus 1.5pt minus 1.5pt
\[
\Rpp\ni \a \leadsto H_{t+1, y,\T_{t,x}^y(\csd),v}(\a)\df \a + \qq_{t+1, y,\T_{t,x}^y(\csd),v}(\a) \,,
\]}%
with inverses $\hat H_{t+1, y,v}\phd$ as in \eqref{inversion}, the balance equation \ref{cross-sys}-(\ref{zze5xb})
can be stated as
{\abovedisplayskip=10pt plus 1.5pt minus 1.5pt\belowdisplayskip=10pt plus 1.5pt minus 1.5pt\belowdisplayshortskip=5pt plus 1.5pt minus 1.5pt 
\[
\pfc_{t,x,\csd}^{ y,v}(u,c) =
{{\hat H}_{t+1, y,\T_{t,x}^y(\csd),v}\Bigl(\bigl(\rho_ y(K_{t}(x,\csd))+1-\dd\bigr){\qq_{t,x,\csd,u}(c)}
+v \,\ee_ y(K_{t}(x,\csd))\Bigr)}\,.
\]}%
Suppose next that the domains of all functions $\qq_{t+1, y,\T_{t,x}^y(\csd),v}\phd$, $ y\in\XXX$, $v \in\EEE$,
include $\Rpp$ and let $\qq_{t+1, y,\T_{t,x}^y(\csd),v}(0)\df\lim_{c\searrow 0}\qq_{t+1, y,\T_{t,x}^y(\csd),v}(c)\,$.
The infimum over all admissible entering wealths in period $t+1$ is
$\qq_{t+1,y,\T_{t,x}^y(\csd),v}(0)$. Suppose that this infimum can be reached, in the sense
that there is a $c^*\ge 0$ (depending on $y$ and $v$) such that 
{\abovedisplayskip=10pt plus 1.5pt minus 1.5pt\belowdisplayskip=10pt plus 1.5pt minus 1.5pt\belowdisplayshortskip=5pt plus 1.5pt minus 1.5pt
\[
\begin{split}
\lim\nolimits_{c\searrow c^*}
\Bigl(\bigl(\rho_ y(K_{t}(x,\csd))+1-\dd\bigr){\qq_{t,x,\csd,u}(c)}
&+v \,\ee_ y(K_{t}(x,\csd))\Bigr)\\
&= \qq_{t+1,\tts y,\tts \T_{t,x}^y(\csd),\tts v}(0) = {H}_{t+1,\tts  y,\tts \T_{t,x}^y(\csd),\tts v}(0)\,,
\end{split}
\]
}%
i.e., assuming that ${\qq_{t,x,\csd,u}\phd}$ is continuous, chosen so that
{\abovedisplayskip=10pt plus 1.5pt minus 1.5pt\belowdisplayskip=10pt plus 1.5pt minus 1.5pt\belowdisplayshortskip=5pt plus 1.5pt minus 1.5pt
\[
\qq_{t,x,\csd,u}(c^*)\df\lim\nolimits_{c\searrow c^*}\qq_{t,x,\csd,u}(c)
= {\qq_{t+1,\tts y,\tts \T_{t,x}^y(\csd),\tts v}(0) - v \,\ee_ y(K_{t}(x,\csd))\over \rho_ y(K_{t}(x,\csd))+1-\dd}\,.
\]
}%
Then $\lim\nolimits_{c\searrow c^*} \pfc_{t,x,\csd}^{ y,v}(u,c)
= {\hat H}_{t+1,\tts y,\tts \T_{t,x}^y(\csd),\tts v}
\bigl({H}_{t+1,\tts y,\tts \T_{t,x}^y(\csd),\tts v}(0)\bigr)=0$, which is possible only if $c^*=\nobreak 0$,
since otherwise the right side of \eqref{new-kern} converges to $0$ as $c\searrow c^*$,
while the left side  converges to $c^*>0$.
Consequently, if one is to insist
that $\qq_{t,x,\csd,u}\phd$ is continuous and non-decreasing and that its domain is contiguous and
extends to $+\infty$ with $\lim_{c\nearrow\infty}\qq_{t,x,\csd,u}(c)=\infty$,
then everywhere in that domain $\qq_{t,x,\csd,u}\phd$ must have a lower bound
given by
{\abovedisplayskip=10pt plus 1.5pt minus 1.5pt\belowdisplayskip=10pt plus 1.5pt minus 1.5pt\belowdisplayshortskip=5pt plus 1.5pt minus 1.5pt
\begin{equation}\label{theta-min}
M_{t,x,\csd}\df \max_{ y\ts\in\ts\XXX,\,v\ts\in\ts\EEE}\
{\qq_{t+1,\tts y,\tts \T_{t,x}^y(\csd),\tts v}(0) - v \,\ee_
y(K_{t}(x,\csd))\over \rho_ y(K_{t}(x,\csd))+1-\dd}\,.
\end{equation}}%
In particular,
{\abovedisplayskip=10pt plus 1.5pt minus 1.5pt\belowdisplayskip=10pt plus 1.5pt minus 1.5pt\belowdisplayshortskip=5pt plus 1.5pt minus 1.5pt
\[
\pfc_{t,x,\csd}^{ y,v}(u,\cdot) \ge {{\hat H}_{t+1, y,v}\Bigl(\bigl(\rho_ y(K_{t,x})+1-\dd\bigr) M_{t,x}
+v \,\ee_ y(K_{t,x})\Bigr)}
\]
}%
everywhere in the domain of $\qq_{t,x,u}\phd$. Of course, this relation is interesting only if 
the right side is strictly positive~-- a situation that is illustrated next. 

Now suppose that, in addition to $\Up(c)=1/c$ and $\q_{t,x,\csd,u}\phd\equiv0$,
the model is also such that $\EEE=\{\eta,0\}$ for some fixed $\eta>0$ (there are only two
employment states, employed and unemployed~-- see Sec.~\ref{sec:KS}).
In~what follows the values of all functions at $0$
are to be understood as the right limits at $0$.
Since no investment takes place in the final period $t=T$, with
$t=T-1$ the lower bound in \eqref{theta-min} becomes $0$, i.e, $\qq_{T-1,x,\csd,u}(0)\ge 0$.
If $\qq_{t,x,\csd,u}(0)\ge 0$ for some $t<T$, then with $v =\eta$ and
$c=0$ the balance equation in \ref{cross-sys}-(\ref{zze5xb}) would give
{\abovedisplayskip=10pt plus 1.5pt minus 1.5pt\belowdisplayskip=10pt plus 1.5pt minus 1.5pt\belowdisplayshortskip=5pt plus 1.5pt minus 1.5pt
\begin{equation}\label{boc-c}
{\pfc_{t,x,\csd}^{ y,\eta}(u,0)}
+ {\qq_{t+1,\tts y,\tts \T_{t,x}^y(\csd),\tts \eta}\bigl(\pfc_{t,x,u}^{ y,\eta}(0)\bigr)}\ge \ee_ y(K_{t,x})\,\eta\,,
\end{equation}}
and with $v =0$ and $c=0$ the same balance equation would give
{\abovedisplayskip=10pt plus 1.5pt minus 1.5pt\belowdisplayskip=10pt plus 1.5pt minus 1.5pt\belowdisplayshortskip=8pt plus 1.5pt minus 1.5pt
\begin{equation}\label{boc-d} 
{\pfc_{t,x,\csd}^{ y,0}(u,0)}
+ {\qq_{t+1,\tts y,\tts\T_{t,x}^y(\csd),\tts 0}\bigl(\pfc_{t,x,\csd}^{ y,0}(u,0)\bigr)}\ge 0\,.
\end{equation}}
With $t=T-1$ the last two relations
become
{\abovedisplayskip=10pt plus 1.5pt minus 1.5pt\belowdisplayskip=10pt plus 1.5pt minus 1.5pt\belowdisplayshortskip=8pt plus 1.5pt minus 1.5pt
\begin{equation}\label{boc-e}
{\pfc_{T-1,x,\csd}^{ y,\eta}(u,0)}\ge \ee_ y(K_{T-1,x})\,\eta\qquad\text{and}\qquad 
{\pfc_{T-1,x,\csd}^{ y,0}(u,0)}\ge 0\,.
\end{equation}
}%
Since ${\pfc_{T-1,x,\csd}^{ y,\eta}(u,0)}$ is strictly positive, due to \eqref{new-kern}
${\pfc_{T-1,x,\csd}^{ y,0}(u,0)}> 0$ is not possible, for otherwise the right side would have a strictly
positive limit as $c\searrow 0$, while the limit of the left side would be $0$.
Since ${\pfc_{T-1,x,\csd}^{ y,0}(u,0)}=0$,
the balance equation \ref{cross-sys}-(\ref{zze5xb}) gives $\qq_{T-1,x,\csd,u}(0)= 0$ for all
$u\in\EEE$. In~particular,
both relations in \eqref{boc-e} are strict identities. Next, setting $t=T-2$ in \eqref{theta-min}
again yields $M_{T-2,x,\csd}=0$, so that $\qq_{T-2,x,\csd,u}(0)\ge 0$ for every $u\in\EEE$.
The balance equation now gives
{\abovedisplayskip=10pt plus 1.5pt minus 1.5pt\belowdisplayskip=10pt plus 1.5pt minus 1.5pt\belowdisplayshortskip=8pt plus 1.5pt minus 1.5pt
\begin{equation*}
\pfc_{T-2,x,\csd}^{ y,\h}(u,0)+\qq_{T-1,\tts y,\tts \T_{T-2,x}^y(\csd),\tts \h}
\bigl(\pfc_{T-2,x,\csd}^{ y,\h}(u,0)\bigr) \ge \ee_y(K_{t}(x,\csd))\h > 0\,,
\end{equation*}}
which is possible only if $\pfc_{T-2,x,\csd}^{ y,\h}(u,0)>0$.
Just as above, due to \eqref{new-kern} the last relation implies $\pfc_{T-2,x,\csd}^{y,0}(u,0)=0$,
so that with $v=0$ the balance equation
\ref{cross-sys}-(\ref{zze5xb}) gives
{\abovedisplayskip=10pt plus 1.5pt minus 1.5pt\belowdisplayskip=10pt plus 1.5pt minus 1.5pt\belowdisplayshortskip=8pt plus 1.5pt minus 1.5pt
\begin{equation*}
\bigl(\rho_y(K_{t}(x,\csd))+1-\dd\bigr)\qq_{T-2,x,\csd,u}(0)=0+\qq_{T-1,\tts
y,\tts \T_{T-2,x}^y(\csd),\tts v}(0)=0\,,
\end{equation*}
}%
with the implication that $\qq_{T-2,x,\csd,u}(0)=0$. By way of induction one can show that
$\qq_{t,x,\csd,u}(0)=0$, $\pfc_{t,x,\csd}^{y,\h}(u,0)>0$, and $\pfc_{t,x,\csd}^{ y,0}(u,0)=0$ for all
$t<T$, $x,y\in\XXX$ and $u\in\EEE=\{\h,0\}$. In particular, capital will never be borrowed and
the consumption level of all employed households will be bounded away from $0$ for all $t<T$. We
stress that these features hold if capital is the only asset and all
households from one of 
the employment states have no income.  

\end{appendices}

\end{document}